\documentclass[a4paper,12pt,twoside,openright]{book}
\usepackage{geometry} 
\usepackage{amsmath}
\usepackage{amssymb}
\usepackage{color}
\usepackage{graphics}
\usepackage{graphicx}
\usepackage{epsfig}
\usepackage{epstopdf}
\usepackage{axodraw}
\usepackage{appendix}
\usepackage{multirow}
\usepackage[normal]{subfigure}
\usepackage[applemac]{inputenc}
\usepackage{rotating}
\usepackage{fancyhdr}
\fancypagestyle{plain}{ %
\fancyhf{} 
}

\usepackage{marvosym}
\usepackage[T1]{fontenc}
\usepackage[light,math]{anttor}
 
\newcommand{\capdef}{}
\newcommand{\mycaption}[2][\capdef]{\renewcommand{\capdef}{#2}
\caption[#1]{{\footnotesize #2}}}

\newcommand{\bwt}{\begin{widetext}}
\newcommand{\ewt}{\end{widetext}}
\newcommand{\be}{\begin{equation}}
\newcommand{\ee}{\end{equation}}
\newcommand{\bdm}{\begin{displaymath}}
\newcommand{\edm}{\end{displaymath}}
\newcommand{\bea}{\begin{eqnarray}}
\newcommand{\eea}{\end{eqnarray}}
\newcommand{\nn}{\nonumber}

\def\eq#1{{Eq.~(\ref{#1})}}
\def\eqs#1#2{{Eqs.~(\ref{#1})--(\ref{#2})}}
\def\fig#1{{Fig.~\ref{#1}}}
\def\figs#1#2{{Figs.~\ref{#1}--\ref{#2}}}
\def\Table#1{{Table~\ref{#1}}}
\def\Tables#1#2{{Tables~\ref{#1}--\ref{#2}}}
\def\sect#1{{Sect.~\ref{#1}}}

\def\app#1{{Appendix~\ref{#1}}}
\def\apps#1#2{{Apps.~\ref{#1}--\ref{#2}}}
\def\vev#1{\left\langle #1\right\rangle}
\def\abs#1{\left| #1\right|}
\def\mod#1{\abs{#1}}
\def\Im{\mbox{Im}\,}
\def\Re{\mbox{Re}\,}
\def\Tr{\mbox{Tr}\,}
\def\det{\mbox{det}\,}

\def\chain#1#2{\mathrel{\mathop{\longrightarrow}\limits^{#1}_{#2}}}

\newcommand{\sepA}{\rule[-0.8cm]{0cm}{1.6cm}}
\newcommand{\sepB}{\rule[-1cm]{0cm}{2cm}}
\newcommand{\sepC}{\rule[-1.2cm]{0cm}{2.4cm}}

\begin{document}

\begin{titlepage}

\title{}
\author{}
\date{}
\begin{center}
\Large{SISSA\\}
\vspace{1cm}
\large{Scuola Internazionale Superiore di Studi Avanzati}
\vspace{4cm}


\huge{{\bf Aspects of Symmetry Breaking in \\ Grand Unified Theories}}
\end{center}

\vspace{4cm}
\begin{center}
\large{Thesis submitted for the degree of \\ Doctor Philosophiae}
\end{center}

\vspace{2.5cm}
\hbox{\hsize=3in \raggedright\parindent=1pt
\vtop {\large{Supervisor:\\ Dr. Stefano Bertolini}}
\hspace{5.cm}
\vtop {\large{Candidate:\\ Luca Di Luzio}}}

\vspace{2.cm}
\begin{center}
{\large{Trieste, September 2011}}
\end{center}

\end{titlepage}

\chapter*{
\centering 
\begin{normalsize}
Abstract
\end{normalsize}}
\begin{quotation}
\noindent 
We reconsider the issue of spontaneous symmetry breaking in $SO(10)$ grand unified theories. 
The emphasis is put on the quest for the minimal Higgs sector leading to a phenomenologically 
viable breaking to the standard model gauge group.
Longstanding results claimed that nonsupersymmetric SO(10) models with just the adjoint representation 
triggering the first stage of the breaking cannot provide a successful gauge unification. 
The main result of this thesis is the observation that this no-go is an artifact of the tree level potential 
and that quantum corrections opens in a natural way the vacuum patterns favoured by gauge coupling unification. 
An analogous no-go, preventing the breaking of $SO(10)$ at the renormalizable level with representations up 
to the adjoint, holds in the supersymmetric case as well. 
In this respect we show that a possible way-out is provided by considering 
the flipped $SO(10)$ embedding of the hypercharge. 
Finally, the case is made for the hunting of the minimal $SO(10)$ theory.
\end{quotation}
\clearpage

\tableofcontents

\chapter*{Foreword}
\addcontentsline{toc}{chapter}{Foreword}
\markboth{\textsc{Foreword}}{\textsc{Foreword}}

This thesis deals with the physics of the 80's. Almost all of the results obtained here could have been achieved by the end of that decade. 
This also means that the field of grand unification is becoming quite old. It dates back in 1974 with the seminal papers of Georgi-Glashow~\cite{Georgi:1974sy} 
and Pati-Salam~\cite{Pati:1974yy}. 
Those were the years just after the foundation of the standard model (SM) of Glashow-Weinberg-Salam~\cite{Glashow:1961tr,Weinberg:1967tq,salam} 
when simple ideas 
(at least simple from our future perspective) seemed to receive an 
immediate confirmation from the experimental data. 

Grand unified theories (GUTs) assume that all the fundamental interactions of the SM (strong and electroweak)
have a common origin. 
The current wisdom is that we live in a broken phase in which 
the world looks $SU(3)_C \otimes U(1)_Q$ invariant to us and the low-energy phenomena are governed by 
strong interactions and electrodynamics. Growing with the energy we start to see the degrees of freedom of a new 
dynamics which can be interpreted as a renormalizable $SU(2)_L \otimes U(1)_Y$ gauge theory spontaneously broken
into $U(1)_Q\footnote{At the time of writing this thesis one of the main ingredients of this theory, the Higgs boson, is still missing 
experimentally. On the other hand a lot of indirect tests suggest that the SM works amazingly well and it 
is exciting that the mechanism of electroweak symmetry breaking is being tested right now at the Large Hadron Collider (LHC).}$. 
Thus, in analogy to the $U(1)_Q \rightarrow SU(2)_L \otimes U(1)_Y$ case, 
one can imagine that at higher energies the SM gauge group $SU(3)_C \otimes SU(2)_L \otimes U(1)_Y$ 
is embedded in a simple group $G$. 

\begin{figure*}[h]
\centering
\includegraphics[width=8.5cm]{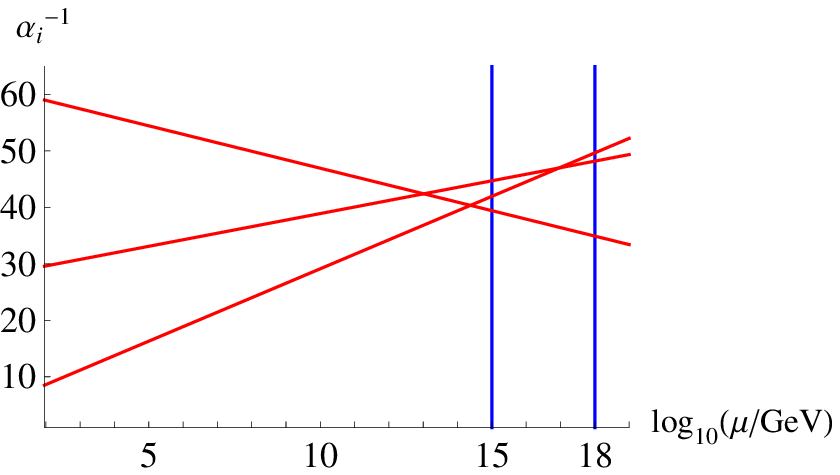}
\mycaption{\label{SMrunintro} One-loop running of the SM gauge couplings assuming the $U(1)_Y$ embedding into $G$.} 
\end{figure*}

The first implication of the grand unification ansatz is that at some mass scale $M_U \gg M_W$ 
the relevant symmetry is $G$ and the $g_3$, $g_2$ and $g'$ coupling constants of $SU(3)_C \otimes SU(2)_L \otimes U(1)_Y$  
merge into a single gauge coupling $g_U$. The rather different values for $g_3$, $g_2$ and $g'$ at low-energy 
are then due to renormalization effects. 
Actually one of the most solid hints in favor of grand unification 
is the fact that the running within the SM shows an approximate convergence 
of the gauge couplings around $10^{15} \ \text{GeV}$ (see e.g.~\fig{SMrunintro}).

This simple idea, though a bit speculative, 
may have a deep impact on the understanding of our low-energy world.
Consider for instance some 
unexplained features of the SM like e.g.~charge quantization or anomaly 
cancellation\footnote{In the SM anomaly cancellation implies charge quantization, 
after taking into account the gauge invariance of the Yukawa couplings~\cite{Geng:1988pr,Minahan:1989vd,Geng:1990nh,Foot:1990uf}. 
This feature is lost as soon as one adds a right-handed neutrino $\nu_R$, unless 
$\nu_R$ is a Majorana particle~\cite{Babu:1989ex}.}. 
They appear just as the natural consequence of starting with an anomaly-free simple group such as $SO(10)$. 

Most importantly grand unification is not just a mere interpretation of our low-energy world,  
but it predicts new phenomena which are correlated with the existing ones. 
The most prominent of these is the instability of matter. The current lower bound on the 
proton lifetime is something like 23 orders of magnitude bigger than the age of the Universe,     
namely $\tau_p \gtrsim 10^{33 \div 34} \ \text{yr}$ depending on the decay channel~\cite{Nakamura:2010zzi}.
This number is so huge that people started to consider baryon number as an exact symmetry of Nature~\cite{Weyl:1929fm,Stueckelberg:1938zz,Wigner:1985ze}.
Nowadays we interpret it as an accidental global symmetry of the standard 
model\footnote{In the SM the baryonic current is anomalous and baryon number violation can arise from
instanton transitions between degenerate $SU(2)_L$ vacua which lead to $\Delta B = \Delta L = 3$ 
interactions for three flavor families~\cite{'tHooft:1976up,'tHooft:1976fv}. 
The rate is estimated to be proportional to $e^{- 2\pi / \alpha_2} \sim e^{-173}$ and thus phenomenologically irrelevant.}. 
This also means that as soon as we extend the SM there is the chance to introduce baryon violating interactions. 
Gravity itself could be responsible for the breaking of baryon number~\cite{Hawking:1979hw}. 
However among all the possible frameworks there is only one of them which predicts a proton lifetime close to its experimental limit 
and this theory is grand unification. 
Indeed we can roughly estimate it by dimensional arguments. The exchange of a baryon-number-violating vector boson 
of mass $M_U$ yields something like  
\be
\tau_p \sim \alpha_U^{-1} \frac{M_U^4}{m_p^5} \, , 
\ee 
and by putting in numbers (we take $\alpha^{-1}_U \sim 40$, cf.~\fig{SMrunintro}) 
one discovers that $\tau_p \gtrsim 10^{33} \ \text{yr}$ corresponds to $M_U \gtrsim 10^{15} \ \text{GeV}$, 
which is consistent with the picture emerging in~\fig{SMrunintro}. Notice that the gauge running is sensitive to the log of the scale. 
This means that a $10\%$ variation on the gauge couplings at the electroweak scale induces a $100\%$ one on $M_U$.  
Were the apparent unification of gauge couplings in the window $10^{15 \div 18} \ \text{GeV}$ just an accident, 
then Nature would have played a bad trick on us. 

Another firm prediction of GUTs are magnetic monopoles~\cite{'tHooft:1974qc,Polyakov:1974ek}. 
Each time a simple gauge group $G$ is broken to a subgroup with a $U(1)$ factor 
there are topologically nontrivial configurations of the Higgs field which leads to stable monopole solutions 
of the gauge potential. 
For instance the breaking of $SU(5)$ generates a monopole with magnetic charge $Q_m = 2\pi / e$ 
and mass $M_m = \alpha_U^{-1} M_U $~\cite{Dokos:1979vu}.
The central core of a GUT monopole contains the fields of the superheavy gauge bosons which mediate 
proton decay, so one expects that baryon number can be violated in baryon-monopole scattering. 
Quite surprisingly it was found~\cite{Callan:1982au,Rubakov:1982fp,Callan:1982ac} that these processes are not suppressed 
by powers of the unification mass, but have a cross section typical of the strong interactions. 

Though GUT monopoles are too massive to be produced at accelerators, 
they could have been produced in the early universe as topological defects arising via the Kibble mechanism~\cite{Kibble:1976sj} 
during a symmetry breaking phase transition. Experimentally one tries to measure their interactions as they 
pass through matter. The strongest bounds on the flux of monopoles come from their interactions with the galactic magnetic field 
($\Phi < 10^{-16} \ \text{cm}^{-2} \, \text{sr}^{-1} \, \text{sec}^{-1}$) and the catalysis of proton decay in compact astrophysical 
objects ($\Phi < 10^{-18 \div 29} \ \text{cm}^{-2} \, \text{sr}^{-1} \, \text{sec}^{-1}$)~\cite{Nakamura:2010zzi}. 

Summarizing the model independent predictions of grand unification are proton decay, magnetic monopoles and charge quantization 
(and their deep connection). 
However once we have a specific model we can do even more. 
For instance the huge ratio between the unification and the electroweak scale, $M_U/M_W \sim 10^{13}$, reminds us about the well established hierarchy 
among the masses of charged fermions and those of neutrinos, $m_f / m_{\nu} \sim 10^{7 \div 13}$. 
This analogy hints to a possible connection between GUTs and neutrino masses.

The issue of neutrino masses caught the attention of particle physicists since a long time ago. 
The model independent way to parametrize them is to consider the SM as an effective field theory by writing all the possible 
operators compatible with gauge invariance.  
Remarkably at the $d=5$ level there is only one operator~\cite{Weinberg:1979sa}  
\be
\label{Weinbergopintro}
\frac{Y_\nu}{\Lambda_L} (\ell^T \epsilon_2 H) C (H^T \epsilon_2 \ell) \, .
\ee
After electroweak symmetry breaking $\vev{H} = v$ and
neutrinos pick up a Majorana mass term 
\be
M_\nu = Y_\nu \frac{v^2}{\Lambda_L} \, .
\ee
The lower bound on the highest neutrino eigenvalue inferred from $\sqrt{\Delta m_{atm}} \sim 0.05 \ \text{eV}$ 
tells us that the scale at which the lepton number is violated is 
\be
\Lambda_L \lesssim Y_\nu \ \mathcal{O}(10^{14 \div 15} \ \text{GeV}) \, .
\ee
Actually there are only three renormalizable ultra-violet (UV) completion of the SM which can give rise to the operator in~\eq{Weinbergopintro}. 
They go under the name of type-I~\cite{Minkowski:1977sc,GellMann:1980vs,Yanagida:1979as,Glashow:1979nm,Mohapatra:1979ia}, 
type-II~\cite{Magg:1980ut,Schechter:1980gr,Lazarides:1980nt,Mohapatra:1980yp} and type-III~\cite{Foot:1988aq} seesaw 
and are respectively obtained by introducing a fermionic singlet $(1,1,0)_F$, 
a scalar triplet $(1,3,+1)_H$ and a fermionic triplet $(1,3,0)_F$. 
These vector-like fields, whose mass can be identified with $\Lambda_L$, couple at the renormalizable level with $\ell$ and $H$ 
so that the operator in~\eq{Weinbergopintro} is generated after integrating 
them out.  
Since their mass is not protected by the chiral symmetry 
it can be super-heavy, thus providing a rationale for the smallness of neutrino masses. 

Notice that this is still an effective field theory language and 
we cannot tell at this level if neutrinos are light because $Y_\nu$ is small or because $\Lambda_L$ is large.
It is clear that without a
theory that fixes the structure of $Y_\nu$ we don't have much to 
say about 
$\Lambda_L$\footnote{The other possibility is that we may probe 
experimentally the new degrees of freedom at the scale $\Lambda_L$ in such a way to reconstruct the theory of neutrino masses. 
This could be the case for left-right symmetric theories~\cite{Mohapatra:1979ia,Mohapatra:1980yp} where $\Lambda_L$ is the scale of the $V+A$ interactions. 
For a recent study of the interplay between LHC signals and neutrinoless double beta decay in the context of left-right scenarios see e.g.~\cite{Tello:2010am}.}.  

As an example of a predictive theory which can fix both $Y_\nu$ and $\Lambda_L$ we can mention $SO(10)$ unification. 
The most prominent feature of $SO(10)$ is that a SM fermion family plus a right-handed neutrino fit 
into a single 16-dimensional spinorial representation. In turn this readily implies that $Y_\nu$ is correlated to 
the charged fermion Yukawas. 
At the same time $\Lambda_L$ can be identified with the $B-L$ generator of $SO(10)$, and
its breaking scale, $M_{B-L} \lesssim M_U$, is subject to the constraints of gauge coupling unification. 

Hence we can say that $SO(10)$ is also a theory of neutrinos, 
whose self-consistency can be tested against complementary observables such as the proton lifetime
and neutrino masses.    

\begin{figure*}[ht]
\centering
\includegraphics[width=8.5cm]{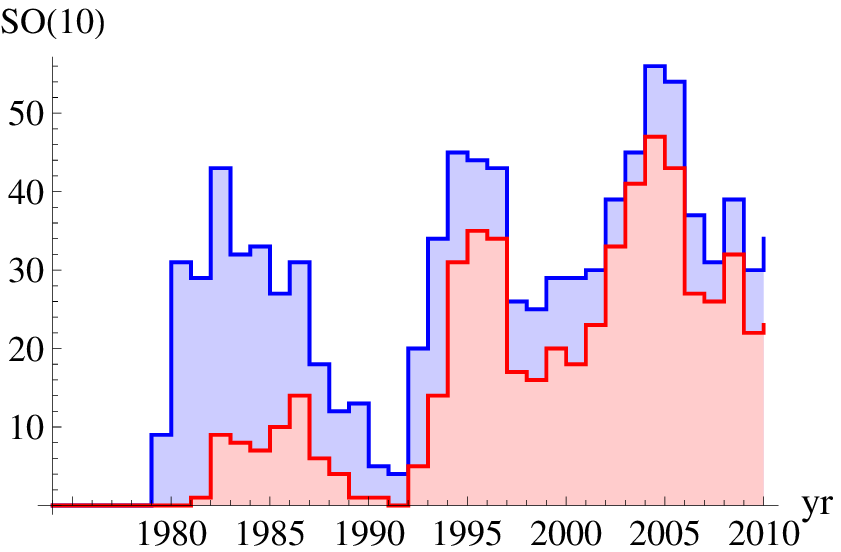}
\mycaption{\label{inspireso10} 
Blue: number of papers per year with the keyword "$SO(10)$" in the title as a function of the years. 
Red: subset of papers with the keyword "supersymmetry" either in the title or in the abstract.
Source: inSPIRE.} 
\end{figure*}

The subject of this thesis will be mainly $SO(10)$ unification. In the arduous attempt of describing the state of the art 
it is crucial to understand what has been done so far. 
In this respect we are facilitated by~\fig{inspireso10},  
which shows the number of $SO(10)$ papers per year from $1974$ to $2010$.

By looking at 
this plot it is possible to reconstruct the following historical phases: 
\begin{itemize}
\item $1974 \div 1986$: Golden age of grand unification. 
These are the years of the foundation in which the fundamental aspects of the theory are worked out. 
The first estimate of the proton lifetime 
yields $\tau_p \sim 10^{31} \ \text{yr}$~\cite{Georgi:1974yf},
amazingly close to the experimental bound
$\tau_p \gtrsim 10^{30} \ \text{yr}$~\cite{Reines:1974pb}. 
Hence the great hope that proton decay is behind the corner. 
\item $1987 \div 1990$: Great depression. 
Neither proton decay nor magnetic monopoles are observed so far. 
Emblematically the last workshop on grand unification is held in $1989$~\cite{Frampton:1989zh}.   
\item $\gtrsim 1991$: SUSY-GUTs. The new data of the Large Electron-Positron collider (LEP) seem to favor low-energy supersymmetry as a candidate for gauge coupling unification. 
From now on almost all the attention is caught by supersymmetry. 
\item $\gtrsim 1998$: Neutrino revolution. Starting from $1998$ experiments begin to show that atmospheric~\cite{Fukuda:1998mi} and solar~\cite{Ahmad:2002jz} 
neutrinos change flavor. 
$SO(10)$ comes back with a rationale for the origin of the sub-eV neutrino mass scale.
\item $\gtrsim 2010$: LHC era. Has supersymmetry something to do with the electroweak scale? 
The lack of evidence for supersymmetry at the LHC would undermine SUSY-GUT scenarios. Back to nonsupersymmetric GUTs? 
\item $\gtrsim 2019$: Next generation of proton decay experiments sensitive to $\tau_p \sim 10^{34 \div 35}$ yr~\cite{Abe:2011ts}. 
The future of grand unification relies heavily on that.
\end{itemize}

Despite the huge amount of work done so far, 
the situation 
does not seem very clear at the moment. 
Especially from a theoretical point of view no model of grand unification emerged as "the" theory.   
The reason can be clearly attributed to the lack of experimental evidence on proton decay. 

In such a situation a good guiding principle in order to discriminate among models and eventually falsify them is given by minimality, 
where minimality deals interchangeably with simplicity, tractability and predictivity.  
It goes without saying that minimality could have nothing to do with our world, but it is anyway the best we can do at the moment. 
It is enough to say that if one wants to have under control all the aspects of the theory 
the degree of complexity of some minimal GUT is already at the edge of the tractability.

Quite surprisingly after 37 years 
there is still no consensus on which is the minimal theory.  
Maybe the reason is also that minimality is not a universal and uniquely defined concept, admitting a number of interpretations. 
For instance it can be understood as a mere simplicity related to the minimum rank of the gauge group. 
This was indeed the remarkable observation of Georgi and Glashow:
$SU(5)$ is the unique rank-4 simple group which contains the SM and has complex representations. 
However nowadays we can say for sure that the Georgi-Glashow model in its original formulation 
is ruled out because it does not unify and neutrinos are 
massive\footnote{Moved by this double issue of the Georgi-Glashow model, two minimal extensions which can cure at the same time 
both unification and neutrino masses have been recently proposed~\cite{Dorsner:2005fq,Bajc:2006ia}.
}.


From a more pragmatic point of view one could instead use predictivity as a measure of minimality. 
This singles out $SO(10)$ as the best candidate. 
At variance with $SU(5)$, the fact that all the SM fermions of one family fit into 
the same representation makes the Yukawa sector of $SO(10)$ much more 
constrained\footnote{Notice that here we do not have in mind flavor symmetries, indeed the GUT symmetry itself already constrains the flavor structure 
just because some particles live together in the same multiplet. 
Certainly one could improve the predictivity by adding additional ingredients like 
local/global/continuous/discrete symmetries on top of the GUT symmetry. 
However, though there is nothing wrong with that, we feel that it would be a no-ending process 
based on assumptions which are difficult to disentangle from the unification idea. 
That is why we prefer to stick as much as possible to the gauge principle without further ingredients.}. 

Actually, if we stick to the $SO(10)$ case, minimality is closely related to the complexity of the symmetry breaking sector.
Usually this is the most challenging and arbitrary aspect of grand unified models. 
While the SM matter nicely fit in three $SO(10)$ spinorial families, 
this synthetic feature has no counterpart in the Higgs sector where 
higher-dimensional representations are usually needed in order to spontaneously break the enhanced 
gauge symmetry down to the SM.

Establishing the minimal Higgs content needed for the GUT breaking is a basic question 
which has been addressed since the early days of the GUT 
program\footnote{ 
Remarkably the general patterns of symmetry breaking in gauge theories with orthogonal and unitary 
groups were already analyzed in 1973/1974 by Li~\cite{Li:1973mq}, contemporarily with the work of Georgi and Glashow.}. 
Let us stress that the quest for the simplest Higgs sector is driven not only by aesthetic criteria but it is also a phenomenologically 
relevant issue related to the tractability and the predictivity of the models.  
Indeed, the details of the symmetry breaking pattern, 
sometimes overlooked in the phenomenological analysis, give further constraints on the low-energy observables such as the proton decay and the effective SM flavor structure. 
For instance in order to assess quantitatively the constraints imposed by gauge coupling unification 
on the mass of the lepto-quarks 
resposible for proton decay 
it is crucial to have the scalar spectrum under control. 
Even in that case some degree of arbitrariness can still persist 
due to the fact that the spectrum can never be fixed completely 
but lives on a manifold defined by the vacuum conditions. 
This also means that if we aim to a falsifiable (predictive) GUT scenario, 
better we start by considering a minimal Higgs 
sector\footnote{As an example of the importance of taking into account the vacuum dynamics 
we can mention the minimal supersymmetric model based on $SO(10)$~\cite{Clark:1982ai,Aulakh:1982sw,Aulakh:2003kg}.
In that case the precise calculation of the mass spectrum~\cite{Aulakh:2002zr,Fukuyama:2004ps,Bajc:2004xe} 
was crucial in order to obtain a detailed fitting of fermion mass parameters and show a tension 
between unification constraints and neutrino masses~\cite{Aulakh:2005mw,Bertolini:2006pe}.}. 

The work done in this thesis can be understood as a general reappraisal of the issue of symmetry breaking in $SO(10)$ GUTs, both in their 
ordinary and supersymmetric realizations. 

We can already anticipate that,   
before considering any symmetry breaking dynamics,
at least 
two Higgs representations are required\footnote{It should be mentioned that a one-step $SO(10) \rightarrow \text{SM}$ breaking 
can be achieved via only one $144_H$ irreducible Higgs representation \cite{Babu:2005gx}. 
However, such a setting requires an extended matter sector, including $45_F$ and $120_F$ multiplets, 
in order to accommodate realistic fermion masses \cite{Nath:2009nf}.}  
by the group theory 
in order to achieve a full breaking of $SO(10)$ to the SM:
\begin{itemize}
\item $16_H$ or $126_H$: they reduce the rank but leave an $SU(5)$ little group unbroken.
\item $45_H$ or $54_H$ or $210_H$: they admit for little groups different from $SU(5) \otimes U(1)$, yielding the SM when intersected with $SU(5)$.
\end{itemize}
While the choice between $16_H$ or $126_H$ is a model dependent issue related to the details of the Yukawa sector, 
the simplest option among $45_H$, $54_H$ and $210_H$ is given by the adjoint $45_H$. 

However, since the early 80's, it has been observed that the vacuum dynamics aligns the adjoint along 
an $SU(5) \otimes U(1)$ direction, making the choice of $16_H$ (or $126_H$) and $45_H$ alone not phenomenologically viable. 
In the nonsupersymmetric case the alignment is only approximate~\cite{Yasue:1980fy,Yasue:1980qj,Anastaze:1983zk,Babu:1984mz},
but it is such to clash with unification constraints 
which do not allow for any $SU(5)$-like intermediate stage, while in the supersymmetric limit 
the alignment is exact due to F-flatness~\cite{Buccella:1981ib,Babu:1994dc,Aulakh:2000sn}, thus never landing to a supersymmetric SM vacuum. 
The focus of the thesis consists in the critical reexamination of these two longstanding no-go for the settings with a $45_H$ 
driving the GUT breaking.

Let us first consider the nonsupersymmetric case. 
We start
by reconsidering the issue of gauge coupling unification in ordinary 
$SO(10)$ scenarios with up to two intermediate mass scales, 
a needed preliminary step 
before entering the details of a specific model. 

After complementing the existing studies 
in several aspects, as the inclusion of the $U(1)$ gauge mixing renormalization at the one- and two-loop 
level and the reassessment of the two-loop beta coefficients, 
a peculiar symmetry breaking pattern 
with just the adjoint representation governing the first stage of the GUT breaking emerges as 
a potentially viable scenario~\cite{Bertolini:2009qj}, contrary to what claimed in the literature~\cite{Deshpande:1992em}.

This brings us to reexamine the vacuum of the minimal conceivable Higgs potential responsible for the 
$SO(10)$ breaking to the SM, containing an adjoint $45_H$ plus a spinor $16_H$. 
As already remarked, a series of studies in the early 80's~\cite{Yasue:1980fy,Yasue:1980qj,Anastaze:1983zk,Babu:1984mz} 
of the $45_H \oplus 16_H$ model indicated that
the only intermediate stages allowed by the scalar sector dynamics were
$SU(5)\otimes U(1)$ for leading $\langle 45_H \rangle$ or $SU(5)$ for  dominant $\langle 16_H \rangle$.
Since an intermediate $SU(5)$-symmetric stage is phenomenologically not allowed, this observation excluded
the simplest $SO(10)$ Higgs sector from realistic consideration.

One of the main results of this thesis is the observation that this no-go "theorem" is actually an artifact of the tree-level potential 
and, as we have shown in Ref.~\cite{Bertolini:2009es} (see also Ref.~\cite{Bertolini:2010ng} for a brief overview), 
the minimization of the one-loop effective potential opens in a natural way 
also the intermediate stages $SU(4)_C \otimes SU(2)_L \otimes U(1)_R$ and 
$SU(3)_C \otimes SU(2)_L \otimes SU(2)_R \otimes U(1)_{B-L}$, 
which are the options favoured by gauge unification.
This result is quite general, since it applies whenever the $SO(10)$ breaking is triggered by the 
$\langle 45_H \rangle$ (while other Higgs representations 
control the intermediate and weak scale stages) and brings back from oblivion 
the simplest scenario of nonsupersymmetric $SO(10)$ unification. 

It is then natural to consider the Higgs system $10_H \oplus 16_H \oplus 45_H$ 
(where the $10_H$ is needed to give mass to the SM fermions at the renormalizable level)
as the potentially minimal $SO(10)$ theory, as advocated long ago by Witten~\cite{Witten:1979nr}.
However, apart from issues related to fermion mixings, the main obstacle with such a model is given by neutrino masses. 
They can be generated radiatively at the two-loop level, but turn out to be too heavy. 
The reason being that the $B-L$ breaking is communicated to 
right-handed neutrinos at the effective level $M_{R} \sim (\alpha_U / \pi)^2 M_{B-L}^2/M_U$ and since $M_{B-L} \ll M_U$ by 
unification constraints, $M_R$ undershoots by several orders of magnitude the value $10^{13 \div 14} \ \text{GeV}$
naturally suggested by the type-I seesaw.

At these point one can consider two possible routes. 
Sticking to the request of Higgs representations with dimensions up to the adjoint one can invoke TeV scale supersymmetry, 
or we can relax this requirement and exchange the $16_H$ with the $126_H$ in the nonsupersymmetric case. 

In the former case the gauge running within the minimal supersymmetric SM (MSSM)
prefers $M_{B-L}$ in the proximity of $M_{U}$ so that one can naturally reproduce the desired range for $M_{R}$, 
emerging from the effective operator $16_F 16_F \overline{16}_H \overline{16}_H / M_P$.

Motivated by this argument, we investigate under which conditions 
an Higgs sector containing only representations up to the adjoint
allows supersymmetric $SO(10)$ GUTs to break spontaneously to the SM. 
Actually it is well known~\cite{Buccella:1981ib,Babu:1994dc,Aulakh:2000sn}
that the relevant superpotential does not support,
at the renormalizable level, a supersymmetric breaking of the $SO(10)$
gauge group to the SM.
Though the issue can be addressed by giving up renormalizability~\cite{Babu:1994dc,Aulakh:2000sn}, 
this option may be rather problematic due to the active role of Planck induced operators in the breaking of the gauge symmetry. 
They introduce an hierarchy in the mass spectrum at the GUT scale which may be an issue for gauge unification, proton decay and neutrino masses. 

In this respect we pointed out~\cite{Bertolini:2010yz} that the minimal Higgs scenario that allows for a renormalizable 
breaking to the SM is obtained considering flipped $SO(10) \otimes U(1)$ with one adjoint 
$45_H$ and two $16_H \oplus \overline{16}_H$ Higgs representations. 

Within the extended $SO(10)\otimes U(1)$ gauge algebra one finds in general three inequivalent embeddings of the SM hypercharge.
In addition to the two solutions with the hypercharge stretching over the $SU(5)$ or the $SU(5)\otimes U(1)$ subgroups of $SO(10)$ 
(respectively dubbed as the ``standard'' and ``flipped'' $SU(5)$ embeddings~\cite{DeRujula:1980qc,Barr:1981qv}), there is a third, 
``flipped'' $SO(10)$ ~\cite{Kephart:1984xj,Rizos:1988jn,Barr:1988yj}, solution inherent to the $SO(10)\otimes U(1)$ case, with a non-trivial projection of the 
SM hypercharge onto the $U(1)$ factor.

Whilst the difference between the standard and the flipped $SU(5)$ embedding is semantical from the $SO(10)$ point of view, the flipped $SO(10)$ case is qualitatively different. In particular, the symmetry-breaking ``power'' of the $SO(10)$ spinor and adjoint representations is boosted with respect to the standard $SO(10)$ case,
increasing the number of SM singlet fields that may acquire non-vanishing vacuum expectation values (VEVs).
This is at the root of the possibility of implementing the gauge symmetry breaking by means of a simple renormalizable Higgs sector.

The model is rather peculiar in the flavor sector and can be 
naturally embedded in a perturbative $E_6$ grand unified scenario above the 
flipped $SO(10) \otimes U(1)$ partial-unification scale. 

On the other hand, sticking to the nonsupersymmetric case with a $126_H$ in place of a $16_H$, 
neutrino masses are generated at the renormalizable level. 
This lifts the problematic $M_{B-L}/M_U$ suppression factor inherent to the $d=5$ effective mass and yields $M_{R}\sim M_{B-L}$, that might be, at least in principle, acceptable. 
As a matter of fact a nonsupersymmetric $SO(10)$ model including $10_H \oplus 45_H \oplus 126_H$ in the Higgs sector has all the ingredients to be the minimal 
realistic version of the theory. 

This option at the time of writing the thesis is subject of ongoing research~\cite{BDLM1}. Some preliminary results are reported 
in the last part of the thesis. 
We have performed the minimization of the $45_H \oplus 126_H$ potential and checked that the vacuum 
constraints allow for threshold corrections leading to phenomenologically
reasonable values of $M_{B-L}$. If the model turned out to lead to a realistic fermionic spectrum 
it would be important then to perform an accurate estimate of the proton decay branching ratios.

The outline of the thesis is the following: the first Chapter is an introduction to the field of grand unification. 
The emphasis is put on the construction of $SO(10)$ 
starting from the SM and passing through 
$SU(5)$ and the left-right symmetric groups.
The second Chapter is devoted to the issue of gauge couplings unification in nonsupersymmetric $SO(10)$. 
A set of tools for a general two-loop analysis of gauge coupling unification, 
like for instance the systematization of the $U(1)$ mixing running and matching, is also collected. 
Then in the third Chapter we consider  the simplest and paradigmatic $SO(10)$ Higgs sector made by 
$45_H \oplus 16_H$. After reviewing the old tree level no-go argument we show, by means of an explicit 
calculation, that the effective potential allows for those patterns which were accidentally excluded at tree level. 
In the fourth Chapter we undertake the analysis of the similar no-go present in supersymmetry with $45_H \oplus 16_H \oplus \overline{16}_H$ 
in the Higgs sector. The flipped $SO(10)$ embedding of the hypercharge is proposed as a way out 
in order to obtain a renormalizable breaking with only representations up to the adjoint. 
We conclude with an Outlook in which we suggest the possible lines of development of the ideas proposed in this thesis. 
The case is made for the hunting of the minimal realistic nonsupersymmetric $SO(10)$ unification.  
Much of the technical details are deferred in a set of Appendices.

\chapter{From the standard model to $SO(10)$}
\label{fromSMtoSO10}

In this chapter we give the physical foundations of $SO(10)$ as a grand unified group, starting from the SM 
and browsing in a constructive way through the Georgi-Glashow $SU(5)$~\cite{Georgi:1974sy} and the left-right symmetric groups 
such as the Pati-Salam one~\cite{Pati:1974yy}. 
This will offer us the opportunity to introduce the fundamental concepts of GUTs, 
as charge quantization, gauge unification, proton decay and the connection with neutrino masses in a simplified and pedagogical way.

The $SO(10)$ gauge group as a candidate for the unification of the elementary 
interactions was proposed long ago by Georgi~\cite{Georgi:1974my} and Fritzsch and Minkowski~\cite{Fritzsch:1974nn}. 
The main advantage of $SO(10)$ with respect to $SU(5)$ grand unification is that 
all the known SM fermions plus three right handed neutrinos fit into three copies of the $16$-dimensional 
spinorial representation of $SO(10)$. In recent years the field received an extra boost due 
to the discovery of non-zero neutrino masses in the sub-eV region. 
Indeed, while in the SM (and similarly in $SU(5)$) there is no rationale for the origin of the extremely small neutrino mass 
scale, the appeal of $SO(10)$ consists in the predictive connection between the local $B-L$ breaking scale (constrained by 
gauge coupling unification somewhat below $10^{16}$ GeV) and neutrino masses around $25$ orders of magnitude below. 
Through the implementation of some variant of the seesaw 
mechanism~\cite{Minkowski:1977sc,GellMann:1980vs,Yanagida:1979as,Glashow:1979nm,Mohapatra:1979ia,Magg:1980ut,Schechter:1980gr,Lazarides:1980nt,Mohapatra:1980yp} the inner structure of $SO(10)$ 
and its breaking makes very natural the appearance of such a small neutrino mass scale. 
This striking connection with neutrino masses is one of the strongest motivations behind $SO(10)$ and it can be traced back to the 
left-right symmetric theories~\cite{Pati:1974yy,Mohapatra:1974gc,Senjanovic:1975rk} 
which provide a direct connection of the smallness of neutrino masses with the non-observation of the $V+A$ interactions~\cite{Mohapatra:1979ia,Mohapatra:1980yp}.

\section{The standard model chiral structure}

The representations of the unbroken gauge symmetry of the world, namely $SU(3)_C \otimes U(1)_Q$, are real.  
In other words, for each colored fermion field 
of a given electric charge we have a fermion field of opposite 
color and charge\footnote{
As is usual in grand unification we use the Weyl notation in which all fermion fields $\psi_L$ are left-handed (LH) four-component spinors. 
Given a $\psi_L$ field transforming as $\psi_L \rightarrow e^{i \sigma \omega} \psi_L$ under the Lorentz group 
($\sigma \omega \equiv \omega^{\mu\nu} \sigma_{\mu\nu}$, $\sigma_{\mu\nu} \equiv \tfrac{i}{2} \left[ \gamma_\mu, \gamma_\nu \right]$ 
and $\left\{ \gamma_\mu, \gamma_\nu \right\} = 2 g_{\mu\nu}$)
an invariant mass term is given by $\psi_L^T C \psi_L$ where $C$ is such that $\sigma^T_{\mu \nu} C = - C \sigma_{\mu \nu}$ 
or (up to a sign) $C^{-1} \gamma_\mu C = -\gamma_\mu^T$. 
Using the following representation for the $\gamma$ matrices 
\be
\gamma^0 = 
\left( 
\begin{array}{cc}
0 & 1 \\
1 & 0
\end{array}
\right) \, ,
\qquad
\gamma^i = 
\left( 
\begin{array}{cc}
0 & \sigma_i \\
-\sigma_i & 0
\end{array}
\right) \, ,
\ee
where $\sigma_i$ are the Pauli matrices,
an expression for $C$ reads $C = i \gamma_2 \gamma_0$, with $C=-C^{-1}=-C^\dag=-C^T$. 

Notice that the mass term is not invariant under the $U(1)$ transformation $\psi_L \rightarrow e^{i\theta} \psi_L$ and in order to avoid 
the breaking of any abelian quantum number carried by $\psi_L$ (such as lepton number or electric charge) 
we can construct $\psi'^T_L C \psi_L$ where for every additive quantum number $\psi'_L$ and $\psi_L$ have opposite charges. 
This just means that if $\psi_L$ is associated with a certain fundamental particle, $\psi'_L$ is associated with its antiparticle.
In order to recast a more familiar notation let us define a field $\psi_R$ by the equation $\overline{\psi}_R \equiv \psi'^T_L C$. 
In therms of the right-handed (RH) spinor $\psi_R$, the mass term can be rewritten as $\overline{\psi}_R \psi_L$.}. If not so we would observe for instance a massless charged fermion field and this is not the case.

More formally, being $g$ an element of a group $G$, a representation $D(g)$ is said to be real (pseudo-real) if it is equal to its conjugate representation $D^*(g)$ 
up to a similarity transformation, namely
\be
\label{realpsrealrep}
S D(g) S^{-1} = D^*(g) \quad \text{for all $g \in G$} \, ,
\ee
whit $S$ symmetric (antisymmetric). A complex representation is neither real nor pseudo-real.

It's easy to prove that $S$ must be either symmetric or antisymmetric. 
Suppose $T_a$ generates a real (pseudo-real) irreducible unitary representation of $G$, $D (g) = \exp{i g_a T_a}$, so that 
\be
\label{defrealpreal}
S T_a S^{-1} = - T^*_a \, .
\ee
Because the $T_a$ are hermitian, we can write
\be
S T_a S^{-1} = - T^T_a \qquad \text{or} \qquad (S^{-1})^T T_a^T S^T  = - T_a \, ,
\ee
which implies 
\be
T_a = (S^{-1})^T S T_a S^{-1} S^T
\ee
or equivalently 
\be
\left[ T_a , S^{-1} S^T \right] = 0 \, .
\ee
But if a matrix commutes with all the generators of an irreducible representation, Schur's Lemma tells us that it is a multiple of the identity, and thus 
\be
S^{-1} S^T = \lambda I \qquad \text{or} \qquad S^T = \lambda S \, . 
\ee
By transposing twice we get back to where we started and thus we must have $\lambda^2 = 1$ and so $\lambda = \pm 1$, i.e.~$S$ must be either symmetric or antisymmetric.

The relevance of this fact for the SM is encoded in the following observation: given a left-handed fermion field $\psi_L$ transforming under some 
representation, reducible or irreducible, $\psi_L \rightarrow D(g) \psi_L$, one can construct a gauge invariant mass term only if the representation is real. 
Indeed, it is easy to verify (by using \eq{realpsrealrep} and the unitarity of $D(g)$) that the mass term $\psi_L^T C S \psi_L$, 
where $C$ denotes the Dirac charge conjugation matrix, is invariant. 
Notice that if the representation were pseudo-real (e.g.~a doublet of $SU(2)$) the mass 
term vanishes because of the antisymmetry of 
$S$\footnote{The relation $C^T = -C$ and the anticommuting 
property of the fermion fields must be also taken into account.}.

The SM is built in such a way that there are no bare mass terms and all the masses stem from the Higgs mechanism. 
Its representations are said to be chiral because they are charged under the $SU(2)_L \otimes U(1)_Y$ chiral 
symmetry in such a way that fermions are massless as long as the chiral symmetry is preserved.
A complex representation of a group $G$ may of course become real when restricted to a subgroup of $G$. This is 
exactly what happens in the $SU(3)_C \otimes SU(2)_L \otimes U(1)_Y \rightarrow SU(3)_C \otimes U(1)_Q$ case.

When looking for a unified UV completion of the SM we would like to keep this feature. Otherwise we should also explain why, 
according to the Georgi's survival hypothesis \cite{Georgi:1979md}, 
all the fermions do not acquire a super-heavy bare mass of the order of the scale at which the unified gauge symmetry is broken. 

\section{The Georgi-Glashow route}

The bottom line of the last section was that a realistic grand unified theory is such that the LH fermions are embedded in a complex representation 
of the unified group (in particular complex under $SU(3)_C \otimes SU(2)_L \otimes U(1)_Y$). 
If we further require minimality (i.e.~rank 4 as in the SM) one reaches the remarkable conclusion \cite{Georgi:1974sy}
that the only simple group with complex representations (which contains 
$SU(3)_C \otimes SU(2)_L \otimes U(1)_Y$ as a subgroup) is $SU(5)$.

Let us consider the fundamental representation of $SU(5)$ and denote it as a $5$-dimensional 
vector $5_i$ ($i=1,\ldots , 5$). It is usual to embed $SU(3)_C \otimes SU(2)_L$ in such a way that the first three 
components of $5$ transform as a triplet of $SU(3)_C$ and the last two components as a doublet of $SU(2)_L$ 
\be
\label{5decomp}
5 = (3,1) \oplus (1,2) \, .
\ee
In the SM we have $15$ Weyl fermions per family with quantum numbers
\be
\label{SMquantnumb}
q \sim (3,2,+\tfrac{1}{6}) \quad \ell \sim (1,2,-\tfrac{1}{2}) \quad u^c \sim (\overline{3},1,-\tfrac{2}{3}) \quad d^c \sim (\overline{3},1,+\tfrac{1}{3}) \quad  e^c \sim (1,1,+1) \, . 
\ee
How to embed these into $SU(5)$? One would be tempted to try with a $15$ of $SU(5)$. Actually from the tensor product 
\be
\label{5t5tp}
5 \otimes 5 = 10_A \oplus 15_S \, ,
\ee  
and the fact that $3 \otimes 3 = \overline{3}_A \oplus 6_S$ one concludes that some of the known quarks should belong to color sextects, which is not the case. 
So the next step is to try with $5 \oplus 10$ or better with $\overline{5} \oplus 10$ since there is no $(3,1)$
in the set of fields in \eq{SMquantnumb}.
The decomposition of $\overline{5}$ under $SU(3)_C \otimes SU(2)_L \otimes U(1)_Y$ is simply 
\be
\label{5bardecomp}
\overline{5} = (\overline{3},1,+\tfrac{1}{3}) \oplus (1,2,-\tfrac{1}{2}) \, ,
\ee
where we have exploited the fact that the hypercharge is a traceless generator of $SU(5)$, which implies the condition $3Y(d^c) + 2Y(\ell) = 0$. 
So, up to a normalization factor, one may choose $Y(d^c) = \tfrac{1}{3}$ and $Y(\ell) = - \tfrac{1}{2}$.  
Then from \eqs{5t5tp}{5bardecomp} we get
\be
10 = (5 \otimes 5)_A 
= (\overline{3},1,-\tfrac{2}{3}) \oplus (3,2,+\tfrac{1}{6}) \oplus (1,1,+1) \, .
\ee
Thus the embedding of a SM fermion family into $\overline{5} \oplus 10$ reads
\be
\overline{5} =
\left( 
\begin{array}{c}
d_1^c \\
d_2^c \\
d_3^c \\
e \\
-\nu
\end{array}
\right) \, ,
\qquad 
10 = 
\left( 
\begin{array}{ccccc}
0 & u_3^c & -u_2^c & u_1 & d_1 \\
-u_3^c & 0 & u_1^c & u_2 & d_2 \\
u_2^c & -u_1^c & 0 & u_3 & d_3 \\
-u_1 & -u_2 & -u_3 & 0 &  e^c \\
-d_1 & -d_2 & -d_3 & -e^c & 0
\end{array}
\right) \, ,
\ee
where we have expressed the $SU(2)_L$ doublets as $q = (u \ d)$ and $\ell = (\nu \ e)$. Notice in particular that 
the doublet embedded in $\overline{5}$ is 
$i \sigma_2 \ell \sim \ell^*$\footnote{Here $\sigma_2$ is the second Pauli matrix and the symbol "$\sim$" stands for the fact 
that $i \sigma_2 \ell$ and $\ell^*$ transform in the same way under $SU(2)_L$.}.

It may be useful to know how the $SU(5)$ generators act of $\overline{5}$ and $10$. From the transformation properties 
\be
\overline{5}^i \rightarrow (U^\dag)^{i}_k \ \overline{5}^k \, , \qquad 10_{ij} \rightarrow U_i^k \, U_j^l \ 10_{kl} \, ,
\ee
where $U = \exp{i T}$ and $T^\dag = T$, we deduce that the action of the generators is
\be
\label{SU5genact}
\delta \, \overline{5}^i = - T^{i}_k \ \overline{5}^k \, , \qquad \delta \, 10_{ij} =  \{ T, 10 \}_{ij} \, .
\ee
Already at this elementary level we can list a set of important features of $SU(5)$ which are typical of any GUT.

\subsection{Charge quantization and anomaly cancellation }

The charges of quarks and leptons are related. 
Let us write the most general electric charge generator compatible with the $SU(3)_C$ invariance and the $SU(5)$ embedding
\be
Q = \text{diag} 
\left(
a, a, a, b, -3a -b
\right) \, ,
\ee
where $\Tr Q = 0$. Then by applying \eq{SU5genact} we find 
\be
Q(d^c) = -a \quad Q(e) = -b \quad Q(\nu) = 3 a + b 
\ee
\be
Q(u^c) = 2 a \quad Q(u) = a + b \quad Q(d) = -(2a + b) \quad Q(e^c) = -3 a \, ,
\ee
so that apart for a global normalization factor the charges do depend just on one parameter, which must be fixed by some 
extra assumption. Let's say we 
require
$Q(\nu) = 0$\footnote{That is needed in order to give mass to the SM fermions with the Higgs mechanism. 
The simplest possibility is given by using an $SU(2)_L$ doublet $H \subset 5_H$ (cf.~\sect{YuksectSU5}) and 
in order to preserve $U(1)_Q$ it must be $Q(\vev{H}) = 0$.  
}, that readily implies 
\be
Q(e^c) = - Q(e) = \tfrac{3}{2} Q(u) = - \tfrac{3}{2} Q(u^c) = -3 Q(d) = 3 Q(d^c) = b \, ,
\ee
i.e.~the electric charge of the SM fermions is a multiple of $2 b$.


Let us consider now the issue of anomalies.
We already know that in the SM all the 
gauge anomalies vanish. This property is preserved in $SU(5)$ since 
$\overline{5}$ and $10$ have equal and opposite anomalies, so that 
the theory is still anomaly free. In order to see this explicitly let us decompose $5$ and $10$ under 
the branching chain $SU(5) \supset SU(4) \otimes U(1)_A \supset SU(3) \otimes U(1)_A \otimes U(1)_B$ 
\begin{align}
& 5 = 1(4) \oplus 4(-1) = 1(4,0) \oplus 1(-1,3) \oplus 3(-1,-1) \, , \\
& 10 = 4(3) \oplus 6(-2) = 1(3,3) \oplus 3 (3,-1) \oplus 3(-2,-2) \oplus \overline{3} (-2,-2) \, ,
\end{align}
where the $U(1)$ charges are given up to a normalization factor.
The anomaly $\mathcal{A}(R)$ relative to a representation $R$ is defined by 
\be
\Tr \{ T_R^a, T_R^b \} T_R^c = \mathcal{A} (R) d^{abc} \, , 
\ee
where $d^{abc}$ is a completely symmetric tensor. 
Then, given the properties 
\be
\mathcal{A} (R_1 \oplus R_1) =  \mathcal{A} (R_1) + \mathcal{A} (R_2) \qquad \text{and} \qquad  \mathcal{A} (\overline{R}) = -  \mathcal{A} (R) \, ,   
\ee
it is enough to compute the anomaly of the $SU(3)$ subalgebra of $SU(5)$, 
\be
\mathcal{A}_{SU(3)} (\overline{5}) = \mathcal{A}_{SU(3)} (\overline{3}) \, , 
\qquad 
\mathcal{A}_{SU(3)} (10) = \mathcal{A}_{SU(3)} (3) + \mathcal{A}_{SU(3)} (3) + \mathcal{A}_{SU(3)} (\overline{3}) \, ,
\ee
in order to conclude that $\mathcal{A} (\overline{5} \oplus 10) = 0$. 

We close this section by noticing that anomaly cancellation and charge quantization are closely related. 
Actually it is not a chance that in the SM anomaly cancellation implies charge quantization, 
after taking into account the gauge invariance of the Yukawa couplings~\cite{Geng:1988pr,Minahan:1989vd,Geng:1990nh,Foot:1990uf,Babu:1989ex}.


\subsection{Gauge coupling unification} 
At some grand unification mass scale $M_U$ the relevant symmetry is $SU(5)$ 
and the $g_3$, $g_2$, $g'$ coupling constants of $SU(3)_C \otimes SU(2)_L \otimes U(1)_Y$ merge into 
one single gauge coupling $g_U$. The rather different values for $g_3$, $g_2$, $g'$ at low-energy are then due to renormalization effects. 

\begin{figure*}[ht]
\centering
\includegraphics[width=8.5cm]{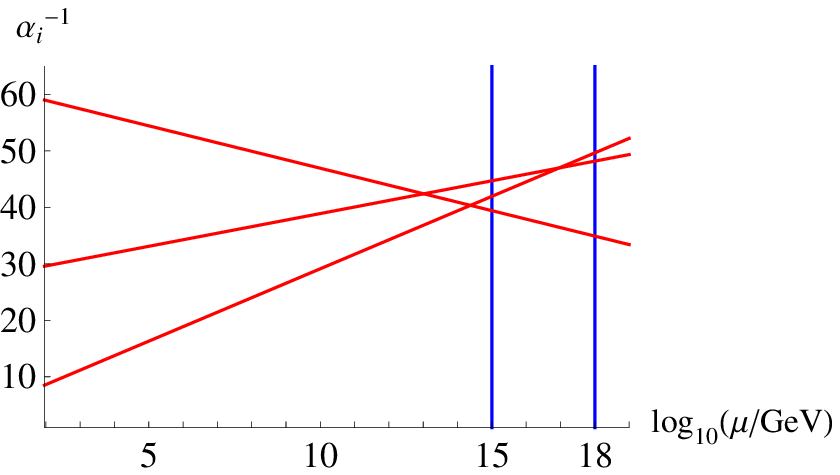}
\mycaption{\label{SMrun} One-loop running of the SM gauge couplings assuming the $U(1)_Y$ embedding into $SU(5)$.} 
\end{figure*}

Before considering the running of the gauge couplings we need to 
fix the relative normalization between $g_2$ and $g'$, which enter the weak interactions
\be
g_2 T_3 + g' Y \, .
\ee
We define 
\be
\label{defzeta}
\zeta = \frac{\Tr Y^2}{\Tr T_3^2} 
\, ,
\ee
so that $Y_1 \equiv \zeta^{-1/2} Y$ is normalized as $T_3$. In a unified theory based on a simple group, the coupling which unifies is then 
($g_1 Y_1 = g' Y$) 
\be
g_1 \equiv \sqrt{\zeta} g' \, .
\ee
Evaluating the normalization over a $\overline{5}$ of $SU(5)$ one finds
\be
\zeta = \frac{3 \left(\tfrac{1}{3}\right)^2 + 2 \left(-\tfrac{1}{2} \right)^2 }{\left(\tfrac{1}{2} \right)^2 + \left(-\tfrac{1}{2} \right)^2} 
= \frac{5}{3} \, ,
\ee
and thus one obtains the tree level matching condition
\be
g_U \equiv g_3 (M_U) = g_2 (M_U) = g_1 (M_U) \, .
\ee
At energies $\mu < M_U$ the running of the fine-structure constants ($\alpha_i \equiv g_i^2/4\pi$) is given by 
\be
\alpha_i^{-1}(t) = \alpha_i^{-1}(0) - \frac{a_i}{2\pi}\ t \,,
\ee
where $t=\log (\mu/\mu_0)$ and the one-loop beta-coefficient for the SM reads $(a_3,a_2,a_1)=(-7,-\tfrac{19}{6},\tfrac{41}{10})$.
Starting from the experimental input values for the (consistently normalized) SM gauge couplings at the scale $M_Z=91.19$ GeV~\cite{Amsler:2008zzb}
\bea
\alpha_1 &=& 0.016946 \pm 0.000006\,, \nn \\
\alpha_2 &=& 0.033812 \pm 0.000021\,, \\
\alpha_3 &=& 0.1176 \pm 0.0020\,,  \nn
\label{alphainormMZintro}
\eea
it is then a simple exercise to perform the one-loop evolution of the gauge couplings assuming just the SM 
as the low-energy effective theory. 
The result is depicted in~\fig{SMrun}

As we can see, the gauge couplings do not unify in the minimal framework, although a small perturbation may suffice 
to restore unification. 
In particular, thresholds effects at the $M_U$ scale (or below) 
may do the job, however depending on the details of the UV completion\footnote{It turns out that threshold corrections are not enough 
in order to restore unification in the minimal Georgi-Glashow $SU(5)$ (see e.g.~Ref.~\cite{Giveon:1991zm}). 
}.  

By now~\fig{SMrun} remains one of the most solid hints in favor of the grand unification idea. 
Indeed, being the gauge coupling evolution sensitive to the $\log$ of the scale, 
it is intriguing that they almost unify in a relatively narrow window, $10^{15 \div 18} \ \text{GeV}$, 
which is still allowed by the experimental lower bound on the proton lifetime and a consistent effective quantum field theory description without gravity.

\subsection{Symmetry breaking}

The Higgs sector of the Georgi-Glashow model spans over the reducible $5_H \oplus 24_H$ representation. These two fields are 
minimally needed in order to break the $SU(5)$ gauge symmetry down to $SU(3)_C \otimes SU(2)_L \otimes U(1)_Y$ and further to $SU(3)_C \otimes U(1)_Q$. 
Let us concentrate on the first stage of the breaking which is controlled by the rank-conserving VEV $\vev{24_H}$.  
The fact that the adjoint preserves the rank is easily seen by considering the action of the Cartan generators on the adjoint vacuum
\be
\delta \vev{24_H}^i_j = \left[ T_{\text{Cartan}}, \vev{24_H} \right]^i_j \, , 
\ee
derived from the transformation properties of the adjoint 
\be
24^i_j \rightarrow (U^\dag)^i_k U^l_j  \, 24^k_l \, .
\ee
Since $\vev{24_H}$ can be diagonalized by an $SU(5)$ transformation 
and the Cartan generators are diagonal by definition, one concludes that the adjoint preserves the Cartan subalgebra. 
The scalar potential is given by 
\be
\label{defV24}
V(24_H) = -m^2 \Tr{24_H^2} + \lambda_1 \left( \Tr{24_H^2} \right)^2 + \lambda_2 \Tr{24_H^4} \, ,
\ee
where just for simplicity we have imposed the discrete symmetry $24_H \rightarrow - 24_H$. 
The minimization of the potential goes as follows. First of all $\vev{24_H}$ is transformed into a real diagonal traceless 
matrix by means of an $SU(5)$ transformation 
\be
\vev{24_H} = \text{diag} (h_1, h_2, h_3, h_4, h_5) \, ,
\ee
where $h_1 + h_2 + h_3 + h_4 + h_5 = 0$. With $24_H$ in the diagonal form, the scalar potential reads
\be
V(24_H) = - m^2 \sum_{i} h_i^2 + \lambda_1 \left( \sum_{i} h_i^2 \right)^2 + \lambda_2 \sum_{i} h_i^4 \, .
\ee 
Since the $h_i$'s are not all independent, we need to use the lagrangian multiplier $\mu$ in order to 
account for the constraint $\sum_i h_i = 0$. The minimization of the potential $V'(24_H) = V(24_H) - \mu \Tr 24_H$ yields
\be 
\frac{\partial V'(24_H)}{\partial h_i} = -2 m^2 h_i + 4 \lambda_1 \left( \sum_{j} h_j^2 \right) h_i + 4 \lambda_2 h_i^3 - \mu = 0 \, .
\ee
Thus at the minimum all the $h_i$'s satisfy the same cubic equation 
\be
4 \lambda_2 x^3 + \left( 4 \lambda_1 a  - 2 m^2 \right) x - \mu = 0 \qquad \text{with} \qquad a = \sum_j h_j^2 \, .
\ee
This means that the $h_i$'s can take at most three different values, $\phi_1$, $\phi_2$ and $\phi_3$, which are the three 
roots of the cubic equation. Note that the absence of the $x^2$ term in the cubic equation implies that 
\be
\phi_1 + \phi_2 + \phi_3 = 0 \, .
\ee
Let $n_1$, $n_2$ and $n_3$ the number of times $\phi_1$, $\phi_2$ and $\phi_3$ appear in $\vev{24_H}$,
\be
\vev{24_H} = \text{diag} (\phi_1, \ldots , \phi_2, \ldots, \phi_3) \qquad \text{with} \qquad n_1 \phi_1 + n_2 \phi_2 + n_3 \phi_3 = 0\, . 
\ee 
Thus $\vev{24_H}$ is invariant under $SU(n_1) \otimes SU(n_2) \otimes SU(n_3)$ transformations. This implies that the most general form of symmetry breaking is 
$SU(n) \rightarrow SU(n_1) \otimes SU(n_2) \otimes SU(n_3)$ as well as possible $U(1)$ factors (total rank is 4) which leave $\vev{24_H}$ invariant.
To find the absolute minimum we have to use the relations 
\be
n_1 \phi_1 + n_2 \phi_2 + n_3 \phi_3 = 0 \qquad \text{and} \qquad \phi_1 + \phi_2 + \phi_3 = 0 
\ee
to compare different choices of $\{n_1,n_2,n_3\}$ in order to get the one with the smallest $V(24_H)$. 
It turns out (see e.g.~Ref.~\cite{Li:1973mq}) that for the case of interest there are two possible patterns for the symmetry breaking
\be
SU(5) \rightarrow SU(3) \otimes SU(2) \otimes U(1) \qquad \text{or} \qquad SU(5) \rightarrow SU(4) \otimes U(1) \, ,
\ee
depending on the relative magnitudes of the parameters $\lambda_1$ and $\lambda_2$.
In particular for $\lambda_1 > 0$ and $\lambda_2 > 0$ the absolute minimum is given by the SM vacuum~\cite{Li:1973mq} and the adjoint VEV reads 
\be
\label{24Hvac}
\vev{24_H} = V\, \text{diag} (2,2,2,-3,-3) \, .
\ee
Then the stability of the vacuum requires 
\be
\lambda_1 \left( \Tr{\vev{24_H}^2} \right)^2 + \lambda_2 \Tr{\vev{24_H}^4} > 0 \qquad \Longrightarrow \qquad \lambda_1 > -\frac{7}{30} \lambda_2
\ee
and the minimum condition 
\be
\frac{\partial V(\vev{24_H})}{\partial V} = 0 \qquad \Longrightarrow \qquad 
60 V \left( -m^2 + 2 V^2 (30 \lambda_1+7 \lambda_2) \right) = 0
\ee
yields 
\be
V^2 = \frac{m^2}{2 (30 \lambda_1+7 \lambda_2)} \, .
\ee
Let us now write the covariant derivative 
\be
D_\mu 24_H = \partial_\mu 24_H + i g \left[ A_\mu, 24_H \right] \, , 
\ee
where $A_\mu$ and $24_H$ are $5 \times 5$ traceless hermitian matrices.
Then from the canonical kinetic term, 
\be
\Tr D_\mu \vev{24_H} D^\mu \vev{24_H}^\dag =  g^2 \Tr \left[ A_\mu, \vev{24_H} \right] \left[ \vev{24_H}, A^\mu \right]
\ee 
and the shape of the vacuum 
\be
\vev{24_H}^i_j = h_j \delta^i_j \, ,
\ee
where repeated indices are not summed,
we can easily extract the gauge bosons mass matrix from the expression
\be
g^2 \left[ A_\mu, \vev{24_H} \right]^i_j \left[ \vev{24_H}, A^\mu \right]^j_i = 
g^2 \left( A_\mu \right)^i_j \left( A^\mu \right)^j_i \left( h_i -h_j \right)^2 \, .
\ee
The gauge boson fields $(A_\mu)^i_j$ having $i=1,2,3$ and $j=4,5$ are massive, $M_X^2 = 25 g^2 V^2$, while $i,j = 1,2,3$ and $i,j=4,5$ are still massless. 
Notice that the hypercharge generator commutes with the vacuum in $\eq{24Hvac}$ and hence the associated gauge boson is massless as well.
The number of massive gauge bosons is then $24 - (8 + 3 +1) = 12$ and their quantum numbers correspond to the coset $SU(5)/SU(3)_C \otimes SU(2)_L \otimes U(1)_Y$. 
Their mass $M_X$ is usually identified with the grand unification scale, $M_U$.

\subsection{Doublet-Triplet splitting}
\label{DTsplit}

The second breaking step, $SU(3)_C \otimes SU(2)_L \otimes U(1)_Y \rightarrow SU(3)_C \otimes U(1)_Q$, 
is driven by a $5_H$ where 
\be
5_H =
\left( 
\begin{array}{c}
T \\
H 
\end{array}
\right) \, ,
\ee
decomposes into a color triplet $T$ and an $SU(2)_L$ doublet $H$. The latter plays the same role of the Higgs doublet of the SM. 
The most general potential containing both $24_H$ and $5_H$ can be written as
\be
V = V(24_H) + V(5_H) + V(24_H, 5_H) \, ,
\ee
where $V(24_H)$ is defined in \eq{defV24}, 
\be
V(5_H) = - \mu^2 \, 5_H^\dag 5_H + \lambda \left( 5_H^\dag 5_H \right)^2 \, ,
\ee
and
\be
V(24_H, 5_H) = \alpha \, 5_H^\dag 5_H \Tr 24_H^2 + \beta \, 5_H^\dag 24_H^2 5_H \, .
\ee
Again we have imposed for simplicity the discrete symmetry $24_H \rightarrow - 24_H$.
It is instructive to compute the mass of the doublet $H$ and the triplet $T$ in the SM vacuum just after the first stage of the breaking
\be
\label{MHMTtree}
M_H^2 = -\mu^2 + (30 \alpha + 9 \beta) V^2 \, , \qquad
M_T^2 = - \mu^2 + (30 \alpha + 4 \beta) V^2 \, .
\ee
The gauge hierarchy $M_X \gg M_W$ requires that the doublet $H$, containing the 
would-be Goldstone bosons eaten by the $W$ and the $Z$ and the physical Higgs boson, live at the $M_W$ scale. 
This is unnatural and can be achieved at the prize of a fine-tuning of one part in $\mathcal{O}(M_X^2 / M_W^2)\sim 10^{26}$ 
in the expression for $M_H^2$. 
If we follow the principle that only the minimal 
fine-tuning needed for the gauge hierarchy is allowed 
then $M_T$ is automatically kept 
heavy\footnote{In some way this is an extension of the Georgi's survival hypothesis for fermions~\cite{Georgi:1979md}, 
according to which the particles do not survive to low energies unless a symmetry forbids their large mass terms. 
This hypothesis is obviously wrong for scalars and must be extended. The extended survival hypothesis (ESH) reads: Higgs scalars (unless protected by some symmetry) 
acquire the maximum mass 
compatible with the pattern of symmetry breaking~\cite{delAguila:1980at}. In practice this corresponds to the requirement 
of the minimal number of fine-tunings to be imposed onto the scalar potential~\cite{Mohapatra:1982aq}.}. 
This goes under the name of doublet-triplet (DT) splitting.
Usually, but not always~\cite{Dvali:1992hc,Dvali:1995hp}, a light triplet is very dangerous for the proton stability since it can couple 
to the SM fermions in such a way that baryon number is not anymore an accidental global 
symmetry of the low-energy lagrangian\footnote{Let us consider for instance the invariants $qqT$ and $q\ell T^*$. There's no way to assign 
a baryon charge to $T$ in such a way that $U(1)_B$ is preserved.}. 

A final comment about the radiative stability of the fine-tuning is in order. 
While supersymmetry helps in stabilizing the hierarchy between $M_X$ and $M_W$ against radiative corrections, 
it does not say much about the origin of this hierarchy. 
Other mechanisms have to be devised to render the hierarchy natural 
(for a short discussion of the solutions proposed so far cf.~\sect{DTsplittingflipped}).
In a nonsupersymmetric scenario one needs  
to compute the mass of the doublet in~\eq{MHMTtree} within a $13$-loop accuracy in order to stabilize the hierarchy.

\subsection{Proton decay}

The theory predicts that protons eventually decay. The most emblematic contribution to proton decay is due to the exchange of super-heavy 
gauge bosons which belong to the coset $SU(5)/SU(3)_C \otimes SU(2)_L \otimes U(1)_Y$.
Let us denote the matter representations of $SU(5)$ as
\be
\label{5bar10index}
\overline{5} = 
\left( 
\psi_\alpha,
\psi_i
\right) \, ,
\qquad 
10 = \left( \psi^{\alpha\beta}, \psi^{\alpha i}, \psi^{ij} \right) \, ,
\ee
where the greek and latin indices run respectively from $1$ to $3$ ($SU(3)_C$ space) and $1$ to $2$ ($SU(2)_L$ space).
Analogously the adjoint $24$ can be represented as
\be
24 = \left( X^\alpha_\beta, X^i_j, X^\alpha_\alpha - \tfrac{3}{2} X^i_i, X^\alpha_i, X^i_\alpha \right) \, ,
\ee
from which we can readily recognize the gauge bosons associated to the SM unbroken generators ($(8,1) \oplus (3,1) \oplus (1,1)$)
and the two super-heavy leptoquark gauge bosons ($(3,2) \oplus (\overline{3},2)$). 
Let us consider now the gauge action of $X^\alpha_i$ on the matter fields
\be
\label{SU5proc}
X^\alpha_i: 
\quad \psi_\alpha \rightarrow \psi_i \ (d^c \rightarrow \nu, e) \, ,
\quad \psi^{\beta i} \rightarrow \psi^{\beta \alpha} \ (d, u \rightarrow u^c) \, ,
\quad \psi^{ij} \rightarrow \psi^{\alpha j} \ (e^c \rightarrow u, d) \, .
\ee
Thus diagrams involving the exchange of a $X^\alpha_i$ boson generate processes like 
\be
\label{SU5bnv}
u d \rightarrow u^c e^c \, , 
\ee
whose amplitude is proportional to the gauge boson propagator. 
After dressing the operator with a spectator quark u, we can have for instance the low-energy process $p \rightarrow \pi^0 e^+$, 
whose decay rate can be estimated by simple dimensional analysis
\be
\Gamma (p \rightarrow \pi^0 e^+) \sim \frac{\alpha_U^2 m_p^5}{M_X^4} \, . 
\ee 
Using $\tau (p \to \pi^0 e^+) > 8.2 \times 10^{33}$ years~\cite{Nakamura:2010zzi}
we extract (for $\alpha^{-1}_{U}= 40$) the naive lower bound on the super-heavy gauge boson mass
\begin{equation}
M_X > 2.3 \times 10^{15} \ \text{GeV}
\end{equation}     
which points directly to the grand unification scale extrapolated by the gauge running (see e.g.~\fig{SMrun}).


Notice that $B-L$ is conserved in the process $p \rightarrow \pi^0 e^+$. 
This selection rule is a general feature of the gauge induced proton decay
and can be traced back to the presence of a global $B-L$ accidental symmetry in the transitions of~\eq{SU5proc} 
after assigning $B-L \ (X^\alpha_i) = 2/3$.


\subsection{Yukawa sector and neutrino masses}
\label{YuksectSU5}

The $SU(5)$ Yukawa lagrangian can be written 
schematically\footnote{More precisely 
$\overline{5}_F Y_5 10_F 5^*_H \equiv  \left(\overline{5}_F\right)^{\alpha x}_m C_{xy} \left(Y_5\right)^{mn} \left(10_F\right)^y_{\alpha\beta n} \left(5^*_H\right)^\beta$ 
and $\epsilon_5 10_F Y_{10} 10_F 5_H \equiv  \epsilon^{\alpha\beta\gamma\delta\epsilon}  \left(10_F\right)^{x}_{\alpha\beta m} C_{xy} \left(Y_{10}\right)^{mn} 
\left(10_F\right)^{y}_{\gamma\delta n} \left(5_H\right)_{\epsilon}$, where $(\alpha, \beta, \gamma, \delta, \epsilon)$, $(m, n)$ and $(x, y)$ are respectively  
$SU(5)$, family and Lorentz indices.} as  
\be
\label{SU5Yuk}
\mathcal{L}_Y = \overline{5}_F Y_5 10_F 5^*_H + \frac{1}{8} \epsilon_5 10_F Y_{10} 10_F 5_H + \text{h.c.} \, ,
\ee
where $\epsilon_5$ is the 5-index Levi-Civita tensor. 
After denoting the $SU(5)$ representations synthetically as
\be
\overline{5}_F = 
\left( 
\begin{array}{c}
d^c \\
\epsilon_2 \ell
\end{array}
\right) 
\qquad 
10_F = 
\left( 
\begin{array}{cc}
\epsilon_3 u^c & q \\
-q^T & \epsilon_2 e^c
\end{array}
\right) 
\qquad 
5_H = 
\left( 
\begin{array}{c}
T \\
H
\end{array}
\right) \, ,
\ee 
where $\epsilon_3$ is the 3-index Levi-Civita tensor and $\epsilon_2 = i \sigma_2$, 
we project \eq{SU5Yuk} over the SM components. This yields  
\be
\label{5bar105bardec}
\overline{5}_F Y_5 10_F 5^*_H = 
\left( d^c \ \ell \epsilon_2^T \right) 
\left( 
\begin{array}{cc}
\epsilon_3 u^c & q \\
-q^T & \epsilon_2 e^c
\end{array}
\right) 
\left( 
\begin{array}{c}
T^* \\
H^*
\end{array}
\right)
\rightarrow 
d^c Y_5 q H^* + \ell Y_5 e^c H^* \, , 
\ee
\be
\frac{1}{8} \epsilon_5 10_F Y_{10} 10_F 5_H \rightarrow \frac{1}{2} u^c \left( Y_{10} + Y_{10}^T \right) q H \, .
\ee
After rearranging the order of the $SU(2)_L$ doublet and singlet fields in the second term of \eq{5bar105bardec}, i.e.~$\ell Y_5 e^c H^* = e^c Y_5^T \ell H^*$,
one gets
\be
\label{btauunif}
Y_d = Y_e^T \qquad \text{and} \qquad Y_u = Y_u^T \, ,
\ee
which shows a deep connection between flavor and the GUT symmetry 
(which is not related to a flavor symmetry).
The first relation in \eq{btauunif} predicts $m_b (M_U) = m_\tau (M_U)$, $m_s (M_U) = m_\mu (M_U)$ and $m_d (M_U) = m_e (M_U)$ at the GUT scale. 
So in order to test this relation one has to run the SM fermion masses starting from their low-energy values. 
While $m_b (M_U) = m_\tau (M_U)$ is obtained in the MSSM
with a typical $20-30\%$ uncertainty~\cite{Babu:1998er}, 
the other two relations are evidently wrong. 
By exploiting the fact that the ratio between $m_d / m_e$ and $m_s / m_\mu$ is essentially independent of renormalization 
effects~\cite{Buras:1977yy}, we get the scale free relation  
\be
m_d / m_s = m_e / m_\mu \, ,
\ee   
which is off by one order of magnitude.

Notice that $m_d = m_e$ comes from the fact that 
the fundamental $\vev{5_H}$ breaks $SU(5)$ down to $SU(4)$ 
which remains an accidental symmetry of the Yukawa sector. 
So one expects that considering higher dimensional representations makes it possible to further break
the remnant $SU(4)$. This is indeed what happens by introducing a $45_H$ which couples to the fermions in the following way~\cite{Georgi:1979df} 
\be
\overline{5}_F 10_F 45^*_H + 10_F 10_F 45_H + \text{h.c.} \, .
\ee
The first operator leads to $Y_d = -3 Y_e$,
so that if both $5_H$ and $45_H$ are present more freedom is available to fit all fermion masses. 
Alternatively one can built an effective coupling~\cite{Ellis:1979fg}
\be
\frac{1}{\Lambda} \overline{5}_F 10_F (\vev{24_H} 5^*_H)_{\overline{45}} \, ,
\ee
which mimics the behavior of the $45_H$. If we take the cut-off to be the planck scale $M_P$, 
this nicely keeps $b-\tau$ unification while corrects the relations among the first two families. 
However in both cases we loose predictivity since we are just fitting $M_d$ and $M_e$ in the extended Yukawa structure. 

Finally what about neutrinos? 
It turns out~\cite{Wilczek:1979et} that the Georgi-Glashow model has an accidental global $U(1)_G$ symmetry
with the charge assignment $G(\overline{5}_F) = -\frac{3}{5}$, $G(10_F) = +\frac{1}{5}$ and $G(5_H) = +\frac{2}{5}$. 
The VEV $\vev{5_H}$ breaks this global symmetry but leaves invariant a linear combination of $G$ and a Cartan generator of $SU(5)$. 
It easy to see that any linear combination of $G + \tfrac{4}{5} Y$, Q, and any color generators is left invariant. The 
extra conserved charge $G + \tfrac{4}{5} Y$ when acting on the fermion fields is just $B-L$.
Thus neutrinos cannot acquire neither a Dirac (because of the field content) nor a Majorana (because of the global $B-L$ symmetry) mass term 
and they remain exactly massless even at the quantum level. 

Going at the non-renormalizable level we can break the accidental $U(1)_G$ symmetry. 
For instance global charges are expected to be violated by gravity and the 
simplest effective operator one can think of is~\cite{Barbieri:1979hc}
\be
\label{MPnuSU5}
\frac{1}{M_P} \overline{5}_F \overline{5}_F 5_H 5_H \, .
\ee
However its contribution to neutrino masses is too much suppressed ($m_\nu \sim \mathcal{O} (M_W^2/M_P)$ $\sim 10^{-5} \ \text{eV}$).  
Thus we have to extend the field content of the theory in order to generate phenomenologically viable neutrino masses. 
Actually, the possibilities are many. 

Minimally one may add an $SU(5)$ singlet fermion field $1_F$. Then, through its renormalizable coupling $\overline{5}_F 1_F 5_H$,
one integrates $1_F$ out and generates an operator similar to that in \eq{MPnuSU5}, but suppressed by the $SU(5)$-singlet mass term which can 
be taken well below $M_P$. 

A slightly different approach could be breaking the accidental $U(1)_G$ symmetry by adding additional scalar representations. 
Let us take for instance a $10_H$ and consider then the new couplings~\cite{Wilczek:1979et} 
\be
\mathcal{L}_{10} \supset f \, \overline{5}_F \overline{5}_F 10_H + M \, 10_H 10_H 5_H \, .
\ee 
Since $G(\overline{5}_F) = -\frac{3}{5}$ and $G(5_H) = +\frac{2}{5}$ there's no way to assign a $G$-charge to $10_H$ in order to preserve $U(1)_G$. 
Thus we expect that loops containing the $B-L$ breaking sources $f$ and $M$ can generate neutrino masses. 

So what is wrong with the two approaches above? In principle nothing. But maybe we should try to do more than getting out what we put in. 
Indeed we are just solving the issue of neutrino masses "ad hoc", without correlations to other phenomena. In addition we do not 
improve unification of minimal $SU(5)$\footnote{An analysis of the thresholds corrections in the Georgi-Glashow model with the addition of the $10_H$ 
indicates that unification cannot be restored.}. 

Guided by this double issue of the Georgi-Glashow model, two minimal extensions which can cure at the same time 
both neutrino masses and unification have been recently proposed 
\begin{itemize}
\item Add a $15_H = (1,3)_H \oplus (6,1)_H \oplus (3,2)_H$~\cite{Dorsner:2005fq}. 
Here $(1,3)_H$ is an Higgs triplet responsible for type-II seesaw.  
The model predicts generically light leptoquarks $(3,2)_H$ and fast proton decay~\cite{Dorsner:2005ii}.
\item Add a $24_F = (1,1)_F \oplus (1,3)_F \oplus (8,1)_F \oplus (3,2)_F \oplus (\overline{3},2)_F$~\cite{Bajc:2006ia}. 
Here $(1,1)_F$ and $(1,3)_F$ are fields responsible respectively for type-I and type-III seesaw. 
The model predicts a light fermion triplet $(1,3)_F$ and fast proton decay~\cite{Bajc:2007zf}.  
\end{itemize}  

Another well motivated and studied extension of the Georgi-Glashow model is given by supersymmetric $SU(5)$~\cite{Dimopoulos:1981zb}. 
In this case the supersymmetrization of the spectrum is enough in order to fix both unification and neutrino masses. 
Indeed, if we do not impose by hand $R$-parity conservation Majorana neutrino masses are automatically generated by lepton number violating interactions~\cite{Hall:1983id}. 

\section{The Pati-Salam route}

In the SM there is an intrinsic lack of left-right symmetry 
without any explanation of the phenomenological facts that neutrino masses are very small and the weak interactions are predominantly $V-A$. 
The situation can be schematically depicted in the following way  
\be
q =
\left( 
\begin{array}{ccc}
u_1 & u_2 & u_3 \\
d_1 & d_2 & d_3 
\end{array}
\right)
\qquad
\ell = 
\left( 
\begin{array}{c}
\nu \\
e 
\end{array}
\right)
\qquad 
\begin{array}{ccc}
d^c = ( d_1^c & d_2^c & d_3^c ) \\
u^c = ( u_1^c & u_2^c & u_3^c )
\end{array}
\qquad
\begin{array}{c}
e^c \\ 
?
\end{array}
\ee
where $q = (3,2,+\tfrac{1}{6})$, $\ell = (1,2,-\tfrac{1}{2})$, $d^c = (\overline{3},1,+\tfrac{1}{3})$, $u^c = (\overline{3},1,-\tfrac{2}{3})$ 
and $e^c = (1,1,+1)$ under $SU(3)_C \otimes SU(2)_L \otimes U(1)_Y$.

Considering the SM as an effective theory, neutrino masses can be generated by a $d=5$ operator~\cite{Weinberg:1979sa} of the type 
\be
\label{Weinbergop}
\frac{Y_\nu}{\Lambda_L} (\ell^T \epsilon_2 H) C (H^T \epsilon_2 \ell) \, ,
\ee
where $\epsilon_2 = i \sigma_2$ and $C$ is the charge-conjugation matrix. After electroweak symmetry breaking, $\vev{H} = v$, neutrinos pick up a Majorana mass term $M_{\nu} \nu^T C \nu$ with 
\be
M_\nu = Y_\nu \frac{v^2}{\Lambda_L} \, .
\ee
The lower bound on the highest neutrino eigenvalue inferred from $\sqrt{\Delta m_{atm}} \sim 0.05 \ \text{eV}$ 
tells us that the scale at which the lepton number is violated is 
\be
\Lambda_L \lesssim Y_\nu \ \mathcal{O}(10^{14 \div 15} \ \text{GeV}) \, .
\ee
Notice that without a theory which fixes the structure of $Y_\nu$ we don't have much to say about $\Lambda_L$. 

Actually, by exploiting the Fierz identity $(\sigma_i)_{ab} (\sigma_i)_{cd} = 2 \delta_{ad} \delta_{cb} - \delta_{ab} \delta_{cd}$,
one finds that the operator in \eq{Weinbergop} can be equivalently written in three different ways
\be
\label{Weinberg3op}
(\ell^T \epsilon_2 H) C (H^T \epsilon_2 \ell) 
= \frac{1}{2} (\ell^T C \epsilon_2 \sigma_i \ell) (H^T \epsilon_2 \sigma_i H)
= - (\ell^T \epsilon_2 \sigma_i H) C (H^T \epsilon_2 \sigma_i \ell) \, .
\ee
Each operator in \eq{Weinberg3op} hints to a different renormalizable UV completion of the SM.  
Indeed one can think those effective operators as the result of the integration of an heavy state 
with a renormalizable coupling of the type
\be
\label{3seesaws}
(\ell^T \epsilon_2 H) C \nu^c \qquad 
(\ell^T C \epsilon_2 \sigma_i \ell) \Delta_i 
\qquad
(\ell^T \epsilon_2 \sigma_i H) C T_i \, ,
\ee
where $\nu^c$, $\Delta_i$ and $T_i$ are a fermionic singlet ($Y=0$), a scalar triplet ($Y=+1$) and a fermionic triplet ($Y=0$).   
Notice that being $\nu^c$, $\Delta_i \oplus \Delta^*_i$ and $T_i$ vector-like states their mass is not protected by the electroweak symmetry and
it can be identified with the scale $\Lambda_L$, 
thus providing a rationale for the smallness of neutrino masses. This goes under the name of seesaw mechanism and the 
three options in \eq{3seesaws} are classified respectively as 
type-I~\cite{Minkowski:1977sc,GellMann:1980vs,Yanagida:1979as,Glashow:1979nm,Mohapatra:1979ia}, 
type-II~\cite{Magg:1980ut,Schechter:1980gr,Lazarides:1980nt,Mohapatra:1980yp} and type-III~\cite{Foot:1988aq} seesaw.  

\subsection{Left-Right symmetry}
\label{leftrightsymmetry}

Guided by the previous discussion on the renormalizable origin of neutrino masses, it is then very natural to 
to fill the gap in the SM by introducing a SM-singlet fermion field $\nu^c$. 
In such a way the spectrum looks more "symmetric" and one can imagine that at higher energies the left-right symmetry is restored, 
in the sense that left and right chirality fermions\footnote{As already stressed we work in a formalism in which 
all the fermions are left-handed four components Weyl spinors. 
The right chirality components are obtained by means of charge conjugation, 
namely $\overline{\psi}_R \equiv \psi^{cT}_L C$ or equivalently $\psi_L^c \equiv C \gamma_0 \psi_R^*$.} 
are assumed to play an identical role prior to some kind of spontaneous symmetry 
breaking.

The smallest gauge group that implement this idea is 
$SU(3)_C \otimes SU(2)_L \otimes SU(2)_R \otimes U(1)_{B-L} \otimes Z_2$~\cite{Pati:1974yy,Mohapatra:1974gc,Senjanovic:1975rk}, 
where $Z_2$ is a discrete symmetry which exchange $SU(2)_L \leftrightarrow SU(2)_R$.  
The field content of the theory can be schematically depicted as 
\be
\label{LRspectrum}
q = 
\left( 
\begin{array}{ccc}
u_1 & u_2 & u_3 \\
d_1 & d_2 & d_3 
\end{array}
\right)
\ \
\ell = 
\left( 
\begin{array}{c}
\nu \\
e 
\end{array}
\right)
\ \ 
q^c = 
\left(
\begin{array}{ccc}
d_1^c & d_2^c & d_3^c \\
-u_1^c & -u_2^c & -u_3^c 
\end{array}
\right)
\ \
\ell^c = 
\left(
\begin{array}{c}
e^c \\ 
-\nu^c
\end{array}
\right) 
\ee
where 
$q = (3,2,1,+\frac{1}{3})$,
$\ell = (1,2,1,-1)$,   
$q^c = (\overline{3},1,2^*,-\frac{1}{3})$,
$\ell^c = (1,1,2^*,+1)$, 
under $SU(3)_C \otimes SU(2)_L \otimes SU(2)_R \otimes U(1)_{B-L}$.
Given this embedding of the fermion fields one readily verifies that the electric charge formula takes the expression
\be
\label{Qformula}
Q = T^3_L + T^3_R + \frac{B-L}{2} \, .
\ee
Next we have to state the Higgs sector. 
In the early days of the development of left-right theories 
the breaking to the SM was minimally achieved by employing the following set of representations: 
$\delta_L = (1,2,1,+1)$, $\delta_R = (1,1,2,+1)$ and $\Phi = (1,2,2^*,0)$~\cite{Pati:1974yy,Mohapatra:1974gc,Senjanovic:1975rk}. 
However, as pointed out in~\cite{Mohapatra:1979ia,Mohapatra:1980yp}, in order to understand the smallness of 
neutrino masses it is better to consider $\Delta_L = (1,3,1,+2)$ and $\Delta_R = (1,1,3,+2)$ in place of $\delta_L$ and 
$\delta_R$. 

Choosing the matrix representation $\Delta_{L,R} = \Delta^i_{L,R} \sigma_i/2$ for the $SU(2)_{L,R}$ adjoint 
and defining the conjugate doublet $\tilde{\Phi} \equiv \sigma_2 \Phi^\ast \sigma_2$, 
the transformation properties for the Higgs fields under $SU(2)_L$ and $SU(2)_R$ read 
\be
\label{transprop}
\Delta_L \rightarrow U_L\, \Delta_L\, U_L^\dag \, , \qquad \Delta_R \rightarrow U_R\, \Delta_R\, U_R^\dag 
\, , \qquad \Phi \rightarrow U_L\, \Phi\, U_R^\dag \, , \qquad \tilde{\Phi} \rightarrow U_L \tilde{\Phi} U_R^\dag \, ,
\ee
and consequently we have
\be
\begin{array}{llll}
\delta_L \Delta_L = \left[ T^3_L , \Delta_L \right] \quad & \delta_L \Delta_R = 0 \quad & \delta_L \Phi = T^3_L \Phi 
\quad & \delta_L \tilde{\Phi} = T^3_L \tilde{\Phi} \\ \\
\delta_R \Delta_L = 0 \quad & \delta_R \Delta_R = \left[ T^3_R , \Delta_R \right] \quad & \delta_R \Phi = - \Phi T^3_R 
\quad & \delta_R \tilde{\Phi} = - \tilde{\Phi} T^3_R \\ \\
\delta_{B-L} \Delta_L = +2 \Delta_L \quad & \delta_{B-L} \Delta_R = +2 \Delta_R \quad & \delta_{B-L} \Phi = 0 
\quad & \delta_{B-L} \tilde{\Phi} = 0 \, .
\end{array} 
\ee
Then, given the expression for the electric charge operator in \eq{Qformula}, 
we can decompose these fields in the charge eigenstates 
\begin{small}
\be
\label{chargeeig}
\Delta_{L,R} = 
\left( 
\begin{array}{cc}
\Delta^+ / \sqrt{2} & \Delta^{++} \\
\Delta^0 & - \Delta^+ / \sqrt{2} 
\end{array}
\right)_{L,R} \, ,
\quad
\Phi = 
\left( 
\begin{array}{cc}
\phi_1^0 & \phi_1^+ \\
\phi_2^- & \phi_2^0 
\end{array}
\right) \, ,
\quad
\tilde{\Phi} = 
\left( 
\begin{array}{cc}
\phi_2^{0\ast} & - \phi_2^+ \\
-\phi_1^- & \phi_1^{0\ast}
\end{array}
\right) \, .
\ee
\end{small}

In order to fix completely the theory one has to specify the action of the $Z_2$ symmetry on the field content. 
There are two phenomenologically viable left-right discrete symmetries: $Z_2^P$ and $Z_2^C$. They are defined as
\be
\label{defZ2}
Z_2^P :
\left\{ 
\begin{array}{ccc}
\psi_L &\longleftrightarrow& \psi_R \\
\Delta_L &\longleftrightarrow& \Delta_R \\
\Phi &\longleftrightarrow& \Phi^\dag \\
W^\mu_L &\longleftrightarrow& W^\mu_R
\end{array}
\right.
\qquad \text{and} \qquad
Z_2^C :
\left\{ 
\begin{array}{ccc}
\psi_L &\longleftrightarrow& \psi_L^c \\
\Delta_L &\longleftrightarrow& \Delta^*_R \\
\Phi &\longleftrightarrow& \Phi^T \\
W^\mu_L &\longleftrightarrow& W^{\mu *}_R
\end{array}
\right. \, .
\ee 
The implications of this two cases differ by the tiny amount of $CP$ violation. 
Indeed when restricted to the fermion fields we can identify $Z_2^P$ and $Z_2^C$ respectively with 
$P: \psi_L \rightarrow \psi_R$ and $C: \psi_L \rightarrow \psi_L^c \equiv C \gamma_0 \psi_R^*$. 
In the former case the Yukawa matrices are hermitian while in the latter they are symmetric. 
So if $CP$ is conserved (real couplings) $Z_2^P$ and $Z_2^C$ lead to the same predictions.

Notice that $Z_2^C$ involves an exchange between spinors with the same chirality. In principle this would 
allow the embedding of $Z_2^C$ into a gauge symmetry which commutes with the Lorentz group. 
The gauging is conceptually important since it protects the symmetry from unknown UV effects. 

Remarkably it turns out that $Z_2^C$ can be identified with a finite gauge transformation 
of $SO(10)$ which, historically, goes under the name of 
D-parity~\cite{Kuzmin:1980yp,Kibble:1982dd,Chang:1983fu,Chang:1984uy,Chang:1984qr}.
The connection with $SO(10)$ motivates our notation in terms of left-handed fermion fields which fits better for the $Z_2^C$ case.

Let us consider now the symmetry breaking sector.  
From~\eq{chargeeig} we deduce that the SM-preserving vacuum directions are 
\be
\vev{\Delta_{L,R}} = 
\left( 
\begin{array}{cc}
0 & 0 \\
v_{L,R} & 0 
\end{array}
\right)
\qquad
\vev{\Phi} = 
\left( 
\begin{array}{cc}
v_1 & 0 \\
0 & v_2 
\end{array}
\right) \, ,
\qquad 
\vev{\tilde{\Phi}} = 
\left( 
\begin{array}{cc}
v^*_2 & 0 \\
0 & v^*_1
\end{array}
\right) \, .
\ee
The minimization of the scalar potential (see e.g.~Appendix B of Ref.~\cite{Mohapatra:1980yp}) shows that beside the expected 
left-right symmetric minimum $v_L = v_R$, we have also the asymmetric one 
\be
\label{asymmetricLRvac}
v_L \neq v_R \, , \qquad v_L v_R = \gamma v_1^2 \, , \qquad (\text{in the approximation} \ v_2 = 0) \, ,
\ee
where $\gamma$ is a combination of parameters of the Higgs potential. 
Since the discrete left-right symmetry is defined to transform $\Delta_L \leftrightarrow \Delta_R$ ($\Delta_L \leftrightarrow \Delta_R^*$) in the case 
of $Z_2^P$ ($Z_2^C$), the VEVs in 
\eq{asymmetricLRvac} breaks it spontaneously. 
Phenomenologically we have to require $v_R \gg v_1 \gg v_L$ which leads to the following breaking pattern
\begin{multline}
SU(3)_C \otimes SU(2)_L \otimes SU(2)_R \otimes U(1)_{B-L} \otimes Z_2  \ \stackrel{v_R}{\longrightarrow}\
SU(3)_C \otimes SU(2)_L \otimes U(1)_Y \\  \ \stackrel{v_1 \gg v_L}{\longrightarrow}\ 
SU(3)_C \otimes U(1)_Q \, ,
\end{multline}
where the gauge hierarchy is set by the gauge boson masses $M_{W_R}, M_{Z_R} \gg M_{W_L}, M_{Z_L}$.
Let us verify this by computing $M_{W_R}$ and $M_{Z_R}$. We start from the covariant derivative
\be
D_\mu \Delta_R = \partial_\mu \Delta_R  + i g_R \left[T^i_R, \Delta_R\right] \left( A^i_R \right)_\mu + i g_{B-L} \frac{B-L}{2} \Delta_R \left( A_{B-L} \right)_\mu \, ,
\ee
and the canonically normalized kinetic term 
\be
\Tr \left( D_\mu \vev{\Delta_R} \right)^\dag D^\mu \vev{\Delta_R} \, ,
\ee
which leads to 
\be
M^2_{W_R} = g_R v_R^2 \, , \qquad M^2_{Z_R} = 2 ( g_R^2 + g_{B-L}^2 ) v_R^2 \, , \qquad M^2_{Y} = 0 \, ,
\ee
where 
\be
W_R^{\pm} = \frac{A_R^1 \mp i A_R^2}{\sqrt{2}} \, , \qquad Z_{R} = \frac{g_{R} A_R^3 + g_{B-L} A_{B-L}}{\sqrt{g_R^2 + g_{B-L}^2}} 
 \, , \qquad Y = \frac{g_{B-L} A_R^3 - g_{R} A_{B-L}}{\sqrt{g_R^2 + g_{B-L}^2}} \, .
\ee
Given the relation $g_Y^{-2} = g_R^{-2} + g_{B-L}^{-2}$\footnote{This relation comes directly from $Y = T^3_R + \frac{B-L}{2}$ 
(cf.~\eq{Qformula}). For a formal proof see~\sect{sec:1Lmatching}.} 
and the $Z_2$ symmetry in~\eq{defZ2} which implies $g_R = g_L \equiv g$, we obtain 
\be
M^2_{Z_R} = \frac{2 g^2}{g^2 - g_Y^2} M^2_{W_R} \sim 2.6 \, M^2_{W_R} \, .
\ee
At the next stage of symmetry breaking ($\vev{\Phi} \neq 0$ and $\vev{\Delta_L} \neq 0$) an analogous calculation yields 
(in the approximation $v_2 = 0$)
\be
M^2_{W_L} = \frac{1}{2} g^2 \left( v_1^2 + 2 v_L^2 \right) \, , \qquad 
M^2_{Z_L} = \frac{1}{2} \left( g^2 + g_Y^2 \right) \left( v_1^2 + 4 v_L^2 \right) \, , \qquad
M^2_{A} = 0 \, ,
\ee 
where 
\be
W_L^{\pm} = \frac{A_L^1 \mp i A_L^2}{\sqrt{2}} \, , \qquad Z_{L} = \frac{g_{L} A_L^3 - g_{Y} A_{Y}}{\sqrt{g_L^2 + g_{Y}^2}} 
 \, , \qquad A = \frac{g_{Y} A_L^3 + g_{L} A_{Y}}{\sqrt{g_L^2 + g_{Y}^2}} \, .
\ee
Notice that in order to preserve $\rho = 1$ at tree level, where 
\be
\rho \equiv \frac{M^2_{W_L}}{M^2_{Z_L}} \frac{g^2 + g_Y^2}{g^2} \, , 
\ee
one has to require $v_L \ll v_1$.  

On the other hand at energy scales between $M_{W_L}$ and $M_{W_R}$, $SU(2)_L \otimes U(1)_Y$ is still preserved and \eq{Qformula} implies 
\be
\Delta T^3_R = - \frac{1}{2} \Delta (B-L) \, .
\ee
Since $\Delta_R$ is an $SU(2)_R$ triplet $\Delta T^3_R = 1$ and we get a violation of $B-L$ by two units. 
Then two classes of $B$ and $L$ violating processes can arise: 
\begin{itemize}
\item $\Delta B = 0$ and $\Delta L = 2$ which imply Majorana neutrinos. 
\item $\Delta B = 2$ and $\Delta L = 0$ which lead to neutron-antineutron oscillations.
\end{itemize}
Let us describe the origin of neutrino masses while postponing the discussion of  
neutron-antineutron oscillations to the next section.

The piece of lagrangian relevant for neutrinos is 
\be
\label{lagneutriniLR}
\mathcal{L}_\nu \supset 
Y_{\Phi} \ell^T C \epsilon_2 \Phi \ell^c + \tilde{Y}_{\Phi} \ell^T C \epsilon_2 \tilde{\Phi} \ell^c 
+ Y_{\Delta} \left( \ell^T C \epsilon_2 \Delta_L \ell + \ell^{cT} C \Delta_R^* \epsilon_2 \ell^c \right) + \text{h.c.} \, ,
\ee
The invariance of \eq{lagneutriniLR} under the $SU(2)_L \otimes SU(2)_R$ might not be obvious. So let us recall that, 
on top of the transformation properties in~\eq{transprop}, 
$\ell \rightarrow U_L\, \ell$, $\ell^c \rightarrow U_R\, \ell^c$, 
and $U_{L,R}^T\, \epsilon_2 = \epsilon_2\, U_{L,R}^\dag$. 
After projecting \eq{lagneutriniLR} on the SM vacuum directions and taking only the pieces relevant to neutrinos we get 
\be
\mathcal{L}_\nu \supset 
Y_{\Phi} \nu^T C \nu^c v_2
+ \tilde{Y}_{\Phi} \nu^T C \nu^c v_1 
+ Y_{\Delta} \left( \nu^T C \nu\, v_L + \nu^{cT} C \nu^c v_R^* \right) + \text{h.c.} \, .
\ee
Let us take for simplicity $v_2 = 0$ and consider real parameters. 
Then the neutrino mass matrix in the symmetric basis $(\nu\ \nu^c)$ reads 
\be
\label{neutrinomassmatrLR}
\left(
\begin{array}{cc}
Y_{\Delta} v_L & \tilde{Y}_{\Phi} v_1 \\
\tilde{Y}_{\Phi}^T v_1 & Y_{\Delta} v_R
\end{array}
\right) \, ,
\ee
and, given the hierarchy $v_R \gg v_1 \gg v_L$, the matrix in \eq{neutrinomassmatrLR} is block-diagonalized by a 
similarity transformation involving the orthogonal matrix 
\be
\label{orthogonalblock}
\left(
\begin{array}{cc}
1 - \frac{1}{2} \rho \rho^T & \rho \\
- \rho^T & 1 - \frac{1}{2} \rho^T \rho
\end{array}
\right) \, ,
\ee
where $\rho = \tilde{Y}_{\Phi} Y_{\Delta} ^{-1} v_1 / v_R$. The diagonalization is valid up to $\mathcal{O}(\rho^2)$ and yields
\be
m_\nu = Y_{\Delta} v_L - \tilde{Y}_{\Phi} Y_{\Delta}^{-1} \tilde{Y}_{\Phi}^T \frac{v_1^2}{v_R} \, . 
\ee
The two contributions go under the name of type-II and type-I seesaw respectively.
From the minimization of the potential\footnote{Even 
without performing the complete minimization we can estimate the induced VEV $v_L$ by looking at the following piece of potential
\be
\label{typeIIinduced}
V \supset - M^2_{\Delta_L} \Tr \Delta_L^\dag \Delta_L + \lambda\, \Tr \Delta_L^\dag \tilde{\Phi} \Delta_R \Phi^\dag \, .
\ee
On the SM-vacuum \eq{typeIIinduced} reads
\be
\vev{V} \supset - M^2_{\Delta_L} v_L^2 + \lambda\, v_L v_R |v_1|^2 \, ,
\ee
and from the extremizing condition with respect to $v_L$ we get 
\be
\label{vLinducedDL}
v_L = \lambda \frac{v_R |v_1|^2}{M^2_{\Delta_L}} \, .
\ee
}
(see \eq{asymmetricLRvac}) one gets $v_L = \gamma v_1^2 / v_R$ and hence the 
effective neutrino mass matrix reads
\be
\label{mnueffective}
m_\nu = \left( Y_{\Delta} \gamma - \tilde{Y}_{\Phi} Y_{\Delta}^{-1} \tilde{Y}_{\Phi}^T \right) \frac{v_1^2}{v_R} \, . 
\ee 
This equation is crucial since it shows a deep connection between the smallness of neutrino masses and the non-observation of $V+A$ 
currents~\cite{Mohapatra:1979ia,Mohapatra:1980yp}. 
Indeed in the limit $v_R \rightarrow \infty$ we recover the $V-A$ structure and $m_\nu$ vanish.  

Nowadays we know that neutrino are massive, but this information is not enough in order to fix the scale $v_R$ because the 
detailed Yukawa structures are unknown. 
In this respect one can adopt two complementary approaches. 
From a pure phenomenological point of view one can hope that 
the $V+A$ interactions are just behind the corner and experiments such us the 
LHC are probing right now the TeV 
region\footnote{It has been pointed out recently~\cite{Tello:2010am} that a low $\mathcal{O}(\text{TeV)}$ left-right symmetry 
scale could be welcome in view of a possible tension between neutrinoless double beta decay signals 
and the upper limit on the sum of neutrino masses coming from cosmology.}. 
Depending on the choice of the discrete left-right symmetry which can be either $Z_2^P$ or $Z_2^C$, 
the strongest bounds on $M_{W_R}$ are given by the $K_L - K_S$ mass difference which 
yields $M_{W_R} \gtrsim 4 \ \text{TeV}$ in the case of $Z_2^P$ and $M_{W_R} \gtrsim 2.5 \ \text{TeV}$ in the case of 
$Z_2^C$~\cite{Maiezza:2010ic,Zhang:2007da}. 

Alternatively one can imagine some well motivated UV completion 
in which the Yukawa structure of the neutrino mass matrix is correlated to that of the charged fermions. 
For instance in $SO(10)$ GUTs it usually not easy to disentangle the highest eigenvalue in \eq{mnueffective} from the top mass. This implies that the scale $v_R$ 
must be very heavy, somewhere close to $10^{14} \ \text{GeV}$. 
As we will see in Chapter~\ref{intermediatescales} 
this is compatible with unification constraints and strengthen the connection between $SO(10)$ and neutrino masses.

\subsection{Lepton number as a fourth color}
\label{leptonnumber4c}

One can go a little step further and imagine a partial unification scenario in which quarks and leptons belong to the same representations. 
The simplest implementation is obtained by collapsing the multiplets in \eq{LRspectrum} in the following way 
\be
\label{PSembedd}
Q = 
\left( 
\begin{array}{cccc}
u_1 & u_2 & u_3 & \nu \\
d_1 & d_2 & d_3 & e
\end{array}
\right)
\qquad 
Q^c = 
\left(
\begin{array}{cccc}
d_1^c & d_2^c & d_3^c & e^c \\
- u_1^c & - u_2^c & - u_3^c & - \nu^c
\end{array}
\right)
\ee  
so that $SU(3)_C \otimes U(1)_{B-L} \subset SU(4)_C$ and the fermion multiplets transform as 
$Q = (4,2,1)$ and $Q^c = (\overline{4},1,2^*)$ under $SU(4)_C \otimes SU(2)_L \otimes SU(2)_R$, 
which is known as the Pati-Salam group~\cite{Pati:1974yy}. 
Even in this case one can attach an extra discrete symmetry which exchange $SU(2)_L \leftrightarrow SU(2)_R$.

The Higgs sector of the model is essentially an extension of that of the left-right symmetric model presented in~\sect{leftrightsymmetry}.  
Indeed we have $\Delta_L = (\overline{10}, 3, 1)$, $\Delta_R = (\overline{10}, 1, 3)$ and $\Phi = (1,2,2^*)$. 
From the decomposition $10 = 6 (+2/3) \oplus 3 (-2/3) \oplus 1(-2)$ under $SU(4)_C \supset SU(3)_C \otimes U(1)_{B-L}$ 
and the expression for the electric charge operator in \eq{Qformula}, we can readily see that $\vev{\Delta_R}$ contains a SM-single direction 
and so the first stage of the breaking is given by
\be
\label{PSbreaking}
SU(4)_C \otimes SU(2)_L \otimes SU(2)_R \ \stackrel{\vev{\Delta_R}}{\longrightarrow}\
SU(3)_C \otimes SU(2)_L \otimes U(1)_Y
\, ,
\ee
while the final breaking to $SU(3)_C \otimes U(1)_Q$ is obtained by means of the bi-doublet VEV $\vev{\Phi}$. 
Analogously to the left-right symmetric case an electroweak triplet VEV $\vev{\Delta_L} \ll \vev{\Phi}$ is induced 
by the Higgs potential and the conclusions about neutrino masses are the same. 

A peculiar feature of the Pati-Salam model is that the proton is stable
in spite of the quark-lepton transitions due to the $SU(4)_C$ interactions. 
Let us consider first gauge interactions. The adjoint of $SU(4)_C$ decomposes as 
$15 = 1(0) \oplus 3(+4/3) \oplus \overline{3}(-4/3) \oplus 8(0)$ 
under $SU(3)_C \otimes U(1)_{B-L}$. In particular the transitions between 
quark and leptons due to $X_{PS} \equiv 3(+\tfrac{4}{3})$ and $\overline{X}_{PS} \equiv \overline{3}(-\tfrac{4}{3})$ 
come from the current interactions 
\be
\label{quarkleptint}
\mathcal{L}_{PS} \supset \frac{g}{\sqrt{2}} \left( X_\mu^{PS} \left[ \overline{u} \gamma^\mu \nu + \overline{d} \gamma^\mu e \right] 
+ \overline{X}_\mu^{PS} \left[ \overline{u^{c}} \gamma^\mu \nu^c + \overline{d^{c}} \gamma^\mu e^c \right] \right) + \text{h.c.}
\ee   
It turns out that \eq{quarkleptint} has an accidental global symmetry $G$, where 
$G(X_{PS})= -\tfrac{2}{3}$, $G(u)=G(d)=+\tfrac{1}{3}$, $G(\nu)=G(e)=+1$, 
$G(\overline{X}_{PS})=+\tfrac{2}{3}$, $G(u^c)=G(d^c)=-\tfrac{1}{3}$, $G(\nu^c)=G(e^c)=-1$. 
$G$ is nothing but $B+L$ when evaluated on the standard fermions. 
Thus, given that $B-L$ is also a (gauge) symmetry, we conclude 
that both $B$ and $L$ are conserved by the gauge interactions. 

The situation regarding the scalar interactions is more subtle. 
Actually in the minimal model there is an hidden discrete symmetry 
which forbids all the $\Delta B = 1$ transitions, like for instance $qqq\ell$ (see e.g.~Ref.~\cite{Mohapatra:1980qe}
)\footnote{Notice that this is just the reverse of the situation with the minimal $SU(5)$ model 
where $\Delta B = 2$ transitions are forbidden.}.  
A simple way to see it is that any operator of the type $qqq\ell \subset QQQQ$ and the $Q^4$ term must be contracted 
with an $\epsilon_{ijkl}$ tensor in order to form an $SU(4)_C$ singlet. However, since the Higgs fields in the minimal model are either 
singlets or completely symmetric in the $SU(4)_C$ space, they cannot mediate $Q^4$ operators.

On the other hand $\Delta B = 2$ transitions like neutron-antineutron oscillations are allowed and 
they proceed through $d=9$ operators of the type~\cite{Mohapatra:1980qe}
\be
\frac{\vev{\Delta_R}}{M^6_{\Delta_R}} (udd) (udd) \, ,
\ee
which are generated by the Pati-Salam breaking VEV $\vev{\Delta_R}$. The fact that $\vev{\Delta_R}$ can be 
pushed down relatively close to the TeV scale without making the proton to decay is phenomenologically interesting, 
since one can hope in testable neutron-antineutron oscillations (for a recent review see Ref.~\cite{Mohapatra:2009wp}).
Present bounds on nuclear instability give $\tau_{N}> 10^{32}$ yr, which
translates into a bound on the neutron oscillation time $\tau_{n-\bar n} > 10^8$ sec.
Analogous limits come from direct reactor oscillations experiments.
This sets a lower bound on the scale of $\Delta B = 2$ nonsupersymmetric ($d=9$) operators
that varies from 10 to 300 TeV depending on model couplings.
Thus neutron-antineutron
oscillations probe scales far below the unification scale. 

\subsection{One family unified}
\label{onefamuni}

The embedding of the left-right symmetric models of the previous sections into a grand unified structure requires the presence of a rank-5 group. 
Actually there are only two candidates which have complex representations and can contain the SM as a subgroup. 
These are $SU(6)$ and $SO(10)$. The former group even though smaller it is somehow 
redundant\footnote{$SU(6)$ as a grand unified group deserves anyway attention especially in its supersymmetric version. 
The reason is that it has an in-built 
mechanism in which the doublet-triplet splitting can be achieved in a very natural way~\cite{Berezhiani:1989bd,Berezhiani:1995dt}. 
The mechanism is based on the fact that the light Higgs doublets arise as pseudo-Goldstone modes of a spontaneously broken accidental global 
$SU(6) \otimes SU(6)$ symmetry of the Higgs superpotential.} 
since the SM fermions would be minimally embedded into 
$\overline{6}_F \oplus 15_F$ which under $SU(5) \otimes U(1)$ decompose as 
\be
\overline{6} = 1(+5) \oplus \overline{5}(-1) \qquad \text{and} \qquad 15 = 5 (-4) \oplus 10(+2) \, , 
\ee
yielding an exotic $5$ on top of the SM fermions.

Thus we are left with $SO(10)$. There are essentially two ways of looking at this unified theory, according to the 
two maximal subalgebras which contain the SM: $SU(5) \otimes U(1)$ and $SO(6) \otimes SO(4)$. 
The latter is locally isomorphic to $SU(4) \otimes SU(2) \otimes SU(2)$. 
The group theory of $SO(10)$ will be the subject of the next section, but let us already anticipate that the spinorial 
16-dimensional representation of $SO(10)$ decomposes in the following way 
$16 = 1(-5) \oplus \overline{5}(+3) \oplus 10(-1)$
under $SU(5) \otimes U(1)$ and 
$16 =  (4,2,1) \oplus (\overline{4},1,2)$ 
under $SU(4)_C \otimes SU(2)_L \otimes SU(2)_R$, 
thus providing a synthesis of both the ideas of Georgi-Glashow and the Pati-Salam. 


\section{$SO(10)$ group theory}

$SO(10)$ is the special orthogonal group of rotations in a 10-dimensional vector space. 
Its defining representation is given by the group of matrices $O$ which leave invariant the 
norm of a 10-dimentional real vector $\phi$. 
Under $O$,
$\phi \rightarrow O \phi$ 
and since $\phi^T \phi$ is invariant $O$ must be orthogonal,
$O O^T = 1$. 
Here special means $\det O = +1$ which selects the group of transformations continuously connected with the identity.
The matrices $O$ may be written in terms of $45$ imaginary generators $T_{ij} = - T_{ji}$, for $i,j = 1, \ldots 10$, as
\be
O = \exp{{\tfrac{1}{2} \epsilon_{ij} T_{ij}}} \, , 
\ee
where $\epsilon_{ij}$ are the parameters of the transformation. A convenient basis for the generators is 
\begin{equation}
\label{SO10genfund}
(T_{ij})_{ab}=-i(\delta_{a[i}\delta_{bj]}) \, ,
\end{equation}
where $a,b,i,j=1,..,10$ and the square bracket stands for anti-symmetrization.
They satisfy the $SO(10)$ commutation 
relations\footnote{These are an higher dimensional generalization of the well known 
$SO(3)$ commutation relations $[J_1, J_2] = i\, J_3$, where $J_1 \equiv T_{23}$, $J_2 \equiv T_{31}$ and $J_3 \equiv T_{12}$. 
Then the right hand side of \eq{SO10cr} takes just into account the antisymmetric nature of $T_{ij}$ and $T_{kl}$.
}
\be
\label{SO10cr}
\left[ T_{ij}, T_{kl} \right] = i ( \delta_{ik}T_{jl} + \delta_{jl}T_{ik} - \delta_{il}T_{jk} - \delta_{jk}T_{il}) \, .
\ee
In oder to study the group theory of $SO(10)$ it is crucial to identify the invariant tensors. 
The conditions $O O^T = 1$ and $\det O = +1$ give rise to two of them. 
The first one is simply the Kronecker tensor $\delta_{ij}$ which is easily proven to be invariant because of $O O^T = 1$, namely
\be
\delta_{ij} \rightarrow O_{ik} O_{jl} \delta_{kl} = O_{ik} O_{jk} = \delta_{ij} \, ,
\ee
while the second one is the 10-index Levi-Civita tensor $\epsilon_{ijklmnopqr}$. Indeed, from the definition of determinant
\be
\det O \, \epsilon_{i'j'k'l'm'n'o'p'q'r'} = O_{i' i} O_{j' j} O_{k' k} O_{l' l} O_{m' m} O_{n' n} O_{o' o} O_{p' p} O_{q' q} O_{r' r} \epsilon_{ijklmnopqr}
\ee
and the fact that $\det O = +1$, we conclude that $\epsilon_{ijklmnopqr}$ is also invariant. 

The irreducible representations of $SO(10)$ can be classified into two categories, single-valued and double-valued 
representations. The single valued representations have the same transformations properties as the ordinary vectors 
in the real $10$-dimensional space and their symmetrized or antisymmetrized tensor products. The doube-valued 
representations, called also spinor representations, trasform like spinors in a 10-dimentional coordinate space. 

\subsection{Tensor representations}

The general $n$-index irreducible representations of $SO(10)$ are built by means of the antisymmetrization 
or symmetrization (including trace subtraction) of the tensor product of $n$-fundamental vectors.
Starting from the 10-dimentional fundamental vector $\phi_i$, whose transformation rule is
\be
\phi_i \rightarrow O_{ij} \phi_j \, ,
\ee
we can decompose the tensor product of two of them in the following way
\be
\phi_i \otimes \phi_j = 
\begin{matrix} 
\underbrace{\frac{1}{2} \left( \phi_i \otimes \phi_j - \phi_j \otimes \phi_i\right)} \\ \phi^A_{ij} 
\end{matrix} 
\begin{matrix} 
+ \underbrace{\frac{1}{2} \left( \phi_i \otimes \phi_j + \phi_j \otimes \phi_i\right) - \frac{\delta_{ij}}{10} \phi_k \otimes \phi_k} \\ \phi^S_{ij}
\end{matrix} 
\begin{matrix} 
+ \underbrace{\frac{\delta_{ij}}{10} \phi_k \otimes \phi_k} \\ S \delta_{ij} 
\end{matrix} \, .
\ee
Since the symmetry properties of tensors under permutation of the indices are not changed by the group transformations, 
the antisymmetric tensor $\phi^A_{ij}$ and the symmetric tensor $\phi^S_{ij}$ clearly do not transform into each other. 
In general one can also separate a tensor in a traceless part and a trace. Because $O$ is orthogonal also 
the traceless property is preserved by the group transformations. 
So we conclude that $\phi^A_{ij}$, $\phi^S_{ij}$ and $S \delta_{ij}$ 
form irreducible representations whose dimensions are respectively $10 (10 - 1) / 2 = 45$, $10 (10 + 1) /2 -1= 54$ and $1$. 
One can continue 
in this way by considering higher order representations and separating each time the symmetric/antisymmetric pieces and subtracting traces. 

However something special happens for $5$-index tensors and the reason has to do with the existence of the invariant $\epsilon_{ijklmnopqr}$  
which induces the following duality map when applied to a $5$-index completely antisymmetric tensor $\phi_{nopqr}$ 
\be
\label{dualitymap}
\phi_{ijklm} \rightarrow \tilde{\phi}_{ijklm} \equiv - \frac{i}{5!} \epsilon_{ijklmnopqr} \phi_{nopqr} \, . 
\ee
This allows us to define the self-dual and the antiself-dual components of $\phi_{ijklm}$ in the following way
\begin{align}
\label{defsigma}
& \Sigma_{ijklm} \equiv \frac{1}{\sqrt{2}} \left( \phi_{ijklm} + \tilde{\phi}_{ijklm}  \right) \, , \\ 
\label{defsigmabar}
& \overline{\Sigma}_{ijklm} \equiv \frac{1}{\sqrt{2}} \left( \phi_{ijklm} - \tilde{\phi}_{ijklm}  \right) \, .  
\end{align}
One verifies that $\tilde{\Sigma}_{ijklm} = \Sigma_{ijklm}$ (self-dual) and $\tilde{\overline{\Sigma}}_{ijklm} = - \overline{\Sigma}_{ijklm}$ (antiself-dual). 
Since the duality property is not changed 
by the group transformations $\Sigma_{ijklm}$ and $\overline{\Sigma}_{ijklm}$ do form irreducible representations whose dimension is 
$\frac{1}{2}\frac{10!}{5! (10-5)!} = 126$.

\subsection{Spinor representations}
\label{spinreps}

We have defined the $SO(10)$ group by those linear transformations on the coordinates $x_1, x_2, \ldots , x_{10}$, such that the quadratic form 
$x_1^2 + x_2^2 + \ldots + x_{10}^2$ is left invariant. If we write this quadratic form as the square of a linear form of $x_i$'s, 
\be
x_1^2 + x_2^2 + \ldots + x_{10}^2 = ( \gamma_1 x_1 + \gamma_2 x_2 + \ldots + \gamma_{10} x_{10} )^2 \, ,
\ee
we have to require 
\be
\label{cliffalg}
\{ \gamma_i, \gamma_j \} = 2 \delta_{ij} \, .
\ee
\eq{cliffalg} goes under the name of Clifford algebra and the $\gamma$'s have to be matrices in order to anticommute with each 
other\footnote{In particular it can be shown that the dimension of the $\gamma$ matrices must be even. 
Indeed from~\eq{cliffalg} we obtain
\be
\gamma_j (\gamma_i \gamma_j + \gamma_j \gamma_i) = 2 \gamma_j 
\qquad \text{or} \qquad 
\gamma_j \gamma_i \gamma_j = \gamma_i \, ,
\ee
with no sum over $j$. Taking the trace we get
\be
\Tr \gamma_j \gamma_i \gamma_j = \Tr \gamma_i \, .
\label{Trgammas1}
\ee
But for the case $i \neq j$ this implies 
\be
\Tr \gamma_j \gamma_i \gamma_j = - \Tr \gamma_i \gamma_j \gamma_j = - \Tr \gamma_i \, ,
\label{Trgammas2}
\ee
and hence, putting together \eqs{Trgammas1}{Trgammas2}, we have $\Tr \gamma_i = 0$. 
On the other hand, $\gamma_i^2 = 1$ implies that the eigenvalues of $\gamma_i$ are either $+1$ or $-1$. 
This means that to get $\Tr \gamma_i = 0$, the number of $+1$ and $-1$ eigenvalues must be the same, i.e.~$\gamma_i$ 
must be even dimensional.}. 

For definiteness let us build an explicit representation of the $\gamma$'s which is valid for $SO(2N)$ 
groups~\cite{Wilczek:1981iz}\footnote{For an alternative approach to the construction of spinor representations 
by means of creation and annihilation operators see e.g.~Ref.~\cite{Mohapatra:1979nn}.}.
We start with $N=1$. Since the Pauli matrices satisfy the Clifford algebra 
\be
\left\{ \sigma_i, \sigma_j \right\} = 2 \delta_{ij} \, ,
\ee
we can choose 
\be
\label{defgamma1}
\gamma_1^{(1)} = \sigma_1 =
\left( 
\begin{array}{cc}
0 & 1 \\
1 & 0
\end{array}
\right)
\qquad \text{and} \qquad
\gamma_2^{(1)} = \sigma_2 =
\left( 
\begin{array}{cc}
0 & -i \\
i & 0
\end{array}
\right) \, .
\ee
Then the case $N > 1$ is constructed by recursion. 
The iteration from $N$ to $N+1$ is defined by
\be
\label{defgamma2}
\gamma_i^{(N+1)} = 
\left( 
\begin{array}{cc}
\gamma_i^{(N)} & 0 \\
0 & - \gamma_i^{(N)}
\end{array}
\right)
\qquad \text{for} \qquad 
i = 1,2,\ldots,2N \, ,
\ee
\be
\label{defgamma3}
\gamma_{2N+1}^{(N+1)} = 
\left( 
\begin{array}{cc}
0 & 1 \\
1 & 0
\end{array}
\right)
\qquad \text{and} \qquad
\gamma_{2N+2}^{(N+1)} =
\left( 
\begin{array}{cc}
0 & -i \\
i & 0
\end{array}
\right) \, .
\ee
Given the fact that the $\gamma_i^{(N)}$ matrices satisfy the Clifford algebra let 
us check explicitly that the $\gamma_i^{(N+1)}$ ones satisfy it as well,
\begin{align}
&& \left\{ \gamma_i^{(N+1)} , \gamma_j^{(N+1)} \right\} = 
\left(
\begin{array}{cc}
\left\{ \gamma_i^{(N)} , \gamma_j^{(N)} \right\} & 0 \\
0 & \left\{ \gamma_j^{(N)} , \gamma_i^{(N)} \right\}
\end{array}
\right) = 
\left(
\begin{array}{cc}
2 \delta_{ij} & 0 \\
0 & 2 \delta_{ij} 
\end{array}
\right) = 
2 \delta_{ij} \, , \\
&& \left\{ \gamma_i^{(N+1)} , \gamma_{2N+1}^{(N+1)} \right\} = 
\left(
\begin{array}{cc}
0 & \gamma_i^{(N)} \\
- \gamma_i^{(N)} & 0 
\end{array}
\right) +
\left(
\begin{array}{cc}
0 & - \gamma_i^{(N)} \\
\gamma_i^{(N)} & 0 
\end{array}
\right) = 0 \, , \\
&& \left( \gamma_{2N+1}^{(N+1)} \right)^2 = 1 \, . 
\end{align}
Analogously one finds
\be
\left\{ \gamma_i^{(N+1)} , \gamma_{2N+2}^{(N+1)} \right\} = 2 \delta_{ij} \, , 
\qquad 
\left\{ \gamma_{2N+1}^{(N+1)} , \gamma_{2N+2}^{(N+1)} \right\} = 0 \, , 
\qquad 
\left( \gamma_{2N+2}^{(N+1)} \right)^2 = 1 \, .
\ee
Now consider a rotation in the coordinate space, 
$x_i' = O_{ik} x_k$, 
where $O$ is an orthogonal matrix. This rotation induces a transformation on the $\gamma_i$ matrix
\be
\gamma_i' = O_{ik} \gamma_k \, .
\ee
Notice that the anticommutation relations remain unchanged, i.e. 
\be
\{ \gamma_i', \gamma_j' \} = O_{ik} O_{jl} \{ \gamma_k, \gamma_l \} = 2 \delta_{ij} \, .
\ee
Because the original set of $\gamma$ matrices form a complete matrix algebra, the new set of $\gamma$ matrices 
must be related to the original set by a similarity transformation, 
\be
\label{vecspinrel}
\gamma_i' = S(O) \gamma_i S^{-1} (O)
\qquad \text{or} \qquad 
O_{ik} \gamma_k = S(O) \gamma_i S^{-1} (O) \, .
\ee
The correspondence $O \rightarrow S(O)$ serves as a $2^N$-dimensional representation of the rotation group which is called spinor representation.
The quantities $\psi_i$, which transform like 
\be
\psi'_i = S(O)_{ij} \psi_j \, , 
\ee
are called spinors. 
For an infinitesimal rotation we can parametrize $O_{ik}$ and $S(O)$ by
\be
O_{ik} = \delta_{ik} + \epsilon_{ik} \quad \text{and} \quad S(O) = 1 + \tfrac{1}{2} i S_{ij} \epsilon_{ij} \, ,
\ee
with $\epsilon_{ik} = - \epsilon_{ki}$. 
Then \eq{vecspinrel} implies 
\be
\label{vecspinrel2}
i \left[ S_{kl}, \gamma_i \right] = \left( \gamma_l \delta_{ik} - \gamma_k \delta_{il} \right) \, , 
\ee
where we have used $\epsilon_{ik} \gamma_k = \epsilon_{lk} \gamma_k \delta_{il} = \tfrac{1}{2} \left( \gamma_k \delta_{il} - \gamma_k \delta_{jl} \right)$.
One can verify that a solution for $S_{kl}$ in \eq{vecspinrel2} is
\be
\label{SO10gen}
S_{kl} = \frac{i}{4} \left[ \gamma_k, \gamma_l \right] \, .
\ee
By expressing the parameter $\epsilon_{kl}$ in terms of rotations angle, one can 
see that $S(O(4\pi)) = 1$\footnote{This is easily 
seen for $SO(3)$. In this case the Clifford algebra is simply given by the three Pauli matrices and a finite transformation 
looks like
\be
S(O(\varphi)) = e^{\frac{i}{2} \sigma_i \varphi_i} = \cos{\frac{|\varphi|}{2}} + i \frac{\sigma_i \varphi_i}{|\varphi|} \sin{\frac{|\varphi|}{2}} \, ,
\ee
where we have defined $\epsilon_{23} \equiv - \varphi_1$, $\epsilon_{13} \equiv - \varphi_2$, $\epsilon_{12} \equiv - \varphi_3$ and 
$|\varphi| = \sqrt{\varphi_1^2 + \varphi_2^2 + \varphi_3^2}$.
}, i.e. $S(O)$ is a double-valued representation. 

However for $SO(2N)$ groups the representation $S(O)$ is not irreducible. 
To see this we construct the chiral projector $\gamma_\chi$ defined by 
\be
\label{chiralproj}
\gamma_\chi = (-i)^N \gamma_1 \gamma_2 \cdots \gamma_{2N} \, . 
\ee
$\gamma_\chi$ anticommutes with $\gamma_i$ since $2 N$ is even\footnote{Notice that this would not be the case for $SO(2N+1)$ groups.}
and consequently we get $\left[ \gamma_\chi , S_{kl} \right] = 0$ (cf.~\eq{SO10gen}).
Thus if $\psi$ transforms as $\psi'_i = S(O)_{ij} \psi_j$, the positive and negative chiral components 
\be
\psi^+ \equiv \frac{1}{2} \left( 1 + \gamma_\chi \right) \psi 
\qquad \text{and} \qquad
\psi^- \equiv \frac{1}{2} \left( 1 - \gamma_\chi \right) \psi 
\ee
transform separately. In other words $\psi^+$ and $\psi^-$ form two irreducible spinor representations 
of dimension $2^{N-1}$. 

Which is the relation between $\psi^+$ and $\psi^-$? In order to address this issue it is necessary to introduce the concept 
of conjugation. Let us consider a spinor $\psi$ of $SO(2N)$. The combination $\psi^T C \psi$ is an $SO(2N)$ invariant 
provided that 
\be
S_{ij}^T C = - C S_{ij} \, . 
\ee 
The conjugation matrix $C$ can be constructed iteratively. We start from $C^{(1)} = i \sigma_2$ for $N=1$ and define
\be
\label{Citer}
C^{(N+1)} = 
\left( 
\begin{array}{cc}
0 & C^{(N)} \\
(-)^{(N+1)} C^{(N)} & 0
\end{array}
\right) \, .
\ee
One can verify that 
\be
\label{gammaC}
(C^{(N)})^{-1} \gamma_i^T C^{(N)} = (-)^{N} \gamma_i  \, . 
\ee
By transposing~\eq{gammaC} and substituting back $\gamma_i^T$ we get 
\be
\left[ \gamma_i , ( ( C^{(N)})^{T})^{-1} C^{(N)} \right] = 0\, .
\ee
Then the Shur's Lemma implies
\be
(( C^{(N)})^{T})^{-1} C^{(N)} = \lambda \, I
\qquad \text{or} \qquad
C^{(N)} = \lambda ( C^{(N)})^{T} \, ,
\ee
which yields $\lambda^2 = 1$. In order to choose between $\lambda = +1$ and $\lambda = -1$ 
one has to apply \eq{Citer}, obtaining
\be
\label{Csymmantisymm}
C^T = (-)^{N(N+1)/2} C \, .
\ee
On the other hand~\eq{chiralproj} and~\eq{gammaC} lead to 
\be
(C^{(N)})^{-1} \gamma_\chi^T C^{(N)} = (-)^{N} \gamma_\chi  \, , 
\ee
which by exploiting $\gamma_\chi^T = \gamma_\chi$ (cf.~again~\eq{chiralproj}) yields
\be
\label{gammachiCrel}
(C^{(N)})^{-1} \gamma_\chi C^{(N)} = (-)^{N} \gamma_\chi  \, . 
\ee
This allows us to write 
\be
\label{conjrel}
(C^{(N)})^{-1} \left( S_{ij} (1+\gamma_\chi) \right)^* C^{(N)} =
(C^{(N)})^{-1} S^*_{ij} (1+\gamma_\chi) C^{(N)} = 
- S_{ij} \left( 1 + (-)^N \gamma_\chi \right) \, .
\ee
where we have also exploited the hermicity of the $\gamma$ matrices.
\eq{conjrel} can be interpreted in the following way: for $SO(2N)$ with $N$ even $\psi^+$ and $\psi^-$ are self-conjugate 
i.e.~real or pseudo-real depending on whether $C$ is symmetric or antisymmetric (cf.~\eq{Csymmantisymm}), 
while for $SO(2N)$ with $N$ odd $\psi^+$ is the conjugate of $\psi^-$. Thus only 
$SO(4k+2)$ can have complex representations and remarkably $SO(10)$ belong to this class. 

\subsubsection{Spinors will be spinors}

We close this section by pointing out a distinctive feature of spinorial representations: 
spinors of $SO(2N)$ decompose into the direct sum of spinors of $SO(2N') \subset SO(2N)$~\cite{Wilczek:1981iz}.
Indeed, since the construction of $\gamma_\chi$ in~\eq{chiralproj} is such that 
\be
\gamma_\chi^{(N+1)} = 
\left(
\begin{array}{cc}
\gamma_\chi^{(N)} & 0 \\
0 & - \gamma_\chi^{(N)}
\end{array}
\right) \, ,
\ee
the positive-chirality spinor $\psi^+$ of $SO(2N + 2M)$ contains $2^{M-1}$ positive-chirality spinors and 
$2^{M-1}$ negative-chirality spinors of $SO(2N)$. More explicitly 
\begin{multline}
\label{spinorswillbespinors}
\psi^+_{SO(2N+2M)} \rightarrow \psi^+_{SO(2N+2M-2)} \ \oplus \ \psi^-_{SO(2N+2M-2)} \\
\rightarrow 2 \times \psi^+_{SO(2N+2M-4)} \ \oplus \ 2 \times \psi^-_{SO(2N+2M-4)} \rightarrow \cdots \\
\rightarrow 2^{M-1} \times \psi^+_{SO(2N)} \ \oplus \ 2^{M-1} \times \psi^-_{SO(2N)} \, .
\end{multline} 
Let us exemplify this important concept in the case of the 16-dimensional positive-chirality spinor of $SO(10)$. 
By taking respectively $(N=3, M=2)$ and $(N=2, M=3)$ we obtain
\begin{itemize}
\item $16 = 2 \times 4^+ \oplus 2 \times 4^-$ under $SO(10) \supset SO(6)$,
\item $16 = 4 \times 2^+ \oplus 4 \times 2^-$ under $SO(10) \supset SO(4)$,
\end{itemize}
where $4^+$ ($4^-$) and $2^+$ ($2^-$) are respectively the positive (negative) chiral components of the $SO(6)$ and $SO(4)$ reducible spinors.
Thus under $SO(10) \supset SO(6) \otimes SO(4)$ the $16$ decomposes as
\be
16 = (4^+, 2^+) \oplus (4^-, 2^-) \, .
\ee
As we will show in~\sect{SMembedding} the Lie algebras $SO(6)$ and $SO(4)$ are locally isomorphic to $SU(4)$ and $SU(2) \otimes SU(2)$. 
This allows us to make the following identifications between the $SO(6)$ and $SU(4)$ representations
\be
4^+ \sim 4 \qquad 4^- \sim \overline{4} \, ,
\ee
and the $SO(4)$ and $SU(2) \otimes SU(2)$ ones
\be
2^+ \sim (2,1) \qquad 2^- \sim (1,2) \, ,
\ee
which justify the decomposition of the $SO(10)$ spinor under the Pati-Salam algebra $SU(4)_C \otimes SU(2)_L \otimes SU(2)_R$ 
as anticipated in~\sect{onefamuni}, namely 
\be
16 = (4,2,1) \oplus (\overline{4},1,2) \, .
\ee
This striking group-theoretic feature of spinors, which under the natural restriction to an orthogonal subgroup decompose 
into several copies of identical spinors of the subgroup, hints to a suggestive 
connection with the repetitive 
structure of the SM families~\cite{Wilczek:1981iz}  
and motivates the study of unification in higher orthogonal groups than 
$SO(10)$~\cite{GellMann:1980vs,Wilczek:1981iz,Senjanovic:1984rw,Bagger:1984rk}. 
To accommodate at least the three observed matter families we must use either $SO(16)$ or $SO(18)$. 
Following the decomposition in~\eq{spinorswillbespinors} we get 
\begin{itemize}
\item $SO(16)$: $\psi^+_{SO(16)} \rightarrow 4 \times \psi^+_{SO(10)} \oplus 4 \times \psi^-_{SO(10)}$ ,
\item $SO(18)$: $\psi^+_{SO(18)} \rightarrow 8 \times \psi^+_{SO(10)} \oplus 8 \times \psi^-_{SO(10)}$ .
\end{itemize} 
However there is a fundamental difference between the two cases above. 
According to the discussion below~\eq{conjrel} only $SO(4k+2)$ groups have complex spinor representations. 
This means that one can write a super-heavy bare mass term for $\psi^+_{SO(16)}$ and it is difficult to explain 
why it should be light. On the other hand no bare mass term can be written 
for $\psi^+_{SO(18)}$, making the last group a more natural choice. 

The obvious difficulty one encounters in this class of models is the overabundance of sequential or mirror families. 
If we decide to embed the SM fermions into three copies of $\psi^+_{SO(10)}$, the remaining families in 
$\psi^+_{SO(10)}$ are called sequential, while those in $\psi^-_{SO(10)}$ are 
mirror\footnote{Mirror fermions have the identical quantum numbers of ordinary fermions under the SM gauge group, except that they have 
opposite handedness. They imply parity restoration at high-energies as proposed long ago by Lee and Yang~\cite{Lee:1956qn}.} 
families. 

It has been pointed out recently~\cite{Martinez:2011ua} that the existence of three (mirror or sequential) 
families is still in accord with the SM, as long as an additional Higgs doublet is also present. 
This however is not enough to allow large orthogonal unification scenarios based on $SO(16)$ or $SO(18)$.

\subsection{Anomaly cancellation}

$SO(10)$ is an anomaly-free group. This important property can be understood from a simple group theoretical argument~\cite{Georgi:1972bb}. 
Let us consider the $SO(10)$ generators $T_{ij}$ in a given arbitrary representation. $T_{ij}$ transforms like an 
antisymmetric tensor in the indices $i$ and $j$. 
Then the anomaly, which is proportional to the invariant tensor 
\be
\Tr \{ T_{ij}, T_{kl} \} T_{mn} \, ,
\ee
must be a linear combination of a product of Kronecker $\delta$'s. 
Furthermore it must be antisymmetric under the exchanges 
$i \leftrightarrow j$, $k \leftrightarrow l$, $m \leftrightarrow n$ and symmetric under the exchange of pairs
$ij \leftrightarrow kl$, $kl \leftrightarrow mn$ and $ij \leftrightarrow mn$.
However the most general form consistent with the antisymmetry in $i \leftrightarrow j$, $k \leftrightarrow l$, $m \leftrightarrow n$ 
\be
\delta_{jk} \delta_{lm} \delta_{ni} - \delta_{ik} \delta_{lm} \delta_{nj} - \delta_{jl} \delta_{km} \delta_{ni} + \delta_{il} \delta_{km} \delta_{nj}   
- \delta_{jk} \delta_{ln} \delta_{mi} + \delta_{ik} \delta_{ln} \delta_{mj} + \delta_{jl} \delta_{kn} \delta_{mi} - \delta_{il} \delta_{kn} \delta_{mj}
\, , \nn
\ee  
is antisymmetric in $ij \leftrightarrow kl$ as well and so it must vanish. The proof fails for $SO(6)$ where the anomaly can be proportional to the invariant 
tensor $\epsilon_{ijklmn}$. Actually this is consistent with the fact that $SO(6)$ is isomorphic to $SU(4)$ which is clearly an anomalous group. 
On the other hand $SO(N)$ is safe for $N > 6$.

\subsection{The standard model embedding}
\label{SMembedding}

From the $SO(10)$ commutation relations in \eq{SO10cr} we find that a complete set of simultaneously commuting generators can be chosen as 
\be
T_{12}, \ T_{34}, \ T_{56}, \ T_{78}, \ T_{90} \, .
\ee
This is also known as the Cartan subalgebra and can be spanned over the left-right group Cartan 
generators 
\be
T^3_C, \ T^8_C, \ T^3_L, \ T^3_R, \ T_{B-L} \, .
\ee
Let us consider the $SO(4) \otimes SO(6)$ maximal subalgebra of $SO(10)$. 
We can span the $SO(4)$ generators over $T_{ij}$ with $i,j = 1, 2, 3, 4$ and the
$SO(6)$ generators over $T_{ij}$ with $i,j = 5, 6, 7, 8, 9, 0$. 
From the $SO(10)$ commutation relations in \eq{SO10cr} one can verify 
that these two sets commute (hence the direct product $SO(4) \otimes SO(6)$).

The next information we need is the notion of local isomorphism for the algebras $SO(4) \sim SU(2) \otimes SU(2)$ and $SO(6) \sim SU(4)$. 
In the $SO(4)$ case we define
\be
T^1_{L,R} \equiv \tfrac{1}{2} \left( T_{23} \pm T_{14} \right) \, , \qquad
T^2_{L,R} \equiv \tfrac{1}{2} \left( T_{31} \pm T_{24} \right) \, , \qquad
T^3_{L,R} \equiv \tfrac{1}{2} \left( T_{12} \pm T_{34} \right) \, ,
\ee
and check by an explicit calculation that 
\be
\left[ T^i_L, T^j_L \right] = i\, \epsilon^{ijk} T^k_L \, , \qquad 
\left[ T^i_R, T^j_R \right] = i\, \epsilon^{ijk} T^k_R \, , \qquad 
\left[ T^i_L, T^j_R \right] = 0 \, .
\ee
Thus $T^i_L$ and $T^i_R$ $(i=1,2,3)$ span respectively the $SU(2)_L$ and the $SU(2)_R$ algebra. 
On the other hand for the $SO(6)$ sector we define 
\be
\begin{array}{ccc}
T^1_C \equiv \frac{1}{2} \left( T_{89} + T_{70} \right) \, ,
& T^2_C \equiv \frac{1}{2} \left( T_{97} + T_{80} \right) \, ,
& T^3_C \equiv \frac{1}{2} \left( T_{09} + T_{87} \right) \, , \\ \\
T^4_C \equiv \frac{1}{2} \left( T_{96} + T_{05} \right) \, ,
& T^5_C \equiv \frac{1}{2} \left( T_{59} + T_{06} \right) \, ,
& T^6_C \equiv \frac{1}{2} \left( T_{67} + T_{85} \right) \, , \\ \\
T^7_C \equiv \frac{1}{2} \left( T_{75} + T_{86} \right) \, ,
& T^8_C \equiv \frac{1}{2\sqrt{3}} \left( 2 T_{65} + T_{78} + T_{09} \right) \, ,
& T^9_C \equiv \frac{1}{2} \left( T_{67} + T_{58} \right) \, , \\ \\
T^{10}_C \equiv \frac{1}{2} \left( T_{75} + T_{68} \right) \, ,
& T^{11}_C \equiv \frac{1}{2} \left( T_{69} + T_{05} \right) \, ,
& T^{12}_C \equiv \frac{1}{2} \left( T_{95} + T_{06} \right) \, , \\ \\
T^{13}_C \equiv \frac{1}{2} \left( T_{89} + T_{07} \right) \, ,
& T^{14}_C \equiv \frac{1}{2} \left( T_{97} + T_{08} \right) \, ,
& T^{15}_C \equiv \frac{1}{\sqrt{6}} \left( T_{65} + T_{87} + T_{90} \right) \, , \nn
\end{array}
\ee
and verify after a tedious calculation that 
\be
\left[ T^i_C , T^j_C \right] = i\, f^{ijk} T^k_C \, ,
\ee
where $f^{ijk}$ are the structure constants of $SU(4)$ (see e.g.~\cite{Lichtenberg:1978pc}). Thus $T^i_C$ $(i = 1, \ldots, 15)$ spans the $SU(4)_C$ algebra 
and, in particular, the $SU(3)_C$ subalgebra is spanned by $T^i_C$ $(i = 1, \ldots, 8)$ while $T^{15}_C$ can be identified with the (normalized) $T_{B-L}$ generator. 
Then the hypercharge and electric charge operators read respectively 
\be
Y = T_R^3 + \sqrt{\frac{2}{3}} T_{B-L} 
= \frac{1}{2} \left( T_{12} - T_{34} \right) + \frac{1}{3} \left( T_{65} + T_{87} + T_{90} \right)
\ee
and 
\be
Q = T_L^3 + Y 
= T_{12} + \frac{1}{3} \left( T_{65} + T_{87} + T_{90} \right) \, .
\ee

\subsection{The Higgs sector}
\label{SO10symmbreak}

As we have seen in the previous sections $SO(10)$ offers a powerful 
organizing principle for the SM matter content whose quantum numbers nicely fit in a 16-dimensional spinorial representation.
However there is an obvious prize to pay: the more one unifies the more one has to work in order to break the enhanced symmetry. 

The symmetry breaking sector can be regarded as the most arbitrary and challenging aspect of GUT models. 
The standard approach is based on the spontaneous symmetry breaking through elementary 
scalars. Though other ways to face the problem may be 
conceived\footnote{For an early attempt of dynamical symmetry breaking in $SO(10)$ see e.g.~\cite{Napoly:1984wt}.} 
the Higgs mechanism remains the most solid one in terms of computability and predictivity. 

\begin{figure*}[h]
\centering
\includegraphics[angle=0,width=16cm]{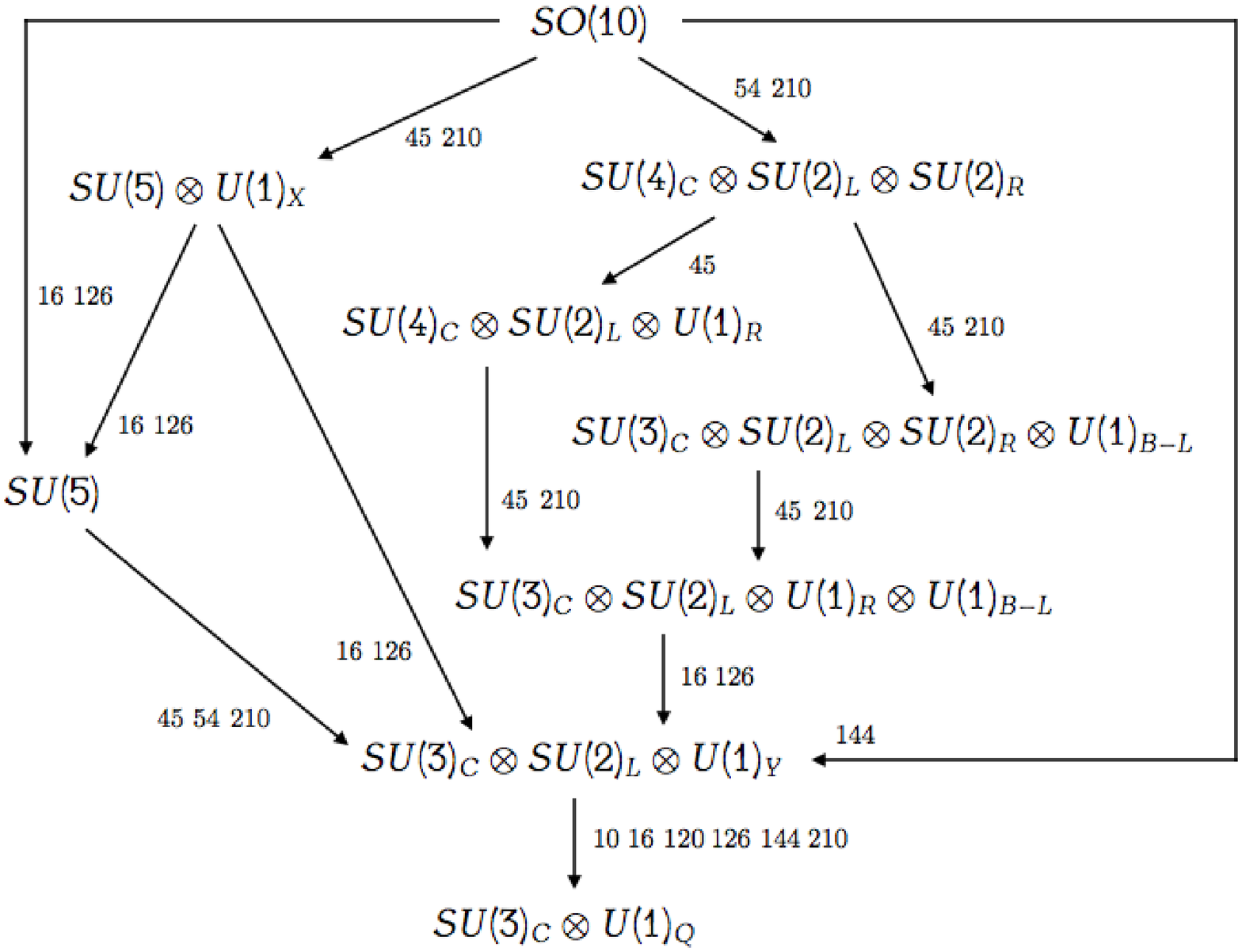}
\mycaption{\label{SO10break}
$SO(10)$ breaking chart with representations up to the $210$. 
$SU(5) \otimes U(1)_X$ can be understood either in the standard or in the flipped realization (cf.~the discussion in~\sect{sec:su5vsflippedsu5} ). 
In the former case $16$ or $126$ breaks it into $SU(5)$, while in the latter into $SU(3)_C \otimes SU(2)_L \otimes U(1)_Y$. 
For simplicity we are neglecting the distinctions due to the discrete left-right symmetry (cf.~\sect{sec:chains} for the discussion 
on the D-parity and~\Table{tab:chains} for an exhaustive account of the intermediate stages). 
}
\end{figure*}

The breaking chart in~\fig{SO10break} shows the possible symmetry stages between $SO(10)$ and $SU(3)_C \otimes U(1)_Q$
with the corresponding scalar representations responsible for the breaking.
That gives an idea of the complexity of the Higgs sector in $SO(10)$ GUTs. 

In view of such a degree of complexity, better we start by considering a minimal Higgs sector.   
Let us stress that the quest for the simplest Higgs sector is driven not only by aesthetic criteria but it is also a phenomenologically 
relevant issue related to tractability and predictivity of the models.  
Indeed, the details of the symmetry breaking pattern, sometimes overlooked in the phenomenological analysis, give further constraints on the low-energy observables such as the proton decay and the effective SM flavor structure.
For instance in order to assess quantitatively the constraints imposed by gauge coupling unification 
on the mass of the lepto-quarks 
resposible for proton decay 
it is crucial to have the scalar spectrum under control\footnote{Even in that case some degree of arbitrariness can still persist 
due to the fact that the spectrum can never be fixed completely 
but lives on a manifold defined by the vacuum conditions.}. 

From the breaking chart in~\fig{SO10break} we conclude that, 
before before considering any symmetry breaking dynamics,
the following representations are required by the group 
theory in order to achieve a full breaking of $SO(10)$ down to the SM:
\begin{itemize}
\item $16_H$ or $126_H$: they reduce the rank by one unit but leave an $SU(5)$ little group unbroken.
\item $45_H$ or $54_H$ or $210_H$: they admit for little groups different from $SU(5) \otimes U(1)$, yielding the SM when intersected with $SU(5)$.
\end{itemize}
It should be also mentioned that a one-step $SO(10) \rightarrow \text{SM}$ breaking can be achieved via only one $144_H$ irreducible Higgs representation \cite{Babu:2005gx}. 
However, such a setting requires an extended matter sector, including $45_F$ and $120_F$ multiplets, in order to accommodate realistic fermion masses \cite{Nath:2009nf}.

As we will see in the next Chapters the dynamics of the spontaneous symmetry breaking imposes further constraints on the 
viability of the options showed in~\fig{SO10break}. On top of that one has to take into account also other phenomenological 
constraints due to the unification pattern, the proton decay and the SM fermion spectrum. 

We can already anticipate at this level that while the choice between $16_H$ or $126_H$ is a model dependent issue related to the details of the 
Yukawa sector (see e.g.~\sect{fermionmasses}), 
the simplest option among $45_H$, $54_H$ and $210_H$ is certainly given by the adjoint $45_H$. 
However, since the early 80's, it has been observed that the vacuum dynamics aligns the adjoint along 
an $SU(5) \otimes U(1)$ direction, making the choice of $16_H$ (or $126_H$) and $45_H$ alone not phenomenologically viable. 
In the nonsupersymmetric case the alignment is only approximate~\cite{Yasue:1980fy,Yasue:1980qj,Anastaze:1983zk,Babu:1984mz},
but it is such to clash with unification constraints (cf.~Chapter~\ref{intermediatescales})  
which do not allow for any $SU(5)$-like intermediate stage, while in the supersymmetric limit 
the alignment is exact due to F-flatness~\cite{Buccella:1981ib,Babu:1994dc,Aulakh:2000sn}, thus never landing to a supersymmetric SM vacuum. 

The critical reexamination of these two longstanding no-go for the setting with the $45_H$ driving the GUT breaking 
will be the subject of Chapters~\ref{thequantumvac} and~\ref{flippedSO10}.

\section{Yukawa sector in renormalizable $SO(10)$}
\label{fermionmasses}

In order to study the $SO(10)$ Yukawa sector, we decompose 
the spinor bilinear 
\be
16 \otimes 16 = 10_S \oplus 120_A \oplus 126_S \, ,
\ee
where $S$ and $A$ denote the symmetric (S) and antisymmetric (A) nature of the bilinear couplings in the family space.
At the renormalizable level we have only three possibilities: $10_H$, $120_H$ and $\overline{126}_H$. Thus the most general $SO(10)$ Yukawa lagrangian is given by 
\be
\label{SO10Yuklag}
\mathcal{L}_Y = 16_F \left( Y_{10} 10_H + Y_{120} 120_H + Y_{126} \overline{126}_H \right) 16_F + \text{h.c.} \, ,
\ee
where $Y_{10}$ and $Y_{126}$ are complex symmetric matrices while $Y_{120}$ is complex 
antisymmetric\footnote{For completeness we report a concise proof of these statements based of the formalism 
used in~\sect{spinreps} and borrowed from Ref.~\cite{Wilczek:1981iz}. 
In a schematic notation we can write a Yukawa invariant term such as those in~\eq{SO10Yuklag} as 
\be
\label{formalSO10Yuk}
(\psi^T C_D C_5 \Gamma_k \psi) \, \Phi_k \, ,
\ee
where $\psi$ is both a Lorentz and an $SO(10)$ spinor (hence the need for $C_D$ and $C_5$ which are respectively 
the Dirac and the $SO(10)$ conjugation matrix). Then $\Gamma_k$ denotes an antisymmetric product of $k$ 
$\gamma$ matrices and $\Phi_k$ is a scalar field 
transforming like an antisymmetric tensor with $k$ indices under $SO(10)$. 
Using the facts that $\psi$ is an eigenstate of $\gamma_\chi$,
$\{\gamma_\chi , \gamma_i \} = 0$, 
$C_5 \gamma_\chi = - \gamma_\chi C_5$ (cf.~\eq{gammachiCrel}) and $\gamma_\chi^T = \gamma_\chi$, 
we deduce that $k$ must be odd (otherwise~\eq{formalSO10Yuk} is zero). This singles out the antisymmetric tensors $\Phi_k$
with $k=1,3,5$, corresponding respectively to dimensions $10$, $\frac{10!}{3! (10-3)!} = 120$ and 
$\frac{10!}{5! (10-5)!} = 252$ (actually the duality map defined in~\eq{dualitymap} is such that only half of these 252 components couples 
to the spinor bilinear). 

Next we consider the constraints imposed by the symmetry properties of the conjugation matrices, 
namely $C_D^T = - C_D$ and $C_5^T = - C_5$ (cf.~\eq{Csymmantisymm}). These yields
\be
\label{proofmanip1}
\psi^T C_D C_5 \Gamma_k \psi = - \psi^T C_D^T \Gamma_k^T C_5^T \psi = 
- \psi^T C_D C_5 (C_5^{-1} \Gamma_k^T C_5) \psi \, ,
\ee
where in the second step we have used the anti-commutation properties of the fermion fields. 
Then, by exploiting the relation $C_5^{-1} \gamma_i^T C_5 = - \gamma_i$ (cf.~\eq{gammaC}), we obtain 
\be
C_5^{-1} (\gamma_1 \cdots \gamma_k)^T C_5 = 
C_5^{-1} \gamma_k^T \cdots \gamma_1^T C_5 = 
(-)^k \gamma_k \cdots \gamma_1 = 
(-)^k (-)^{k(k-1)/2} \gamma_1 \cdots \gamma_k \, ,
\ee  
which plugged into~\eq{proofmanip1} implies
\be
\psi^T C_D C_5 \Gamma_k \psi = (-)^{k(k-1)/2+k+1} \psi^T C_D C_5 \Gamma_k \psi \, .
\ee
Hence for $k=1,3$ the invariant in~\eq{formalSO10Yuk} is symmetric in the flavor space of $\psi$, 
while for $k=2$ is antisymmetric.
}.

It should be mentioned that $10_H$ and $120_H$ are real representation from the $SO(10)$ point of 
view\footnote{This can be easily seen from the fact that the $SO(10)$ generators in the fundamental representation are both imaginary and antisymmetric 
(cf.~\eq{SO10genfund}). This implies $T_a = -T^\ast_a$ which corresponds to the definition of real representation in \eq{defrealpreal} with $S=1$.}. 
In spite of that the components of $10_H$ and $120_H$ can be chosen either real or complex. In the latter case we have 
$10_H \neq 10_H^\ast$ and $120_H \neq 120_H^\ast$, which means that the complex conjugate fields differ from the original ones by some extra charge. 
Actually both the components are allowed in the Yukawa lagrangian, since they transform in the same way under 
$SO(10)$\footnote{Alternatively one can imagine a complex $10$ as the linear combination of two real $10$'s, 
i.e.~$10 \equiv \tfrac{1}{\sqrt{2}} (10_1 + i 10_2)$. This should make clearer the origin of the new structures in \eq{SO10Yuklagcomplex}.}, 
and thus we have
\be
\label{SO10Yuklagcomplex}
\mathcal{L}_Y = 16_F \left( Y_{10} 10_H + \tilde{Y}_{10} 10_H^\ast + Y_{120} 120_H + \tilde{Y}_{120} 120_H^\ast + Y_{126} \overline{126}_H \right) 16_F + \text{h.c.} \, .
\ee
For instance complex scalars are a must in supersymmetry where the fundamental objects are chiral superfields 
made of Weyl fermions and complex scalars. However in supersymmetry we never see the couplings $\tilde{Y}_{10}$ 
and $\tilde{Y}_{120}$ because of the holomorphic properties of the superpotential. 
Even without supersymmetry there could be the phenomenological need, as we are going to see soon, of having 
either a complex $10_H$ or a complex $120_H$. In this case the new structures in \eq{SO10Yuklagcomplex} are 
still there, unless an extra symmetry which forbids them is imposed. 

In order to understand the implications of having a complex $10_H$, let us decompose it under the subgroup $SU(4)_C \otimes SU(2)_L \otimes SU(2)_R$
\be
10 = (1,2,2) \oplus (6,1,1) \, . 
\ee
In particular the bi-doublet can be further decomposed under $SU(3)_C \otimes SU(2)_L \otimes U(1)_Y$, 
yielding $(1,2,2) = (1,2,+\tfrac{1}{2}) \equiv H_u \oplus (1,2,-\tfrac{1}{2}) \equiv H_d$. 
Now if $10_H = 10_H^\ast$ we have $H_u^\ast = H_d$ as in the SM, 
while if $10_H \neq 10_H^\ast$ then $H_u^\ast \neq H_d$ as much as in the MSSM or in the two-higgs doublet model (2HDM).

To simplify a bit the discussion let us assume that 
we are either in the supersymmetric case or in the nonsupersymmetric one with an extra symmetry which forbids 
$\tilde{Y}_{10}$ and $\tilde{Y}_{120}$, so that~\eq{SO10Yuklag} applies with complex 
bi-doublets ($H_u^\ast \neq H_d$). 
The remaining representations in \eq{SO10Yuklag} decompose as 
\begin{align}
& 16 = (4,2,1) \oplus (\overline{4},1,2) \, , \\
& 120 = (1,2,2) \oplus (10,1,1) \oplus (\overline{10},1,1) \oplus (6,3,1) \oplus (6,1,3) \oplus (15,2,2) \, , \\
& \overline{126} = (6,1,1) \oplus (10,3,1) \oplus (\overline{10},1,3) \oplus (15,2,2) \, ,
\end{align}
under the Pati-Salam group
and thus 
the fields which can develop a SM-invariant VEV are $(10,3,1)$, $(\overline{10},1,3)$, $(1,2,2)$ and $(15,2,2)$. 
With the exception of the last one we already encountered these 
representations in the context of the Pati-Salam model (cf.~\sect{leptonnumber4c}). 
Let us also fix the following notation for the SM-invariant VEVs 
\be
v_L \equiv \vev{(\overline{10},3,1)_{126}} \, , \qquad 
v_R \equiv \vev{(10,1,3)_{126}} \, , 
\ee
\be
v_{10}^{u,d} \equiv \vev{(1,2,2)_{10}^{u,d}} \, , \qquad
v_{126}^{u,d} \equiv \vev{(15,2,2)_{126}^{u,d}} \, , 
\ee
\be
v_{{120}_1}^{u,d} \equiv \vev{(1,2,2)_{120}^{u,d}} \, , \qquad 
v_{{120}_{15}}^{u,d} \equiv \vev{(15,2,2)_{120}^{u,d}} \, .
\ee
Given the embedding of a SM fermion family into $(4,2,1) \oplus (\overline{4},1,2)$ (c.f.~\eq{PSembedd})
one finds the following fermion mass sum rule after the electroweak symmetry breaking
\begin{align}
\label{Mu3Yuk}
& M_u = Y_{10} v_{10}^u + Y_{126} v_{126}^u 
+ Y_{120} ( v_{120_1}^u + v_{120_{15}}^u ) \\
\label{Md3Yuk}
& M_d = Y_{10} v_{10}^d + Y_{126} v_{126}^d
+ Y_{120} ( v_{120_1}^d + v_{120_{15}}^d )  \\
\label{Me3Yuk}
& M_e = Y_{10} v_{10}^d -3 Y_{126} v_{126}^d
+ Y_{120} ( v_{120_1}^d -3 v_{120_{15}}^d )  \\
\label{MD3Yuk}
& M_D = Y_{10} v_{10}^u -3 Y_{126} v_{126}^u
+ Y_{120} ( v_{120_1}^u -3 v_{120_{15}}^u )  \\
\label{MR3Yuk}
& M_R = Y_{126} v_R \\
\label{ML3Yuk} 
& M_L = Y_{126} v_L 
\end{align}
where $M_D$, $M_R$ and $M_L$ enter the neutrino mass matrix defined on the symmetric basis $(\nu, \nu^c)$ 
\be
\label{seesawmatrix}
\left(
\begin{array}{cc}
M_L & M_D \\
M_D^T & M_R
\end{array}
\right) \, .
\ee 
\eqs{Mu3Yuk}{ML3Yuk} follow from the SM decomposition\footnote{For a formal proof see e.g.~\cite{Mohapatra:1979nn}.}, 
but it is maybe worth of a 
comment the $-3$ factor in front of $\vev{(15,2,2)}$ for the leptonic components $M_e$ and $M_D$. 
That is understood by looking at the Pati-Salam invariant 
\be
(4,2,1) \vev{(15,2,2)} (\overline{4},1,2) \, .
\ee
The adjoint of $SU(4)_C$ is a traceless hermitian matrix, 
so the requirement of an $SU(3)_C \otimes U(1)_Q$ preserving vacuum implies the following shape for $\vev{(15,2,2)}$
\be
\vev{(15,2,2)} \propto \text{diag} (1,1,1,-3) \otimes 
\left(
\begin{array}{cc}
0 & v_u \\
v_d & 0
\end{array} 
\right) \, ,
\ee
which leads to an extra $-3$ factor for leptons with respect to quarks. Conversely $\vev{(1,2,2)}$ 
preserves the symmetry between quarks and leptons.

In order to understand the implications of the sum-rule in~\eqs{Mu3Yuk}{ML3Yuk} 
it is useful to estimate the magnitude of the VEVs appearing there: $v_R$ is responsible for the rank reduction of $SO(10)$ 
and gauge unification constrains its value to be around (or just below) the unification scale $M_U$, 
then all the bi-doublets can develop a VEV (collectively denoted as $v$) which is at most of the order of the 
electroweak scale, 
while $v_L$ is a small $\mathcal{O}(M_W^2/M_U)$ VEV
induced by the scalar potential\footnote{In the contest of $SO(10)$ this was pointed out for the first time in Ref.~\cite{Lazarides:1980nt}.} 
in analogy to what happens in the left-right symmetric models (cf.~\sect{leftrightsymmetry}).

Thus, given the hierarchy $v_R \gg v \gg v_L$, \eq{seesawmatrix} can be block-diagonalized (cf.~\eq{orthogonalblock}) 
and the light neutrino mass matrix is very well approximated by
\be
\label{Mnu3Yuk}
M_\nu = M_L - M_D M_R^{-1} M_D^T \, ,
\ee
where the first and the second term are the type-II and type-I seesaw contributions already encountered in~\sect{leftrightsymmetry}.

Which is the minimum number of Higgs representations needed in the Yukawa sector in order to have a realistic theory?
With only one Higgs representation at play there is no fermion mixing, since one Yukawa matrix can be always diagonalized 
by rotating the $16_F$ fields, so at least two of them must be present. 
Out of the six combinations (see e.g.~\cite{Senjanovic:2006nc}): 
\begin{enumerate}
\item $10_H \oplus \overline{126}_H$
\item $120_H \oplus \overline{126}_H$
\item $10_H \oplus 120_H$
\item $10_H \oplus 10_H$
\item $120_H \oplus 120_H$ 
\item $\overline{126}_H \oplus \overline{126}_H$ 
\end{enumerate}
the last three can be readily discarded since they predict wrong mass relations, namely $M_d = M_e$ (case 4), $M_d = -3 M_e$ (case 6), while in case 5 the 
antisymmetry of $Y_{120}$ implies $m_1 = 0$ (first generation) and $m_2 = - m_3$ (second and third generation). 
Notice that in absence of $\overline{126}_H$ (case 3) neutrinos are Dirac and their mass is related to that of charged leptons 
which is clearly wrong. In order to cure this one has to introduce the bilinear $\overline{16}_H \overline{16}_H$ which plays effectively the role 
of $\overline{126}_H$ (cf.~\sect{neutrinostellus} for a discussion of this case in the context of the Witten mechanism~\cite{Witten:1979nr,Bajc:2004hr,Bajc:2005aq}). 
Though all the cases 1, 2 and 3 give rise to well defined Yukawa sectors, for definiteness we are going to analyze in more detail just the first one.  

\subsection{$10_H \oplus \overline{126}_H$ with supersymmetry}
\label{10p126barsusy} 
 
This case has been the most studied especially in the context of the minimal supersymmetric 
version, featuring $210_H \oplus 126_H \oplus \overline{126}_H \oplus 10_H$ in the Higgs sector~\cite{Clark:1982ai,Aulakh:1982sw,Aulakh:2003kg}.  
The effective mass sum-rule in~\eqs{Mu3Yuk}{ML3Yuk} can be rewritten in the following way
\begin{align}
M_u    &= Y_{10} v_u^{10} +  Y_{126} v_u^{126} \,,\nonumber\\
M_d    &= Y_{10} v_d^{10} +  Y_{126} v_d^{126} \,,\nonumber\\
M_e &= Y_{10} v_d^{10} -3 Y_{126} v_d^{126} \,, \label{eq:mass-relations}\\
M_D    &= Y_{10} v_u^{10} -3 Y_{126} v_u^{126} \,,\nonumber\\
M_R    &= Y_{126} v_R \,, \nonumber \\
M_L    &= Y_{126} v_L \,, \nonumber
\end{align}
and, exploiting the symmetry of $Y_{10}$ and $Y_{126}$, the neutrino mass matrix reads
\begin{equation}\label{eq:Mnubis}
M_\nu = M_L - M_D M_R^{-1} M_D \,.
\end{equation}
In the recent years this model received a lot of attention\footnote{For a set of references on the subject see~\cite{Fukuyama:2002ch,Goh:2003sy,Fukuyama:2003hn,Goh:2003hf,Fukuyama:2004xs,Bajc:2004fj,Dutta:2004wv,Goh:2004fy,Aulakh:2004hm,Fukuyama:2004ti,Aulakh:2005sq,Bertolini:2005qb,Babu:2005ia,Dutta:2005ni,Bajc:2005qe}.} due to the observation~\cite{Bajc:2002iw} that 
the dominance of type-II seesaw leads to a nice correlation 
between the large atmospheric mixing in the leptonic sector and the 
convergence of the bottom-quark and tau-lepton masses at the unification 
scale ($b-\tau$ unification) 
which is a phenomenon occurring in the MSSM up to $20-30\%$ corrections~\cite{Babu:1998er}. 

Another interesting prediction of the model is $\theta_{13} \sim 10^\circ$~\cite{Goh:2003sy}, 
in agreement with the recent data released by the T2K collaboration~\cite{Abe:2011sj}.

The correlation between $b-\tau$ unification and large atmospheric mixing can 
be understood with a simple two generations argument. 
Let us assume $M_\nu = M_L$ in \eq{eq:Mnubis}, then we get 
\be
\label{mnumdmme}
M_\nu \propto M_d - M_e \, .
\ee
In the the basis in which charged leptons are diagonal and 
for small down quark mixing $\epsilon$, \eq{mnumdmme} is approximated by 
\be
M_\nu \propto 
\left( 
\begin{array}{cc}
m_s - m_\mu & \epsilon \\
\epsilon & m_b - m_\tau 
\end{array}
\right) \, ,
\ee  
and, being the $22$ entry the largest one, maximal atmospheric mixing requires a cancellation between $m_b$ and $m_\tau$. 

For a more accurate analysis~\cite{Bertolini:2006pe} it is convenient to
express the $Y_{10}$ and $Y_{126}$ Yukawa matrices in terms of
$M_e$ and $M_d$, and substitute them in the expressions for
$M_u,M_D$ and $M_\nu$:
\begin{align}
M_u &= f_u \left[  (3+r)M_d +  (1-r) M_e \right] \,,
\label{eq:Mu}\\[1ex]
M_D &= f_u \left[ 3(1-r)M_d + (1+3r) M_e \right] \,,
\label{eq:MD}
\end{align}
where
\begin{equation}
f_u = \frac{1}{4} \, \frac{v_u^{10}}{v_d^{10}} \,,\qquad
r = \frac{v_d^{10}}{v_u^{10}} \, \frac{v_u^{126}}{v_d^{126}} \,.
\label{eq:fu-r}
\end{equation}
The neutrino mass matrix is obtained as
\begin{equation}
M_\nu = f_\nu \left[ (M_d - M_e) +
\xi \, \frac{M_D}{f_u}(M_d-M_e)^{-1}\frac{M_D}{f_u} \right] \,,
\label{eq:Mnu2}
\end{equation}
with
\begin{equation}
f_\nu = \frac{1}{4} \, \frac{v_L}{v_d^{126}} \,,\qquad
\xi = - \frac{\left(4 f_u v_d^{126}\right)^2}{v_L v_R} \,.
\label{eq:fnu-xi}
\end{equation}
%
In what follows we denote diagonal mass matrices by $\hat m_x$,
$x=u,d,e,\nu$, with eigenvalues corresponding to the
particle masses, i.e.~being real and positive.
We choose a basis where the down-quark matrix is diagonal: $M_d =
\hat m_d$. In this basis $M_e$ is a general complex symmetric
matrix, that can be written as $M_e = W_e^\dagger \hat m_e
W_e^*$, where $W_e$ is a general unitary matrix. 
Without loss of
generality $f_u$ and $f_\nu$ can be taken to be real and
positive. Hence, the independent parameters are given by 3 down-quark
masses, 3 charged lepton masses, 3 angles and 6 phases in $W_e$,
$f_u, f_\nu$, together with two complex parameters $r$ and $\xi$:
21 real parameters in total, among which 8 phases.
Using Eqs.~(\ref{eq:Mu}), (\ref{eq:MD}), and (\ref{eq:Mnu2}) all observables
(6 quark masses, 3 CKM angles, 1 CKM phase, 3 charged
lepton masses, 2 neutrino mass-squared differences, the mass of the
lightest neutrino, and 3 PMNS angles, 19 quantities altogether)
can be calculated in terms of these input parameters.

Since we work in a basis where the down-quark mass matrix is diagonal
the CKM matrix is given by the unitary matrix diagonalizing the
up-quark mass matrix up to diagonal phase matrices:
\begin{equation}
\label{hatmu}
\hat m_u = W_u M_u W_u^T
\end{equation}
with
\begin{equation}\label{eq:Vckm}
W_u = \mathrm{diag}(e^{i\beta_1},
e^{i\beta_2}, e^{i\beta_3}) \, V_\mathrm{CKM} \,
\mathrm{diag}(e^{i\alpha_1}, e^{i\alpha_2}, 1) \,,
\end{equation}
where $\alpha_i,\beta_i$ are unobservable phases at low energy. The neutrino mass
matrix given in Eq.~(\ref{eq:Mnu2}) is diagonalized by $\hat
m_\nu = W_\nu M_\nu W_\nu^T$, and the PMNS matrix is determined by
$W_e^* W_\nu^T = \hat D_1 V_\mathrm{PMNS} \hat D_2$, where $\hat D_1$
and $\hat D_2$ are diagonal phase matrices similar to those in~\eq{eq:Vckm}.

Allowing an arbitrary Higgs sector it is possible to obtain a good fit of the SM flavor structure~\cite{Bertolini:2006pe}. 
However, after including the constraints of the vacuum in the minimal supersymmetric version of the 
theory~\cite{Aulakh:2002zr,Fukuyama:2004ps,Bajc:2004xe}, one finds~\cite{Aulakh:2005mw,Bertolini:2006pe} 
an irreducible incompatibility between the fermion mass spectrum and the unification constraints. 
The reason can be traced back in the proximity between the unification scale and the seesaw scale, at odds with the lower bound 
on the neutrino mass scale implied by the oscillation phenomena. 

The proposed ways out consist in invoking a split supersymmetry spectrum~\cite{Bajc:2008dc} or resorting to a non-minimal 
Higgs sector~\cite{Bertolini:2004eq,Grimus:2006bb,Aulakh:2007ir,Aulakh:2011at}, 
but they hardly pair the appeal of the minimal setting.
In this respect it is interesting to notice that without supersymmetry gauge unification exhibits naturally the 
required splitting between the seesaw and the GUT scales. This is one of the motivations behind the study of 
the $10_H \oplus \overline{126}_H$ system in the absence of supersymmetry.

\subsection{$10_H \oplus \overline{126}_H$ without supersymmetry}

In the nonsupersymmetric case it would be natural to start with a real $10_H$. 
However, as pointed out in Ref.~\cite{Bajc:2005zf} (see also~\cite{Babu:1992ia} for an earlier reference), 
this option is not phenomenologically viable.  
The reason is that one predicts $m_t \sim m_b$, at list when working in the two heaviest generations limit with real parameter and 
in the sensible approximation $\theta_q = V_{cb} = 0$. It is 
instructive to reproduce this statement with the help of the parametrization given in~\sect{10p126barsusy}.

Let us start from \eq{eq:Mu} and apply $W_u$ (from the left) and $W_u^T$ (from the right).  
Then, taking into account \eq{hatmu} and the choice of basis $M_d =\hat m_d$, we get
\be
\label{WMWT}
\hat m_u = f_u \left[ (3+r) \hat{m}_d W_u W_u^T +  (1-r) W_u M_\ell W_u^T \right] \, .
\ee
Next we make the following approximations:
\begin{itemize}
\item $W_u W_u^T \sim 1$ (real approximation)
\item $W_u \sim V_\mathrm{CKM}$ (real approximation)
\item $V_\mathrm{CKM} \sim 1$ (for the 2nd and 3th generation and in the limit $V_{cb} \sim V_{ts} \sim A \lambda^2 \sim 0$)
\item $W_u M_\ell W_u^T \sim \hat m_\ell$ (for the self-consistency of~\eq{WMWT} in the limits above)
\end{itemize}
which lead to the system
\begin{align}
m_c \sim f_u \left[ (3+r) m_s +  (1-r) m_\mu \right] \,, \\
m_t \sim f_u \left[ (3+r) m_b +  (1-r) m_\tau \right] \,.
\end{align}
It is then a simple algebra to substitute back $r$ and find the relation
\be
\label{fuapprox}
f_u \sim \frac{1}{4} \frac{m_c (m_\tau -m_b) - m_t (m_\mu - m_s)}{m_s m_\tau - m_\mu m_b} \sim \frac{1}{4} \frac{m_t}{m_b} \, .
\ee
On the other hand a real $10_H$ predicts $|v^u_{10}| = |v^d_{10}|$ and hence from~\eq{eq:fu-r} $f_u = \frac{1}{4}$. 
More quantitatively, considering the nonsupersymmetric running for the fermion masses 
evaluated at $2 \times 10^{16}$~\cite{Xing:2007fb}, one gets $f_u \sim 22.4$, which is off by a factor of $\mathcal{O}(100)$.    

This brief excursus shows that the $10_H$ must be complex. 
In such a case the fermion mass sum-rule reads
\begin{align}
\label{nonsusysumrule}
M_u    &= Y_{10} v_u^{10} + \tilde{Y}_{10} v_d^{10*} +  Y_{126} v_u^{126} \,,\nonumber\\
M_d    &= Y_{10} v_d^{10} + \tilde{Y}_{10} v_u^{10*} +  Y_{126} v_d^{126} \,,\nonumber\\
M_e &= Y_{10} v_d^{10} + \tilde{Y}_{10} v_u^{10*} -3 Y_{126} v_d^{126} \,, \\
M_D    &= Y_{10} v_u^{10} + \tilde{Y}_{10} v_d^{10*} -3 Y_{126} v_u^{126} \,,\nonumber\\
M_R    &= Y_{126} v_R \,, \nonumber \\
M_L    &= Y_{126} v_L \,. \nonumber
\end{align}
The three different Yukawa sources would certainly weaken the predictive power of the model. So the proposal in Ref.~\cite{Bajc:2005zf} was to 
impose a Peccei-Quinn (PQ) symmetry~\cite{Peccei:1977hh,Peccei:1977ur} which forbids the coupling 
$\tilde{Y}_{10}$, thus mimicking a supersymmetric Yukawa sector (see also Ref.~\cite{Babu:1992ia}). 
The following charge assignment: $PQ(16_F) = \alpha$, $PQ(10_H) = - 2\alpha$ and $PQ(\overline{126}_H) = - 2\alpha$ would suffice. 

In this case $\vev{\overline{126}_H}$ is responsible both for $U(1)_{R} \otimes U(1)_{B-L} \rightarrow U(1)_Y$ 
and the $U(1)_{PQ}$ breaking. 
However, since it cannot break the rank of $SO(10) \otimes U(1)_{PQ}$ by two units, a global linear combination of 
$U(1)_{PQ} \otimes U(1)_{Y_\perp}$ (where $Y_\perp$ is the generator orthogonal to $Y$) survives at the electroweak scale.
This remnant global symmetry is subsequently broken by the VEV of the electroweak doublets, that is 
phenomenological unacceptable since it would give rise to a 
visible axion~\cite{Weinberg:1977ma,Wilczek:1977pj} which is experimentally excluded. 

Actually astrophysical and cosmological limits prefers the PQ breaking scale in the window $10^{9 \div 12} \ \text{GeV}$ (see e.g.~\cite{Raffelt:2006cw}). 
It is therefore intriguing to link the $B-L$ breaking scale responsible for neutrino masses and the PQ breaking one in the same model. 
This has been proposed long ago in~\cite{Mohapatra:1982tc} and advocated again in~\cite{Bajc:2005zf}. 
What is needed is another representation charged under the PQ symmetry 
in such a way that it is decoupled from the SM fermions and 
which breaks $U(1)_{PQ}$ completely at very high scales. 

In summary the PQ approach is very physical and well motivated since it does not just forbid a coupling 
in the Yukawa sector making it more "predictive", but correlates $SO(10)$ with other two relevant questions: it offers 
the axion as a dark matter candidate and it solves the strong CP problem 
predicting a zero $\overline{\theta}$\footnote{This is true as long as we ignore gravity~\cite{Kamionkowski:1992mf}.}. 

However one should neither discard pure minimal $SO(10)$ solutions with the SM as the effective low-energy theory.   
Notice that in the PQ case we are in the presence of a 2HDM which is more than what required by the extended 
survival hypothesis (cf.~the discussion in~\sect{DTsplit}) in order to set the gauge hierarchy. 
Indeed two different fine-tunings are needed in order to get two light 
doublets\footnote{The situation is different in supersymmetry where the minimal fine-tuning in the doublet sector makes both $H_u$ and $H_d$ light.}.  

Thus we could minimally consider the sum-rule in \eq{nonsusysumrule} with either $v_d^{10} = v_d^{126} = 0$ or $v_u^{10} = v_u^{126} = 0$. 
The first option leads to a clearly wrong conclusion, i.e. $M_d = M_e$. So we are left with the second one which implies 
\begin{align}
\label{nonsusysumrulevd}
M_u    &= \tilde{Y}_{10} v_d^{10*} \,,\nonumber\\
M_d    &= Y_{10} v_d^{10} +  Y_{126} v_d^{126} \,,\nonumber\\
M_e &= Y_{10} v_d^{10} -3 Y_{126} v_d^{126} \,, \\
M_D    &= \tilde{Y}_{10} v_d^{10*} \,,\nonumber\\
M_R    &= Y_{126} v_R \,, \nonumber \\
M_L    &= Y_{126} v_L \,, \nonumber 
\end{align}
and
\begin{equation}\label{eq:Mnunonsup}
M_\nu = M_L - M_D M_R^{-1} M_D \,.
\end{equation}
Notice that in the case of type-I seesaw the strong hierarchy due to $M_D = M_u$ must by undone by $M_R$ which remains proportional to $M_d - M_e$.  
More explicitly, in the case of type-I seesaw, one finds 
\be
\label{sumrule}
M_\nu = 4 M_u \left( M_d - M_e \right)^{-1} M_u \frac{v_d^{126}}{v_R} \, .
\ee
Though a simple two generations argument with real parameters shows that \eq{sumrule} could lead to an incompatibility with the data, 
a full preliminary three generations study indicates that this is not the case~\cite{Ketan}.

\subsection{Type-I vs type-II seesaw}
\label{typeIvstypeII}

Here we would like to comment about the interplay between type-I and type-II seesaw in \eq{Mnu3Yuk}. 
In a supersymmetric context one generally expects these two contributions to be comparable. 
As we have previously seen (see~\sect{10p126barsusy}) the dominance of type-II seesaw leads to a nice connection between the large atmospheric 
mixing and $b-\tau$ unification and one would like to keep this 
feature\footnote{See e.g.~Ref.~\cite{Melfo:2010gf} for a supersymmetric $SO(10)$ model in which the type-II seesaw dominance can be realized.}.
On the other hand without supersymmetry the $b-\tau$ convergence is far from being obtained. 
For instance the running within the SM yields $m_b = 1.00 \pm 0.04 \ \text{GeV}$ and $m_\tau = 1685.58 \pm 0.19 \ \text{MeV}$ 
at the scale $2 \times 10^{16} \ \text{GeV}$~\cite{Xing:2007fb}.
Thus in the nonsupersymmetric case the dominance of type-II seesaw would represent a serious issue. 

In this respect it is interesting to note that the type-II seesaw contribution can be naturally subdominant in nonsupersymmetric $SO(10)$.  
The reason has to do with the left-right asymmetrization of the scalar spectrum in the presence of intermediate symmetry breaking 
stages\footnote{For a similar phenomenon occurring in the context of left-right symmetric theories see Ref.~\cite{Chang:1985en}.}. 
Usually the unification pattern is such that the mass of the $SU(2)_R$ triplet $\Delta_R \subset \overline{126}_H$ responsible for the $B-L$ breaking 
is well below the GUT scale $M_U$. The reason is that $M_{\Delta_R}$ must be fine-tuned at the level of the intermediate scale VEV $v_R \equiv \vev{\Delta_R}$.
Then, unless there is a discrete left-right 
symmetry\footnote{As we will see in Chapter~\ref{intermediatescales} this can be the case if the $SO(10)$ symmetry breaking is due to either 
a $54_H$ or a $210_H$.} 
which locks $M_{\Delta_R} = M_{\Delta_L}$, the mass of the $SU(2)_L$ triplet $\Delta_L \subset \overline{126}_H$, 
remains automatically at $M_U$. 
On the other hand the induce VEV $v_L \equiv \vev{\Delta_L}$ is given by (cf.~e.g.~\eq{vLinducedDL})
\be
v_L = \lambda \frac{v_R}{M^2_{\Delta_L}} v^2 \, .
\ee
where $\lambda$ and $v$ denote a set of parameters of the scalar potential and an electroweak VEV respectively.  
So we can write
\be
\label{vRvL}
v_R \sim M_{B-L} \, , \qquad v_L \sim \lambda \left( \frac{M_{B-L}}{M_U} \right)^2 \left( \frac{v^2}{M_{B-L}} \right) \, ,
\ee 
which shows that type-II seesaw is suppressed by a factor $(M_{B-L} / M_U)^2$ with respect to type-I.

\section{Proton decay}

The contributions to the proton decay can be classified according to the dimension of the baryon violating 
operators appearing in the lagrangian. Since the external states are fermions and because 
of the color structure the proton decay operators arise first at the $d=6$ level. 
Sometimes the source of the baryon violation is hidden in a $d=5$ or a $d=4$ operator involving also scalar fields. 
These operators are successively dressed with the exchange of other states in order to get effectively the $d=6$ ones. 

The so-called $d=6$ gauge contribution is the most important in nonsupersymmetric GUTs. 
In particular if the mass of the lepto-quarks which mediate these operators is constrained by the running then the major uncertainty 
comes only from fermion mixing. There is also another class of $d=6$ operators coming from the Higgs sector 
but they are less important and more model dependent. 

The supersymmetrization of the scalar spectrum gives rise to $d=5$ and $d=4$ baryon and lepton number violating operators 
which usually lead to a strong enhancement of the proton decay amplitudes, though they are very model dependent. 

In the next subsections we will analyze in more detail just the gauge contribution while we will briefly pass through all the other ones. 
We refer the reader to the reviews~\cite{Langacker:1980js,Nath:2006ut,Senjanovic:2009kr} 
for a more accurate account of the subject.

\subsection{$d=6$ (gauge)}
\label{d6gauge}

Following the approach of Ref.~\cite{FileviezPerez:2004hn},
we start by listing all the possible $d=6$ baryon number violating operators due to the exchange of a vector boson 
and invariant under $SU(3)_C \otimes SU(2)_L \otimes U(1)_Y$~\cite{Weinberg:1979sa,Wilczek:1979hc,Sakai:1981pk}
\begin{align}
\label{O1}
& \textit{O}^{B-L}_I= k^2_1
\ \epsilon_{ijk} \ \epsilon_{\alpha \beta} 
\ \overline{u}_{i a}^c \ \gamma^{\mu} \ {q_{j \alpha a}}   \
\overline{e}_b^c \ \gamma_{\mu} \ {q_{k \beta b}} \, , \\ 
\label{O2}
& \textit{O}^{B-L}_{II}= k^2_1
\ \epsilon_{ijk} \ \epsilon_{\alpha \beta}
\ \overline{u}_{i a}^c \ \gamma^{\mu} \ {q_{j \alpha a}}   \
\overline{d}^c_{k b} \ \gamma_{\mu} \ {\ell_{\beta b}} \, ,\\
\label{O3}
& \textit{O}^{B-L}_{III}= k^2_2
\ \epsilon_{ijk} \ \epsilon_{\alpha \beta}
\ \overline{d}_{i a}^c \ \gamma^{\mu} \ {q_{j \beta a}}   \
\overline{u}_{k b}^c \ \gamma_{\mu} \ {\ell_{\alpha b}} \, , \\
\label{O4}
& \textit{O}^{B-L}_{IV}= k^2_2
\ \epsilon_{ijk} \ \epsilon_{\alpha \beta}
\ \overline{d}^c_{i a} \ \gamma^{\mu} \ {q_{j \beta a}}   \
\overline{\nu}_b^c \ \gamma_{\mu} \ {q_{k \alpha b}} \, .
\end{align}
In the above expressions $k_1= g_{U}/ {\sqrt 2} M_{X}$ 
and $k_2= g_{U}/ {\sqrt 2} M_{Y}$, where 
$M_{X}$, $M_{Y} \sim M_{U}$ 
and $g_{U}$ are the masses of the superheavy gauge bosons
and the gauge coupling at the unification scale. 
The indices $i, j, k = 1, 2, 3$ are referred to $SU(3)_C$, $\alpha, \beta = 1, 2$ to $SU(2)_L$ and 
$a$ and $b$ are family indices. The fields $q= ( u,d)$ and $\ell = ( \nu, e)$ are $SU(2)_L$ doublets.

The effective operators $\textit{O}^{B-L}_I$ and
$\textit{O}^{B-L}_{II}$
appear when we integrate out the superheavy gauge field 
$X=(3,2,-\tfrac{5}{6})$.
This is the case in theories based on the gauge group $SU(5)$.
Integrating out $Y=(3,2,+\tfrac{1}{6})$ we obtain the 
operators $\textit{O}^{B-L}_{III}$ and $\textit{O}^{B-L}_{IV}$. 
This is the case of flipped 
$SU(5)$ theories~\cite{DeRujula:1980qc,Barr:1981qv}, while 
in $SO(10)$ models both the lepto-quarks $X$ and $Y$ are present. 

Using the operators listed above, we can write the effective operators in the physical basis
for each decay channel~\cite{FileviezPerez:2004hn}
\begin{align}
\label{Oec}
& \textit{O}(e_{\alpha}^c, d_{\beta})= c(e^c_{\alpha},
d_{\beta}) \ \epsilon_{ijk} 
\ \overline{u}^c_i \ \gamma^{\mu} \ {u_j} \ \overline{e}^c_{\alpha} \
\gamma_{\mu} \ {d_{k \beta}} \, , \\
\label{Oe}
& \textit{O}(e_{\alpha}, d^c_{\beta})= c(e_{\alpha}, d^c_{\beta}) \ \epsilon_{ijk} \ 
\overline{u}^c_i \ \gamma^{\mu} \ {u_j} \ \overline{d}^c_{k \beta} \
\gamma_{\mu} \ {e_{\alpha}} \, , \\
\label{On}
& \textit{O}(\nu_l, d_{\alpha}, d^c_{\beta} )= c(\nu_l, d_{\alpha}, d^c_{\beta}) \
\epsilon_{ijk} \ \overline{u}^c_i \ \gamma^{\mu} \ {d_{j \alpha}}
\ \overline{d}^c_{k \beta} \ \gamma_{\mu} \ {\nu_l} \, , \\
\label{OnC}
& \textit{O}(\nu_l^c, d_{\alpha}, d^c_{\beta} )= 
c(\nu_l^c, d_{\alpha}, d^c_{\beta}) \
\epsilon_{ijk} \
\overline{d}_{i \beta}^c \ \gamma^{\mu} \ {u_j} \ \overline{\nu}_l^c \
\gamma_{\mu} \ {d_{k \alpha}} \, ,
\end{align}
where
\begin{align}
\label{cec}
& c(e^c_{\alpha}, d_{\beta})= k_1^2 [V^{11}_1 V^{\alpha \beta}_2 + ( V_1 V_{UD})^{1
\beta}( V_2 V^{\dagger}_{UD})^{\alpha 1}] \, , \\
\label{ce}
& c(e_{\alpha}, d_{\beta}^c) = k^2_1  \ V^{11}_1 V^{\beta \alpha}_3 +  \ k_2^2 \
(V_4 V^{\dagger}_{UD} )^{\beta 1} ( V_1 V_{UD} V_4^{\dagger} V_3)^{1 \alpha} \, , \\
\label{cnu} 
& c(\nu_l, d_{\alpha}, d^c_{\beta})= k_1^2 \ ( V_1 V_{UD} )^{1 \alpha}
( V_3 V_{EN})^{\beta l} + \ k_2^2 \ V_4^{\beta \alpha}( V_1 V_{UD}
V^{\dagger}_4 V_3 V_{EN})^{1l} \, , \\
\label{cnuc}
& c(\nu_l^c, d_{\alpha}, d^c_{\beta})= k_2^2 [( V_4 V^{\dagger}_{UD} )^{\beta
 1} ( U^{\dagger}_{EN} V_2)^{l \alpha }+ V^{\beta \alpha}_4
 (U^{\dagger}_{EN} V_2 V^{\dagger}_{UD})^{l1}] \, , 
\end{align}
with $\alpha = \beta \neq 2$. In the equations above we have defined the fermion mixing matrices as: 
$V_1= U_c^{\dagger} U$, $V_2=E_c^{\dagger}D$,
$V_3=D_c^{\dagger}E$, $V_4=D_c^{\dagger} D$, $V_{UD}=U^{\dagger}D$,
$V_{EN}=E^{\dagger}N$ and $U_{EN}= {E^C}^{\dagger} N^C$, 
where $U,D,E$ define the Yukawa coupling diagonalization so that 
\begin{align}
& U^T_c \ Y_U \ U = Y_U^{diag} \, , \\
& D^T_c \ Y_D \ D = Y_D^{diag} \, ,\\
& E^T_c \ Y_E \ E = Y_E^{diag} \, , \\
& N^T \ Y_N \ N = Y_N^{diag} \, . 
\end{align}    
Further, one may write $V_{UD}=U^{\dagger}D=K_1 V_{CKM} K_2$,
where $K_1$ and $K_2$ are diagonal matrices containing respectively three and two
phases. 
Similarly, $V_{EN}=K_3 V^D_{PMNS} K_4$ in the case
of Dirac neutrinos, or $V_{EN}=K_3 V^M_{PMNS}$ in the Majorana case.

From this brief excursus we can see that the 
theoretical predictions of the proton lifetime from the gauge $d=6$ operators 
require the knowledge of the quantities
$k_1$, $k_2$, $V^{1b}_1$, $V_2$, $V_3$, $V_4$ and $U_{EN}$. In addition
we have three (four) diagonal matrices containing phases in the case of Majorana (Dirac) neutrino.

Since the gauge $d=6$ operators conserve $B-L$ the nucleon decays into a meson and an antilepton.  
Let us write the decay rates for the different channels. 
We assume that in the proton decay experiments one can not 
distinguish the flavor of the neutrino and the 
chirality of charged leptons in the exit
channel. 
Using the Chiral Lagrangian techniques 
(see e.g.~\cite{Claudson:1981gh}), the decay rates of 
the different channels due to the gauge $d=6$ operators are~\cite{FileviezPerez:2004hn}
\begin{small}
\begin{multline}
\label{A1}
\Gamma(p \to K^+\overline{\nu}) \\
		= \frac{(m_p^2-m_K^2)^2}{8\pi m_p^3 f_{\pi}^2} A_L^2 
\left|\alpha\right|^2 
\sum_{i=1}^3 \left|\frac{2m_p}{3m_B}D \ c(\nu_i, d, s^c) 
+ [1+\frac{m_p}{3m_B}(D+3F)] c(\nu_i,s, d^c)\right|^2  \, ,
\end{multline}
\end{small}
\begin{small}
\begin{align}
\label{A2}
& \Gamma(p \to \pi^+ \overline{\nu})
		=\frac{m_p}{8\pi f_{\pi}^2}  A_L^2 \left|\alpha
		\right|^2 (1+D+F)^2 
\sum_{i=1}^3 \left| c(\nu_i, d, d^c) \right|^2 \, , \\
\label{A3}
& \Gamma(p \to \eta \, e_{\beta}^+) 
		= {(m_p^2-m_\eta^2)^2\over 48 \pi f_\pi^2 m_p^3}
A_L^2 \left|\alpha \right|^2 (1+D-3 F)^2 
\left\{ \left| c(e_{\beta},d^c)\right|^2 + \left|c(e^c_{\beta}, d)\right|^2 \right\} \, , \\
\label{A4}
& \Gamma (p \to K^0 e_{\beta}^+) 
		= {(m_p^2-m_K^2)^2\over 8 \pi f_\pi^2 m_p^3}  A_L^2
		\left|\alpha\right|^2 [1+{m_p\over m_B} (D-F)]^2
\left\{ \left|c(e_{\beta},s^c)\right|^2 +  \left|c(e^c_{\beta},s)\right|^2\right\} \, , \\
\label{A5}
& \Gamma(p \rightarrow \pi^0 e_{\beta}^+)
           = \frac{m_p}{16\pi f_{\pi}^2} A_L^2 \left|\alpha\right|^2
		(1+D+F)^2 \left\{ \left|c(e_{\beta},d^c)\right|^2 + 
		\left|c(e^c_{\beta},d)\right|^2 \right\} \, , 
\end{align}
\end{small}
where $\nu_i= \nu_e, \nu_{\mu}, \nu_{\tau}$ and $e_{\beta}= e, \mu$.
In the equations above $m_B \sim 1.15 \ \text{MeV}$ is the average baryon mass $m_B \sim m_\Sigma \sim m_\Lambda$, 
$f_\pi \sim 131 \ \text{MeV}$ is the pion decay constant, 
$D \sim 0.80$ and $F \sim 0.47$ are low-energy constants of the Chiral Lagrangian which can be obtained 
from the analysis of semileptonic hyperon decays~\cite{Cabibbo:2003cu}
and $\alpha \sim - 0.0112 \ \text{GeV}^3$ is a proton-to-vacuum matrix element parameter extracted via Lattice QCD techniques~\cite{Aoki:2008ku}.
Finally $A_L \sim 1.4$ takes into account the renormalization 
from $M_Z$ to $1$ GeV. 

In spite of the complexity and the model-dependency of the branching ratios in~\eqs{A1}{A5} 
the situation becomes much more constrained in the presence of symmetric Yukawas, 
relevant for realistic $SO(10)$ models based on $10_H \oplus \overline{126}_H$ 
in the Yukawa sector. 
In that case we get the following relations for the mixing matrices: 
$U_c = U K_u$, $D_c = D K_d$ and $E_c = E K_e$, where $K_u$, $K_d$ and $K_e$ 
are diagonal matrices involving phases.
These relations lead to the remarkable prediction~\cite{FileviezPerez:2004hn} 
\be
\label{testSO10}
k_1 = \frac{Q_1^{1/4}}{\left[ |A_1|^2 |V^{11}_{CKM}|^2 + |A_2|^2 |V^{12}_{CKM}|^2 \right]^{1/4}} \, ,
\ee
where 
\be
Q_1 = \frac{8 \pi m_p^3 f_\pi^2 \Gamma (p \rightarrow K^+ \overline{\nu})}{(m_p^2 - m_K^2)^2 A_L^2 |\alpha|^2} \, ,
\quad 
A_1 = \frac{2 m_p}{3 m_B} D \, ,
\quad 
A_2 = 1 + \frac{m_p}{3 m_B} (D + 3F) \, .
\ee
Notice that the expression for $k_1 = g_{U}/ {\sqrt 2} M_{X}$ is independent from unknown 
mixing matrices and CP violating phases, while the values of $g_{U}$ and $M_{X}$ are subject to 
gauge coupling unification constraints. This is a clear example of how to test a (nonsupersymmetric) $SO(10)$ model with 
$10_H \oplus \overline{126}_H$ in the Yukawa sector through the decay channel $\Gamma (p \rightarrow K^+ \overline{\nu})$.

\begin{figure*}[ht]
\centering
\includegraphics[width=8.5cm]{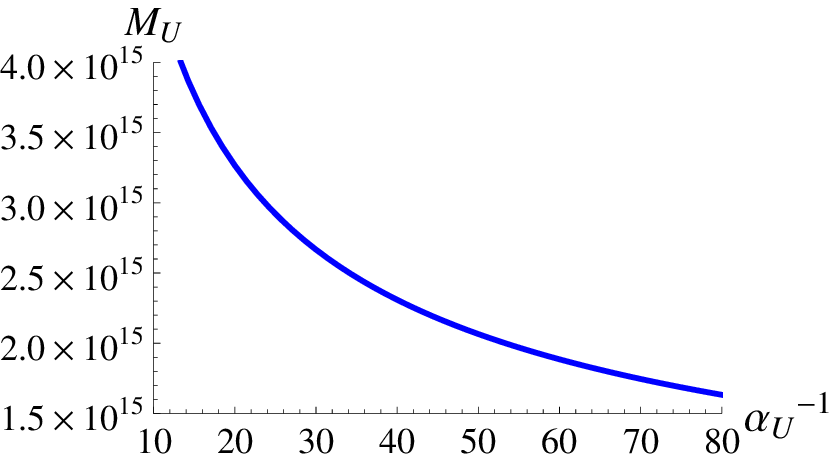}
\mycaption{\label{protonbound}
Naive lower bound on the superheavy gauge boson mass $M_U$ as a function of $\alpha^{-1}_{U}$. 
}
\end{figure*}

We close this subsection with a naive model-independent estimate for the 
mass of the superheavy gauge bosons $M_X \sim M_Y \sim M_U$.
Approximating the inverse lifetime of the proton in the following way (cf.~the real computation in \eqs{A1}{A5})
\begin{equation}
\label{appinvplt}
\Gamma_p \sim \alpha_{U}^2 \ \frac{m_p^5}{M_U^4}
\end{equation}     
and using $\tau (p \to \pi^0 e^+) \ > \ 8.2 \times 10^{33}$ yr~\cite{Nakamura:2010zzi},
one finds the naive lower bound
\begin{equation}
M_U \ > 2.3 \times 10^{15} \ \ \textrm{GeV} \, ,
\end{equation}     
where we fixed $\alpha^{-1}_{U}= 40$. 
The bound on $M_U$ as a function of $\alpha^{-1}_{U}$ is plotted in~\fig{protonbound}.

\subsection{$d=6$ (scalar)}

In nonsupersymmetric scenarios the next-to-leading contribution 
to the decay of the proton comes from the Higgs induced $d=6$ operators. 
In this case the proton decay is mediated by scalar leptoquarks $T=(3,1,-\tfrac{1}{3})$. 
For definiteness let us illustrate the case of minimal $SU(5)$ with just one scalar 
leptoquark. 
In this model the scalar 
leptoquark lives in the $5_H$ representation together 
with the SM Higgs. The relevant interactions for 
proton decay can be written in the following way~\cite{Nath:2006ut}
\begin{multline}
\label{pdecaytrip}
\mathcal{L}_T \ = \ \epsilon_{ijk} \ \epsilon_{\alpha \beta} \ q_{i \alpha}^T \ C \ \underline{A} \ q_{j \beta} \ T_k  \ 
+ \ {u^c_{i}}^T \ C \ \underline{B} \ e^c \ T_i \\
+ \ \epsilon_{\alpha \beta} \ q_{i \alpha}^T \ C \ \underline{C} \ \ell_{\beta} \ T_i^* 
\ + \ \epsilon_{ijk} \ {u^c_{i}}^T \ C \ \underline{D} \ d_{j}^c \ T_i^* \  + \ h.c. 
\end{multline}
In the above equation we have used the same notation as in the previous subsection.
The matrices $\underline{A}$, $\underline{B}$, $\underline{C}$ 
and $\underline{D}$ are linear combinations of the Yukawa 
couplings in the theory and the possible contributions 
coming from higher-dimensional operators. In the minimal $SU(5)$ 
we have the following relations: $\underline{A}=\underline{B}=Y_U$, 
and $\underline{C}=\underline{D}= Y_D=Y_E^T$.
Now, using the above interactions we can write the Higgs $d=6$ 
effective operators for proton decay~\cite{Nath:2006ut}
\begin{align}
& \textit{O}_H (d_{\alpha}, e_{\beta}) =  a(d_{\alpha}, e_{\beta}) \ u^T \ L \ C \ d_{\alpha} \ u^T \ L \ C e_{\beta} \, , \\
& \textit{O}_H (d_{\alpha}, e_{\beta}^c) =  a(d_{\alpha}, e_{\beta}^c)\ u^T \ L \ C \ d_{\alpha} \ 
{e^c_{\beta}}^{\dagger} \ L \ C {u^c}^*  \, , \\
& \textit{O}_H (d_{\alpha}^c, e_{\beta}) =  a(d_{\alpha}^c, e_{\beta}) \ {d^c_{\alpha}}^{\dagger} \ L \ C \ {u^c}^* 
\ u^T \ L \ C e_{\beta} \, , \\
& \textit{O}_H (d_{\alpha}^c, e_{\beta}^c) =  a(d_{\alpha}^c, e_{\beta}^c) \ {d^c_{\alpha}}^{\dagger} \ L \ C \ {u^c}^* 
\ {e^c_{\beta}}^{\dagger} \ L \ C^{-1} {u^c}^* \, , \\
& \textit{O}_H (d_{\alpha}, d_{\beta}, \nu_i)  =  a(d_{\alpha}, d_{\beta}, \nu_i) \ u^T \ L \ C \ d_{\alpha} \ d_{\beta}^T \ L \ C\ \nu_i \, , \\
& \textit{O}_H (d_{\alpha}, d_{\beta}^c, \nu_i)  =  a(d_{\alpha}, d_{\beta}^c, \nu_i) \ {d^c_{\beta}}^{\dagger} \ L \ C \ {u^c}^* 
\ d_{\alpha}^T \ L \ C^{-1}\ \nu_i \, ,  
\end{align}
where
\begin{align}
& a(d_{\alpha}, e_{\beta}) = \frac{1}{M_T^2} \ (U^T ( \underline{A} + \underline{A}^T) D)_{1 \alpha} 
\ (U^T \underline{C} E)_{1 \beta} \, , \\
& a(d_{\alpha}, e_{\beta}^c) = \frac{1}{M_T^2} \ (U^T (\underline{A}+ \underline{A}^T) D)_{1 \alpha} 
\ (E_c^{\dagger} \underline{B}^{\dagger} U_c^*)_{\beta 1} \, , \\
& a(d_{\alpha}^c, e_{\beta}) = \frac{1}{M_T^2} \ (D_c^{\dagger} \underline{D}^{\dagger} U_c^*)_{\alpha 1} 
\ (U^T \underline{C} E)_{1 \beta} \, , \\
& a(d_{\alpha}^c, e_{\beta}^c) = \frac{1}{M_T^2} \ (D_c^{\dagger} \underline{D}^{\dagger} U_c^*)_{\alpha 1} 
\ (E_c^{\dagger} \underline{B}^{\dagger} U_c^*)_{\beta 1} \, , \\
& a(d_{\alpha}, d_{\beta}, \nu_i)= \frac{1}{M_T^2} (U^T ( \underline{A} + \underline{A}^T) D)_{1 \alpha} 
\ (D^T \underline{C} N)_{\beta i} \, , \\
& a (d_{\alpha}, d_{\beta}^c, \nu_i) =\frac{1}{M_T^2} \ (D_c^{\dagger} \underline{D}^{\dagger} U_c^*)_{\beta 1} \ 
(D^T \underline{C} N)_{\alpha i} \, .
\end{align}
Here $L=(1- \gamma_5)/2$, $M_T$ is the triplet mass, $\alpha=\beta=1,2$ are $SU(2)_L$ indices and $i=1,2,3$ are $SU(3)_C$ indices. 

The above analysis exhibits that the Higgs $d=6$ contributions are quite model dependent, and because of this 
it is possible to suppress them in specific models of fermion masses. For instance, we can 
set to zero these contributions by the  constraints  
$\underline{A}_{ij}=-\underline{A}_{ji}$ and $\underline{D}_{ij}=0$, 
except for $i=j=3$. 

Also in this case 
we can make 
a naive model-independent estimation for the mass of the 
scalar leptoquark using the experimental lower 
bound on the proton lifetime. 
Approximating the inverse lifetime of the proton in the following way
\begin{equation}
\Gamma_p \sim |Y_u Y_d|^2 \ \frac{m_p^5}{M_T^4}
\end{equation}     
and taking $\tau (p \to \pi^0 e^+) \ > \ 8.2 \times 10^{33}$ yr~\cite{Nakamura:2010zzi},
we find the naive lower bound
\begin{equation}
M_T \ > \ 4.5 \times 10^{11} \ \ \textrm{GeV} \, .
\end{equation}     
This bound tells us that the 
triplet Higgs has to be heavy, unless some special condition on the matrices in~\eq{pdecaytrip} 
is fulfilled (see e.g.~\cite{Dvali:1992hc,Dvali:1995hp}). 
Therefore since the 
triplet Higgs lives with the SM Higgs in the same multiplet 
we have to look for a doublet-triplet splitting 
mechanism\footnote{Cf.~\sect{DTsplittingflipped} for a short overview of the mechanisms proposed so far.}.

\subsection{$d=5$}

In the presence of supersymmetry new $d=5$ operators of the type
\be
\frac{1}{M_T} q \, q \, \tilde{q} \, \tilde{\ell} \qquad \text{and} \qquad \frac{1}{M_T} u^c \, u^c \, \tilde{d}^c \, \tilde{e}^c  
\ee
are generated via colored triplet Higgsino exchange with mass $M_T$~\cite{Sakai:1981pk,Weinberg:1981wj}. These operators can be subsequently 
dressed at one-loop with an electroweak gaugino (gluino or wino) or higgsino leading to the standard $q q q \ell$ and $u^c u^c d^c e^c$ operators. 
Since the amplitude turns out to be suppressed just by the product $M_T \tilde{m}$, where $\tilde{m}$ is the soft scale, 
this implies a generic enhancement of the proton decay rate with respect to the 
ordinary $d=6$ operators. 

Another peculiarity of $d=5$ operators is that the dominant decay mode is $p \rightarrow K^+ \overline{\nu}_\mu$
which differs from the standard nonsupersymmetric mode $p \rightarrow \pi^0 e^+$. 
A simple symmetry argument shows the reason: the operators $\hat{q}_i \hat{q}_j \hat{q}_k \hat{\ell}_l$ 
and $\hat{u}^c_i \hat{u}^c_j \hat{d}^c_k \hat{e}^c_l$ (where $i,j,k,l=1,2,3$ are family indices and color and weak indices are implicit) 
must be invariant under $SU(3)_C$ and $SU(2)_L$. This means that their color and weak indices must be antisymmetrized. 
However since this operators are given by bosonic superfields, they must be totally symmetric under interchange of all indices. 
Thus the first operator vanishes for $i=j=k$ and the second vanishes for $i=j$. 
Hence a second or third generation member must exist in the final state.  

In minimal supersymmetric $SU(5)$~\cite{Dimopoulos:1981zb} the coefficient of the baryon number violating operator 
$q q q \ell$ can be schematically written as (see e.g.~\cite{Bajclectures})
\be
\label{qqqlceoff}
\frac{\alpha_3}{4 \pi} \frac{Y_{10} Y_5}{M_T} \frac{m_{\tilde{g}}}{m^2_{\tilde{q}}} \, ,
\ee 
where we have assumed the dominance of the gluino exchange and that the sfermion masses ($m_{\tilde{q}}$) are bigger than the gluino one ($m_{\tilde{g}}$), 
while $Y_{10}$ and $Y_5$ are couplings of the Yukawa superpotential. 
Though there could be a huge enhancement of the proton decay rate which brought to the claim that 
minimal supersymmetric $SU(5)$ was ruled out~\cite{Hisano:1992jj,Murayama:2001ur}, 
a closer look to the uncertainties at play makes this claim much more weaker~\cite{Bajc:2002pg}: 
\begin{itemize}
\item The Yukawa couplings in \eq{qqqlceoff} are not directly related to those of the SM, since in minimal $SU(5)$ one needs to 
take into account non-renormalizable operators in order to break the relation $M_d = M_e^T$, and thus they can conspire to suppress the decay mode~\cite{EmmanuelCosta:2003pu}. 
\item A similar suppression could also originate from the soft sector even after including the constraints coming from flavor violating effects~\cite{Bajc:2002bv}.  
\item Last but not least the mass of the triplet $M_T$ is constrained by the running only in the renormalizable version of the theory \cite{Murayama:2001ur}. As soon 
as non-renormalizable operators (which are anyway needed for fermion mass relations) are included this is not true anymore~\cite{Bajc:2002pg}. 
In this respect it is remarkable that even in the worse case scenario of the renormalizable theory the recent accurate three-loop 
analysis in Ref.~\cite{Martens:2010nm}
increases by about one order of magnitude the upper bound on $M_T$ due to the running constraints.     
\end{itemize}
Thus the bottom-line is that minimal supsersymmetric $SU(5)$ is still a viable theory and more input on the experimental side is needed in order to say 
something accurate on proton decay.  

\subsection{$d=4$}

This last class of operators originates from the $R$-parity violating superpotential of the MSSM
\be
W_{RPV} = \mu_i\, \hat{\ell}_i \hat{h}_u + \lambda_{ijk}\, \hat{\ell}_i \hat{\ell}_i \hat{e}^c_i 
+ \lambda^{'}_{ijk}\, \hat{q}_i \hat{\ell}_j \hat{d}^c_k + \lambda^{''}_{ijk}\, \hat{u}^c_i \hat{d}^c_j \hat{d}^c_k \, .
\ee 
Notice that $\lambda^{''}$ violates baryon number while $\mu$, $\lambda$ and $\lambda'$ violate lepton number. So for instance we have the 
following interactions in the $R$-parity violating lagrangian 
\be
\mathcal{L}_{RPV} \supset \lambda^{'}_{ijk}\, q_i \ell_j \tilde{d^c_k} + \lambda^{''}_{ijk}\, u^c_i d^c_j \tilde{d^c_k} + \text{h.c.} \, .
\ee 
The tree-level exchange of $\tilde{d^c}$ generates the baryon violating operator $q \ell u^{c\dag} d^{c\dag}$ with a coefficient which can be written schematically as 
\be
\lambda^{'} \lambda^{''} / m^2_{\tilde{d^c}} \, . 
\ee
Barring cancellations in the family structure of this coefficient and assuming a TeV scale soft spectrum, 
the proton lifetime implies the generic bound~\cite{Smirnov:1996bg}
\be
\label{lambdasbound}
\lambda^{'} \lambda^{''} \lesssim 10^{-26} \, .
\ee
It's easy to see that the $R$-parity violating operators are generated in supersymmetric GUTs unless special conditions are fulfilled. 
For instance in $SU(5)$ the effective trilinear couplings originate from the operator 
\be
\Lambda_{ijk}\, \hat{\overline{5}}_i \hat{\overline{5}}_j \hat{10}_k \, , 
\ee 
which leads to $\lambda = \frac{1}{2} \lambda^{'} = \lambda^{''} = \Lambda$.
Analogously in $SO(10)$ the $R$-parity violating trilinears stem from the operator
\be
\label{RPVSO10}
\frac{\Lambda_{ijk}}{M_P} \hat{16}_i \hat{16}_j \hat{16}_k \vev{\hat{16}_H} \, .
\ee
If one doesn't like small numbers such as in \eq{lambdasbound} the standard approach is to impose a $Z_2$ matter parity which forbids 
the baryon and lepton number violating operators~\cite{Dimopoulos:1981zb}. 
A more physical option in $SO(10)$ is instead suggested by \eq{RPVSO10}. Actually it seems that as soon as $SO(10)$ is preserved the $R$-parity violating trilinears 
are not generated. In order to better understand this point let us rephrase the $R$-parity in the following language~\cite{Martin:1992mq}
\be
R_P = (-)^{3(B-L)+2S} \, ,
\ee 
where the spin quantum number $S$ is irrelevant as long as the Lorentz symmetry is preserved. 
Then, since $B-L$ is a local generator of $SO(10)$, it is enough to embed the SM fermions in 
representations with odd $B-L$ (e.g.~$16, \ldots$) and the Higgs doublets in representations with even $B-L$ (e.g.~$10, 120, 126, 210, \ldots$)
in order to ensure exact $R$-parity conservation.  
After the $SO(10)$ breaking the fate of $R$-parity depends on the order parameter responsible for the $B-L$ breaking. 
Employing either a $16_H$ or a $126_H$ for the rank reduction of $SO(10)$ the action of the $R_P$ operator on their VEV is
\be
R_P \vev{16_H} = - \vev{16_H} \qquad \text{or} \qquad R_P \vev{126_H} = \vev{126_H} \, . 
\ee
In the latter case the $R$-parity is preserved by the vacuum and becomes an exact symmetry of the MSSM. 
This feature makes supersymmetric $SO(10)$ models with $126_H$ very appealing~\cite{Aulakh:2000sn}. 
On the other hand with a $16_H$ at play the amount of R-parity 
violation is dynamically controlled by the parameter $M_{B-L}/M_P$, where $\vev{16_H} \sim M_{B-L}$. Though conceptually 
interesting it is fair to say that in $SO(10)$ it is unnatural to have the $B-L$ breaking scale 
much below the unification scale both from the point of view of unification constraints and 
neutrino masses\footnote{As we will see in Chapter~\ref{intermediatescales} when the GUT breaking 
is driven either by a $45_H$ or a $210_H$ there are vacuum configurations such that $M_{B-L}$ 
can be pulled down till to the TeV scale without conflicting with unification constraints. 
On the other hand the issue of neutrino masses with a low $M_{B-L}$ is more serious. 
One has either to invoke a strong fine-tuning in the Yukawa sector or extend the 
theory with an $SO(10)$ singlet (see e.g.~\cite{Malinsky:2005bi}).}.


\chapter{Intermediate scales in nonsupersymmetric $SO(10)$ unification}
\label{intermediatescales}

The purpose of this chapter is to review the constraints enforced by gauge unification
on the intermediate mass scales in the nonsupersymmetric $SO(10)$ GUTs, a needed
preliminary step for assessing the structure of
the multitude of the different breaking patterns
before entering the details of a specific model.
Eventually, our goal is to envisage and examine scenarios potentially
relevant for the understanding of the low energy matter spectrum.
In particular those setups that, albeit
nonsupersymmetric, may exhibit a predictivity comparable to that of the minimal supersymmetric 
$SO(10)$~\cite{Clark:1982ai,Aulakh:1982sw,Aulakh:2003kg}, scrutinized at length in the last few 
years.


The constraints imposed by the absolute neutrino
mass scale on the position of the $B-L$ threshold, together with the proton decay bound on the unification scale $M_{U}$,  provide a
discriminating tool among the many $SO(10)$ scenarios and the corresponding breaking patterns.
These were studied at length in the 80's and early 90's, and detailed surveys
of two- and three-step $SO(10)$ breaking chains
(one and two intermediate thresholds respectively)
are found in Refs.~\cite{Gipson:1984aj,Chang:1984qr,Deshpande:1992au,Deshpande:1992em}.

We perform a systematic survey of $SO(10)$ unification with two intermediate stages.
In addition to updating the analysis to present day data,
this reappraisal is motivated by
(a) the absence of $U(1)$ mixing in previous studies, both at one- and two-loops in the gauge coupling renormalization,
(b) the need for additional Higgs multiplets at some intermediate stages,
and
(c) a reassessment of the two-loop beta coefficients reported in the literature.

The outcome of our study is the emergence of sizeably different features in some of the breaking patterns as compared to the existing results.
This allows us to rescue previously excluded scenarios.
All that before considering the effects of
threshold corrections~\cite{Dixit:1989ff,Mohapatra:1992dx,Lavoura:1993su},
that are unambiguously assessed only when the details of a specific model are worked out. 
Eventually we will comment on the impact of threshold effects in the Outlook of the thesis.

It is remarkable that the chains corresponding to the minimal $SO(10)$ setup with the smallest Higgs representations ($10_H$, $45_H$ and $\overline{16}_H$, or $\overline{126}_H$
in the renormalizable case) and the smallest number of parameters in the
Higgs potential, are still viable. The complexity of this nonsupersymmetric scenario is comparable to that of the minimal supersymmetric $SO(10)$ model, what makes it worth of detailed consideration.

In \sect{sec:chains} we set the framework of the analysis.
\sect{sec:2Lrge} provides a collection of the tools needed for a two-loop study of grand unification.
The results of the numerical study are reported and scrutinized in \sect{sec:results}.
Finally, the relevant one- and two-loop $\beta$-coefficients are detailed
in Appendix \ref{app:2Lbeta}.

\section{Three-step $SO(10)$ breaking chains}
\label{sec:chains}

The relevant $SO(10)\to G2\to G1\to SM$ symmetry breaking chains with two intermediate
gauge groups $G2$ and $G1$ are listed in Table~\ref{tab:chains}. Effective two-step chains are obtained
by joining two of the high-energy scales, paying attention to the possible
deviations from minimality of the scalar content in the remaining intermediate stage
(this we shall discuss in \sect{sec:2loopgauge}).

For the purpose of comparison we follow closely the notation of Ref.~\cite{Deshpande:1992em},
where $P$ denotes the unbroken D-parity~\cite{Kuzmin:1980yp,Kibble:1982dd,Chang:1983fu,Chang:1984uy,Chang:1984qr}.
For each step the Higgs representation responsible for the
breaking is given.

The breakdown of the lower intermediate symmetry $G1$ to the SM gauge group is driven either by the
$16$- or $126$-dimensional Higgs multiplets  $\overline{16}_{H}$ or  $\overline{126}_{H}$.
An important feature of the scenarios with $\overline{126}_{H}$
is the fact that in such a case a potentially realistic $SO(10)$ Yukawa
sector can be constructed already at the renormalizable level (cf.~\sect{fermionmasses}).
Together with $10_{H}$ all the
effective Dirac Yukawa couplings as well as the Majorana mass matrices at the SM level
emerge from the contractions of the matter bilinears $16_{F}16_{F}$ with $\overline{126}_H$
or with $\overline{16}_H \overline{16}_H/\Lambda$, where $\Lambda$
denotes the scale (above $M_{U}$)
at which the effective dimension five Yukawa couplings arise.

\begin{table}[h]
\centering
\begin{tabular}{lllll}
\hline
Chain  &  & G2 & & G1
\\
\hline
{\rm I:}  &
           $ \chain{}{210}  $ & $ \{4_C 2_L 2_R\} $
         & $ \chain{}{\Lambda^{45}}  $ & $ \{3_C 2_L 2_R 1_{B-L}\} $
\\
{\rm II:} &
           $ \chain{}{54}  $ & $ \{4_C 2_L 2_R P\} $
         & $ \chain{}{\Lambda^{210}}   $ & $ \{3_C 2_L 2_R 1_{B-L} P\} $
\\
{\rm III:}  &
           $ \chain{}{ 54 }  $ & $ \{4_C 2_L 2_R P\} $
         & $ \chain{}{\Lambda^{45}}  $ & $ \{3_C 2_L 2_R 1_{B-L}\} $
\\
{\rm IV:}  &
           $ \chain{}{ 210 }  $ & $ \{3_C 2_L 2_R 1_{B-L} P\} $
         & $ \chain{}{\Lambda^{45}}  $ & $ \{3_C 2_L 2_R 1_{B-L}\} $
\\
{\rm V:}  &
           $ \chain{}{ 210 }  $ & $ \{4_C 2_L 2_R\} $
         & $ \chain{}{\Sigma_R^{45}}  $ & $ \{4_C 2_L 1_R\} $
\\
{\rm VI:}  &
           $ \chain{}{ 54 }  $ & $ \{4_C 2_L 2_R P\} $
         & $ \chain{}{\Sigma_R^{45}}  $ & $ \{4_C 2_L 1_R\} $
\\
{\rm VII:}  &
           $ \chain{}{ 54 }  $ & $ \{4_C 2_L 2_R P\} $
         & $ \chain{}{\lambda^{210}}  $ & $ \{4_C 2_L 2_R\} $
\\
{\rm VIII:}  &
           $ \chain{}{ 45 }  $ & $ \{3_C 2_L 2_R 1_{B-L}\} $
         & $ \chain{}{\Sigma_R^{45}}   $ & $ \{3_C 2_L 1_R 1_{B-L}\} $
\\
{\rm IX:}  &
           $ \chain{}{ 210 }  $ & $ \{3_C 2_L 2_R 1_{B-L} P\} $
         & $ \chain{}{\Sigma_R^{45}}  $ & $ \{3_C 2_L 1_R 1_{B-L}\} $
\\
{\rm X:}  &
           $ \chain{}{ 210 }  $ & $ \{4_C 2_L 2_R\} $
         & $ \chain{}{\sigma_R^{210}} $ & $ \{3_C 2_L 1_R 1_{B-L}\} $
\\
{\rm XI:}  &
           $ \chain{}{ 54 }  $ & $ \{4_C 2_L 2_R P\} $
         & $ \chain{}{\sigma_R^{210}}  $ & $ \{3_C 2_L 1_R 1_{B-L}\} $
\\
{\rm XII:}  &
           $ \chain{}{ 45 }  $ & $ \{4_C 2_L 1_R\} $
         & $ \chain{}{\Lambda^{45}}  $ & $ \{3_C 2_L 1_R 1_{B-L}\} $
\\[1ex]
\hline
\end{tabular}
\mycaption{Relevant $SO(10)$ symmetry breaking chains via two intermediate
gauge groups G1 and G2. For each step the representation of the Higgs multiplet 
responsible for the breaking is given in $SO(10)$ or intermediate symmetry group notation (cf.~\Table{tab:submultiplets}).
The breaking to the SM group $3_C2_L1_Y$ is obtained via a $16$ or $126$ Higgs representation. 
}
\label{tab:chains}
\end{table}

D-parity is a discrete symmetry acting as charge conjugation in a left-right
symmetric context~\cite{Kuzmin:1980yp,Kibble:1982dd}, 
and as that it plays the role of a left-right symmetry (it enforces for instance equal left and right gauge
couplings).
$SO(10)$ invariance then implies exact D-parity (because D belongs
to the $SO(10)$ Lie algebra). D-parity may be spontaneously broken by D-odd Pati-Salam (PS)
singlets contained in 210 or 45 Higgs representations. Its breaking can therefore be decoupled
from the $SU(2)_R$ breaking, allowing for different left and right gauge couplings~\cite{Chang:1983fu,Chang:1984uy}.

\begin{table*}[h]
\centering
\begin{tabular}{cccccc}
\hline
& \multicolumn{4}{c}{Surviving Higgs multiplets in $SO(10)$ subgroups} & \\
\cline{2-5}
\multicolumn{1}{c}{$SO(10)$}&
\multicolumn{1}{c}{$\{4_C2_L 1_R\}$} &
\multicolumn{1}{c}{$ \{4_C2_L 2_R\}$} &
\multicolumn{1}{c}{$\{3_C 2_L 2_R 1_{B-L}\}$} &
\multicolumn{1}{c}{$\{3_C 2_L 1_R 1_{B-L}\}$} &
\multicolumn{1}{c}{Notation}\\
\hline
10 & $(1,2, {+\frac{1}{2}})$ & $(1,2,2)$ & $(1,2,2,0)$ &
$(1,2,{+\frac{1}{2}},0)$ &
$\phi^{10}$\\
$\overline{16}$ & $(4,1, +{\frac{1}{2}})$ & $(4,1,2)$ & $(1,1,2,{-\frac{1}{2}})$ &
$(1,1,+{\frac{1}{2}},{-\frac{1}{2}})$ &
$\delta^{16}_R$\\
$\overline{16}$ &  & $(\overline{4},2,1)$ & $(1,2,1,+{\frac{1}{2}})$ &  &
$\delta^{16}_L$\\
$\overline{126}$ & $(15,2,+\frac{1}{2})$  & $(15,2,2)$  & $(1,2,2,0)$ &
$(1,2,{+\frac{1}{2}},0)$ &
$\phi^{126}$\\
$\overline{126}$ &$(10,1,1)$  & $(10,1,3)$ & $(1,1,3,-1)$ & $(1,1,1,-1)$ &
$\Delta^{126}_R$\\
$\overline{126}$ &  & $(\overline{10},3,1)$ & $(1,3,1,1)$ & &
$\Delta^{126}_L$\\
45 & $(15,1,0)$ & $(15,1,1)$ &  & &
$\Lambda^{45}$\\
210 &  & $(15,1,1)$ &  & &
$\Lambda^{210}$\\
45 &  & $(1,1,3)$ &$(1,1,3,0)$  & &
$\Sigma^{45}_R$\\
45 &  & $(1,3,1)$ &$(1,3,1,0)$  & &
$\Sigma^{45}_L$\\
210 &  & $(15,1,3)$ &  & &
$\sigma^{210}_R$\\
210 &  & $(15,3,1)$ &  & &
$\sigma^{210}_L$\\
210 &  & $(1,1,1)$ &  & &
$\lambda^{210}$\\
\hline
\end{tabular}
\mycaption{Scalar multiplets contributing to the running of the
gauge couplings for a given $SO(10)$ subgroup according to minimal fine tuning.
The survival of $\phi^{126}$ (not required by minimality) is needed by
a realistic leptonic mass spectrum, as discussed in the text
(in the $3_C 2_L 2_R 1_{B-L}$ and $3_C 2_L 1_R 1_{B-L}$ stages only one linear combination
of $\phi^{10}$ and $\phi^{126}$ remains).
The $U(1)_{B-L}$ charge is given, up to a factor $\sqrt{3/2}$, by $(B-L)/2$
(the latter is reported in the table). For the naming of the
Higgs multiplets we follow the notation of Ref.~\cite{Deshpande:1992em}
with the addition of $\phi^{126}$.
When the D-parity (P) is unbroken the particle content must be left-right symmetric.
D-parity may be broken via P-odd Pati-Salam singlets in $45_H$ or $210_H$.}
\label{tab:submultiplets}
\end{table*}

The possibility of decoupling the D-parity breaking from the scale of right-handed interactions is a cosmologically
relevant issue. On the one hand baryon asymmetry cannot arise in a left-right symmetric
($g_L=g_R$) universe~\cite{Kuzmin:1980yp}.
On the other hand, the spontaneous breaking of a discrete symmetry, such as D-parity,
creates domain walls that, if massive enough (i.e. for intermediate mass scales) do not disappear, overclosing the universe~\cite{Kibble:1982dd}.
These potential problems may be overcome either
by confining D-parity at the GUT scale or by invoking inflation. The latter solution implies
that domain walls are formed above the reheating temperature, enforcing a lower bound
on the D-parity breaking scale of $10^{12}$ GeV. Realistic $SO(10)$ breaking
patterns must therefore include this constraint.

\subsection{The extended survival hypothesis}
\label{sec:ESH}

Throughout all three stages of running we assume that the scalar spectrum obeys the so
called extended survival hypothesis (ESH)~\cite{delAguila:1980at}
which requires that {\it at every stage of the symmetry breaking chain only those scalars
are present that develop a vacuum expectation value (VEV)
at the current or the subsequent levels of the spontaneous symmetry breaking}.
ESH is equivalent to the requirement of the minimal number of
fine-tunings to be imposed onto the scalar potential~\cite{Mohapatra:1982aq}
so that all the symmetry breaking steps are performed at the desired scales.

On the technical side one should identify all the Higgs multiplets needed by the
breaking pattern under consideration and keep them
according to the gauge symmetry down to the scale of their VEVs.
This typically pulls down a large number of scalars in scenarios where
$\overline{126}_{H}$ provides the $B-L$ breakdown.

On the other hand, one must take into account that the role of $\overline{126}_{H}$ is twofold: in addition to triggering the $G1$ breaking it plays a relevant role
in the Yukawa sector
where it provides the necessary breaking of
the down-quark/charged-lepton mass degeneracy (cf.~\eq{eq:mass-relations}).
For this to work one needs a reasonably large admixture of the $\overline{126}_{H}$
component in the effective electroweak doublets. Since $(1,2,2)_{10}$ can mix
with $(15,2,2)_{\overline{126}}$ only below the Pati-Salam breaking scale, both fields
must be present at the Pati-Salam level (otherwise the scalar doublet mass matrix does not provide large enough components of both these multiplets in the light Higgs fields).

Note that the same argument applies also to the $4_{C}2_{L}1_{R}$ intermediate stage
when one must retain the doublet component of $\overline{126}_{H}$, namely $(15,2,+\frac{1}{2})_{\overline{126}}$,
in order for it to eventually admix with  $(1,2,+\frac{1}{2})_{10}$
in the light Higgs sector.
On the other hand, at the $3_C 2_L 2_R 1_{B-L}$ and $3_C 2_L 1_R 1_{B-L}$ stages,
the (minimal) survival of only one combination of the $\phi^{10}$
and $\phi^{126}$ scalar doublets (see Table \ref{tab:submultiplets}) is compatible with the Yukawa sector constraints because the degeneracy between the quark and lepton spectra has already been smeared-out by the Pati-Salam breakdown.

In summary, potentially realistic renormalizable Yukawa textures in settings with
well-separated $SO(10)$ and Pati-Salam breaking scales
call for an additional fine tuning in the Higgs sector.
In the scenarios with $\overline{126}_{H}$,
the $10_{H}$ bidoublet $(1,2,2)_{10}$,
included in Refs~\cite{Gipson:1984aj,Chang:1984qr,Deshpande:1992au,Deshpande:1992em},
must be paired at the $4_{C}2_{L}2_{R}$ scale with an
extra $(15,2,2)_{\overline{126}}$ scalar bidoublet
(or $(1,2,+\frac{1}{2})_{10}$ with $(15,2,+\frac{1}{2})_{\overline{126}}$ at the $4_{C}2_{L}1_{R}$ stage).
This can affect the running of the gauge couplings in chains I, II, III, V, VI, VII, X, XI
and XII.

For the sake of comparison with previous
studies~\cite{Gipson:1984aj,Chang:1984qr,Deshpande:1992au,Deshpande:1992em}
we shall not include the $\phi^{126}$ multiplets in the first part of the analysis.
Rather, we shall comment on their relevance for gauge unification in \sect{sec:extrahiggs}.

\section{Two-loop gauge renormalization group equations}
\label{sec:2Lrge}

In this section we report, in order to fix a consistent notation, the two-loop renormalization group equations
(RGEs) for the gauge couplings.
We consider a gauge group of the form 
$ U(1)_{1} \otimes ... \otimes U(1)_{N}\otimes G_1\otimes ... \otimes G_{N'}$, where $G_i$ are simple groups.

\subsection{The non-abelian sector}
\label{sec:2Lnonabelian}

Let us focus first on the non-abelian sector  corresponding to $G_1\otimes ... \otimes G_{N'}$  and defer the full treatment of the effects due to the extra $U(1)$ factors to section \ref{sec:U1mix}.
Defining $t=\log (\mu/\mu_0)$ we write
\be
\frac{dg_p}{dt} = g_p\ \beta_p
\label{rge}
\ee
where $p=1,...,N'$ is the gauge group label. Neglecting for the time being the abelian components,
the $\beta$-functions for the $G_1\otimes ...\otimes G_{N'}$ gauge couplings  read at two-loop level~\cite{Jones:1974mm,Caswell:1974gg,Jones:1981we,
Machacek:1983tz,Machacek:1983fi,Machacek:1984zw}
\begin{align}
\label{Gp2loops}
\beta_p &= \frac{g_p^2}{(4\pi)^2} \left\{
    - \frac{11}{3} C_2(G_p) + \frac{4}{3}\kappa S_2(F_p) + \frac{1}{3} \eta S_2(S_p) \right. \nn \\ 
& + \frac{g_p^2}{(4\pi)^2} \left[- \frac{34}{3} \left( C_2(G_p) \right)^2 + \left( 4 C_2(F_p) + \frac{20}{3} C_2(G_p) \right) \kappa S_2(F_p) \right. \nn \\
& + \left. \left( 4 C_2(S_p) + \frac{2}{3} C_2(G_p) \right) \eta S_2(S_p) \right] \nn \\
&+ \left. \frac{g_q^2}{(4\pi)^2} 4 \Big[  \kappa C_2(F_q) S_2(F_p) + \eta C_2(S_q) S_2(S_p) \Big] - \frac{2\kappa}{(4\pi)^2} Y_4(F_p)  \!\right\} \, ,
\end{align}
where $\kappa=1,\frac{1}{2}$ for Dirac and Weyl fermions respectively.
Correspondingly, $\eta=1, \frac{1}{2}$ for complex and real scalar fields. The sum over $q\neq p$ corresponding to contributions to $\beta_{p}$ from the other gauge sectors labelled by $q$ is understood.
Given a fermion $F$ or a scalar $S$ field that transforms according to the
representation $R=R_1\otimes ... \otimes R_{N'}$, where $R_p$ is an irreducible
representation of the group $G_p$ of dimension $d(R_p)$, the factor $S_2(R_p)$ is defined by
\be
S_2(R_p) \equiv T(R_p)\frac{d(R)}{d(R_p)}\,,
\label{S2}
\ee
where $T(R_p)$ is the Dynkin index of the representation $R_p$. The corresponding Casimir
eigenvalue is then given by
\be
C_2(R_p) d(R_p) = T(R_p) d(G_p)\,,
\label{C2}
\ee
where $d(G)$ is the dimension of the group.
In \eq{Gp2loops} the first row represents the one-loop contribution while the other
terms stand for the two-loop corrections, including that induced by Yukawa interactions.
The latter is accounted for in terms of a factor
\be
Y_4(F_p) = \frac{1}{d(G_p)} \Tr \left[ C_2(F_p) YY^\dagger \right],
\label{Y4}
\ee
where the ``general'' Yukawa coupling
\be
Y^{abc}\ \overline{\psi}_a \psi_b\ h_c \ + \ h.c.
\label{Yukawa}
\ee
includes family as well as group indices.
The coupling in \eq{Yukawa} is written in terms of four-component Weyl spinors $\psi_{a,b}$
and a scalar field $h_c$ (be complex or real). The trace includes the sum over all relevant fermion and
scalar fields.

\vspace*{3ex}
\subsection{The abelian couplings and $U(1)$ mixing}
\label{sec:U1mix}

In order to include the abelian contributions to \eq{Gp2loops} at two loops and the
one- and two-loop effects of $U(1)$ mixing~\cite{Holdom:1985ag}, let us write the most general
interaction of $N$ abelian gauge bosons $A^\mu_b$ and a set of
Weyl fermions $\psi_f$ as
\be
\overline{\psi_f} \gamma_\mu Q^r_f\psi_f g_{rb} A^\mu_b\,.
\label{mincoupl}
\ee
The gauge coupling constants $g_{rb}$, $r,b=1,...,N$, couple $ A^\mu_b$ to the fermionic
current $J^r_\mu = \overline{\psi}_f \gamma_\mu Q^r_f\psi_f$. The $N\times N$ gauge coupling
matrix $g_{rb}$ can be diagonalized by two independent rotations: one acting on the $U(1)$ charges
$Q^r_f$ and the other on the gauge boson fields $A^\mu_b$.
For a given choice of the charges,
$g_{rb}$ can be set in a triangular form ($g_{rb}=0$ for $r>b$) by the gauge boson rotation. The resulting
$N(N+1)/2$ entries are observable couplings.

Since $F^a_{\mu\nu}$ in the abelian case is itself gauge invariant, the most general kinetic part of the lagrangian reads at the renormalizable level
\be
-\frac{1}{4} F^a_{\mu\nu}  F^{a\mu\nu}  - \frac{1}{4} \xi_{ab} F^a_{\mu\nu}  F^{b\mu\nu}\,,
\label{FmunuFmunu}
\ee
where $a\neq b$ and $\mod{\xi_{ab}} < 1$.
A non-orthogonal rotation of the fields $A^\mu_a$ may be performed to set the gauge kinetic term
in a canonical diagonal form. Any further orthogonal rotation of the gauge fields will preserve
this form. Then, the renormalization prescription may be conveniently chosen to
maintain at each scale the kinetic terms canonical and diagonal on-shell while renormalizing
accordingly the gauge coupling matrix $g_{rb}$~\footnote{Alternatively
one may work with off-diagonal kinetic terms while keeping the
gauge interactions diagonal~\cite{Luo:2002iq}.}.
Thus, even if at one scale $g_{rb}$ is diagonal, in general non-zero off-diagonal
entries are generated by renormalization effects. One shows~\cite{delAguila:1988jz}
that in the case
the abelian gauge couplings are at a given scale diagonal {\em and} equal (i.e. there is a $U(1)$ unification),
there may exist a (scale independent) gauge field basis such that the abelian interactions
remain to all orders diagonal along the RGE trajectory~\footnote{Vanishing of the commutator of the $\beta$-functions and their derivatives is needed~\cite{delAguila:1995rb}.}.

In general, the renormalization of the abelian part of the gauge interactions
is determined by
\be
\frac{dg_{rb}}{dt} = g_{ra} \beta_{ab}\,,
\label{U1rge}
\ee
where, as a consequence of gauge invariance,
\be
\beta_{ab} = \frac{d}{dt}\left(\log Z^{1/2}_3 \right)_{ab}\,.
\label{U1beta}
\ee
with $Z_3$ denoting the gauge-boson wave-function renormalization matrix.
In order to further simplify the notation
it is convenient to introduce the ``reduced'' couplings~\cite{delAguila:1988jz}
\be
g_{kb} \equiv Q^r_k g_{rb}\,,
\label{redcoupl}
\ee
that evolve according to
\be
\frac{dg_{kb}}{dt} = g_{ka} \beta_{ab}\,.
\label{U1rgered}
\ee
The index $k$ labels the fields (fermions and scalars) that carry $U(1)$ charges.

In terms of the reduced couplings the $\beta$-function that governs the $U(1)$
running up to two loops is given
by~\cite{Jones:1974mm,Caswell:1974gg,Jones:1981we}
\bea
\label{bU12loops}
\beta_{ab} &=& \frac{1}{(4\pi)^2} \left\{
      \frac{4}{3}\kappa\ g_{fa}g_{fb} + \frac{1}{3}\eta\ g_{sa}g_{sb}  \right. \\
&+& \left.  \frac{4}{(4\pi)^2} \Big[
    \kappa \left( g_{fa}g_{fb}g_{fc}^2 + g_{fa}g_{fb} g_q^2 C_2(F_q) \right) + \eta \left( g_{sa}g_{sb}g_{sc}^2 + g_{sa}g_{sb} g_q^2 C_2(S_q) \right) \Big] \right. \nn \\
    &-& \left. \frac{2\kappa}{(4\pi)^2}   \Tr \left[g_{fa}g_{fb}\ YY^\dagger \right] \right\}\,, \nn
\eea
where repeated indices are summed over, labelling fermions ($f$), scalars ($s$)
and $U(1)$ gauge groups ($c$). The terms proportional to the quadratic Casimir $C_2(R_p)$ represent the two-loop
contributions of the non abelian components $G_q$ of the gauge group to the $U(1)$ gauge coupling
renormalization.

Correspondingly, using the notation of \eq{redcoupl}, an additional two-loop term that represents the renormalization of the non abelian gauge couplings
induced at two loops by the $U(1)$ gauge fields is to be added
to \eq{Gp2loops}, namely
\be
\Delta \beta_p = \frac{g_p^2}{(4\pi)^4}
                4 \Big[ \kappa\ g_{fc}^2 S_2(F_p) +  \eta\ g_{sc}^2 S_2(S_p)\Big]\,.
\label{bpU12loops}
\ee
In \eqs{bU12loops}{bpU12loops}, we use the abbreviation $f\equiv F_p$ and $s\equiv S_p$ and, as before, $\kappa=1,\frac{1}{2}$ for Dirac and Weyl fermions, while $\eta=1,\frac{1}{2}$ for complex and real scalar fields respectively.

\subsection{Some notation}
\label{sec:2Lnotation}

When at most one $U(1)$ factor is present, and neglecting the Yukawa contributions,
the two-loop RGEs can be conveniently written as
\be
\frac{d\alpha_i^{-1}}{dt} = - \frac{a_i}{2\pi} - \frac{b_{ij}}{8\pi^2}\alpha_j\,,
\label{alpharge}
\ee
where $\alpha_i=g_i^2/4\pi$. The $\beta$-coefficients $a_i$ and $b_{ij}$ for the
relevant $SO(10)$ chains are given  in Appendix \ref{app:2Lbeta}.

Substituting the one-loop solution for $\alpha_j$ into the right-hand side of \eq{alpharge}
one obtains
\be
\alpha_i^{-1}(t) - \alpha_i^{-1}(0) = - \frac{a_i}{2\pi}\ t + \frac{\tilde b_{ij}}{4\pi}
                                     \log\left(1- \omega_j t\right)\,,
\label{alpha2loops}
\ee
where $\omega_j=a_j \alpha_j(0)/(2\pi)$ and $\tilde b_{ij} = {b_{ij}}/{a_j}$ .
The analytic solution in (\ref{alpha2loops}) holds at two loops
(for $\omega_j t < 1$) up to higher order
effects. A sample of the rescaled $\beta$-coefficients $\tilde b_{ij}$ is given,
for the purpose of comparison with previous results,
in Appendix~\ref{app:2Lbeta}.

We shall conveniently write the $\beta$-function in \eq{bU12loops},
that governs the abelian mixing, as
\be
\beta_{ab} = \frac{1}{(4\pi)^2}\ g_{sa}\ \gamma_{sr}\ g_{rb}\,,
\label{bU1gamma}
\ee
where $\gamma_{sr}$ include both one- and two-loop contributions.
Analogously, the non-abelian beta function in \eq{Gp2loops},
including the $U(1)$ contribution in \eq{bpU12loops},
is conveniently written as
\be
\beta_{p} = \frac{g_{p}^2}{(4\pi)^2}\ \gamma_{p}\,.
\label{bU1gammanonabel}
\ee
The $\gamma_{p}$ functions for the $SO(10)$ breaking chains considered in this work are reported
in Appendix~\ref{app:U1mix}.

Finally, the Yukawa term in \eq{Y4}, and correspondingly in \eq{bU12loops}, can be written as
\be
Y_4(F_p) =  y_{pk}\Tr \left(Y_kY_k^\dagger \right)\,,
\label{Ycpk}
\ee
where $Y_{k}$ are the ``standard'' $3\times 3$ Yukawa matrices in the family space labelled by the flavour index $k$. The trace is taken over family indices and $k$ is summed over the different Yukawa
terms present at each stage of $SO(10)$ breaking. The coefficients $y_{pk}$ are given
explicitly in Appendix \ref{app:Yukawa}

\subsection{One-loop matching}
\label{sec:1Lmatching}

The matching conditions between effective theories in the framework of
dimensional regularization have been derived in \cite{Weinberg:1980wa,Hall:1980kf}.
Let us consider first a simple gauge group $G$ spontaneously broken into subgroups $G_p$.
Neglecting terms involving logarithms of mass ratios which are assumed
to be subleading (massive states clustered near the threshold),
the one-loop matching for the gauge couplings can be written as
\be
\alpha_{p}^{-1} - \frac{C_2(G_p)}{12\pi} = \alpha_G^{-1} - \frac{C_2(G)}{12\pi}\,.
\label{2LmatchGUT}
\ee
Let us turn to the case when several non-abelian simple groups $G_p$
(and at most one $U(1)_X$)
spontaneously break whilst preserving a $U(1)_Y$ charge. The conserved $U(1)$ generator $T_Y$
can be written in terms of the relevant generators of the various Cartan subalgebras (and of the consistently
normalized $T_X$)
as
\be
T_Y = p_i T_i\,,
\label{oneU1}
\ee
where $\sum p_i^2 = 1$, and $i$ runs over the relevant $p$ (and $X$) indices.
The matching condition is then give 
by\footnote{This is easily proven at tree level~\cite{Georgi:1977wk}. 
Let us imagine that the gauge symmetry is spontaneously broken by the VEV of 
an arbitrary set of scalar fields $\vev{\phi}$, such that $T_Y \vev{\phi} = 0$. 
Starting from the covariant derivative 
\be
D_\mu \phi = \partial_\mu \phi + i g_i T_i \left( A_\mu \right)_i \phi \, ,
\ee
we derive the gauge boson mass matrix
\be
\mu^2_{ij} = g_i g_j \vev{\phi}^\dag T_i T_j \vev{\phi} \, ,
\ee 
which has a zero eigenvector corresponding to the massless gauge field $A^Y_\mu = q_i \left( A_\mu \right)_i$, where 
\be
\mu^2_{ij} q_j = 0 \qquad \text{with} \qquad \sum_j q_j^2 = 1 \, . 
\ee
It's easy to see then that the components of $q$ are 
\be
q_i = N p_i / g_i \qquad \text{with} \qquad N \equiv \left( \sum_i \left( p_i / g_i  \right)^2 \right)^{-\tfrac{1}{2}} \, ,
\ee
and from the coupling of $A^Y_\mu$ to fermions 
\be
g_i \left(A_\mu\right)_i \overline{\psi} \gamma^\mu T_i \psi = 
g_i q_i A^Y_\mu \overline{\psi} \gamma^\mu T_i \psi + \ldots =
N p_i A^Y_\mu \overline{\psi} \gamma^\mu T_i \psi + \ldots =
N A^Y_\mu \overline{\psi} \gamma^\mu T_Y \psi + \ldots \, ,
\ee
we conclude that 
\be
g_Y = N \equiv \left( \sum_i \left( p_i / g_i  \right)^2 \right)^{-\tfrac{1}{2}} \, .
\ee
}
\be
\alpha_Y^{-1} = \sum_i p_i^2 \left(\alpha_i^{-1} - \frac{C_2(G_i)}{12\pi}\right)\,,
\label{2LmatchNOmix}
\ee
where for $i=X$, if present, $C_2=0$.

Consider now the breaking of $N$ copies of $U(1)$ gauge factors to a subset of $M$ elements $U(1)$ (with $M<N$).
Denoting by $T_n$ ($n=1,...,N$) and by $\widetilde T_m$ ($m=1,...,M$)
their properly normalized generators we have
\be
\widetilde T_m = P_{mn} T_n
\label{U1charges}
\ee
with the orthogonality condition $P_{mn}P_{m'n} = \delta_{mm'}$. Let us denote
by $g_{na}$ ($n,a=1,...,N$) and by $\widetilde g_{mb}$ ($m,b=1,...,M$) the matrices
of abelian gauge couplings above and below the breaking scale respectively.
By writing the abelian gauge boson mass matrix in the broken vacuum and by identifying
the massless states, we find the following matching condition
\be
(\widetilde g \widetilde g^T )^{-1} = P \left(g g^T \right)^{-1} P^T\,.
\label{2LU1match}
\ee
Notice that \eq{2LU1match} depends on the chosen basis for the $U(1)$ charges (via $P$)
but it is invariant under orthogonal rotations of the gauge boson fields ($gO^TOg^T=gg^T$).
The massless gauge bosons $\widetilde A_m^\mu$ are given in terms of $A_n^\mu$ by
\be
\widetilde A_m^\mu = \left[ \widetilde g^T P \left(g^{-1}\right)^T \right]_{mn} A_n^\mu\,,
\label{masslessA}
\ee
where $m=1,...,M$ and $n=1,...,N$.

The general case of a gauge group $ U(1)_{1} \otimes ... \otimes U(1)_{N}\otimes G_1\otimes ... \otimes G_{N'}$
spontaneously broken to $U(1)_{1} \otimes ... \otimes U(1)_{M}$ with $M\leq N+N'$
is taken care of by replacing $(gg^T)^{-1}$  in \eq{2LU1match}
with the block-diagonal $(N+N')\times (N+N')$ matrix
\be
(G G^T)^{-1} = \mbox{Diag}\left[(g g^T)^{-1},g_p^{-2} - \frac{C_2(G_p)}{48\pi^2} \right]
\label{2Lmatchgen}
\ee
thus providing, together with the extended \eq{U1charges}
and \eq{2LU1match}, a generalization of \eq{2LmatchNOmix}.

\section{Numerical results}
\label{sec:results}

At one-loop, and in absence of the $U(1)$ mixing, the gauge
RGEs are not coupled and the unification constraints can be studied analytically.
When two-loop effects are included (or at one-loop more than one $U(1)$ factor is present) there is no closed solution and one must solve the system of coupled equations, matching all stages between the weak and unification scales, numerically.
On the other hand (when no $U(1)$ mixing is there)
one may take advantage of the analytic formula in \eq{alpha2loops}.
The latter turns out to provide, for the cases here studied, a very good approximation to the numerical solution. The discrepancies with the numerical integration do not
generally exceed the $10^{-3}$ level.

We perform a scan over the relevant breaking scales $M_{U}$, $M_{2}$ and $M_{1}$ and the value of the grand unified coupling $\alpha_{U}$ and
impose the matching with the SM gauge couplings at the $M_Z$ scale
requiring a precision at the per mil level.
This is achieved by minimizing the parameter
\be
\delta=\sqrt{\sum_{i=1}^{3}\left(\frac{\alpha_i^{\rm th} -\alpha_i}{\alpha_i}\right)^2} \ ,
\label{deltaMZ}
\ee
where $\alpha_i$ denote the experimental values at $M_Z$
and $\alpha_i^{\rm th}$ are the renormalized couplings obtained from unification.

The input values for the (consistently normalized) gauge SM couplings
at the scale $M_Z=91.19$ GeV are~\cite{Amsler:2008zzb}
\bea
\alpha_1 &=& 0.016946 \pm 0.000006\,, \nn \\
\alpha_2 &=& 0.033812 \pm 0.000021\,, \\
\alpha_3 &=& 0.1176 \pm 0.0020\,,  \nn
\label{alphainormMZ}
\eea
corresponding to the electroweak scale parameters
\bea
\alpha^{-1}_{em} &=& 127.925 \pm 0.016\,, \nn \\
\sin^2\theta_W &=& 0.23119 \pm 0.00014\,.
\label{alphasinthetaMZ}
\eea
All these data refer to the modified minimally subtracted ($\overline{\mbox{MS}}$) quantities
at the $M_Z$ scale.

For $\alpha_{1,2}$ we shall consider only the central values while we resort to scanning over the whole $3\sigma$  domain for $\alpha_3$ when a stable solution is not found.

The results, i.e. the positions of the intermediate scales $M_{1}$, $M_{2}$ and $M_{U}$ shall be reported in terms of decadic logarithms of their values in units of GeV, i.e.
$n_1 = \log_{10}({M_1}/{\text{GeV}})$,
$n_2 = \log_{10}({M_2}/{\text{GeV}})$,
$n_U = \log_{10}({M_U}/{\text{GeV}})$.
In particular, $n_U$, $n_2$
are given as functions of $n_1$ for each breaking pattern
and for different approximations in the loop expansion. Each of the breaking patterns is further supplemented by the relevant range of the values of $\alpha_U$.

\subsection{$U(1)_R \otimes U(1)_{B-L}$ mixing}
\label{sec:U1RU1Xmixing}

The chains VIII to XII
require consideration of the mixing between the two $U(1)$ factors.
While $U(1)_{R}$ and $U(1)_{B-L}$ do emerge
with canonical diagonal kinetic
terms, being the remnants of the breaking of non-abelian groups,
the corresponding gauge couplings are at the onset different in size.
In general, no {\em scale independent}
orthogonal rotations of charges and gauge fields exist that diagonalize
the gauge interactions to all orders along the RGE trajectories.
According to the discussion in \sect{sec:2Lrge}, off-diagonal gauge
couplings arise at the one-loop level that must be accounted for
in order to perform the matching at the $M_1$ scale with
the standard hypercharge. The preserved direction in the $Q^{R,B-L}$ charge
space is given by
\be
Q^Y = \sqrt{\frac{3}{5}} Q^R + \sqrt{\frac{2}{5}} Q^{B-L}\,,
\label{QYdef}
\ee
where
\begin{equation}
Q^R = T_{3R}\;\;\; \text{ and } \;\;\;
Q^{B-L} = \sqrt{\frac{3}{2}}\ \left(\frac{B-L}{2}\right)\,.
\label{QRQXdefs}
\end{equation}
The matching of the gauge couplings is then obtained from \eq{2LU1match}
\be
g_Y^{-2} = P\left(g g^T\right)^{-1} P^T\,,
\label{g_Ymatch}
\ee
with
\be
P = \left(\sqrt{\frac{3}{5}},\ \sqrt{\frac{2}{5}} \right)
\label{P_RX}
\ee
and
\be
g=
\left(\begin{array}{cc}
g_{R,R}&g_{R,B-L}\\[1.1ex]
g_{B-L,R}&g_{B-L,B-L}\\
\end{array}\right)\,.
\label{g_RX}
\ee


When neglecting the off-diagonal terms, \eq{g_Ymatch} reproduces the matching condition
used in Refs.~\cite{Gipson:1984aj,Chang:1984qr,Deshpande:1992au,Deshpande:1992em}.
For all other cases, in which only one $U(1)$ factor is present,
the matching relations can be
read off directly from \eq{2LmatchGUT} and \eq{2LmatchNOmix}.

\subsection{Two-loop results (purely gauge)}
\label{sec:2loopgauge}

\begin{figure*}
 \centering
 \subfigure[\ Chain Ia]
   {\includegraphics[width=5cm]{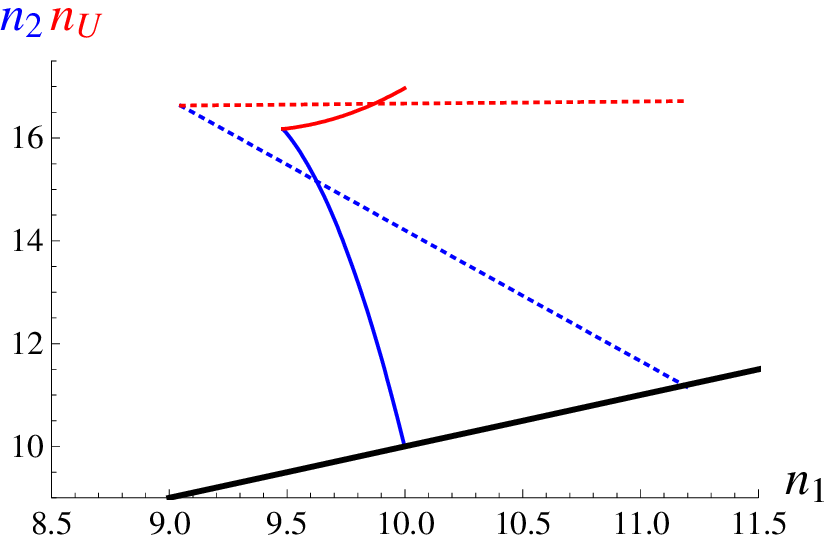}}
 \vspace{2mm}
 \subfigure[\ Chain IIa]
   {\includegraphics[width=5cm]{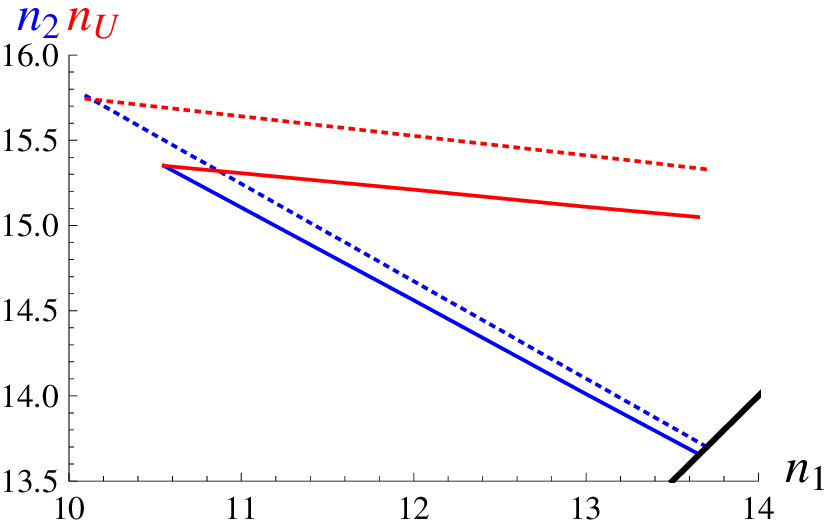}}
 \vspace{2mm}
 \subfigure[\ Chain IIIa]
   {\includegraphics[width=5cm]{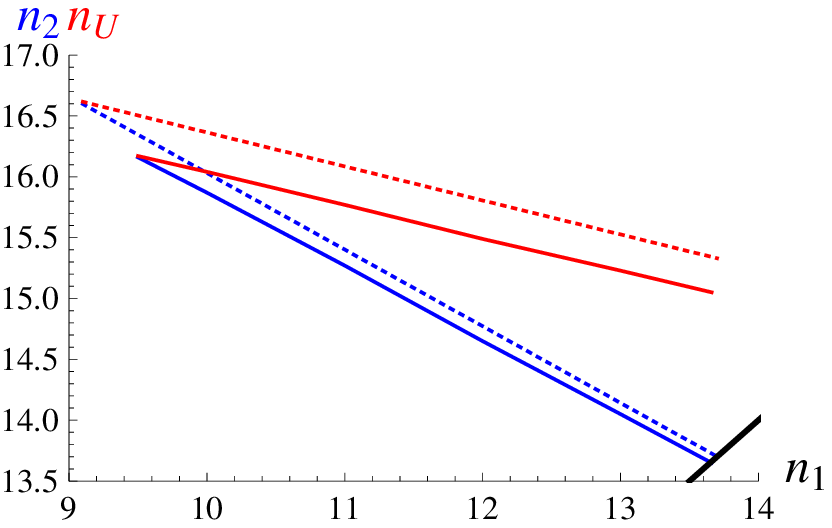}}
 \vspace{2mm}
\subfigure[\ Chain IVa]
   {\includegraphics[width=5cm]{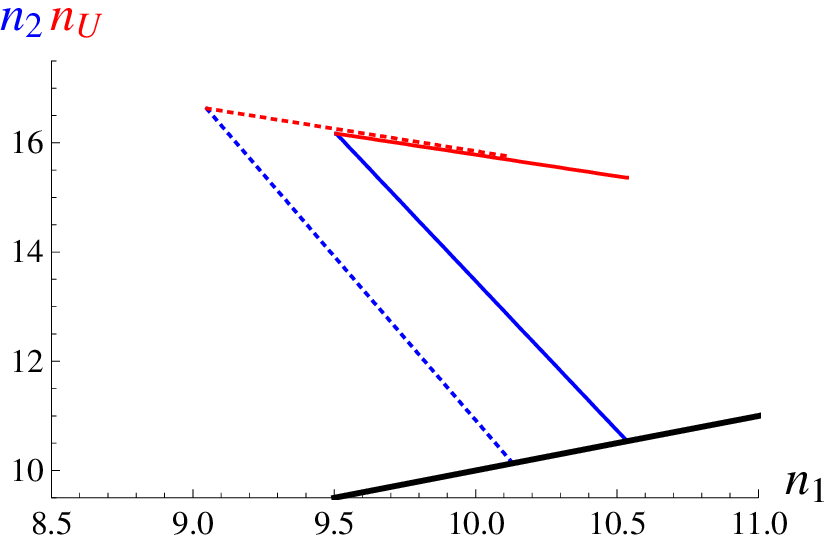}}
 \vspace{2mm}
 \subfigure[\ Chain Va]
   {\includegraphics[width=5cm]{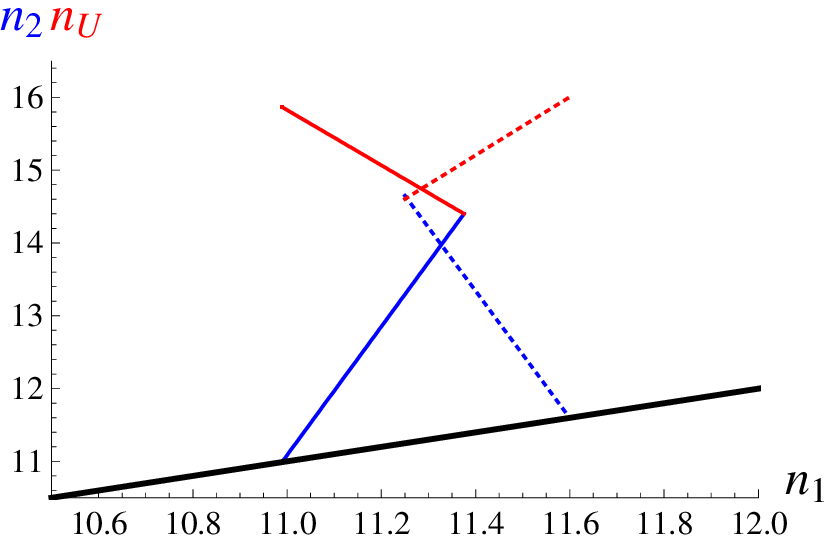}}
 \vspace{2mm}
 \subfigure[\ Chain VIa]
   {\includegraphics[width=5cm]{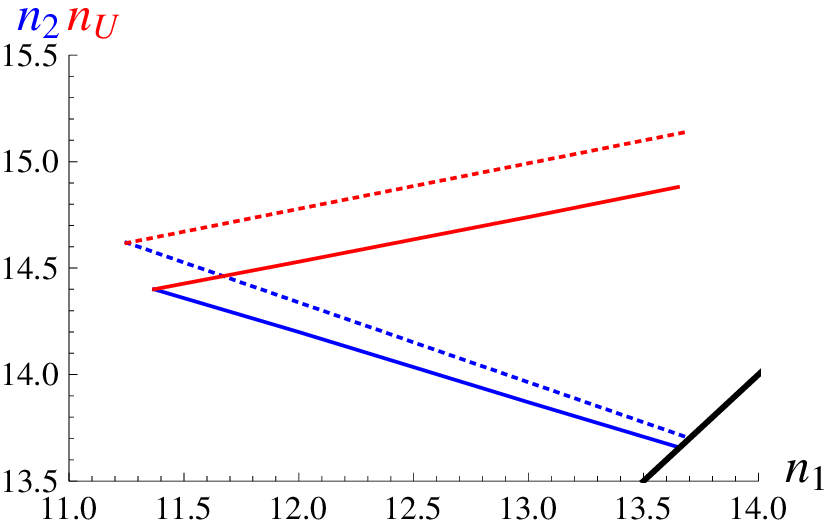}}
 \vspace{2mm}
 \subfigure[\ Chain VIIa]
   {\includegraphics[width=5cm]{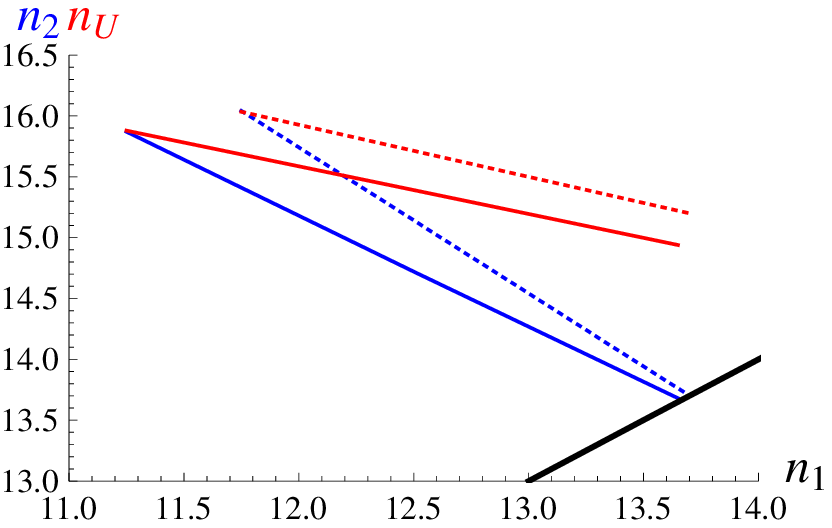}}
 \vspace{2mm}
 \subfigure[\ Chain VIIIa]
   {\includegraphics[width=5cm]{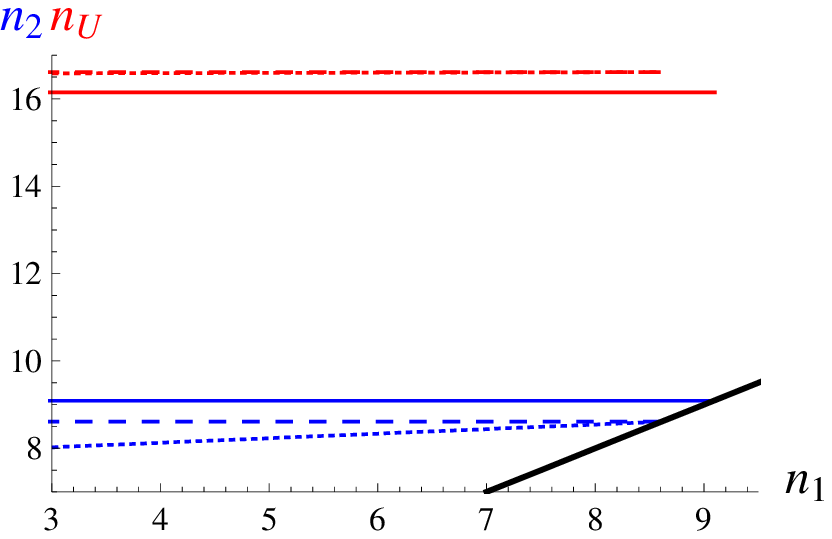}}
 \vspace{2mm}
 \subfigure[\ Chain IXa]
   {\includegraphics[width=5cm]{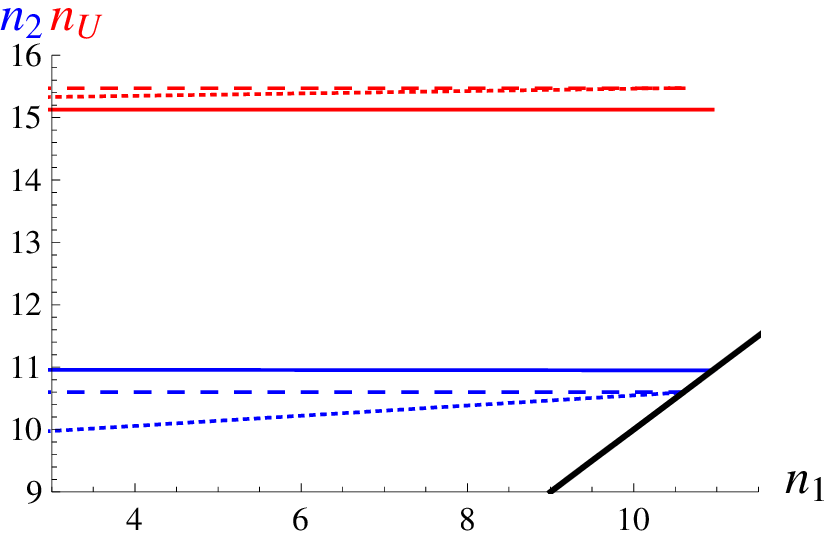}}
 \vspace{2mm}
 \subfigure[\ Chain XIa]
   {\includegraphics[width=5cm]{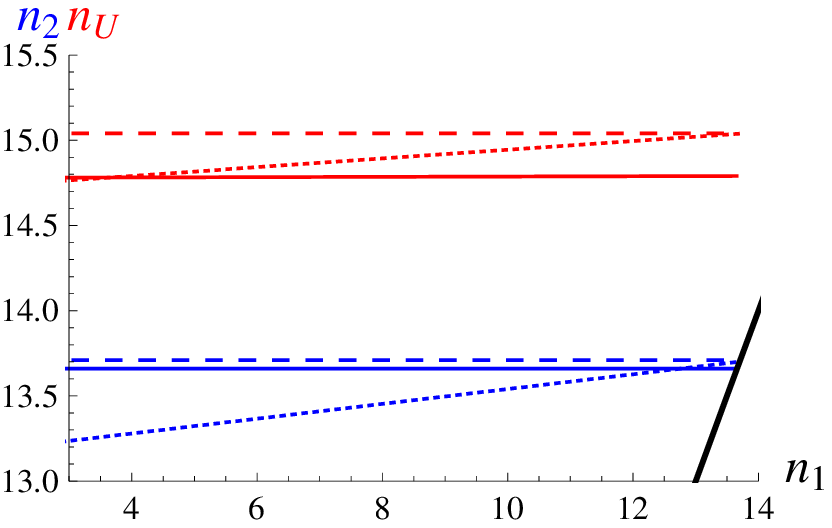}}
 \vspace{2mm}
 \subfigure[\ Chain XIIa]
   {\includegraphics[width=5cm]{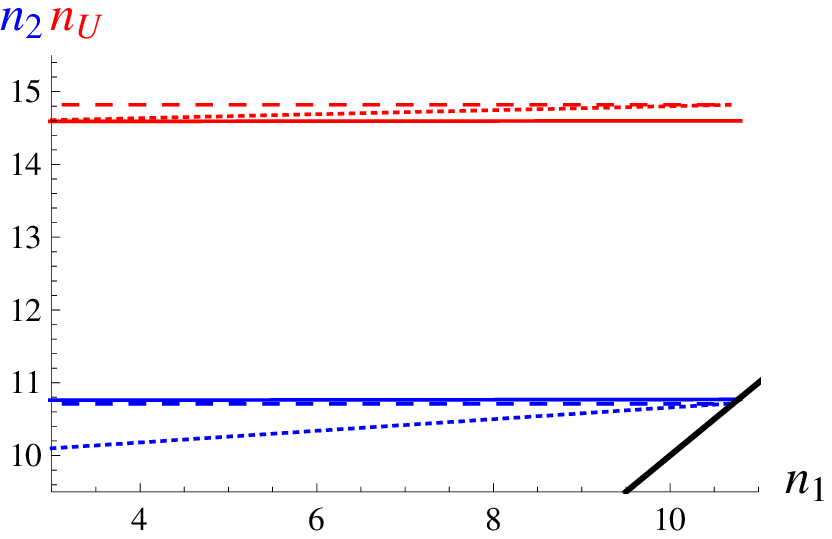}}
 \vspace{2mm}
\mycaption{The values of $n_U$ (red/upper branches)
and $n_2$ (blue/lower branches) are shown as functions
of $n_1$ for the pure gauge running in the $\overline{126}_H$ case.
The bold black line bounds the region $n_1 \leq n_2$.
From chains Ia to VIIa the short-dashed lines represent the result of one-loop running
while the solid ones correspond to the two-loop solutions. For chains
VIIIa to XIIa  the short-dashed lines represent the one-loop results without the $U(1)_{R}\otimes U(1)_{B-L}$ mixing,
the long-dashed lines account for the complete one-loop results,
while the solid lines represent the two-loop solutions.
The scalar content at each stage corresponds to that
considered in Ref.~\cite{Deshpande:1992em},
namely to that reported in \Table{tab:submultiplets} without the $\phi^{126}$ multiplets.
For chains I to VII the two-step $SO(10)$ breaking consistent with minimal fine tuning is recovered in the $n_2 \to n_U$ limit. No solution is found for chain Xa.}
\label{fig:1to12a}
\end{figure*}

\begin{figure*}
 \centering
 \subfigure[\ Chain Ib]
   {\includegraphics[width=5cm]{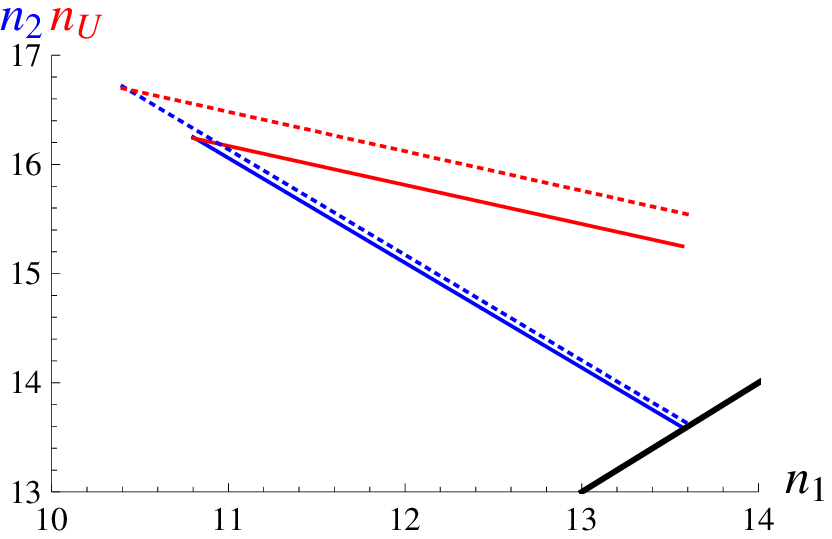}}
 \vspace{2mm}
 \subfigure[\ Chain IIb]
   {\includegraphics[width=5cm]{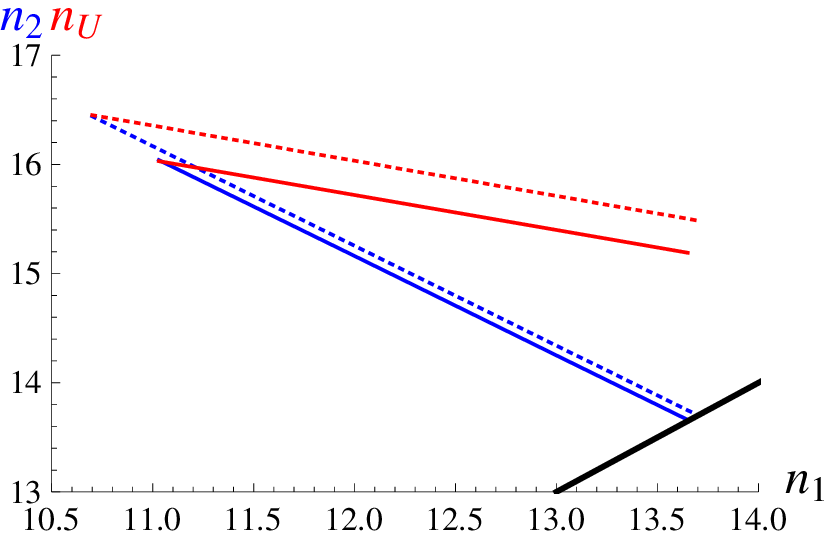}}
 \vspace{2mm}
 \subfigure[\ Chain IIIb]
   {\includegraphics[width=5cm]{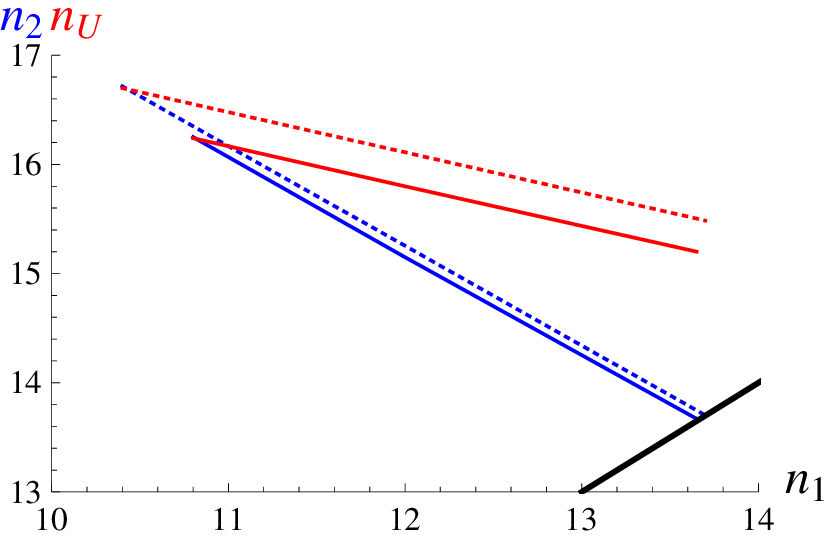}}
 \vspace{2mm}
\subfigure[\ Chain IVb]
   {\includegraphics[width=5cm]{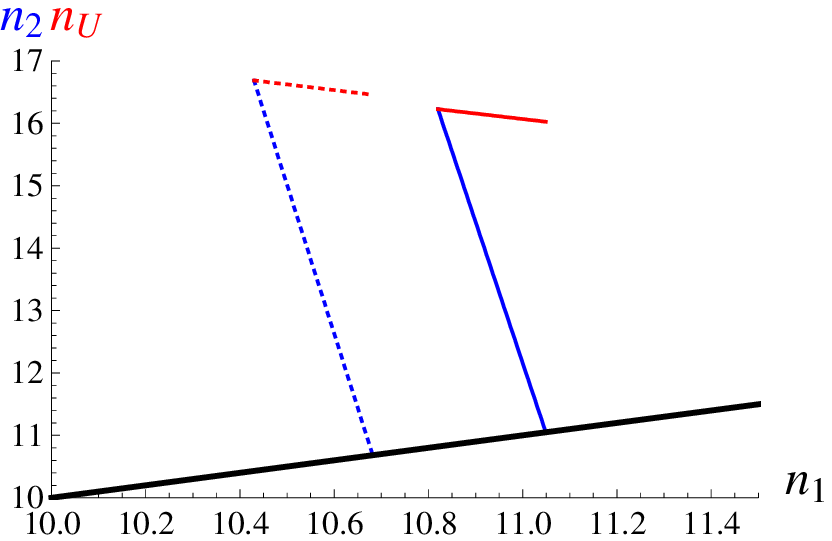}}
 \vspace{2mm}
 \subfigure[\ Chain Vb]
   {\includegraphics[width=5cm]{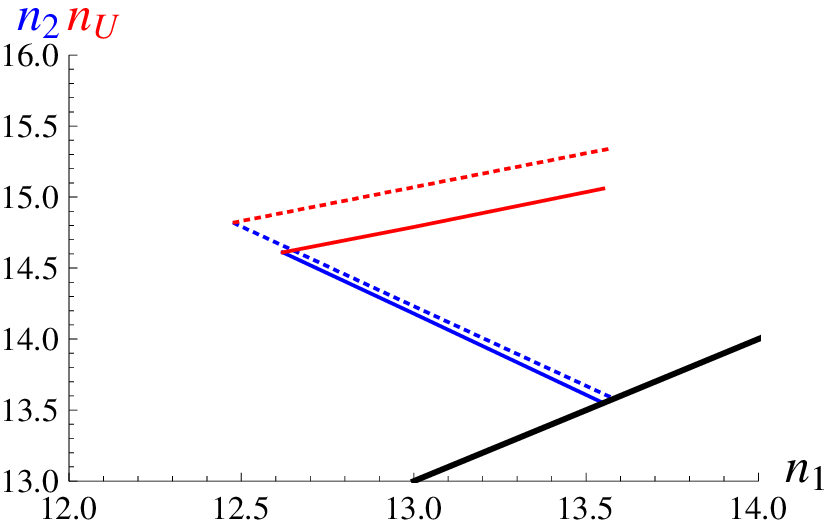}}
 \vspace{2mm}
 \subfigure[\ Chain VIb]
   {\includegraphics[width=5cm]{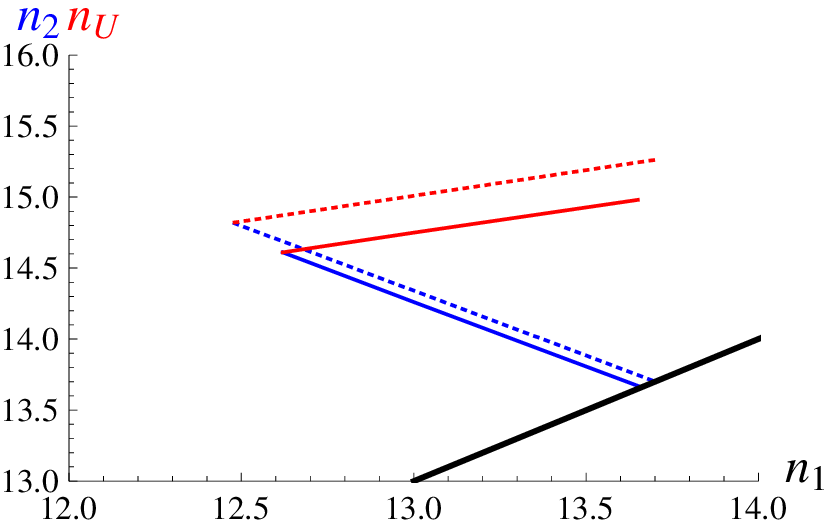}}
 \vspace{2mm}
 \subfigure[\ Chain VIIb]
   {\includegraphics[width=5cm]{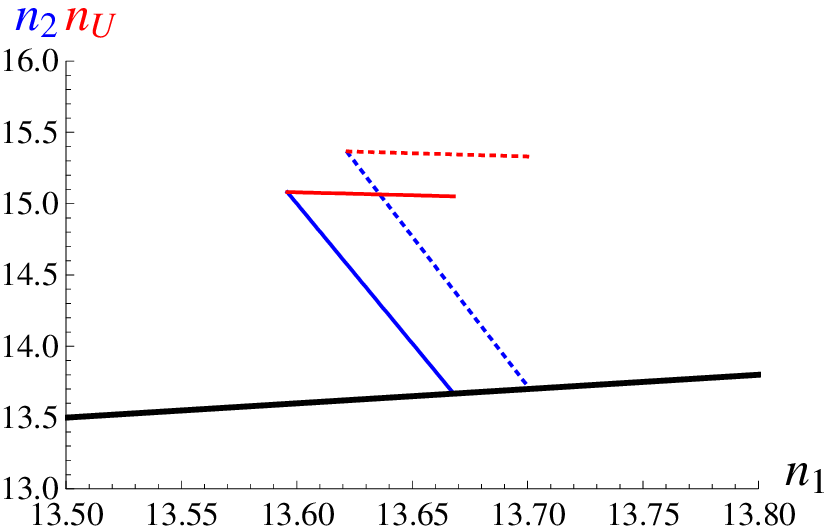}}
 \vspace{2mm}
 \subfigure[\ Chain VIIIb]
   {\includegraphics[width=5cm]{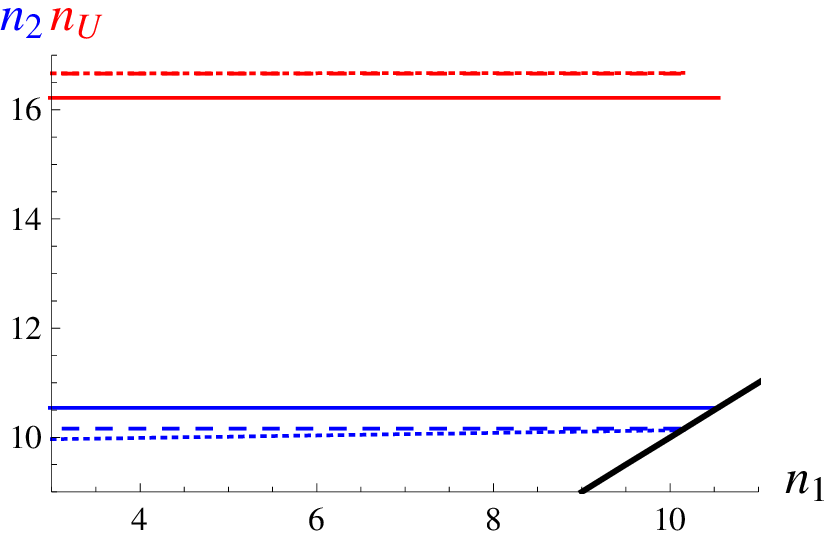}}
 \vspace{2mm}
 \subfigure[\ Chain IXb]
   {\includegraphics[width=5cm]{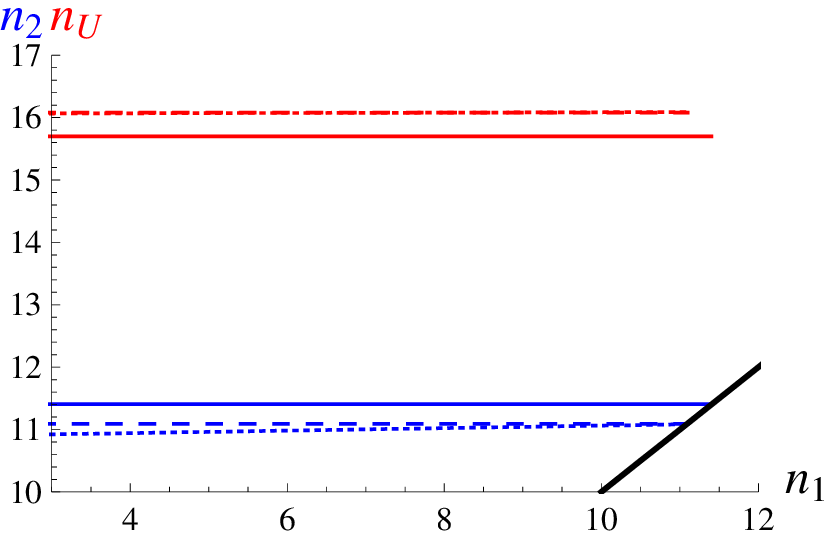}}
 \vspace{2mm}
\subfigure[\ Chain Xb]
   {\includegraphics[width=5cm]{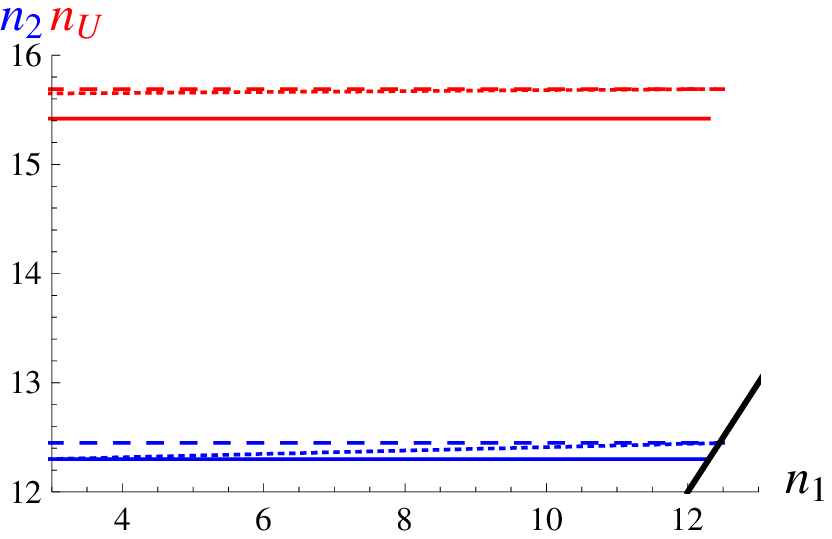}}
 \vspace{2mm}
 \subfigure[\ Chain XIb]
   {\includegraphics[width=5cm]{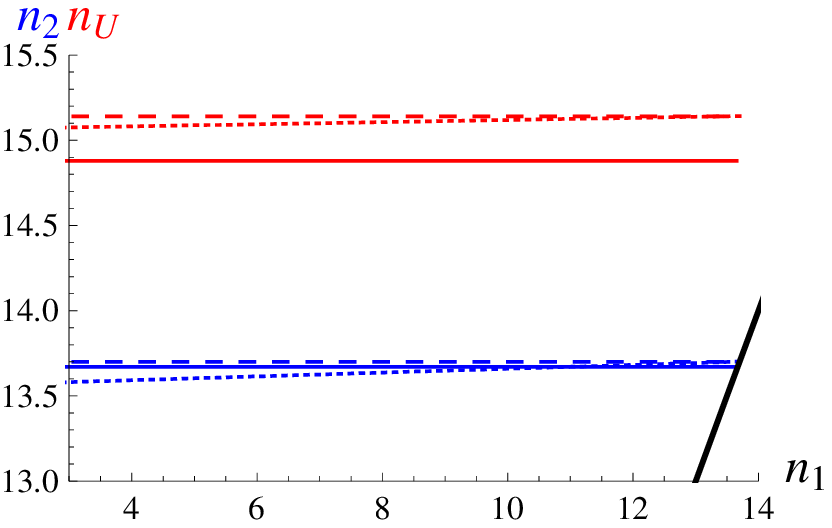}}
 \vspace{2mm}
 \subfigure[\ Chain XIIb]
   {\includegraphics[width=5cm]{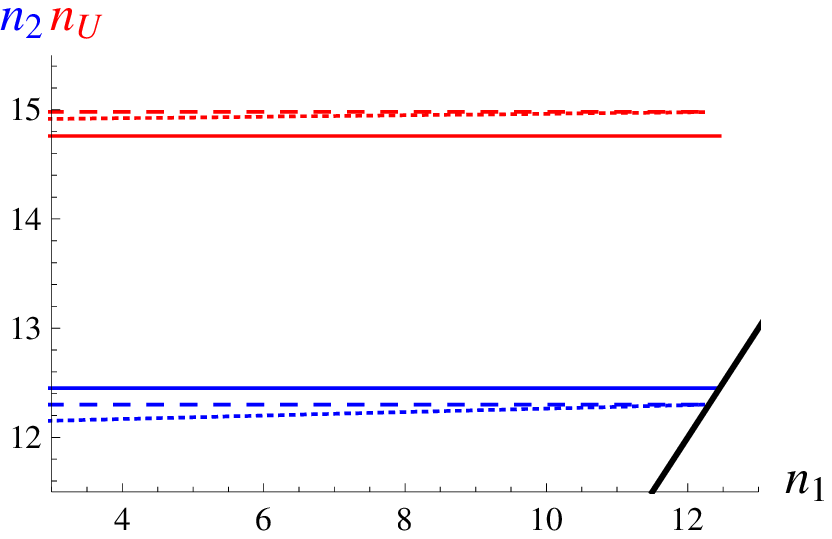}}
 \vspace{2mm}
\mycaption{Same as in \fig{fig:1to12a} for the $\overline{16}_H$ case.
}
\label{fig:1to12b}
\end{figure*}

The results of the numerical analysis are organized as follows:
\fig{fig:1to12a} and \fig{fig:1to12b} show the values of $n_U$ and $n_2$
as functions of $n_1$ for the pure gauge running (i.e. no Yukawa interactions),
in the $\overline{126}_H$ and $\overline{16}_H$ case respectively. The differences between the patterns for the $\overline{126}_H$ and
$\overline{16}_H$ setups depend on the substantially different scalar content.
The shape and size of the various contributions (one-loop, with and without $U(1)$ mixing,
and two-loops) are compared in each figure.
The dissection of the RGE results shown in the figures allows us
to compare our results with those
of Refs.~\cite{Gipson:1984aj,Chang:1984qr,Deshpande:1992au,Deshpande:1992em}.

\Table{tab:alphaU} shows the two-loop values of $\alpha_U^{-1}$ in the
allowed region for $n_1$.
The contributions of the additional $\phi^{126}$ multiplets,
and the Yukawa terms are discussed separately in \sect{sec:extrahiggs}
and \sect{sec:yukawaterms}, respectively.
With the exception of a few singular cases detailed therein, these effects turn out to be generally subdominant.

As already mentioned in the introduction, two-loop precision
in a GUT scenario makes sense once (one-loop) thresholds effects are coherently
taken into account, as their effect may become comparable if not
larger than the two loop itself (the argument becomes stronger as the number
of intermediate scales increases). On the other hand, there is no control
on the spectrum unless a specific model is studied in details.
The purpose of this chapter is to set the stage for such a study
by reassessing and updating the general constraints and patterns
that $SO(10)$ grand unification enforces on the spread of intermediate
scales.

The one and two-loop $\beta$-coefficients used in the present study are
reported in Appendix \ref{app:2Lbeta}. \Table{tab:betaChang} in the appendix shows
the reduced $\widetilde b_{ij}$ coefficients for those cases where we are at variance
with Ref.~\cite{Chang:1984qr}.

One of the largest effects in the comparison with
Refs.~\cite{Gipson:1984aj,Chang:1984qr,Deshpande:1992au,Deshpande:1992em}
emerges at one-loop and it is due to the implementation of the $U(1)$ gauge mixing
when $U(1)_R \otimes U(1)_{B-L}$ appears as an intermediate stage of the $SO(10)$
breaking\footnote{The lack of abelian gauge mixing in Ref.~\cite{Deshpande:1992em}
was first observed in Ref.~\cite{Lavoura:1993ut}.}.
This affects chains VIII to XII, and it exhibits itself in the exact (one-loop)
flatness of $n_2$, $n_U$ and $\alpha_U$ as functions of $n_1$.

The rationale for such a behaviour is quite simple.
When considering
the gauge coupling renormalization in the $3_C 2_L 1_R 1_{B-L}$ stage,
no effect at one-loop appears in the non-abelian $\beta$-functions
due to the abelian gauge fields. On the other hand,
the Higgs fields surviving at the $3_C 2_L 1_R 1_{B-L}$ stage, responsible
for the breaking to $3_C 2_L 1_Y$, are (by construction) SM singlets.
Since the SM one-loop $\beta$-functions are not affected by their presence,
the solution found for $n_2$, $n_U$ and $\alpha_U$ in the $n_1 = n_2$ case
holds for $n_1 < n_2$ as well.
Only by performing correctly the mixed $U(1)_R \otimes U(1)_{B-L}$
gauge running and the consistent matching with $U(1)_Y$
one recovers the expected $n_1$ flatness of the GUT solution.

In this respect, it is interesting to notice that the absence of $U(1)$ mixing in
Refs.~\cite{Gipson:1984aj,Chang:1984qr,Deshpande:1992au,Deshpande:1992em}
makes the argument for the actual possibility of a light (observable) $U(1)_R$ gauge boson
an ``approximate" statement (based on the approximate flatness of the solution).

One expects this feature to break at two-loops. The $SU(2)_L$ and $SU(3)_C$ $\beta$-functions
are affected at two-loops directly by the abelian gauge bosons via \eq{bpU12loops}
(the Higgs multiplets
that are responsible for the $U(1)_R\otimes U(1)_{B-L}$ breaking do not enter through the Yukawa interactions).
The net effect on the non-abelian gauge running is related to the difference between the
contribution of the $U(1)_R$ and $U(1)_{B-L}$ gauge bosons  and that of the standard hypercharge.
We checked that such a difference is always a small fraction (below 10\%) of the
typical two-loop contributions to the $SU(2)_L$ and $SU(3)_C$  $\beta$-functions.
As a consequence, the $n_1$ flatness of the GUT solution is at a very high accuracy 
($10^{-3}$) preserved at two-loops as well, as the inspection of the relevant chains
in \figs{fig:1to12a}{fig:1to12b} shows.

Still at one-loop we find a sharp disagreement in the $n_1$ range of chain XIIa,
with respect to the result of Ref.~\cite{Deshpande:1992em}.
The authors find $n_1 < 5.3$, while
strictly following their procedure and assumptions we find $n_1 < 10.2$
(the updated one- and two-loop results are given in \fig{fig:1to12a}k).
As we shall see, this difference brings chain XIIa back among the potentially realistic ones.

\begin{table}
\centering
\begin{tabular}{lcclc}
\hline
{\rm Chain}  &  {\rm $\alpha^{-1}_U$}   && {\rm Chain}  &  {\rm $\alpha^{-1}_U$}  \\
\hline
Ia  & $[45.5,46.4]$ && Ib  & $[45.7,44.8]$
\\
IIa  & $[43.7,40.8]$ && IIb  & $[45.3,44.5]$
\\
IIIa  & $[45.5,{40.8}]$ && IIIb  & $[45.7,{44.5}]$
\\
IVa  & $[45.5,43.4]$ && IVb  & $[45.7,45.1]$
\\
Va  & $[45.4,44.1]$ && Vb  & $[44.3,44.8]$
\\
VIa  & $[44.1,41.0]$ && VIb  & $[44.3,44.2]$
\\
VIIa  & $[45.4,41.1]$ && VIIb  & $[44.8,44.4]$
\\
VIIIa  & $45.4$ && VIIIb  & $45.6$
\\
IXa  & $42.8$ && IXb  & $44.3$
\\
Xa  & $ $ && Xb  & $44.8$
\\
XIa  & $38.7$ && XIb  & $41.5$
\\
XIIa  & $44.1$ && XIIb  & $44.3$
\\
\hline
\end{tabular}

\mycaption{Two-loop values of $\alpha^{-1}_U$ in the allowed region for $n_1$.
From chains I to VII, $\alpha^{-1}_U$ is $n_1$ dependent and its range is given in square brackets for the minimum (left) and the maximum (right) value of $n_1$ respectively.
For chains VIII to XII, $\alpha^{-1}_U$ depends very weekly on $n_1$ (see the discussion on $U(1)$ mixing in the text). No solution is found for chain Xa.}
\label{tab:alphaU}
\end{table}

As far as two-loop effects are at stakes, their relevance is generally related to the
length of the running involving the largest non-abelian groups.
On the other hand, there are chains where $n_2$ and $n_U$ have a strong dependence
on $n_1$ (we will refer to them as to ``unstable" chains) and where two-loop
corrections affect substantially the one-loop results.
Evident examples of such unstable chains are Ia, IVa, Va, IVb, and VIIb.
In particular, in chain Va the two-loop effects flip
the slopes of $n_{2}$ and $n_{U}$, that implies a sharp change in the allowed region for $n_1$.
It is clear that when dealing with these breaking chains any
statement about their viability should account for the details of the thresholds
in the given model.

In chains VIII to XII (where the second intermediate stage is $3_C 2_L 1_R 1_{B-L}$,
two-loop effects are mild
and exhibit the common behaviour
of lowering the GUT scale ($n_U$) while raising (with the exception of Xb and XIa,b)
the largest intermediate scale ($n_2$).
The mildness of two-loop corrections (no more that one would a-priori expect) is strictly
related to the ($n_1$) flatness of the GUT solution discussed before.

Worth mentioning are the limits $n_2\sim n_U$ and $n_1\sim n_2$.
While the former is equivalent to neglecting the first stage $G2$
and to reducing effectively the three breaking steps to just two
(namely $SO(10)\rightarrow G1\rightarrow SM$) with a minimal fine tuning in the scalar sector, care must be taken of the latter.
One may naively expect that the chains with the same G2
should exhibit for $n_1\sim n_2$ the same numerical behavior ($SO(10)\rightarrow G2\rightarrow SM$), thus
clustering the chains (I,V,X), (II,III,VI,VII,XI) and (IV,IX).
On the other hand, one must recall that the existence of G1 and its breaking
remain encoded in the G2 stage through the Higgs scalars that are responsible
for the G2$\to$G1 breaking. This is why the chains with the same G2 are
not in general equivalent in the $n_1\sim n_2$ limit.
The numerical features of the degenerate patterns (with $n_2\sim n_U$) can be
crosschecked among the different chains by direct
inspection of \figs{fig:1to12a}{fig:1to12b} and \Table{tab:alphaU}.

In any discussion of viability of the various scenarios the main attention is paid to
the constraints emerging from the proton decay.
In nonsupersymmetric GUTs this process is dominated by baryon number violating
gauge interactions, inducing at low energies
a set of effective $d = 6$ operators that conserve $B-L$ (cf.~\sect{d6gauge}).
In the $SO(10)$ scenarios we consider here
such gauge bosons are integrated out at the unification scale and therefore
proton decay constrains $n_U$ from below. 
Considering the naive estimate of the inverse lifetime of the proton in~\eq{appinvplt},
the present experimental limit $\tau (p \to \pi^0 e^+) \ > \ 8.2 \times 10^{33}$ yr~\cite{Nakamura:2010zzi}
yields $n_U \gtrsim 15.4$, for $\alpha^{-1}_{U}= 40$.
Taking the results in \figs{fig:1to12a}{fig:1to12b} and \Table{tab:alphaU} at face value the chains VIab, XIab, XIIab, Vb and VIIb should be excluded from realistic considerations.

On the other hand, one must recall that once a specific model is scrutinized
in detail there can be large threshold corrections in the
matching~\cite{Dixit:1989ff,Mohapatra:1992dx,Lavoura:1993su},
that can easily move the unification scale by a few orders of magnitude (in both directions).
In particular, as a consequence of the spontaneous breaking of accidental would-be
global symmetries of the scalar
potential, pseudo-Goldstone modes (with
masses further suppressed with respect to the expected threshold range) may appear in the scalar spectrum, leading to
potentially large RGE effects~\cite{Aulakh:1982sw}.
Therefore, we shall follow a conservative approach in interpreting the limits on the
intermediate scales coming from a simple threshold clustering. These limits, albeit
useful for a preliminary survey, may not be sharply used to exclude marginal
but otherwise well motivated scenarios.

Below the scale of the $B-L$ breaking, processes that violate separately the barion or the lepton
numbers emerge.
In particular,
$\Delta B = 2$ effective interactions give rise to the phenomenon of neutron oscillations
(for a recent review see Ref.~\cite{Mohapatra:2009wp}).
Present bounds on nuclear instability give $\tau_{N}> 10^{32}$ years, which
translates into a bound on the neutron oscillation time $\tau_{n-\bar n} > 10^8$ sec.
Analogous limits come from direct reactor oscillations experiments.
This sets a lower bound on the scale of $\Delta B = 2$ nonsupersymmetric ($d = 9$) operators
that varies from 10 to 300 TeV depending on model couplings.
Thus, neutron-antineutron
oscillations probe scales far below the unification scale. In a supersymmetric
context the presence of $\Delta B = 2$
$d = 7$ operators softens the dependence on the $B-L$ scale and for the present bounds
the typical limit goes up to about $10^7$ GeV.

Far more reaching in scale sensitivity are the $\Delta L = 2$
neutrino masses emerging from the see-saw mechanism.
At the $B-L$ breaking scale the $\Delta^{126}_R$ ($\delta^{16}_R$) scalars
acquire $\Delta L = 2$ ($\Delta L = 1$) VEVs
that give a Majorana mass to the right-handed neutrinos. Once the latter are integrated out,
$d = 5$ operators of the form $(\ell^T \epsilon_2 H) C (H^T \epsilon_2 \ell)/\Lambda_L$
generate light Majorana neutrino states in the low
energy theory.

In the type-I seesaw,  the neutrino mass matrix $m_{\nu}$ is proportional
to $Y_D M_R^{-1} Y_D^Tv^{2}$ where the largest entry
in the Yukawa couplings is typically of the order of the top quark one and $M_{R}\sim M_{1}$.
Given a neutrino mass above the limit obtained from atmospheric neutrino oscillations and below the eV, one infers a (loose) range 
$10^{13}\ \mbox{GeV} < M_1 < 10^{15}\ \mbox{GeV}$.
It is interesting to note that the lower bound pairs with the cosmological limit
on the D-parity breaking scale (see \sect{sec:chains}).

In the scalar-triplet induced (type-II) seesaw the evidence of the neutrino mass
entails a lower bound on the VEV of the heavy $SU(2)_{L}$ triplet in $\overline{126}_{H}$. 
This translates into an upper bound on the
mass of the triplet that depends on the structure of the relevant Yukawa coupling.
If both type-I as well as type-II contribute to the light neutrino mass,
the lower bound on the $M_{1}$ scale may then be weakened by the interplay between
the two contributions. Once again this can be quantitatively assessed only when the
vacuum of the model is fully investigated.

Finally, it is worth noting that if the $B-L$ breakdown is driven by $\overline{126}_{H}$,
the elementary triplets couple to the Majorana currents at the renormalizable level and $m_{\nu}$
is directly sensitive to the position of the $G1\to SM$ threshold $M_{1}$.
On the other hand, the $n_{1}$-dependence of $m_{\nu}$ is loosened in the b-type of chains
due to the non-renormalizable nature of the relevant effective operator
$16_{F}16_{F}\overline{16}_{H}\overline{16}_{H}/\Lambda$, where the effective
scale $\Lambda>M_{U}$ accounts for an extra suppression.

With these considerations at hand, the constraints from proton decay and the see-saw
neutrino scale favor the chains II, III and VII, which all share $4_C 2_L 2_R P$
in the first $SO(10)$ breaking stage~\cite{Bajc:2005zf}.
On the other hand, our results rescue from oblivion
other potentially interesting scenarios
that, as we shall expand upon shortly, are worth of in depth consideration.
In all cases, the bounds on the $B-L$ scale enforced by the see-saw neutrino mass
excludes the possibility of observable $U(1)_R$ gauge bosons.

\subsection{The $\phi^{126}$ Higgs multiplets}
\label{sec:extrahiggs}

As mentioned in \sect{sec:ESH},
in order to ensure a rich enough Yukawa sector in realistic models
there may be the need to keep more than one $SU(2)_{L}$ Higgs doublet at
intermediate scales, albeit at the price of an extra fine-tuning.
A typical example is the
case of a relatively low Pati-Salam breaking scale where
one needs at least a pair of $SU(2)_{L}\otimes SU(2)_{R}$ bidoublets with different $SU(4)_{C}$ quantum numbers to
transfer the information about the PS breakdown into the matter sector. Such
additional Higgs multiplets are those labelled by $\phi^{126}$ in \Table{tab:submultiplets}.

\Table{tab:phi126chains} shows the effects of including $\phi^{126}$ at the
$SU(4)_{C}$ stages of the relevant breaking chains. The two-loop
results at the extreme values of the intermediate scales, with and without
the $\phi^{126}$ multiplet, are compared.
In the latter case the complete functional dependence among
the scales is given in \fig{fig:1to12a}. Degenerate patterns
with only one effective intermediate stage are easily crosschecked among the different chains in \Table{tab:phi126chains}.

In most of the cases, the numerical results do not exhibit a sizeable dependence on the additional $(15,2,2)_{\overline{126}}$ (or $(15,2,+\frac{1}{2})_{\overline{126}}$)
scalar multiplets. The reason can be read off
\Table{tab:phi126betas} in Appendix \ref{app:2Lbeta} and it rests on an
accidental approximate coincidence of the $\phi^{126}$ contributions to the $SU(4)_{C}$ and $SU(2)_{L,R}$ one-loop beta coefficients
(the same argument applies to the $4_{C}2_{L}1_{R}$ case).

Considering for instance the $4_{C}2_{L}2_{R}$ stage, one obtains
$\Delta a_{4}=\frac{1}{3} \times 4\times T_{2}(15)=\frac{16}{3}$,
and $\Delta a_{2}=\frac{1}{3}\times 30 \times T_{2}(2)=5$, that only slightly affects
the value of $\alpha_U$ (when the PS scale is low enough), but has generally
a negligible effect on the intermediate scales.

\begin{table}
\centering
\begin{tabular}{lcccc}
\hline
{\rm Chain}     & {\rm $n_1$}  & {\rm $n_2$}  & {\rm $n_U$}  & {\rm $\alpha^{-1}_U$}
\\
\hline
Ia              & [9.50, 10.0] & [16.2, 10.0] & [16.2, 17.0] & [45.5, 46.4]
\\
                & [8.00, 9.50] & [10.4, 16.2] & [18.0, 16.2] & [30.6, {45.5}]
\\
IIa             & [10.5, 13.7] & [15.4, 13.7] & [15.4, 15.1] & [43.7, 40.8]
\\
                & [10.5, 13.7] & [15.4, 13.7] & [15.4, 15.1] & [43.7, 37.6]
\\
IIIa            & [9.50, 13.7] & [16.2, 13.7] & [16.2, 15.1] & [45.5, {40.8}]
\\
                & [9.50, 13.7] & [16.2, 13.7] & [16.2, 15.1] & [45.5, {37.6}]
\\
Va              & [11.0, 11.4] & [11.0, 14.4] & [15.9, 14.4] & [45.4, 44.1]
\\
                & [10.1, 11.2] & [10.1, 14.5] & [16.5, 14.5] & [32.5, 40.8]
\\
VIa             & [11.4, 13.7] & [14.4, 13.7] & [14.4, 14.9] & [44.1, 41.0]
\\
                & [11.2, 13.7] & [14.5, 13.7] & [14.5, 14.9] & [40.8, 38.1]
\\
VIIa            & [11.3, 13.7] & [15.9, 13.7] & [15.9, 14.9] & [45.4, 41.1]
\\
                & [10.5, 13.7] & [16.5, 13.7] & [16.5, 15.0] & [33.3, 38.1]
\\
XIa             & [3.00, 13.7] & [13.7, 13.7] & [14.8, 14.8] & [38.7, 38.7]
\\
                & [3.00, 13.7] & [13.7, 13.7] & [14.8, 14.8] & [36.0, 36.0]
\\
XIIa            & [3.00, 10.8] & [10.8, 10.8] & [14.6, 14.6] & [44.1, 44.1]
\\
                & [3.00, 10.5] & [10.5, 10.5] & [14.7, 14.7] & [39.8, 39.8]
\\

\hline
\end{tabular}
\mycaption{Impact of the additional multiplet $\phi^{126}$
(second line of each chain) on those chains that contain the gauge
groups $4_C2_L2_R$ or $4_C2_L1_R$ as intermediate stages,
and whose breaking to the SM is obtained via a $\overline{126}_H$ representation.
The values of $n_2$, $n_U$ and $\alpha^{-1}_U$ are showed for
the minimum and maximum values allowed for $n_1$ by the two-loop analysis.
Generally the effects on the intermediate scales
are below the percent level,
with the exception of chains Ia and Va that are most sensitive to
variations of the $\beta$-functions.}
\label{tab:phi126chains}
\end{table}

An exception to this argument is observed in chains
Ia and Va that, due to their $n_{2,U}(n_1)$ slopes, are most sensitive to variations of the $\beta$-coefficients. In particular, the inclusion of $\phi^{126}$ in the Ia chain flips at two-loops
the slopes of $n_2$ and $n_U$ so that the limit $n_2 = n_U$ (i.e. no G2 stage) is obtained for the maximal value of $n_1$ (while the same happens for the minimum $n_1$ if there is no $\phi^{126}$).

\begin{figure}
 \centering
 \subfigure[\ Chain Ia]
   {\includegraphics[width=5cm]{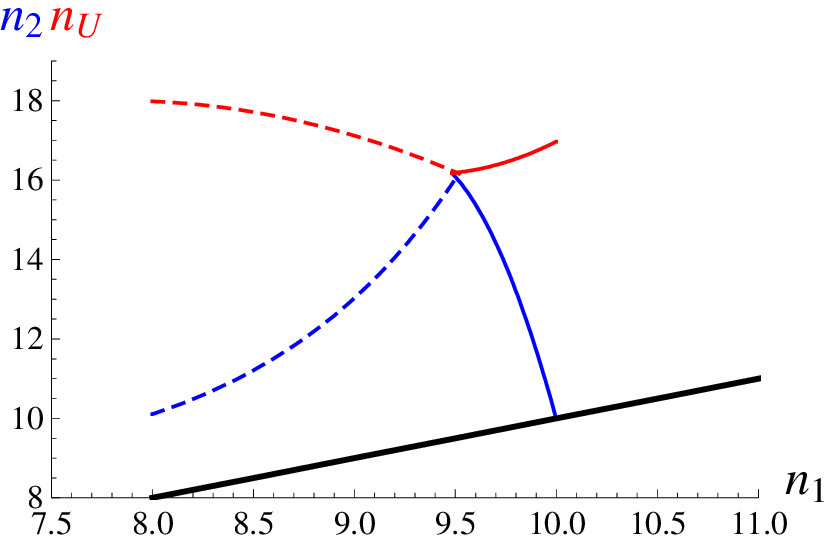}}
 \vspace{2mm}
 \subfigure[\ Chain Va]
   {\includegraphics[width=5cm]{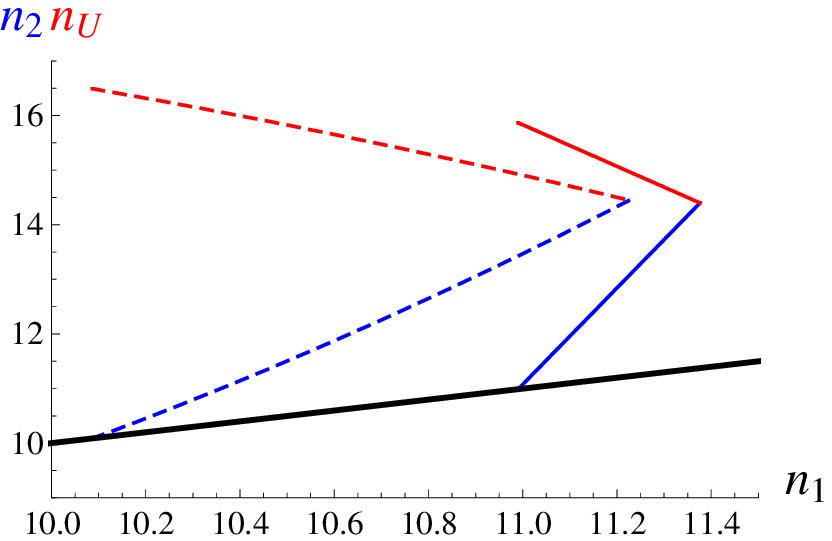}}
 \vspace{2mm}
 \subfigure[\ Chain VIIa]
   {\includegraphics[width=5cm]{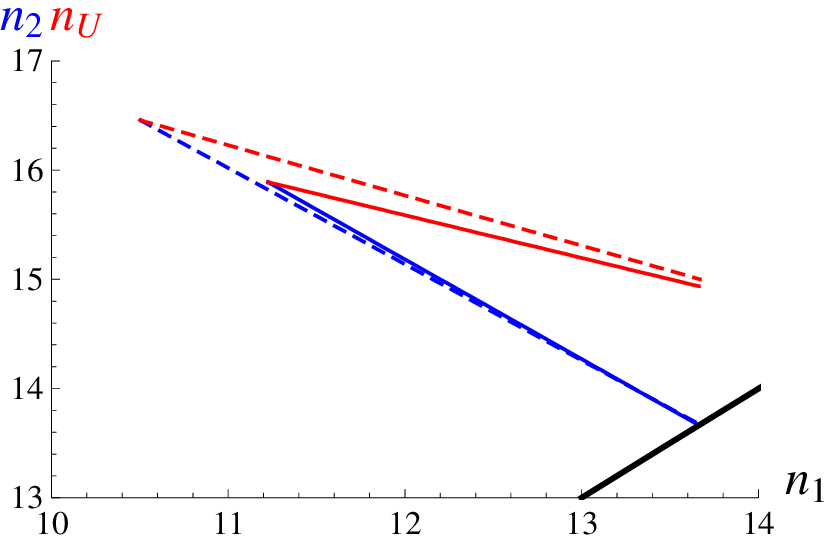}}
\mycaption{Example of chains with sizeable $\phi^{126}$ effects (long-dashed curves)
on the position of the intermediate scales.
The solid curves represent the two-loop results in \fig{fig:1to12a}.
The most dramatic effects appear in the chain Ia, while moderate scale shifts
affect chain Va (both ``unstable" under small variations of the $\beta$-functions).
Chain VIIa, due to the presence of two PS stages, is the only "stable''
chain with visible $\phi^{126}$ effects.
}
\label{fig:phi126}
\end{figure}

\fig{fig:phi126} shows three template cases where the $\phi^{126}$ effects are visible.
The highly unstable Chain Ia shows, as noticed earlier, the largest effects.
In chain Va the
effects of $\phi^{126}$ are moderate. Chain VII is the only "stable" chain that
exhibits visible effects on the intermediate scales.
This is due to the presence of two full-fledged PS stages.

\subsection{Yukawa terms}
\label{sec:yukawaterms}


The effects of the Yukawa couplings can be at leading order approximated by constant negative shifts of the one-loop $a_{i}$ coefficients $a_{i}\to a_{i}'=a_{i}+\Delta a_{i} $ with
\begin{equation}
\Delta a_{i}=-\frac{1}{(4\pi)^{2}}y_{ik}\text{Tr}\,Y_{k}\,Y_{k}^{\dagger}\,.
\end{equation}
The impact of $\Delta a_{i}$ on the position of the unification scale and the value of the unified coupling can be simply estimated by considering the running induced by the Yukawa couplings from a scale $t$ up to the unification point ($t=0$). The one-loop result for the change of the intersection of the curves corresponding to
$\alpha_{i}^{-1}(t)$ and $\alpha_{j}^{-1}(t)$ reads (at the leading order in
$\Delta a_{i}$):
\begin{equation}
\Delta t_{U}= 2\pi \frac{\Delta a_{i}-\Delta a_{j}}{(a_{i}-a_{j})^{2}}\left[\alpha_{j}^{-1}(t_{})-\alpha_{i}^{-1}(t_{})\right]+\ldots
\end{equation}
and
\be
\Delta \alpha_{U}^{-1} = \frac{1}{2}\left[\frac{\Delta a_{i}+\Delta a_{j}}{a_{i}-a_{j}}-\frac{(a_{i}+a_{j})(\Delta a_{i}-\Delta a_{j})}{(a_{i}-a_{j})^{2}}\right]
\left[\alpha_{j}^{-1}(t_{})-\alpha_{i}^{-1}(t_{})\right]+\ldots
\label{eq:deltaalphaUinv}
\ee
for any $i\neq j$.
For simplicity we have neglected the changes in the $a_{i}$ coefficients due to crossing intermediate thresholds.
It is clear that for a common change $\Delta a_{i}=\Delta a_j$ the unification scale is not affected,
while a net effect remains on $\alpha_{U}^{-1}$.
In all cases, the leading contribution is always proportional to
$\alpha_{j}^{-1}(t_{})-\alpha_{i}^{-1}(t_{})$ (this holds exactly for
$\Delta t_U$).

In order to assess quantitatively such effects we shall consider first the SM stage that accounts for a large part of the running in all realistic chains.
The case of a low $n_{1}$ scale
leads, as we explain in the following, to comparably smaller effects.
The impact of the Yukawa interactions on the gauge RGEs is readily estimated assuming only the up-type Yukawa contribution to be sizeable and constant,
namely $\text{Tr}\,Y_{U}\,Y_{U}^{\dagger}\sim 1$.
This yields $\Delta a_{i}\sim -6\times 10^{-3}y_{iU}$, where
the values of the $y_{iU}$ coefficients are given in Table~\ref{tab:Yukawas}.
For $i=1$ and $j=2$ one obtains $\Delta a_{1}\sim -1.1\times 10^{-2}$ and $\Delta a_{2}\sim -0.9\times 10^{-2}$ respectively.
Since $a_{1}^{SM}=\frac{41}{10}$ and $a_{2}^{SM}=-\frac{19}{6}$, the first term in (\ref{eq:deltaalphaUinv}) dominates and one finds $\Delta \alpha_{U}^{-1}\sim 0.04$.
For a typical value of $\alpha_{U}^{-1}\sim 40$ this translates into $\Delta \alpha_{U}^{-1}/\alpha_{U}^{-1}\sim 0.1\%$.
The impact on $t_U$ is indeed tiny, namely $\Delta n_{U}\sim -1\times 10^{-2}$.
In both cases the estimated effect agrees to high accuracy
with the actual numerical behavior we observe.

The effects of the Yukawa interactions emerging at intermediate scales (or of a non-negligible $\text{Tr }Y_{D}\,Y_{D}^{\dagger}$ in a two Higgs doublet settings with large $\tan\beta$)
can be analogously accounted for. As a matter of fact, in the $SO(10)$ type of  models
$\text{Tr }Y_{N}\,Y_{N}^{\dagger}\sim \text{Tr }Y_{U}\,Y_{U}^{\dagger}$
due to the common origin of $Y_{U}$ and $Y_{N}$.
The unified structure of the Yukawa sector yields therefore
homogeneous $\Delta a_{i}$ factors
(see the equality of $\sum_{k}y_{ik}$ in \Table{tab:Yukawas}).
This provides the observed large suppression of the Yukawa effects on threshold scales and unification compared to typical two-loop gauge contributions.

In summary, the two-loop RGE effects due to Yukawa couplings on the magnitude of the unification scale (and intermediate thresholds)
and the value of the GUT gauge coupling turn out to be very small. Typically we observe
negative shifts at the per-mil level in both $n_U$
and $\alpha_{U}$, with no relevant impact on the gauge-mediated proton decay rate.

\subsection{The privilege of being minimal}
\label{sec:discussion}

With all the information at hand we can finally approach an assessment of the
viability of the various scenarios.
As we have argued at length, we cannot discard a marginal
unification setup without a detailed information on the fine threshold structure.

Obtaining this piece of information involve the study of the vacuum
of the model, and for $SO(10)$ GUTs this is in general a most challenging task.
In this respect supersymmetry helps: the superpotential is holomorphic and the couplings in the renormalizable case are limited to at most cubic terms;
the physical vacuum is constrained by GUT-scale $F$- and $D$-flatness
and supersymmetry may
be exploited to studying the fermionic rather than the scalar spectra.

It is not surprising that for nonsupersymmetric $SO(10)$, only a few detailed studies of the Higgs potential and the related threshold effects
(see for instance
Refs.~\cite{Yasue:1980fy,Harvey:1981hk,Anastaze:1983zk,Babu:1984mz,Abud:1989uu})
are available.
In view of all this and of the intrinsic predictivity related to minimality, the relevance of
carefully scrutinizing the simplest scenarios is hardly overstressed.

The most economical $SO(10)$ Higgs sector includes the adjoint $45_{H}$, that provides the breaking of the GUT symmetry, either $\overline{16}_{H}$ or $\overline{126}_{H}$, responsible for the subsequent $B-L$ breaking, and $10_{H}$, participating to the electroweak symmetry breaking.
The latter is needed together with $\overline{16}_{H}$ or $\overline{126}_{H}$
in order to obtain realistic patterns for the fermionic masses and mixing.
Due to the properties of the adjoint representation this scenario exhibits a minimal
number of parameters in the Higgs potential.
In the current notation such a minimal nonsupersymmetric $SO(10)$ GUT
corresponds to the chains VIII and XII.

From this point of view, it is quite intriguing that our analysis of the gauge unification constraints improves the standing of these chains (for XIIa dramatically) with respect to
existing studies.
In particular, considering the renormalizable setups ($\overline{126}_H$), we find
for chain VIIIa, $n_1 \leq 9.1$, $n_U=16.2$ and $\alpha^{-1}_U=45.4$
(to be compared to $n_1 \leq 7.7$ given in Ref.~\cite{Deshpande:1992em}).
This is due to the combination of the updated weak scale data and two loop running effects.
For chain XIIa we find $n_1 \leq 10.8$, $n_U=14.6$ and $\alpha^{-1}_U=44.1$, showing a dramatic (and pathological) change from $n_{1} \leq 5.3$ obtained in \cite{Deshpande:1992em}.
Our result sets the $B-L$ scale nearby the needed scale for realistic light neutrino masses.

We observe non-negligible two-loop effects for the chains VIIIb and XIIb
($\overline{16}_H$) as well.
For chain VIIIb we obtain $n_1 \leq 10.5$, $n_U=16.2$ and $\alpha^{-1}_U=45.6$
(that lifts the $B-L$ scale while preserving $n_{U}$ well above the proton decay bound \eq{pdecaybound}).
A similar shift in $n_{1}$ is observed in chain XIIb where we find
$n_1 \leq 12.5$, $n_U=14.8$ and $\alpha^{-1}_U=44.3$.
As we have already stressed one should not too readily discard $n_U=14.8$ as being incompatible with the proton decay bound. We have verified that reasonable GUT threshold patterns exist that easily lift $n_{U}$ above the experimental bound.
For all these chains D-parity is broken at the GUT scale thus avoiding any cosmological issues
(see the discussion in \sect{sec:chains}).

As remarked in \sect{sec:2loopgauge}, the limit $n_1 = n_2$ leads
to an effective two-step $SO(10)\to \mbox{G2}\to \mbox{SM}$ breaking with a non-minimal set of surviving scalars at the G2 stage. As a consequence,
the unification setup for the minimal scenario can be recovered (with the needed minimal fine tuning) by considering the limit $n_2 = n_U$ in those
chains among I to VII where G1 is either $3_C 2_L 2_R 1_{B-L}$ or $4_C 2_L 1_R$
(see \Table{tab:chains}). From inspection of \figs{fig:1to12a}{fig:1to12b} and of \Table{tab:alphaU}, one reads the following results:
for $SO(10)\chain{}{45} 3_C 2_L 2_R 1_{B-L} \to \mbox{SM}$
we find
\begin{itemize}
\item $n_1 = 9.5$, $n_U=16.2$ and $\alpha^{-1}_U=45.5$ (case $a$),
\item $n_1 = 10.8$ $n_U=16.2$ and $\alpha^{-1}_U=45.7$ (case $b$),
\end{itemize}
while for $SO(10)\chain{}{45} 4_C 2_L 1_R \to \mbox{SM}$
\begin{itemize}
\item $n_1 = 11.4$, $n_U=14.4$ and $\alpha^{-1}_U=44.1$ (case $a$),
\item $n_1 = 12.6$, $n_U=14.6$ and $\alpha^{-1}_U=44.3$ (case $b$).
\end{itemize}
We observe that the patterns are quite similar to those of the non-minimal setups
obtained from chains VIII and XII in the $n_1=n_2$ limit.
Adding the $\phi^{126}$ multiplet, as required by a realistic
matter spectrum in case $a$, does not modify the scalar content in the $3_C 2_L 2_R 1_{B-L}$ case:
only one linear combination of the $10_H$ and $\overline{126}_H$ bidoublets (see \Table{tab:submultiplets})
is allowed by minimal fine tuning. On the other hand, in the $4_C 2_L 1_R$ case,
the only sizeable effect is a shift on the unified coupling constant, namely $\alpha^{-1}_U=40.7$
(see the discussion in \sect{sec:extrahiggs}).

In summary, in view of realistic thresholds effects at the GUT (and $B-L$) scale and of
a modest fine tuning in the see-saw neutrino mass, we consider both scenarios
worth of a detailed investigation.

\chapter{The quantum vacuum of the minimal $SO(10)$ GUT}
\label{thequantumvac}

\section{The minimal SO(10) Higgs sector}
\label{sect:minimalSO10}

In this chapter we consider a nonsupersymmetric $SO(10)$ setup
featuring the minimal Higgs content sufficient to trigger the spontaneous breakdown of the GUT symmetry down to the
standard electroweak model.
Minimally, the scalar sector spans over a reducible $45_{H}\oplus 16_{H}$ representation. The adjoint $45_{H}$ and the spinor $16_{H}$ multiplets contain three SM singlets that may acquire GUT
scale VEVs.


As we have seen in Chapter \ref{intermediatescales}
the phenomenologically favored scenarios allowed by gauge
coupling unification correspond to a three-step breaking along one of the following directions:
\begin{eqnarray}
\label{chainVIII}
SO(10)&\stackrel{M_U}{\longrightarrow}
& 3_{C}\, 2_{L}\, 2_{R}\, 1_{B-L}\ \stackrel{M_I}{\longrightarrow}\
3_{C}\, 2_{L}\, 1_{R}\, 1_{B-L}
\ \stackrel{M_{B-L}}{\longrightarrow}\ \mbox{SM}
\,,\\[1ex]
\label{chainXII}
SO(10)&\stackrel{M_U}{\longrightarrow}
&4_{C}\, 2_{L}\, 1_{R}\ \stackrel{M_I}{\longrightarrow}\
3_{C}\, 2_{L}\, 1_{R}\, 1_{B-L}
\ \stackrel{M_{B-L}}{\longrightarrow}\ \mbox{SM}
\,,
\end{eqnarray}
where the first two breaking stages at $M_U$ and $M_I$ are driven by the $45_H$ VEVs, while the breaking to the SM at the intermediate scale $M_{B-L}$ 
is controlled by the $16_H$. 
The constraints coming from gauge unification are such that $M_U \gg M_I > M_{B-L}$.
In particular, even without proton decay limits, any
intermediate $SU(5)$-symmetric stage is excluded. 
On the other hand,
a series of studies in the early 1980's of the $45_H\oplus 16_H$ model \cite{Yasue:1980fy,Yasue:1980qj,Anastaze:1983zk,Babu:1984mz} indicated that
the only intermediate stages allowed by the scalar sector dynamics were
the flipped $SU(5)\otimes U(1)$ for leading $45_H$ VEVs or the standard $SU(5)$ GUT
for  dominant $16_H$ VEV.
This observation excluded
the simplest $SO(10)$ Higgs sector from realistic consideration.

In this chapter we show that the exclusion of the breaking patterns
in \eqs{chainVIII}{chainXII} is an artifact of the tree level potential. As a matter of fact,
some entries of the scalar hessian
are accidentally
over-constrained at the tree level. A number of scalar interactions that, by a simple inspection of the relevant global symmetries and their explicit breaking, are expected to contribute to these critical entries, are not effective at the tree level.

On the other hand, once quantum corrections are considered,
contributions of $O(M_U^2/16\pi^2)$ induced on these entries
open in a natural way
all group-theoretically allowed vacuum configurations.
Remarkably enough, the study of the one-loop effective potential
can be consistently carried out just for the critical tree level hessian entries (that correspond to specific pseudo-Goldstone boson masses).
For all other states in the scalar spectrum, quantum corrections
remain perturbations of the tree level results and do not affect
the discussion of the vacuum pattern.


Let us emphasize that the issue we shall be dealing with is inherent to all nonsupersymmetric $SO(10)$ models with one adjoint $45_H$ governing the first breaking step. Only one additional scalar representation interacting with the adjoint is sufficient to demonstrate conclusively our claim.
In this respect, the choice of the $SO(10)$ spinor to trigger the intermediate symmetry breakdown is a mere convenience and a similar line of reasoning can be devised for the scenarios in which $B-L$ is broken for instance by a 126-dimensional $SO(10)$ tensor.

We shall therefore study the structure of the vacua of a $SO(10)$ Higgs potential with only the $45_{H} \oplus 16_{H}$ representation at play.
Following the common convention, we define $16_{H} \equiv \chi$ and denote by $\chi_{+}$ and $\chi_{-}$ the multiplets transforming as positive and negative chirality components of the reducible 32-dimensional $SO(10)$ spinor representation respectively.
Similarly, we shall use the symbol $\Phi$ (or the derived $\phi$ for the components in the natural basis, cf.~\app{app:so10algebra}) for the adjoint Higgs representation $45_{H}$.


\subsection{The tree-level Higgs potential}
The most general renormalizable tree-level scalar potential which
can be constructed out of $45_H$ and $16_H$ reads (see for instance
Refs.~\cite{Li:1973mq,Buccella:1980qb}):
\be
V_0=V_{\Phi}+V_{\chi}+V_{\Phi\chi} \, ,
\label{potentialV0}
\ee
where, according to the notation in \app{app:so10algebra},
\bea
\label{potentialV45}
V_{\Phi}&=&
-\frac{\mu^2}{2}\Tr\Phi^2 + \frac{a_1}{4}(\Tr\Phi^2)^2 + \frac{a_2}{4}\Tr\Phi^4 \, , \\
\label{potentialV16}
V_{\chi}&=&
-\frac{\nu^2}{2}\chi^\dag\chi
+\frac{\lambda_1}{4}(\chi^\dag\chi)^2
 +\frac{\lambda_2}{4}(\chi_+^\dag\Gamma_j\chi_-)(\chi_-^\dag\Gamma_j\chi_+) \,\nn
 \eea
and
\be
\label{potentialV4516}
V_{\Phi\chi}=
\alpha(\chi^\dag\chi)\Tr\Phi^2+\beta\chi^\dag\Phi^2\chi
+\tau\chi^\dag\Phi\chi \, .
\ee
The mass terms and coupling constants above are real by hermiticity.
The cubic $\Phi$ self-interaction is absent
due the zero trace of the $SO(10)$ adjoint representation.
For the sake of simplicity, all tensorial indices have been suppressed.

\subsection{The symmetry breaking patterns}
\label{sec:breakingpatterns}

\subsubsection{The SM singlets}
\label{sec:SMsinglet}

There are in general three SM singlets in the $45_H\oplus16_H$ representation of $SO(10)$. 
Labeling the field components according to
$3_{C}\, 2_{L}\, 2_{R}\, 1_{B-L}$ (where the $U(1)_{B-L}$ generator is $(B-L)/2$), the SM singlets
reside in the $(1,1,1,0)$ and $(1,1,3,0)$ submultiplets of $45_{H}$
and in the $(1,1,2,+\tfrac{1}{2})$ component of $16_{H}$.
We denote their VEVs as
\begin{align}
\label{vevs}
&\vev{(1,1,1,0)}\equiv \omega_{B-L}, \nn \\[0.5ex]
&\vev{(1,1,3,0)}\equiv \omega_{R},  \\[0.5ex]
&\vev{(1,1,2,+\tfrac{1}{2})}\equiv \chi_{R}, \nn
\end{align}
where $\omega_{B-L,R}$ are real and $\chi_{R}$ can be taken real by
a phase redefinition of the $16_H$.
Different VEV configurations
trigger the spontaneous breakdown of the $SO(10)$ symmetry into a number of subgroups. Namely, for $\chi_{R}= 0$ one finds
\begin{align}
\label{vacua}
&\omega_{R}= 0,\, \omega_{B-L}\neq 0\; : & 3_C 2_L 2_R 1_{B-L} \nn \\[0.5ex]
&\omega_{R}\neq 0,\, \omega_{B-L}= 0\; : & 4_{C} 2_L 1_R \nn \\[0.5ex]
&\omega_{R}\neq 0,\, \omega_{B-L}\neq 0\; : & 3_C 2_L 1_R 1_{B-L}   \\[0.5ex]
&\omega_{R}=-\omega_{B-L}\neq 0\; : & \mbox{flipped}\, 5'\, 1_{Z'} \nn \\[0.5ex]
&\omega_{R}=\omega_{B-L}\neq 0\; :  & \mbox{standard}\, 5\, 1_{Z} \nn
\end{align}
with $5\, 1_{Z}$ and $5'\, 1_{Z'}$ standing for the two different
embedding of the $SU(5) \otimes U(1)$ subgroup into $SO(10)$, i.e. standard and ``flipped'' respectively (see the discussion at the end of the section).

When $\chi_{R}\neq 0$ all intermediate gauge symmetries are spontaneously broken down to the SM group, with the exception of the last case which maintains the standard $SU(5)$ subgroup unbroken and will no further be considered.

The classification in \eq{vacua} depends on the phase conventions used in the parametrization of the SM singlet subspace of $45_{H} \oplus 16_{H}$.
The statement that $\omega_{R}=\omega_{B-L}$ yields the standard $SU(5)$
vacuum while  $\omega_{R}=-\omega_{B-L}$ corresponds to the flipped setting defines a particular basis in this subspace
(see \sect{sec:su5vsflippedsu5}).
The consistency of any chosen framework is then verified against the corresponding Goldstone boson spectrum.

The decomposition of the $45_H$ and $16_H$ representations with respect to the relevant $SO(10)$ subgroups is detailed in Tables \ref{tab:16decomp} and \ref{tab:45decomp}.

\renewcommand{\arraystretch}{1.3}
\begin{table}
\centering
\begin{tabular}{lllllccc}
\hline \hline
 $4_C\,2_L\,2_R $
& $4_C\,2_L\,1_R $
& $3_C\,2_L\,2_R\,1_{B-L} $
& $3_C\,2_L\,1_R\,1_{B-L} $
& $3_C\,2_L\,1_Y $
& $5$ 
& $5'\,1_{Z'}$ 
& $1_{Y'}$ 
\\
\hline
$\left({4,2,1} \right)$
& $\left({4,2,0} \right)$
& $\left({ 3,2,1},+\frac{1}{6} \right)$
& $\left({ 3,2},0,+\frac{1}{6} \right)$
& $\left({ 3,2},+\frac{1}{6} \right)$
& $\left.{ 10}\right.$
& $\left({ 10},+1 \right)$
& $\left. +\frac{1}{6} \right.$
\\
\null
&
& $\left({ 1,2,1},-\frac{1}{2} \right)$
& $\left({ 1,2},0,-\frac{1}{2} \right)$
& $\left({ 1,2},-\frac{1}{2} \right)$
& $\left.{ \overline{5}} \right.$
& $\left({ \overline{5}},-3 \right)$
& $\left. -\frac{1}{2} \right.$
\\
$\left({ \overline{4},1,2} \right)$
& $\left({ \overline{4},1,+\frac{1}{2}} \right)$
& $\left({ \overline{3},1,2},-\frac{1}{6} \right)$
& $\left({ \overline{3},1},+\frac{1}{2},-\frac{1}{6} \right)$
& $\left({ \overline{3},1},+\frac{1}{3} \right)$
& $\left.{ \overline{5}} \right.$
& $\left({ 10},+1 \right)$
& $\left. -\frac{2}{3} \right.$
\\
\null
& $\left({ \overline{4},1,-\frac{1}{2}} \right)$
&
& $\left({ \overline{3},1},-\frac{1}{2},-\frac{1}{6} \right)$
& $\left({ \overline{3},1},-\frac{2}{3} \right)$
& $\left.{ 10} \right.$
& $\left({ \overline{5}},-3 \right)$
& $\left.  +\frac{1}{3} \right.$
\\
\null
&
& $\left({ 1,1,2},+\frac{1}{2} \right)$
& $\left({ 1,1},+\frac{1}{2},+\frac{1}{2} \right)$
& $\left({ 1,1},+1 \right)$
& $\left.{ 10} \right.$
& $\left({ 1},+5 \right)$
& $\left. 0 \right.$
\\
\null
&
&
& $\left({ 1,1},-\frac{1}{2},+\frac{1}{2} \right)$
& $\left({ 1,1},0 \right)$
& $\left.{ 1} \right.$
& $\left({ 10},+1 \right)$
& $\left. +1 \right.$
\\
\hline \hline
\end{tabular}
\mycaption{Decomposition of the spinorial representation  $16$ with respect to the various  $SO(10)$ subgroups. The definitions and normalization of the abelian charges are given in the text.}
\label{tab:16decomp}
\end{table}

\renewcommand{\arraystretch}{1.3}
\begin{table*}
\centering
\begin{tabular}{lllllccc}
\hline \hline
 $4_C\,2_L\,2_R $
& $4_C\,2_L\,1_R $
& $3_C\,2_L\,2_R\,1_{B-L} $
& $3_C\,2_L\,1_R\,1_{B-L} $
& $3_C\,2_L\,1_Y $
& $5$ 
& $5'\,1_{Z'}$ 
& $1_{Y'}$ 
\\
\hline
 $\left({ 1,1,3} \right)$
& $\left({ 1,1},+1 \right)$
& $\left({ 1,1,3},0 \right)$
& $\left({ 1,1},+1,0 \right)$
& $\left({ 1,1},+1 \right)$
& $\left.{ 10} \right.$
& $\left({ 10},-4 \right)$
& $\left. +1 \right.$
\\
\null
& $\left({ 1,1},0 \right)$
&
& $\left({ 1,1},0,0 \right)$
& $\left({ 1,1},0 \right)$
& $\left.{ 1} \right.$
& $\left({ 1},0 \right)$
& $\left. 0 \right.$
\\
\null
& $\left({ 1,1},-1 \right)$
&
& $\left({ 1,1},-1,0 \right)$
& $\left({ 1,1},-1 \right)$
& $\left.{ \overline{10}} \right.$
& $\left({ \overline{10}},+4 \right)$
& $\left. -1 \right.$
\\
$\left({ 1,3,1} \right)$
& $\left({ 1,3},0 \right)$
& $\left({ 1,3,1},0 \right)$
& $\left({ 1,3},0,0 \right)$
& $\left({ 1,3},0 \right)$
& $\left.{ 24} \right.$
& $\left({ 24},0 \right)$
& $\left. 0 \right.$
\\
$\left({ 6,2,2} \right)$
& $\left({ 6,2},+\frac{1}{2} \right)$
& $\left({ 3,2,2},-\frac{1}{3} \right)$
& $\left({ 3,2},+\frac{1}{2},-\frac{1}{3} \right)$
& $\left({ 3,2},\frac{1}{6} \right)$
& $\left.{ 10} \right.$
& $\left({ 24},0 \right)$
& $\left. -\frac{5}{6} \right.$
\\
\null
& $\left({ 6,2},-\frac{1}{2} \right)$
&
& $\left({ 3,2},-\frac{1}{2},-\frac{1}{3} \right)$
& $\left({ 3,2},-\frac{5}{6} \right)$
& $\left.{ 24} \right.$
& $\left({ 10},-4 \right)$
& $\left. +\frac{1}{6} \right.$
\\
\null
&
& $\left({ \overline{3},2,2},+\frac{1}{3} \right)$
& $\left({ \overline{3},2},+\frac{1}{2},+\frac{1}{3} \right)$
& $\left({ \overline{3},2},+\frac{5}{6} \right)$
& $\left.{ 24} \right.$
& $\left({ \overline{10}},+4 \right)$
& $\left. -\frac{1}{6} \right.$
\\
\null
&
&
& $\left({ \overline{3},2},-\frac{1}{2},+\frac{1}{3} \right)$
& $\left({ \overline{3},2},-\frac{1}{6} \right)$
& $\left.{ \overline{10}} \right.$
& $\left({ 24},0 \right)$
& $\left. +\frac{5}{6} \right.$
\\
$\left({ 15,1,1} \right)$
& $\left({ 15,1},0 \right)$
& $\left({ 1,1,1},0 \right)$
& $\left({ 1,1},0,0 \right)$
& $\left({ 1,1},0 \right)$
& $\left.{ 24} \right.$
& $\left({ 24},0 \right)$
& $\left. 0 \right.$
\\
\null
&
& $\left({ 3,1,1},+\frac{2}{3} \right)$
& $\left({ 3,1},0,+\frac{2}{3} \right)$
& $\left({ 3,1},+\frac{2}{3} \right)$
& $\left.{ \overline{10}} \right.$
& $\left({ \overline{10}},+4 \right)$
& $\left. +\frac{2}{3} \right.$
\\
\null
&
& $\left({ \overline{3},1,1},-\frac{2}{3} \right)$
& $\left({ \overline{3},1},0,-\frac{2}{3} \right)$
& $\left({ \overline{3},1},-\frac{2}{3} \right)$
& $\left.{ 10} \right.$
& $\left({ 10},-4 \right)$
& $\left. -\frac{2}{3} \right.$
\\
\null
&
& $\left({ 8,1,1},0 \right)$
& $\left({ 8,1},0,0 \right)$
& $\left({ 8,1},0 \right)$
& $\left.{ 24} \right.$
& $\left({ 24},0 \right)$
& $\left. 0 \right.$
\\
\hline \hline
\end{tabular}
\mycaption{Same as in Table \ref{tab:16decomp} for the $SO(10)$ adjoint ($45$) representation.}
\label{tab:45decomp}
\end{table*}

\subsubsection{The L-R chains}
\label{sec:L-Rchains}

According to the analysis in Chapter \ref{intermediatescales},
the potentially viable breaking chains fulfilling the basic gauge
unification constraints (with a minimal $SO(10)$ Higgs sector)
correspond to the settings:
\be
\omega_{B-L}\gg \omega_{R} > \chi_{R}\ : \quad
SO(10)\to 3_{C}2_{L}2_{R}1_{B-L}\to 3_{C}2_{L}1_{R}1_{B-L}\to 3_{C}2_{L}1_{Y}
\ee
and
\be
\omega_{R}\gg \omega_{B-L}\ > \chi_{R}\ : \quad 
SO(10)\to 4_{C}2_{L}1_{R}\to 3_{C}2_{L}1_{R}1_{B-L}\to 3_{C}2_{L}1_{Y} \, .
\ee
As remarked in \sect{sec:2loopgauge}, the cases
$\chi_{R} \sim \omega_{R}$ or $\chi_{R} \sim \omega_{B-L}$ lead to effective two-step $SO(10)$ breaking patterns with a non-minimal
set of surviving scalars at the intermediate scale.
On the other hand, a truly two-step setup can be recovered (with a minimal fine tuning) by considering the cases where
$\omega_{R}$ or $\omega_{B-L}$ exactly vanish. Only the explicit study
of the scalar potential determines which of the textures are allowed.


\subsubsection{Standard SU(5) versus flipped SU(5)}
\label{sec:su5vsflippedsu5}

There are in general two distinct SM-compatible embeddings  of  $SU(5)$
into $SO(10)$~\cite{DeRujula:1980qc,Barr:1981qv}.
They differ in one generator of the $SU(5)$ Cartan algebra and
therefore in the $U(1)_Z$ cofactor.

In the ``standard'' embedding, the weak hypercharge operator
$
Y=T^{3}_R+T_{B-L}
$
belongs to the $SU(5)$ algebra and the orthogonal Cartan generator
$Z$ (obeying $ [T_{i},Z]=0$ for all $T_{i}\in SU(5)$) is given by
$
Z =-4T^{3}_R+6T_{B-L}
$.

In the ``flipped'' $SU(5)'$ case, the right-handed isospin
assignment of quark and leptons into the $SU(5)'$ multiplets
is turned over so that the ``flipped'' hypercharge generator reads
$Y'=-T^{3}_R+T_{B-L}$.
Accordingly, the additional $U(1)_{Z'}$ generator reads
$Z' =4T^{3}_R+6T_{B-L}$,
such that $ [T_{i},Z']=0$ for all $T_{i}\in SU(5)'$.
Weak hypercharge is then given by
$Y=(Z'-Y')/5$.

\Tables{tab:16decomp}{tab:45decomp} show
the standard and flipped $SU(5)$ decompositions of the spinorial and adjoint $SO(10)$ representations respectively.

The two $SU(5)$ vacua in \eq{vacua} differ
by the texture of the adjoint representation VEVs:
in the standard $SU(5)$ case they are aligned with the $Z$ operator
while they match the $Z'$ structure
in the flipped $SU(5)'$ setting (see \app{app:explicitgenerators} for
an explicit representation).
\vskip 1mm

\section{The classical vacuum}
\label{sect:classicalvacuum}

\subsection{The stationarity conditions}

By substituting \eq{vevs} into \eq{potentialV0} the vacuum manifold reads
\begin{multline}
\vev{V_0}=-2\mu^2(2\omega_R^2 + 3\omega_{B-L}^2) + 4a_1(2\omega_R^2 +3\omega_{B-L}^2)^2 \\
+\frac{a_2}{4}(8\omega_R^4 + 21\omega_{B-L}^4 + 36\omega_R^2\omega_{B-L}^2) 
-\frac{\nu^2}{2}\chi_R^2 + \frac{\lambda_1}{4}\chi_R^4 +4\alpha\chi_R^2(2\omega_R^2 +3\omega_{B-L}^2) \\
+\frac{\beta}{4}\chi_R^2(2\omega_R + 3\omega_{B-L})^2 -\frac{\tau}{2}\chi_R^2(2\omega_R + 3\omega_{B-L})
\end{multline}
The corresponding three stationary conditions can be conveniently written as
\be
\frac{1}{8}\left(\frac{\partial \vev{V_0}}{\partial\omega_R}
-\frac{2}{3}\frac{\partial \vev{V_0}}{\partial\omega_{B-L}}\right) = 0 \, ,
\quad 
\omega_{B-L}\frac{\partial \vev{V_0}}{\partial\omega_R}
-\omega_R \frac{2}{3}\frac{\partial \vev{V_0}}{\partial\omega_{B-L}} = 0 \, ,
\quad 
\frac{\partial \vev{V_0}}{\partial\chi_R} = 0 \, ,
\ee
which lead respectively to 
\begin{multline}
[ -\mu^2+4a_1(2\omega_R^2 + 3\omega_{B-L}^2)
+ \frac{a_2}{4}
(4\omega_R^2+7\omega_{B-L}^2-2\omega_{B-L}\omega_R)+2\alpha\chi_R^2 ] \\
\times (\omega_R-\omega_{B-L}) = 0 \, ,
\label{eqstatmu}
\end{multline}
\begin{align}
&[ - 4a_2(\omega_R+\omega_{B-L})\omega_R\omega_{B-L}
-  \beta\chi_R^2(2\omega_R + 3\omega_{B-L})
+\tau\chi_R^2 ](\omega_R-\omega_{B-L}) = 0 \, ,
\label{eqstat0} \\[1ex]
& [ -\nu^2+\lambda_1\chi_R^2+8\alpha(2\omega_R^2 + 3\omega_{B-L}^2)
+ \frac{\beta}{2}(2\omega_R + 3\omega_{B-L})^2-\tau(2\omega_R + 3\omega_{B-L}) ] \chi_R = 0 \, .
\label{eqstatnu}
\end{align}
We have chosen linear combinations that factor out the
uninteresting standard $SU(5)\otimes U(1)_Z$ solution, namely $\omega_R=\omega_{B-L}$.

In summary, when $\chi_R = 0$, \eqs{eqstatmu}{eqstat0} allow for four possible vacua:
\begin{itemize}
\item{$\omega = \omega_R = \omega_{B-L}$ (standard $5\, 1_{Z}$)}
\item{$\omega = \omega_R = -\omega_{B-L}$ (flipped $5'\, 1_{Z'}$)}
\item{$\omega_R=0$ and $\omega_{B-L} \neq 0$ ($3_C\, 2_L\, 2_R\, 1_{B-L}$)}
\item{$\omega_R \neq 0$ and $\omega_{B-L} = 0$ ($4_C 2_L 1_R$)}
\end{itemize}

As we shall see, the last two options are not {\em tree level} minima.
Let us remark that for $\chi_R \neq 0$,
\eq{eqstat0} implies naturally a correlation
among the $45_H$ and $16_H$ VEVs,
or a fine tuned relation between $\beta$ and $\tau$, depending on the stationary solution.
In the cases $\omega_R=-\omega_{B-L}$, $\omega_R=0$ and $\omega_{B-L} = 0$
one obtains $\tau = \beta\omega$, $\tau = 3\beta\omega_{B-L}$ and $\tau = 2\beta\omega_R$ respectively. Consistency with the scalar mass spectrum
must be verified in each case.

\subsection{The tree-level spectrum}

The gauge and scalar spectra corresponding to the SM vacuum configuration
(with non-vanishing VEVs in $45_{H}\oplus 16_{H}$) are detailed
in \app{app:Treemasses}.

The scalar spectra obtained in various limits of the tree-level Higgs potential, corresponding to the appearance of accidental global symmetries, are derived in \apps{45only}{4516trivialinteraction}.
The emblematic case $\chi_R=0$ is scrutinized in \app{4516spectrumchi0}.

\subsection{Constraints on the potential parameters}
\label{furtherconstraints}

The parameters (couplings and VEVs) of the scalar potential are constrained by the requirements
of boundedness and the absence of tachyonic states, ensuring that
the vacuum is stable and the stationary points correspond to physical minima.

Necessary conditions for vacuum stability are derived
in \app{boundedpot}. In particular, on the $\chi_R=0$ section
one obtains
\begin{equation}
\label{strongestboundednessconstraints}
a_1 > -\tfrac{13}{80} a_2
\, .
\end{equation}
Considering the general case, the absence of tachyons in the scalar spectrum yields among else
\be
\label{bound130810}
a_2 < 0 \, , \qquad -2<\omega_{B-L}/\omega_R<-\tfrac{1}{2} \,.
\ee
The strict constraint on $\omega_{B-L}/\omega_R$
is a consequence of the tightly correlated form of the tree-level
masses of the $(8,1,0)$ and $(1,3,0)$ submultiplets of $45_{H}$,
labeled according to the SM ($3_C\, 2_L\, 1_Y$) quantum numbers, namely
\begin{align}
\label{310PGBmass}
M^2(1,3,0) & =
2 a_2 (\omega_{B-L} - \omega _R) (\omega_{B-L} + 2 \omega _R) \, , \\[1ex]
\label{810PGBmass}
M^2(8,1,0) & = 
2 a_2 (\omega _R - \omega_{B-L}) (\omega _R + 2 \omega_{B-L}) \,,
\end{align}
that are simultaneously positive only if \eq{bound130810} is enforced.
For comparison with previous studies, let us remark that in the $\tau=0$ limit (corresponding to an extra $Z_2$ symmetry
$\Phi\to -\Phi$) the intersection of the constraints from \eq{eqstat0}, \eqs{310PGBmass}{810PGBmass} and
the mass eigenvalues of the
$(1,1,1)$ and
$(3,2,1/6)$ states, yields
\be
\label{boundYasue}
a_2 < 0 \, ,
\quad  -1 \leq \omega_{B-L}/\omega_R \leq -\tfrac{2}{3}
\, ,
\ee
thus recovering the results of
Refs.~\cite{Yasue:1980fy,Yasue:1980qj,Anastaze:1983zk,Babu:1984mz}.

In either case, one concludes by inspecting the scalar mass spectrum
that flipped $SU(5)'\otimes U(1)_{Z'}$ is for $\chi_R=0$
the only solution admitted by \eq{eqstat0} consistent with the
constraints in \eq{bound130810} (or \eq{boundYasue}).
For $\chi_R\neq 0$,
the fine tuned possibility of having or $\omega_{B-L}/\omega_R \sim -1$ such that $\chi_R$ is obtained
at an intermediate scale fails to reproduce the SM couplings (see e.g.~\sect{sec:2loopgauge}).
Analogous and obvious conclusions hold for
$\omega_{B-L} \sim \omega_R \sim \chi_R \sim M_U$ and for
$\chi_R\gg \omega_{R,B-L}$
 (standard $SU(5)$ in the first stage).

This is the origin of the common knowledge that nonsupersymmetric $SO(10)$ settings with the adjoint VEVs driving the gauge symmetry breaking are not phenomenologically viable. In particular, a large
hierarchy between the $45_H$ VEVs, that would set the stage for
consistent unification patterns, is excluded.

The key question is: why are the masses of the states in \eqs{310PGBmass}{810PGBmass} so tightly correlated?
Equivalently, why do they depend on $a_2$ only?

\section{Understanding the scalar spectrum}
\label{sect:understanding}

A detailed comprehension of the patterns in the scalar spectrum
may be achieved by understanding the correlations between mass textures and the symmetries of the scalar potential. In particular, the
appearance of accidental global symmetries in limiting cases
may provide the rationale for the dependence of mass eigenvalues from specific couplings.
To this end we classify the most interesting cases, providing
a counting of the would-be Goldstone bosons (WGB)
and pseudo Goldstone bosons (PGB) for each case.
A side benefit of this discussion is a consistency check of the
explicit form of the mass spectra.

\subsection{45 only with $a_2=0$}
\label{45a2}

Let us first consider the potential generated by $45_H$, namely
$V_\Phi$ in \eq{potentialV0}.
When $a_2=0$,
i.e. when only trivial $45_H$ invariants (built off moduli) are considered, the scalar potential
exhibits an enhanced global symmetry: $O(45)$.
The spontaneous symmetry breaking (SSB) triggered by the $45_H$ VEV
reduces the global symmetry to $O(44)$.
As a consequence, 44 massless states are expected in the scalar spectrum.
This is verified explicitly in \app{45only}.
Considering the case of the $SO(10)$ gauge symmetry broken to the flipped $SU(5)'\otimes U(1)_{Z'}$, $45-25=20$ WGB, with the quantum numbers
of the coset $SO(10)/SU(5)'\otimes U(1)_{Z'}$ algebra,
decouple from the physical spectrum
while, $44-20=24$ PGB remain, whose
mass depends on the explicit breaking term $a_2$.

\subsection{16 only with $\lambda_2=0$}
\label{16lambda2}

We proceed in analogy with the previous discussion.
Taking $\lambda_{2}=0$ in $V_\chi$
enhances the global symmetry to $O(32)$.
The spontaneous breaking of $O(32)$ to $O(31)$
due to the $16_H$  VEV leads to
31 massless modes, as it is explicitly seen in \app{16only}.
Since the gauge $SO(10)$ symmetry is broken by $\chi_R$ to the standard $SU(5)$, $45-24=21$ WGB, with the quantum numbers
of the coset $SO(10)/SU(5)$ algebra,
decouple from the physical spectrum,
while $31-21=10$ PGB do remain.
Their masses depend on the explicit
breaking term $\lambda_2$.

\subsection{A trivial 45-16 potential $(a_2=\lambda_2=\beta=\tau=0)$}
\label{therelevantone}

When only trivial invariants (i.e. moduli) of both $45_H$ and $16_H$ are considered,
the global symmetry of $V_{0}$ in \eq{potentialV0}
is $O(45)\otimes O(32)$.
This symmetry is spontaneously broken into $O(44)\otimes O(31)$ by the $45_H$ and $16_H$ VEVs
yielding 44+31=75 GB in the scalar spectrum (see \app{4516justnorms}).
Since in this case, the gauge $SO(10)$ symmetry is broken to the SM gauge group,
$45-12=33$ WGB, with the quantum numbers
of the coset $SO(10)/SM$ algebra,
decouple from the physical spectrum,
while $75-33=42$ PGB remain.
Their masses are generally expected to receive contributions
from the explicitly breaking terms $a_2$, $\lambda_2$, $\beta$ and $\tau$.

\subsection{A trivial 45-16 interaction $(\beta=\tau=0)$}
\label{45-16betatau}

Turning off just the $\beta$ and $\tau$ couplings still allows for
independent global rotations of the $\Phi$ and $\chi$ Higgs fields.
The largest global symmetries are those determined
by the $a_2$ and $\lambda_2$ terms
in $V_{0}$, namely $O(10)_{45}$ and $ O(10)_{16}$, respectively.
Consider the spontaneous breaking to global flipped
$SU(5)'\otimes U(1)_{Z'}$ and the standard $SU(5)$
by the $45_H$ and $16_H$ VEVs, respectively.
This setting gives rise to  $20 + 21 = 41$ massless scalar modes.
The gauged $SO(10)$ symmetry is broken
to the SM group so that 33 WGB decouple from the physical spectrum.
Therefore, 41-33=8 PGB remain,
whose masses receive contributions from the explicit
breaking terms $\beta$ and $\tau$.
All of these features are readily verified by inspection of the scalar mass spectrum in \app{4516trivialinteraction}.

\subsection{A tree-level accident}

The tree-level masses of the crucial $(1,3,0)$ and $(8,1,0)$ multiplets belonging to the $45_H$
depend only on the parameter $a_2$ but {\em not} on the other parameters expected (cf.~\ref{therelevantone}),
namely $\lambda_2$, $\beta$ and $\tau$.

While the $\lambda_2$ and $\tau$ terms cannot obviously contribute
at the tree level to $45_H$ mass terms,
one would generally expect a contribution from the $\beta$ term, proportional to $\chi_R^2$.
Using the parametrization $\Phi=\sigma_{ij}\phi_{ij}/4$, where the $\sigma_{ij}$  ($i,j\in\{1,..,10\}$, $i\neq j$) matrices
represent the $SO(10)$ algebra on the 16-dimensional spinor basis
(cf.~\app{app:so10algebra}), one obtains a $45_H$ mass term of the form
\be
\label{45massbeta}
\frac{\beta}{16}\chi_R^2\ (\sigma_{ij})_{16\beta}(\sigma_{kl})_{\beta 16}\ \phi_{ij}\phi_{kl} \, .
\ee
The projection of the $\phi_{ij}$ fields onto the
$(1,3,0)$ and $(8,1,0)$ components lead, as we know, to vanishing
contributions.

This result can actually be understood on general grounds
by observing that
the scalar interaction in \eq{45massbeta} has the same structure as the
gauge boson mass from the covariant-derivative interaction with the $16_H$, cf.~\eq{fielddepmass16}.
As a consequence, no tree-level mass contribution from the $\beta$ coupling can be generated  for the $45_H$
scalars carrying the quantum numbers of the standard $SU(5)$ algebra.
This behavior can be again verified by inspecting the relevant scalar spectra in \app{app:scalarspectrum}.

The above considerations provide a clear rationale for the
accidental tree level constraint on $\omega_{B-L}/\omega_R$,
that holds independently on the size of $\chi_R$.

On the other hand, we should expect the $\beta$ and $\tau$ interactions to contribute $O(M_U/4\pi)$ terms to the masses of $(1,3,0)$ and $(8,1,0)$ at the quantum level.
Similar contributions should also arise from the gauge interactions,
that break explicitly the independent global transformations on
the $45_H$ and $16_H$ discussed in the previous subsections.

\begin{figure}[h]
\centering
\includegraphics[width=6.5cm]{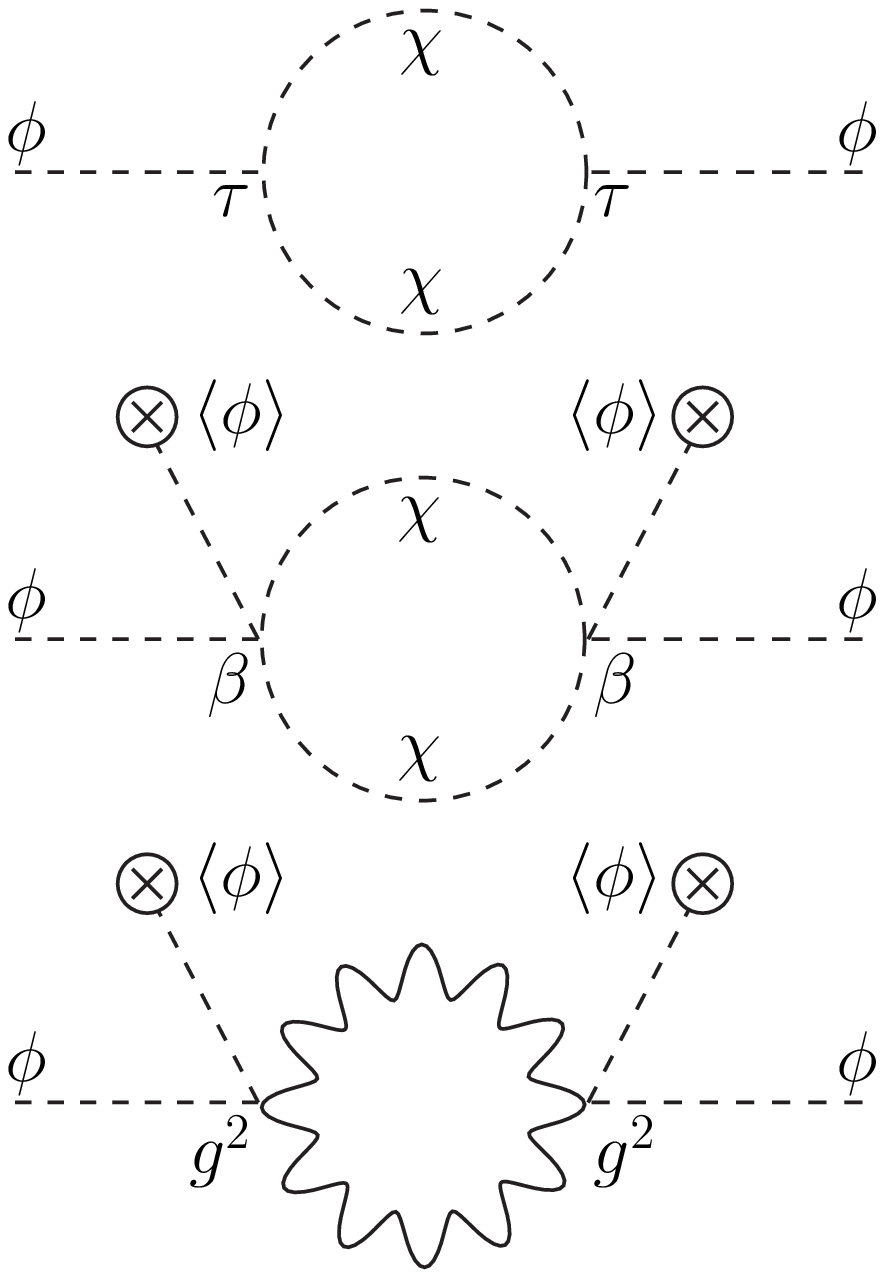}
\mycaption{\label{graphs} Typical one-loop diagrams that induce for
$\vev{\chi}=0$, $O(\tau/4\pi, \beta\vev{\phi}/4\pi, g^2\vev{\phi}/4\pi)$
renormalization to the mass of $45_H$ fields at the unification scale.
They are relevant for the PGB states, whose tree level mass is proportional to $a_2$.}
\end{figure}

The typical one-loop self energies, proportional to the $45_H$ VEVs,
are diagrammatically depicted in \fig{graphs}.
While the exchange of $16_{H}$ components is crucial,
the $\chi_R$ is not needed to obtain the large mass shifts.
In the phenomenologically allowed unification patterns it gives actually negligible contributions.

It is interesting to notice that the
$\tau$-induced mass corrections do not depend on
the gauge symmetry breaking, yielding an $SO(10)$
symmetric contribution to all scalars in $45_{H}$.

One is thus lead to the conclusion that any result based on the particular shape of the tree-level $45_H$ vacuum is drastically affected at the quantum level. Let us emphasize that although one may in principle avoid the $\tau$-term by means of e.g.~an extra $Z_{2}$ symmetry, no symmetry can forbid the $\beta$-term and the gauge loop contributions.

In case one resorts to $126_H$, in place of $16_H$,
for the purpose of $B-L$ breaking,
the more complex tensor
structure of the class of $126_{H}^{\dagger}45_H^{2}126_{H}$ quartic invariants in the scalar potential may admit tree-level
contributions to the states $(1,3,0)$ and $(8,1,0)$ proportional to $\vev{126_H}$.
On the other hand, as mentioned above, whenever
$\vev{126_H}$ is small on the unification scale, the same considerations apply, as for the $16_H$ case.

\subsection{The $\chi_{R}=0$ limit}
\label{sec:chi0limit}

From the previous discussion it is clear that
the answer to the question whether the non-$SU(5)$ vacua are allowed at the quantum level is independent
on the specific value of the $B-L$ breaking VEV ($\chi_R\ll M_U$ in potentially realistic cases).

In order to simplify the study of the scalar potential beyond the classical level it is therefore convenient (and sufficient) to consider the $\chi_R = 0$ limit.

When $\chi_R =0$ the mass matrices of the $45_H$ and $16_H$ sectors
are not coupled.
The stationary equations in \eqs{eqstatmu}{eqstat0} lead to the
four solutions
\begin{itemize}
\item{$\omega = \omega_R = \omega_{B-L}$\quad ($5\, 1_{Z}$)}
\item{$\omega = \omega_R = -\omega_{B-L}$\quad ($5'\, 1_{Z'}$)}
\item{$\omega_R=0$ and $\omega_{B-L} \neq 0$\quad ($3_C\, 2_L\, 2_R\, 1_{B-L}$)}
\item{$\omega_R \neq 0$ and $\omega_{B-L} = 0$\quad ($4_C 2_L 1_R$)}
\end{itemize}
In what follows, we will focus our discussion on the last three cases only.

It is worth noting that the tree level spectrum in the $\chi_{R}=0$ limit is not
directly obtained from the general formulae given in \app{4516spectrum}, since \eq{eqstatnu} is trivially satisfied for $\chi_{R}=0$.
The corresponding scalar mass spectra are derived and discussed in \app{4516spectrumchi0}. Yet again, it is apparent that the non $SU(5)$ vacuum configurations exhibit unavoidable tachyonic states in the scalar spectrum.

\section{The quantum vacuum}
\label{sec:quantumvacuum}

\subsection{The one-loop effective potential}
\label{sec:1loopeffpot}

We shall compute the relevant one-loop corrections to the tree level
results by means of the one-loop effective potential (effective action
at zero momentum) \cite{Coleman:1973jx}.
We can formally write
\begin{equation}
V_{\rm eff}=V_{0}+\Delta V_s+\Delta V_f+\Delta V_g \, ,
\label{Veff}
\end{equation}
where $V_0$ is the tree level potential and $\Delta V_{s,f,g}$ denote the quantum contributions induced by scalars, fermions and gauge bosons respectively.
In dimensional regularization with the modified minimal subtraction
($\overline{MS}$) and in the Landau gauge, they are given by
\begin{align}
& \Delta V_s(\phi ,\chi ,\mu)= 
\frac{\eta}{64\pi^2}
\Tr\left[W^4(\phi ,\chi)\left(\log\frac{W^2(\phi ,\chi)}{\mu^2}
-\frac{3}{2}\right)\right] \, , \\
& \Delta V_f(\phi ,\chi ,\mu)= 
\frac{-\kappa}{64\pi^2}
\Tr\left[M^4(\phi ,\chi)\left(\log\frac{M^2(\phi ,\chi)}{\mu^2}
-\frac{3}{2}\right)\right] \, , \\
& \Delta V_g(\phi ,\chi ,\mu)= 
\frac{3}{64\pi^2}
\Tr\left[\mathcal{M}^4(\phi ,\chi)\left(\log\frac{\mathcal{M}^2(\phi ,\chi)}{\mu^2}
-\frac{5}{6}\right)\right] \, ,
\end{align}
with $\eta=1(2)$ for real (complex) scalars and $\kappa=2(4)$ for
Weyl (Dirac) fermions. $W$, $M$ and $\mathcal{M}$ are the functional scalar, fermion and gauge boson mass matrices respectively, as obtained from the tree level potential.

In the case at hand, we may write the functional scalar mass matrix,
$W^2(\phi,\chi)$ as a 77-dimensional hermitian matrix, with a lagrangian term
\be
\label{defbasis}
\frac{1}{2}\psi^\dag W^2 \psi \, ,
\ee
defined on the vector basis $\psi=(\phi,\chi,\chi^\ast)$. More explicitly, $W^{2}$ takes the block form
\be
W^2(\phi,\chi)=\left(
\begin{array}{ccc}
V_{\phi\phi} & V_{\phi\chi} & V_{\phi\chi^\ast} \\
V_{\chi^{\ast}\phi} & V_{\chi^{\ast}\chi} & V_{\chi^{\ast}\chi^{\ast}} \\
V_{\chi\phi} & V_{\chi\chi} & V_{\chi\chi^{\ast}}
\end{array}
\right) \, ,
\label{W2matrix}
\ee
where the subscripts denote the derivatives of the scalar potential with respect to the set of fields $\phi$, $\chi$ and $\chi^\ast$.
In the one-loop part of the effective potential $V\equiv V_0$.

We neglect the fermionic component of the effective potential since there are no fermions at the GUT scale (we assume that the right-handed (RH) neutrino mass is substantially lower than the unification scale).

The functional gauge boson
mass matrix, $\mathcal{M}^2(\phi ,\chi)$ is given in
\app{app:Treemasses},
\eqs{fielddepmass45}{fielddepmass16}.

\subsection{The one-loop stationary equations}
\label{sec:1loopstationary}

The first derivative of the one-loop part of the effective potential, with respect to the scalar field component $\psi_a$, reads
\be
\label{der1deltaV}
\frac{\partial\Delta V_s}{\partial\psi_a}=
\frac{1}{64\pi^2}\Tr\left[\left\{W^2_{\psi_a},
  W^2\right\} 
\left(\log\frac{W^2}{\mu^2}-\frac{3}{2}\right)
+  W^2W^2_{\psi_a}\right] \,
\ee
where the symbol $W^{2}_{\psi_{a}}$ stands for the partial derivative of $W^{2}$ with respect to $\psi_{a}$.
Analogous formulae hold for $\partial\Delta V_{f,\, g}/\partial\psi_a$.
The trace properties ensure that \eq{der1deltaV} holds independently on whether $W^{2}$ does commute with its first derivatives or not.

The calculation of the loop corrected stationary equations
due to gauge bosons and scalar exchange is straightforward
(for $\chi_R=0$ the $45_H$ and $16_H$ blocks decouple in \eq{W2matrix}). On the other hand, the corrected equations are quite cumbersome and we do not explicitly report them here. It is enough to say that
the quantum analogue of \eq{eqstat0} admits analytically the same solutions as we had at the tree level. Namely, these are
$\omega_R = \omega_{B-L}$, $\omega_R = -\omega_{B-L}$, $\omega_R=0$ and $\omega_{B-L} = 0$,
corresponding respectively to the
standard $5\, 1_Z$, flipped $5'\, 1_{Z'}$, $3_C 2_L 2_R 1_{B-L}$ and $4_C 2_L 1_R$ preserved subalgebras.

\subsection{The one-loop scalar mass}
\label{sec:1loopspectrum}

In order to calculate the second derivatives of the one-loop contributions to $V_{\rm eff}$ it is in general necessary to take into account the commutation properties of $W^{2}$ with its derivatives that enter as a series of nested commutators. The general expression can be written as
\begin{small}
\begin{multline}
\frac{\partial^2\Delta V_s}{\partial\psi_a\partial\psi_b}=
\frac{1}{64\pi^2}\Tr \Big[ W^2_{\psi_a}W^2_{\psi_b}
+W^2W^2_{\psi_a\psi_b} 
+\left[\left\{W^2_{\psi_a\psi_b}, W^2\right\}
+ \left\{W^2_{\psi_a},W^2_{\psi_b}\right\}\right]
\left(\log\frac{W^2}{\mu^2}-\frac{3}{2}\right) \\[1ex]
+  \sum_{m=1}^{\infty}(-1)^{m+1}\frac{1}{m}\sum_{k=1}^{m}{m\choose k}\left\{ W^{2},W^2_{\psi_a}\right\} 
\left[W^{2},..\left[W^{2},W^2_{\psi_b}\right]..\right]\left(W^{2}-1\right)^{m-k}
\Bigr]
\label{der2deltaV}
\end{multline}
\end{small}
where the commutators in the last line are taken $k-1$ times.
Let us also remark that, although not apparent, the RHS of \eq{der2deltaV} can be shown to be symmetric under $a\leftrightarrow b$, as it should be. In specific cases (for instance when the nested commutators vanish or they can be rewritten as powers of a certain matrix commuting with $W$) the functional mass evaluated on the vacuum may take a closed form.

\subsubsection{Running and pole mass}
\label{sec:polemass}

The effective potential is a functional computed
at zero external momenta.
Whereas the stationary equations allow for the localization of the new
minimum (being the VEVs translationally invariant),
the mass shifts obtained from \eq{der2deltaV} define the
running masses
$\overline{m}_{ab}^2$
\be
\overline{m}_{ab}^2 \equiv
\frac{\partial^2 V_{\rm eff}(\phi)}{\partial\psi_a\partial\psi_b}\Big |_{\vev{\psi}} =
m_{ab}^2 + \Sigma_{ab}(0)
\label{runningmass}
\ee
where $m_{ab}^2$ are the renormalized masses and $\Sigma_{ab}(p^2)$
are the $\overline{MS}$ renormalized self-energies.
The physical (pole) masses $M_a^2$ are then obtained as a solution
to the equation
\be
\mbox{det}\left[p^2\delta_{ab}-\left(\overline{m}_{ab}^2 +
\Delta\Sigma_{ab}(p^2)
\right)\right] = 0
\label{invprop}
\ee
where
\be
\Delta\Sigma_{ab}(p^2) = \Sigma_{ab}(p^2) - \Sigma_{ab}(0)
\label{DeltaS}
\ee
For a given eigenvalue
\be
M_a^2 = \overline{m}_{a}^2
+ \Delta\Sigma_a (M_a^2)
\label{physmasses}
\ee
gives the physical mass. The gauge and scheme dependence in \eq{runningmass} is canceled by the relevant contributions from \eq{DeltaS}. In particular, infrared
divergent terms in \eq{runningmass} related to the presence
of massless WGB in the Landau gauge cancel in \eq{physmasses}.

Of particular relevance is the case when $M_a$ is substantially smaller than
the (GUT-scale) mass of the particles that contribute to $\Sigma(0)$.
At $\mu=M_U$, in the $M_{a}^2 \ll M_U^2$ limit, one has
\be
\Delta\Sigma_{a}(M_a^2) = O(M_a^4/M_U^2)\ .
\label{DeltaS-PGB}
\ee
In this case the running mass computed from \eq{runningmass} contains the leading gauge independent corrections.
As a matter of fact, in order to study the vacua of the potential in \eq{Veff},
we need to compute the zero momentum mass corrections just to those states
that are tachyonic at the tree level and whose corrected mass turns out
to be of the order of $M_U/4\pi$.

We may safely neglect the one loop corrections
for all other states with masses of order $M_U$.
It is remarkable, as we shall see, that for $\chi_R=0$ the relevant corrections to the masses of the critical PGB states can be obtained from \eq{der2deltaV} with vanishing commutators.

\subsection{One-loop PGB masses}
\label{subsec:1Lmatching}

The stringent tree-level
constraint on the ratio $\omega_{B-L}/\omega_R$, coming from the positivity
of the $(1,3,0)$ and $(8,1,0)$ masses, follows from the fact that some scalar masses depend only on the parameter $a_2$.
On the other hand, the discussion on the would-be global symmetries of the scalar potential shows that in general their mass should depend on other terms in the scalar potential, in particular $\tau$ and $\beta$.

A set of typical one-loop diagrams contributing $O(\vev{\phi}/4\pi)$ renormalization to the masses of $45_H$ states is depicted in \fig{graphs}. As we already pointed out the $16_H$ VEV does not play any role in the leading GUT scale corrections (just the interaction between $45_H$ and $16_H$, or with the massive gauge bosons is needed).
Therefore we henceforth work in the strict $\chi_R=0$ limit, that
simplifies substantially the calculation.
In this limit the scalar mass matrix in \eq{W2matrix} is block diagonal (cf.~\app{4516spectrumchi0}) and the leading corrections from the one-loop effective potential are encoded in the $V_{\chi^{\ast}\chi}$
sector.

More precisely, we are interested in the corrections to those
$45_H$ scalar states whose tree level mass depends only on $a_2$
and have the quantum numbers of the preserved non-abelian algebra
(see \sect{45a2} and \app{4516spectrumchi0}).
It turns out that focusing  to this set of PGB states
the functional mass matrix $W^2$ and its first derivative do commute for $\chi_R=0$  and \eq{der2deltaV} simplifies accordingly. This allows us to compute the relevant mass corrections in a closed form.

The calculation of the EP running mass
from \eq{der2deltaV} leads for the states $(1,3,0)$
and $(8,1,0)$ at $\mu=M_U$ to the mass shifts
\begin{multline}
\label{310onthevac}
\Delta M^2(1,3,0) = \\ 
\frac{ \tau^2
+\beta^2(2\omega_R^2-\omega_R\omega_{B-L}+2\omega_{B-L}^2)
+g^4 \left(16 \omega _R^2+\omega_{B-L} \omega _R+19 \omega_{B-L}^2\right) }{4\pi^2} \, , 
\end{multline}
\begin{multline}
\label{810onthevac}
\Delta M^2(8,1,0) = \\ 
\frac{ \tau^2
+\beta^2(\omega_R^2-\omega_R\omega_{B-L}+3\omega_{B-L}^2)
+g^4 \left(13 \omega _R^2+\omega_{B-L} \omega _R+22 \omega_{B-L}^2\right) }{4\pi^2} \, ,
\end{multline}
where the sub-leading (and gauge dependent) logarithmic terms
are not explicitly reported.
For the vacuum configurations of interest we find the results reported
in \app{app:1Lmasses}. In particular,  we obtain
\begin{itemize}
\item{
$\omega = \omega_R = -\omega_{B-L}$\quad ($5'\, 1_{Z'}$):
\begin{align}
M^{2}(24,0) =
-4 a_2 \omega^2 +
\frac{\tau ^2 + (5 \beta ^2 + 34 g^4) \omega ^2}{4 \pi ^2}\,,
\label{m240}
\end{align}
}
\item{
$\omega_R=0$ and $\omega_{B-L} \neq 0$\quad  ($3_C 2_L 2_R 1_{B-L}$):
\begin{align}
& M^{2}(1,3,1,0) = M^{2}(1,1,3,0) = 
2 a_2 \omega_{B-L}^2
+\frac{\tau ^2 + (2 \beta ^2 + 19 g^4) \omega_{B-L}^2}{4 \pi ^2} \, , 
\label{m1310} \\
& M^{2}(8,1,1,0) = - 4 a_2 \omega_{B-L}^2 +
\frac{\tau ^2 + (3 \beta ^2 + 22 g^4) \omega_{B-L}^2}{4 \pi ^2}\,,
\label{m8110}
\end{align}
}
\item{
$\omega_R \neq 0$ and $\omega_{B-L} = 0$\quad  ($4_C 2_L 1_R$):
\begin{align}
& M^{2}(1,3,0)  = - 4 a_2 \omega_R^2 +
\frac{\tau ^2 + (2 \beta ^2 + 16 g^4) \omega _R^2}{4 \pi ^2}  \label{m130}\,, \\
& M^{2}(15,1,0) = 2 a_2 \omega_R^2 +
\frac{\tau ^2 + (\beta ^2 + 13 g^4) \omega _R^2}{4 \pi ^2}\,.
\label{m1510}
\end{align}
}
\end{itemize}
In the effective theory language
\eqs{m240}{m1510}
can be interpreted as the one-loop GUT-scale matching due to the decoupling of the massive $SO(10)/G$ states where $G$ is the preserved gauge group.
These are the only relevant one-loop corrections needed in order
to discuss the vacuum structure of the model.

It is quite apparent that a consistent scalar mass spectrum can be obtained
in all cases, at variance with the tree level result.

In order to fully establish the existence of the non-$SU(5)$ minima at the quantum level one
should identify the regions of the parameter space supporting the desired vacuum configurations and estimate their depths. We shall address these issues in the next section.

\subsection{The one-loop vacuum structure}
\label{sec:1loopvacuum}

\subsubsection{Existence of the new vacuum configurations}
\label{subsec:newvacua}

The existence of the different minima of the one-loop effective potential is related to the values
of the parameters  $a_2$, $\beta$, $\tau$ and $g$ at the scale $\mu=M_U$. For the flipped $5'\, 1_{Z'}$ case it is sufficient,
as one expects, to assume the tree level condition $a_2<0$.
On the other hand, from \eqs{m1310}{m1510} we obtain
\begin{itemize}
\item{
$\omega_R=0$ and $\omega_{B-L} \neq 0$\quad  ($3_C 2_L 2_R 1_{B-L}$):
\be
- 8 \pi ^2 a_2 <
\frac{\tau ^2}{\omega_{B-L}^2} + 2 \beta ^2 + 19 g^4 \, ,
\label{pm13loop}
\ee
}
\item{
$\omega_R \neq 0$ and $\omega_{B-L} = 0$\quad  ($4_C 2_L 1_R$):
\be
- 8 \pi ^2 a_2 <
\frac{\tau ^2}{\omega _R^2} + \beta ^2 + 13 g^4 \, .
\label{pm81loop}
\ee
}
\end{itemize}
Considering for naturalness $\tau \sim \omega_{Y,R}$, \eqs{pm13loop}{pm81loop} imply $|a_2| < 10^{-2}$.
This constraint remains within the natural perturbative range for
dimensionless couplings.
While all PGB states whose mass is proportional to $-a_2$
receive large positive loop corrections,
quantum corrections are numerically irrelevant for all of the
states with GUT scale mass. On the same grounds we may safely neglect the multiplicative $a_2$ loop corrections induced by the $45_H$ states on the PGB masses.

\subsubsection{Absolute minimum}
\label{subsec:absoluteminimum}

It remains to show that the non $SU(5)$ solutions may actually be absolute minima of the potential.
To this end
it is necessary to consider the one-loop corrected stationary equations and calculate the vacuum energies in the relevant cases.
Studying the shape of the one-loop effective potential is a numerical
task. On the other hand, in the approximation of
neglecting at the GUT scale the logarithmic corrections, we
may reach non-detailed but definite conclusions.
For the three relevant vacuum configurations we obtain:
\begin{itemize}
\item{
$\omega = \omega_R = -\omega_{B-L}$\quad ($5'\, 1_{Z'}$)
\begin{small}
\begin{align}
& V(\omega, \chi_R=0) =
-\frac{3 \nu ^4}{16 \pi ^2}
+ \left(\frac{5 \alpha  \nu ^2}{\pi ^2}+\frac{5 \beta  \nu ^2}{16
   \pi ^2}-\frac{5 \tau ^2}{16 \pi ^2}\right) \omega ^2 \\[1ex]
& + \left(-100 a_1-\frac{65 a_2}{4}+\frac{600 a_1^2}{\pi ^2}-\frac{45 a_1 a_2}{\pi ^2}-\frac{645 a_2^2}{32 \pi ^2}+\frac{100 \alpha
   ^2}{\pi ^2}+\frac{25 \alpha  \beta }{2 \pi ^2}+\frac{65 \beta ^2}{64 \pi ^2}-\frac{5 g^4}{2 \pi ^2}\right) \omega ^4 \, , \nn
\end{align}
\end{small}
}
\item{
$\omega_R=0$ and $\omega_{B-L} \neq 0$\quad ($3_C 2_L 2_R 1_{B-L}$)
\begin{small}
\begin{align}
& V(\omega_{B-L}, \chi_R=0) =
-\frac{3 \nu ^4}{16 \pi ^2}
+ \left(\frac{3 \alpha  \nu ^2}{\pi ^2}+\frac{3 \beta  \nu
   ^2}{16 \pi ^2}-\frac{3 \tau ^2}{16 \pi ^2}\right) \omega_{B-L}^2 \\[1ex]
& + \left(-36 a_1 -\frac{21 a_2}{4} +\frac{216 a_1^2}{\pi ^2}+\frac{33 a_1 a_2}{\pi ^2}+\frac{45 a_2^2}{32 \pi ^2}+\frac{36 \alpha
   ^2}{\pi ^2}+\frac{9 \alpha  \beta }{2 \pi ^2}+\frac{21 \beta ^2}{64 \pi ^2}-\frac{15 g^4}{16 \pi ^2}\right)\omega_{B-L}^4 \, , \nn
\end{align}
\end{small}
}
\item{
$\omega_R \neq 0$ and $\omega_{B-L} = 0$\quad ($4_C 2_L 1_R$)
\begin{small}
\begin{align}
& V(\omega_R, \chi_R=0) =
-\frac{3 \nu ^4}{16 \pi ^2}
+\left(\frac{2 \alpha  \nu ^2}{\pi ^2}+\frac{\beta  \nu ^2}{8 \pi
   ^2}-\frac{\tau ^2}{8 \pi ^2}\right) \omega _R^2 \\[1ex]
& +\left(-16 a_1-2 a_2+\frac{96 a_1^2}{\pi ^2}+\frac{42 a_1 a_2}{\pi ^2}+\frac{147 a_2^2}{32 \pi ^2}+\frac{16 \alpha ^2}{\pi
   ^2}+\frac{2 \alpha  \beta }{\pi ^2}+\frac{\beta ^2}{8 \pi ^2}-\frac{7 g^4}{16 \pi ^2}\right)\omega _R^4 \, . \nn
\end{align}
\end{small}
}
\end{itemize}
A simple numerical analysis reveals that for natural values of the dimensionless couplings and GUT mass parameters any of the qualitatively different vacuum configurations may be a global minimum of the one-loop effective potential in a large domain of the  parameter space.

This concludes the proof of existence of all of the group-theoretically
allowed vacua.
Nonsupersymmetric $SO(10)$ models
broken at $M_U$ by the $45_{H}$ SM preserving VEVs, do exhibit at the quantum level the full spectrum of intermediate symmetries. This is
crucially relevant for those chains that, allowed by gauge unification, are accidentally excluded by the tree level potential.

\chapter{SUSY-$SO(10)$ breaking with small representations}
\label{flippedSO10}

\section{What do neutrinos tell us?}
\label{neutrinostellus}

In Chapter~\ref{thequantumvac} we showed that quantum effects solve the long-standing issue of the incompatibility between the dynamics of the simplest 
nonsupersymmetric $SO(10)$ Higgs sector spanning over $45_H \oplus 16_H$ and gauge coupling unification. 

In order to give mass to the SM fermions at the renormalizable level one has to minimally add a $10_H$. 
So it would be natural to consider the Higgs sector $10_H\oplus 16_H\oplus 45_H$ as a candidate for the 
minimal $SO(10)$ theory, as advocated long ago by Witten~\cite{Witten:1979nr}.  
However the experimental data accumulated since the $1980$ tell us that such an Higgs sector cannot work. 
It is anyway interesting to review the general idea, especially as far as concerns the generation of neutrino masses. 

First of all with just one $10_H$ the Yukawa lagrangian is
\be
\label{diracneu}
\mathcal{L}_Y = Y_{10} 16_F 16_F 10_H + \text{h.c.} \, ,
\ee
which readily implies $V_{CKM} = 1$, since $Y_{10}$ can be always diagonalized by a rotation in the flavor space of the $16_F$. 
However this is not a big issue. It would be enough to add a second $10_H$ or even better a $120_H$ which can break the 
down-quark/charged-lepton symmetry (cf.~\eqs{Mu3Yuk}{MD3Yuk}).   

The most interesting part is about neutrinos. In order to give a Majorana mass to the RH neutrinos $B-L$ must be broken by two units. 
Since $B-L \vev{16_H^*} = - 1$ this means that we have to couple the bilinear $16_H^* 16_H^*$ to $16_F 16_F$. 
Such a $d=5$ operator can be generated radiatively due to the exchange of GUT states~\cite{Witten:1979nr}.  

\begin{figure*}[ht]
\centering
\includegraphics[width=8.5cm]{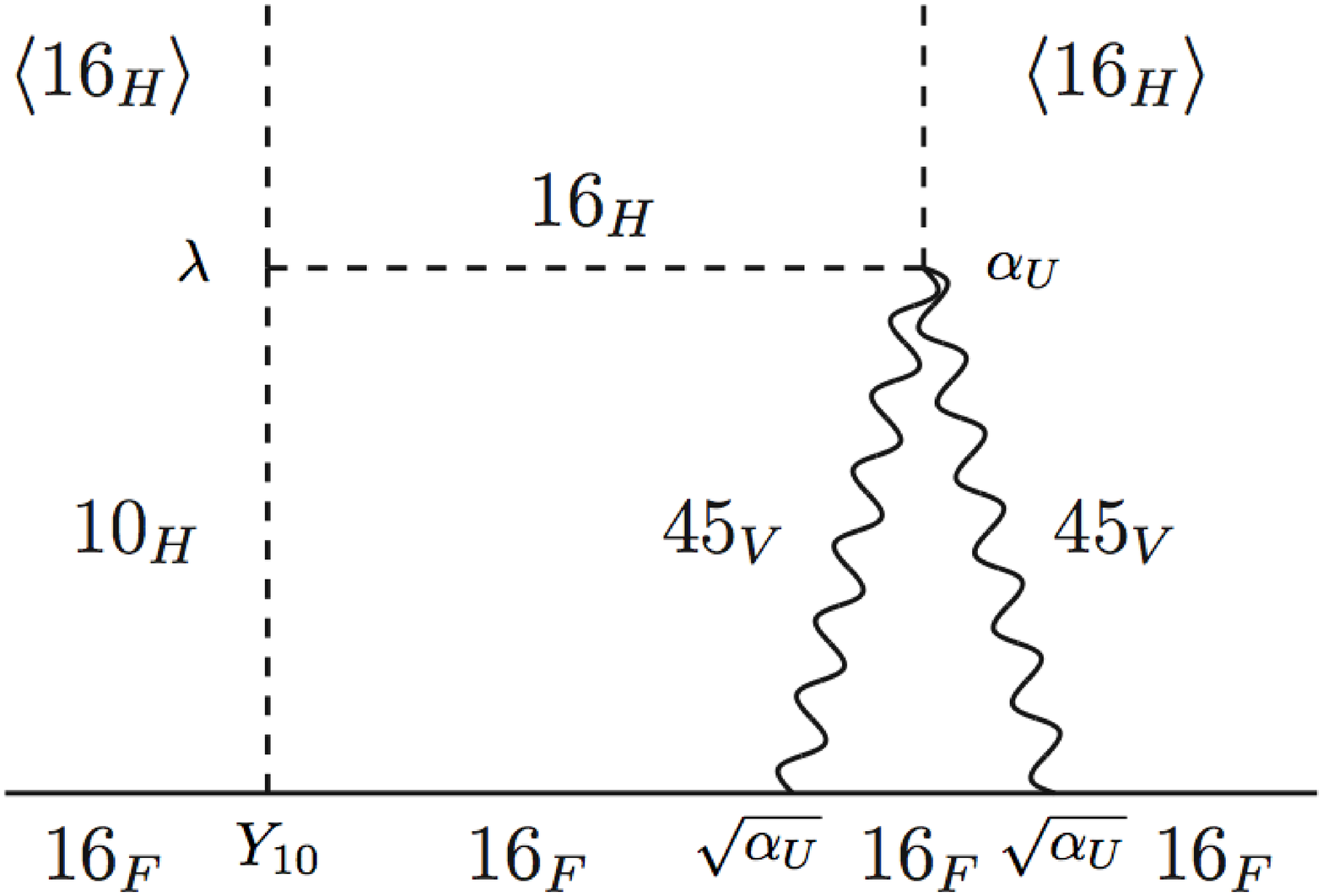}
\mycaption{\label{witten2loop}Two-loop diagram responsible for neutrino masses in the Witten mechanism. Figure taken from~\cite{Bajc:2004hr}.} 
\end{figure*}

Effectively the bilinear $16_H^* 16_H^*$ can be viewed as a $126_H^*$. So we are looking for states which 
can connect the matter bilinear $16_F 16_F$ with an effective $126_H^*$. 
Since $10 \otimes 45 \otimes 45 \supset 126$ a possibility is given by the combination $10_H 45_V 45_V$ (where $45_V$ are the $SO(10)$ gauge bosons). 
Indeed the $10_H$ and the $45_V$'s can be respectively attached to the matter bilinear via Yukawa ($Y_{10}$) and gauge ($g_U$) interactions; 
and to the bilinear $16_H^* 16_H^*$ via 
scalar potential couplings ($\lambda$) and again gauge interactions. The topology of the diagram is such that this 
happens for the first time at the two-loop level (see e.g.~\fig{witten2loop}). 

Notice that the same diagram generates also a 
Majorana mass term for LH neutrinos, while a Dirac mass term arises from the Yukawa lagrangian in~\eq{diracneu}.
At the leading order the contribution to the RH Majorana, Dirac and LH Majorana neutrino mass matrices (respectively $M_R$, $M_D$ and $M_L$) 
is estimated to be 
\be
M_R \sim Y_{10} \, \lambda \left( \frac{\alpha_U}{\pi} \right)^2 \frac{\chi_R^2}{M_U} \, , \qquad
M_D \sim Y_{10} v^u_{10} \, , \qquad
M_L \sim Y_{10} \, \lambda \left( \frac{\alpha_U}{\pi} \right)^2 \frac{\chi_L^2}{M_U} \, , 
\ee
where $\chi_R$ is $B-L$ breaking VEV of the $16_H$ in the $SU(5)$ singlet direction, $v_u^{10} = \vev{(1,2,+\tfrac{1}{2})_{10}}$ and 
$\chi_L = \vev{(1,2,+\tfrac{1}{2})_{16^*}}$ are instead electroweak VEVs.
After diagonalizing the full $6 \times 6$ neutrino mass matrix 
\be
\left( 
\begin{array}{cc}
M_L & M_D \\
M_D^T & M_R
\end{array}
\right) \, ,
\ee
defined on the symmetric basis $(\nu, \nu^c)$, we get the usual type-II and type-I contributions to the $3 \times 3$ light neutrinos mass matrix
\be
m_\nu = M_L - M_D M_R^{-1} M_D^T \, .
\ee
The type-II seesaw is clearly too small ($M_L \sim Y_{10} \, 10^{-6} \ \text{eV}$) while the type-I is naturally too 
big\footnote{Accidentally gravity would be responsible for a contribution of the same order of magnitude. 
Indeed if we take the Plank scale as the cut-off of the $SO(10)$ theory 
we find a $d=5$ effective operators of the type $16_F 16_F 16_H^* 16_H^* / M_P$, which leads to $M_R \sim \chi_R^2/M_P$. 
The analogy comes from the fact that $(\alpha_U/\pi)^{-2} M_U \sim M_P$.}
\be
M_D M_R^{-1} M_D^T \sim Y_{10} \, \lambda^{-1} \left( \frac{\alpha_U}{\pi} \right)^{-2} \frac{M_U \, (v_{10}^u)^2}{\chi_R^2} \sim Y_{10} \, 10^6 \ \text{eV} \, .
\ee
For the estimates we have taken $\lambda \sim 1$, $\alpha_U / \pi \sim 10^{-2}$, $\chi_L \sim v_{10}^u \sim 10^2 \ \text{GeV}$, $M_U \sim 10^{15} \ \text{GeV}$ 
and $\chi_R \sim 10^{13} \ \text{GeV}$, where $\chi_R \ll M_U$ by unification constraints. 

The only chance in order to keep neutrino masses below $1 \ \text{eV}$ is either to push $\chi_R \sim M_U$ or 
to take $Y_{10} \sim 10^{-6}$. The first option is unlikely in nonsupersymmetric $SO(10)$ because of unification 
constraints\footnote{In supersymmetry $\chi_R \sim M_U$, but then the two-loop diagram in \fig{witten2loop} would disappear due to the non-renormalization theorems 
of the superpotential. This brought the authors of Refs.~\cite{Bajc:2004hr,Bajc:2005aq} to reconsider the Witten mechanism 
in the context of split-supersymmetry~\cite{ArkaniHamed:2004fb,Giudice:2004tc,ArkaniHamed:2004yi}.}, 
while the second one is forbidden by the fact that $M_u = M_D$, in the simplest case with just the $10_H$ in the Yukawa sector (cf.~e.g.~\eqs{Mu3Yuk}{MD3Yuk}) . 

As we have already anticipated the reducible representation $10_H \oplus 120_H$ is needed in the Yukawa sector in order to generate non trivial mixing and break the 
down-quark/charged-lepton symmetry. Interestingly this system would also allow for a disentanglement between $M_u$ and $M_D$ 
(cf.~e.g.~\eqs{Mu3Yuk}{MD3Yuk}) and a fine-tuning in order to suppress neutrino masses is in principle conceivable.  
However the Higgs sector $10_H \oplus 16_H \oplus 45_H \oplus 120_H$ starts to deviate from minimality and maybe 
there is a better option to be considered. 

The issue can be somewhat alleviated by considering a $126_H$ in place of a $16_H$ in the Higgs sector, since in such a case the neutrino masses is generated at the renormalizable level by the term $16_{F}^{2} 126^*_H$. This lifts the problematic $\chi_R /M_U$ suppression factor inherent to the $d=5$ effective mass and yields 
$M_{R} \sim \chi_R \sim M_{B-L}$, that might be, at least in principle, acceptable. 
This scenario, though conceptually simple, 
involves a detailed one-loop analysis of the scalar potential governing the dynamics of the $10_H \oplus 45_H \oplus 126_H$ 
Higgs sector and is subject of an ongoing investigation~\cite{BDLM1}. We will briefly mention some preliminary results in the Outlook of the 
thesis. 

On the other hand it would be also nice to have a viable Higgs sector with only representations up to the adjoint. 
This is not possible in ordinary $SO(10)$, but what about the supersymmetric case?
Invoking TeV-scale supersymmetry (SUSY), the qualitative picture changes dramatically. Indeed, the gauge running within the MSSM prefers $M_{B-L}$ in the proximity of $M_{U}$ and, hence, the Planck-suppressed $d=5$ RH neutrino mass operator {$16_{F}^{2}\overline{16}_{H}^{2}/M_{P}$, available whenever $16_H\oplus \overline{16}_H$ is present in the} Higgs sector, can naturally reproduce the desired range for $M_{R}$. 

This well known fact motivates us to re-examin the issue of the breaking of SUSY-$SO(10)$ in the presence of small representations. 

\section{SUSY alignment: a case for flipped $SO(10)$}

In the presence of supersymmetry one would naively say that the minimal Higgs sector that suffices to break $SO(10)$ 
to the SM is given by $45_H \oplus 16_H \oplus \overline{16}_H$. 
Let us recall that both $16_H$ as well as  $\overline{16}_H$ are required in order to retain SUSY below the GUT scale.

However, it is well known \cite{Buccella:1981ib,Babu:1994dc,Aulakh:2000sn}
that the relevant superpotential does not support,
at the renormalizable level, a supersymmetric breaking of the $SO(10)$
gauge group to the SM. This is due to the constraints on the vacuum manifold
imposed by the $F$- and $D$-flatness conditions which, apart from linking
the magnitudes of the $SU(5)$-singlet ${16}_H$ and ${\overline{16}_H}$
vacuum expectation values ({VEVs}), make the adjoint {VEV $\vev{45_{H}}$}
aligned to {$\vev{16_H\overline{16}_{H}}$}.
As a consequence, an $SU(5)$ subgroup of the initial $SO(10)$ gauge symmetry remains unbroken. In this respect, a renormalizable Higgs sector with $126_H\oplus \overline{126}_H$ in place of $16_H\oplus \overline{16}_H$ suffers from the same ``$SU(5)$ lock''~\cite{Aulakh:2000sn},
{because also in $\overline{126}_{H}$ the SM singlet direction is $SU(5)$-invariant.}

This issue {can} be addressed by giving up renormalizability~\cite{Babu:1994dc,Aulakh:2000sn}. However, this option may be rather problematic since it introduces a delicate interplay between physics at two different scales, $M_{U}\ll M_{P}$, with the consequence of splitting the GUT-scale thresholds over several orders of magnitude around $M_{U}$. This may affect proton decay as well as the SUSY gauge unification, and {may} force the $B-L$ scale below the GUT scale. The latter is harmful for the setting with $16_H\oplus \overline{16}_H$ relying on a $d=5$ RH neutrino mass operator. The models with $126_H\oplus \overline{126}_H$ are also prone to trouble with gauge unification, due to the {number of large} Higgs multiplets spread around the GUT-scale.

Thus, in none of the cases above the simplest conceivable $SO(10)$ Higgs sector spanned over the lowest-dimensionality irreducible representations (up to the adjoint) seems to offer a natural scenario for realistic model building.
Since the option of a simple GUT-scale Higgs dynamics involving small representations governed by a simple renormalizable superpotential is particularly attractive, we aimed at studying the
conditions under which the seemingly ubiquitous {$SU(5)$ lock} can be {overcome}, while keeping only spinorial and adjoint $SO(10)$ representations.

Let us emphasize that the assumption that the gauge symmetry breaking is driven by the renormalizable part of the Higgs superpotential does not clash with the fact that, in models with $16_H\oplus \overline{16}_H$, the neutrino masses are generated at the non-renormalizable level, and other fermions may be sensitive to physics beyond the GUT scale.
As far as symmetry breaking is concerned, Planck induced $d\geq 5$ effective interactions are irrelevant perturbations in this picture.

The simplest attempt to breaking the $SU(5)$ lock by doubling either $16_H\oplus \overline{16}_H$ or $45_H$ in order to relax the $F$-flatness constraints is easily shown not to work. In the former case, there is only one SM singlet field direction associated to each of the $16_H\oplus \overline{16}_H$ pairs. Thus, $F$-flatness makes the VEVs in $45_H$ align along this direction regardless of the number of $16_{H}\oplus \overline{16}_{H}$'s contributing to the relevant $F$-term, $\partial W/\partial 45_{H}$ (see for instance Eq. (6) in ref. \cite{Aulakh:2000sn}). Doubling the number of $45_H$'s does not help either.
Since there is no mixing among the 45's besides the mass term,
$F$-flatness aligns both $\vev{45_H}$ in the $SU(5)$ direction of  $16_H\oplus \overline{16}_H$.
For three (and more) adjoints a mixing term of the form $45_{1}45_{2}45_{3}$
is allowed, but it turns out to be irrelevant to the minimization so that the alignment is maintained.

From this brief excursus one might conclude that, as far as the Higgs content is considered, the price for tractability and predictivity
is high on SUSY $SO(10)$ models,
as the desired group-theoretical simplicity
of the Higgs sector, with representations up to the
adjoint, appears not viable.

In this chapter, we point out that all these issues are alleviated if one considers a flipped variant of the SUSY $SO(10)$ unification.
In particular, we shall show that the flipped $SO(10)\otimes U(1)$ scenario~\cite{Kephart:1984xj,Rizos:1988jn,Barr:1988yj} offers an attractive option to break the gauge symmetry to the SM at the renormalizable level by means of a quite simple Higgs sector,
namely a couple of  $SO(10)$  spinors $16_{1,2}\oplus \overline{16}_{1,2}$ and one adjoint $45_{H}$.

Within the extended $SO(10)\otimes U(1)$ gauge algebra one finds in general three inequivalent embeddings of the SM hypercharge.
In addition to the two solutions with the hypercharge stretching over the $SU(5)$ or the $SU(5)\otimes U(1)$ subgroups of $SO(10)$ (respectively dubbed as the ``standard'' and ``flipped'' $SU(5)$ embeddings), there is a third, ``flipped'' $SO(10)$, solution inherent to the $SO(10)\otimes U(1)$ case, with a non-trivial projection of the SM hypercharge onto the $U(1)$ factor.

Whilst the difference between the standard and the flipped $SU(5)$ embedding is semantical from the $SO(10)$ point of view, the flipped $SO(10)$ case is qualitatively different. In particular, the symmetry-breaking ``power'' of the $SO(10)$ spinor and adjoint representations is boosted with respect to the standard $SO(10)$ case,
increasing the number of SM singlet fields that may acquire non vanishing VEVs.
Technically, flipping allows for a pair of SM singlets in each of the $16_H$ and $\overline{16}_H$ ``Weyl'' spinors, together with four SM singlets within $45_H$.
This is at the root of the possibility of implementing the gauge symmetry breaking by means of a simple renormalizable Higgs sector.
Let us just remark that, if renormalizability is not required, the
breaking can be realized without the adjoint Higgs field, see for instance the flipped $SO(10)$ model with an additional anomalous $U(1)$ of Ref. \cite{Maekawa:2003wm}.

Nevertheless, flipping is not per-se sufficient to cure the $SU(5)$ lock of standard $SO(10)$ with $16_H\oplus \overline{16}_H\oplus 45_H$ in the Higgs sector. Indeed, the adjoint does not reduce the rank and the bi-spinor, in spite of the two qualitatively different SM singlets involved, can lower it only by a single unit, leaving a residual $SU(5)\otimes U(1)$ symmetry (the two SM singlet directions in the
$16_H$ still retain an $SU(5)$ algebra as a little group).
Only when two pairs of $16_H\oplus \overline{16}_H$ (interacting via $45_H$) are introduced the two pairs of SM singlet VEVs in the spinor multiplets may not generally be aligned and the little group is reduced to the SM.

Thus, the simplest renormalizable SUSY Higgs model that can provide the spontaneous breaking of the $SO(10)$ GUT symmetry to the SM by means of Higgs representations not larger than the adjoint, is the flipped $SO(10)\otimes U(1)$ scenario with two copies of the $16\oplus \overline{16}$ bi-spinor supplemented by the adjoint $45$. Notice further that in the flipped embedding the spinor representations include also weak doublets that may trigger the electroweak symmetry breaking
and allow for renormalizable Yukawa interactions with the chiral matter fields distributed in the flipped embedding over $16\oplus 10\oplus 1$.

Remarkably, the basics of the mechanism we advocate can be embedded in an underlying non-renormalizable $E_{6}$ Higgs model featuring a pair of $27_H\oplus \overline{27}_H$ and the adjoint $78_H$.

Technical similarities apart, there is, however, a crucial difference between the $SO(10)\otimes U(1)$ and $E_{6}$ scenarios, that is related  to the fact that the Lie-algebra of $E_{6}$ is larger than that of $SO(10)\otimes U(1)$.  It has been shown long ago \cite{Buccella:1987kc} that the renormalizable SUSY $E_{6}$ Higgs model spanned on a single copy of $27_H\oplus \overline{27}_H\oplus 78_H$ leaves an $SO(10)$ symmetry unbroken. Two pairs of $27_H \oplus \overline{27}_H$ are needed to reduce the rank by two units.
In spite of the fact that the two SM singlet directions in the $27_H$ are exactly those of the ``flipped'' $16_H$, the little group of the SM singlet directions 
$\vev{27_{H_1} \oplus \overline{27}_{H_1} \oplus 27_{H_2} \oplus \overline{27}_{H_2}}$ and $\vev{78_H}$
remains at the renormalizable level $SU(5)$, as we will explicitly show. 

Adding non-renormalizable adjoint interactions allows for a disentanglement of the $\vev{78_H}$, such that the little group is reduced to the SM. Since a one-step $E_6$ breaking is phenomenologically problematic as mentioned earlier, we argue for a two-step breaking, via flipped $SO(10)\otimes U(1)$, with the $E_6$ scale near the Planck scale.

In summary, we make the case for an anomaly free flipped $SO(10)\otimes U(1)$ partial unification scenario. We provide a detailed discussion of the symmetry breaking pattern obtained within the minimal flipped $SO(10)$ SUSY Higgs model and consider its possible $E_{6}$ embedding. We finally present an elementary discussion of the flavour structure offered by these settings.

\section{The GUT-scale little hierarchy}
\label{littlehierarchy}

In supersymmetric $SO(10)$ models with just $45_H \oplus 16_H \oplus \overline{16}_H$ governing the GUT breaking, one way to obtain the misalignment between the adjoint and the spinors is that of invoking new physics at the Planck scale, parametrized in a model-independent way by a tower of effective operators suppressed by powers of $M_P$.

What we call the ``GUT-scale little hierarchy" is the hierarchy induced in the GUT spectrum by $M_{U}/M_{P}$ suppressed effective operators, which may split the GUT-scale thresholds over several orders of magnitude.
In turn this may be highly problematic for proton stability and the gauge unification in low energy SUSY scenarios (as discussed for instance in Ref. \cite{Chacko:1998jz}). It may also jeopardize the neutrino mass generation in the seesaw scheme. We briefly review the
relevant issues here.

\subsection{GUT-scale thresholds and proton decay}

In Ref.~\cite{Babu:1998wi} the emphasis is set on a class of neutrino-mass-related operators which turns out to be particularly dangerous for proton stability in scenarios with a non-renormalizable GUT-breaking sector.
The relevant interactions can be schematically written as
\begin{multline}
W_Y
\supset \frac{1}{M_P} 16_F\ g\  16_F 16_H 16_H + \frac{1}{M_P} 16_F\ f\ 16_F \overline{16}_H \overline{16}_H \\
\supset \frac{v_R}{M_P} \left( Q\ g\ L\ \overline{T} + Q\ f\ Q\ T \right) \, ,
\end{multline}
where $g$ and $f$ are matrices in the family space, $v_R \equiv |\vev{16_H}| = |\vev{\overline{16}_H}|$ and $T$ ($\overline{T}$) is the color triplet (anti-triplet)
contained in the $\overline{16}_H$ ($16_H$). Integrating out the color triplets, whose mass term is labelled $M_T$, one obtains the following effective superpotential involving fields belonging to $SU(2)_L$ doublets
\be
W_{eff}^{L} = \frac{v_R^2}{M_P^2 M_{T}} \left( u^T F d' \right) \left( u^T G V' \ell - d'^T G V' \nu' \right) \, ,
\label{Peffop}
\ee
where $u$ and $\ell$ denote the physical left-handed up quarks and charged lepton superfields in the basis in which neutral gaugino interactions are flavor diagonal. The $d'$ and $\nu'$ fields are related to the physical down quark and light neutrino fields $d$ and $\nu$ by $d'=V_{CKM} d$ and $\nu' = V_{PMNS} \nu$. In turn $V' = V^\dag_u V_\ell$, where $V_u$ and $V_\ell$ diagonalize the left-handed up quark and charged lepton mass matrices respectively. The $3 \times 3$ matrices $(G,F)$ are given by $(G,F) = V^T_u (g,f) V_u$.

By exploiting the correlations between the $g$ and $f$ matrices and the matter masses and mixings and by taking into account the uncertainties related to the low-energy SUSY spectrum, the GUT-thresholds and the hadronic matrix elements, the authors of Ref.~\cite{Babu:1998wi} argue that the effective operators in \eq{Peffop} lead to a proton lifetime
\be
\Gamma^{-1} (\overline{\nu} K^+) \sim \left( 0.6 - 3 \right) \times 10^{33} \ \text{yrs} \, ,
\label{plife}
\ee
at the verge of the current
experimental lower bound of $0.67 \times 10^{33}$ years \cite{Amsler:2008zzb}.
In obtaining \eq{plife} the authors 
assume that the color triplet masses cluster about the GUT scale, $M_T \approx \vev{16_{H}}\sim\vev{45_{H}}\equiv M_U$. On the other hand, in scenarios where at the renormalizable level $SO(10)$ is broken to $SU(5)$ and the residual $SU(5)$ symmetry is broken to SM by means of non-renormalizable operators, the effective scale of the $SU(5)$ breaking physics is  typically suppressed by $\vev{16_{H}}/M_{P}$ or $\vev{45_{H}}/M_{P}$ with respect to $M_U$. As a consequence, the $SU(5)$-part of the colored triplet higgsino spectrum is effectively pulled down to the $M_{U}^{2}/M_{P}$ scale, in a clash with proton stability.

\subsection{GUT-scale thresholds and one-step unification}
The ``delayed'' residual $SU(5)$ breakdown has obvious implications for
the shape of the gauge coupling unification pattern.
Indeed, the gauge bosons associated to the $SU(5)/SM$ coset, together with the relevant
part of the Higgs spectrum, tend to be uniformly shifted \cite{Babu:1994dc}
by a factor $M_U / M_P \sim 10^{-2}$ below the scale of the $SO(10)/SU(5)$
gauge spectrum, that sets the unification scale, $M_U$.
These thresholds may jeopardize the successful one-step gauge unification pattern favoured by the TeV-scale SUSY extension of the SM (MSSM).

\subsection{GUT-scale thresholds and neutrino masses}

With a non-trivial interplay among several GUT-scale thresholds \cite{Babu:1994dc} one may in principle end up with a viable gauge unification pattern. Namely, the threshold effects in different SM gauge sectors may be such that unification is preserved at a larger scale. In such a case the $M_{U}/M_{P}$ suppression is at least partially undone. This, in turn, is unwelcome for the neutrino mass scale because the VEVs entering the $d=5$ effective operator responsible for the RH neutrino Majorana mass 
term $16_{F}^{2}\overline{16}_{H}^{2}/M_{P}$
are raised accordingly and thus $M_R \sim M_U^{2}/{M_P}$ tends to overshoot the upper limit $M_{R}\lesssim 10^{14}$ GeV implied by the light neutrino masses generated by the seesaw mechanism.

Thus, although the Planck-induced operators can provide a key to overcoming the $SU(5)$ lock of the minimal SUSY 
$SO(10) \rightarrow SU(3)_C \otimes SU(2)_L \otimes U(1)_Y$ Higgs model with $16_{H}\oplus \overline{16}_{H}\oplus 45_{H}$, such an
effective scenario is prone to failure when addressing the measured proton stability and light neutrino phenomenology.

\section{Minimal flipped $SO(10)$ Higgs model}
\label{sect:minimalflippedSO10}

As already anticipated in the previous sections, in a standard $SO(10)$ framework with a Higgs sector
built off the lowest-dimensional representations (up to the adjoint), it is rather difficult to achieve a phenomenologically viable symmetry breaking pattern even admitting multiple copies of each type of multiplets. Firstly, with a single $45_{H}$ at play, at the renormalizable-level the little group of all SM singlet VEVs is $SU(5)$ regardless of the number of $16_{H}\oplus \overline{16}_{H}$ pairs.
The reason is that one can not get anything more than an $SU(5)$ singlet out of a number of $SU(5)$ singlets. The same is true with a second $45_{H}$ added into the Higgs sector because there is no renormalizable mixing among the two $45_{H}$'s apart from the mass term that, without loss of generality, can be taken diagonal.
With a third adjoint Higgs representation at play a cubic $45_145_2 45_3$ interaction is allowed.
However, due to the total antisymmetry of the invariant and to the fact that the adjoints commute on the SM vacuum,
the cubic term does not contribute to the F-term equations~\cite{Babu:1993we}.
This makes the simple flipped $SO(10)\otimes U(1)$ model proposed in this work a framework worth of consideration.
For the sake of completeness, let us also recall that admitting Higgs representations
larger than the adjoint a renormalizable $SO(10)\to $ SM breaking can be devised with the Higgs sector of the form $54{_H} \oplus 45{_H} \oplus 16{_H} \oplus \overline{16}{_H}$ \cite{Buccella:2002zt}, or $54{_H}\oplus 45{_H} \oplus 126{_H} \oplus \overline{126}{_H}$ \cite{Aulakh:2000sn} for a renormalizable seesaw.

In Tables \ref{tab:standvsflipbreak} and \ref{tab:E6break} we collect
a list of the supersymmetric vacua that are obtained in the basic $SO(10)$ Higgs
models and their $E_6$ embeddings by considering a set of Higgs
representations of the dimension of the adjoint and smaller, with all SM singlet VEVs turned on. The cases of a renormalizable (R) or non-renormalizable (NR) Higgs potential are compared.
We quote reference papers where results relevant for the present study were obtained without any aim of exhausting the available literature.
The results without reference are either verified by us or follow by comparison with other cases and rank counting.
The main results of this study are shown in boldface.

\renewcommand{\arraystretch}{1.3}
\begin{table}[h]
\centering
\begin{tabular}{lllll}
\hline \hline
& \multicolumn{2}{l}{Standard $SO(10)$} & \multicolumn{2}{l}{Flipped $SO(10) \otimes U(1)$}
\\
\hline
Higgs superfields & R & NR & R & NR
\\
\hline
$16 \oplus \overline{16}$ & $SO(10)$ & $SU(5)$ & $SO(10) \otimes U(1)$ & $SU(5) \otimes U(1)$
\\
\null
$ 2 \times \left(16 \oplus \overline{16} \right)$ & $SO(10)$ & $SU(5)$ & $SO(10) \otimes U(1)$ & SM
\\
\null
$45 \oplus 16 \oplus \overline{16}$ & $SU(5)$ \cite{Buccella:1981ib} & SM \cite{Babu:1994dc} & $SU(5) \otimes U(1)$ & $\text{SM} \otimes U(1)$
\\
\null
$45 \oplus 2 \times \left( 16 \oplus \overline{16} \right) $ & $SU(5)$ & SM & \bf SM & SM
\\
\hline \hline
\end{tabular}
\mycaption{Comparative summary of supersymmetric vacua left invariant by the SM singlet VEVs in various combinations of spinorial and adjoint
Higgs representations of
standard $SO(10)$ and flipped $SO(10) \otimes U(1)$.
The results for a renormalizable (R) and a non-renormalizable (NR) Higgs superpotential are respectively listed.
}
\label{tab:standvsflipbreak}
\end{table}

\renewcommand{\arraystretch}{1.3}
\begin{table}[h]
\centering
\begin{tabular}{lll}
\hline \hline
Higgs superfields & R & NR
\\
\hline
$27 \oplus \overline{27}$ & $E_6$ & $SO(10)$
\\
\null
$2 \times \left( 27 \oplus \overline{27} \right)$ & $E_6$ & $SU(5)$
\\
\null
$78 \oplus 27 \oplus \overline{27}$ & $SO(10)$ \cite{Buccella:1987kc} & $\text{SM} \otimes U(1)$
\\
\null
$78 \oplus 2 \times \left( 27 \oplus \overline{27} \right)$ & $\mathbf{SU(5)}$ & SM
\\
\hline \hline
\end{tabular}
\mycaption{Same as in Table \ref{tab:standvsflipbreak} for the $E_6$ gauge group with fundamental and adjoint Higgs representations.
}
\label{tab:E6break}
\end{table}

We are going to show that by considering a non-standard hypercharge embedding in $SO(10) \otimes U(1)$ (flipped $SO(10)$) the breaking
to the SM is achievable at the renormalizable level with
$45{_H} \oplus 2 \times \left( 16{_H} \oplus \overline{16}{_H} \right)$ Higgs fields. Let us stress that what we require is that the GUT symmetry breaking is driven by the renormalizable part of the superpotential,
while Planck suppressed interactions may be relevant for the fermion
mass spectrum, in particular for the neutrino sector.


\subsection{Introducing the model}
\subsubsection{Hypercharge embeddings in $SO(10)\otimes U(1)$}
\label{Yembeds}

The so called flipped realization of the  $SO(10)$ gauge symmetry
requires an additional $U(1)_{X}$ gauge factor in order to provide an extra degree of freedom for the SM hypercharge identification.
For a fixed embedding of the $SU(3)_{C}\otimes SU(2)_{L}$ subgroup within $SO(10)$, the SM hypercharge can be generally spanned over the three remaining Cartans generating the abelian $U(1)^{3}$ subgroup of the $SO(10)\otimes U(1)_{X}/(SU(3)_{C}\otimes SU(2)_{L})$ coset.
There are two consistent implementations of the SM hypercharge
within the $SO(10)$ algebra (commonly denoted by standard and flipped $SU(5)$), while a third one becomes available due to the presence of $U(1)_{X}$.

In order to discuss the different embeddings
we find useful to consider two bases for the $U(1)^{3}$ subgroup.
Adopting the traditional left-right (LR) basis corresponding to the $SU(3)_{C}\otimes SU(2)_{L}\otimes SU(2)_{R}\otimes U(1)_{B-L}$ subalgebra of $SO(10)$, one can span the SM hypercharge on the generators of $U(1)_{R}\otimes U(1)_{B-L}\otimes U(1)_{X}$:
\be
\label{YPS}
Y=\alpha T_{R}^{(3)}+\beta(B-L)+\gamma X.
\ee
The normalization of the $T_{R}^{(3)}$ and $B-L$ charges is chosen so that the decompositions of the spinorial and vector representations of $SO(10)$ with respect to  $SU(3)_{C}\otimes SU(2)_{L}\otimes U(1)_{R}\otimes U(1)_{B-L}$ read
\begin{align}
16 &= (3,2;0,+\tfrac{1}{3})\oplus (\overline{3},1;+\tfrac{1}{2},-\tfrac{1}{3})\oplus (\overline{3},1;-\tfrac{1}{2},-\tfrac{1}{3}) 
\oplus  (1,2;0,-1)\oplus (1,1;+\tfrac{1}{2},+1) \nn \\ 
& \oplus (1,1;-\tfrac{1}{2},+1)\nn\;,\\
10 &= (3,1;0,-\tfrac{2}{3})\oplus (\overline{3},1;0,+\tfrac{2}{3})
\oplus (1,2;+\tfrac{1}{2},0)\oplus (1,2;-\tfrac{1}{2},0) \label{LRdecompositions} \;, 
\end{align}
which account for the standard $B-L$ and $T_{R}^{(3)}$ assignments.

Alternatively, considering the $SU(5)\otimes U(1)_{Z}$ subalgebra of $SO(10)$, we identify the 
$U(1)_{Y'}\otimes U(1)_{Z}\otimes U(1)_{X}$ subgroup of $SO(10)\otimes U(1)_{X}$, and equivalently write:
\be
\label{YSU5}
Y=\tilde\alpha Y'+\tilde\beta Z+\tilde\gamma X\;,
\ee
where $Y'$ and $Z$ are normalized
so that the $SU(3)_{C}\otimes SU(2)_{L}\otimes U(1)_{Y'}\otimes U(1)_{Z}$ analogue of eqs.~(\ref{LRdecompositions}) reads:
\begin{align}
16 &= (3,2;+\tfrac{1}{6},+1)\oplus (\overline{3},1;+\tfrac{1}{3},-3)\oplus (\overline{3},1;-\tfrac{2}{3},+1)
\oplus (1,2;-\tfrac{1}{2},-3)\oplus (1,1;+1,+1) \nn \\
& \oplus (1,1;0,+5)\nn\;,\\
10 &= (3,1;-\tfrac{1}{3},-2)\oplus (\overline{3},1;+\tfrac{1}{3},+2)
\oplus (1,2;+\tfrac{1}{2},-2)\oplus (1,2;-\tfrac{1}{2},+2) \label{SU5decompositions} \;.
\end{align}
In both cases,  the $U(1)_X$ charge has been conveniently fixed to $X_{16}=+1$ for the spinorial representation (and thus $X_{10}=-2$ and also $X_{1}=+4$ for the $SO(10)$ vector and singlet, respectively;
this is also the minimal way to obtain an anomaly-free $U(1)_{X}$, that
allows $SO(10)\otimes U(1)_{X}$ to be naturally embedded into $E_{6}$).

It is a straightforward exercise to show that in order to accommodate the SM quark multiplets with  quantum numbers $Q=(3,2,{+}\frac{1}{6})$, $u^{c}=(\overline{3},1,-\frac{2}{3})$ and $d^{c}=(\overline{3},1,+\frac{1}{3})$ there are only three solutions.

On the $U(1)^3$ bases of \eq{YPS} (and \eq{YSU5}, respectively) one obtains,
\be\label{standardabc}
\alpha=1\,,\beta=\tfrac{1}{2}\,,\gamma=0\,, \qquad \left( \tilde\alpha=1\,,\tilde\beta=0\,,\tilde\gamma=0 \right)\,,
\ee
which is nothing but the ``standard'' embedding of the SM matter into $SO(10)$. Explicitly, $Y=T_{R}^{(3)}+\tfrac{1}{2}(B-L)$ in the LR basis (while $Y=Y'$ in the $SU(5)$ picture).

The second option is characterized by
\be\label{flippedSU5abc}
\alpha=-1\,, \beta=\tfrac{1}{2}\,,\gamma=0\,, \qquad \left(  \tilde\alpha=-\tfrac{1}{5}\,,\tilde\beta=\tfrac{1}{5}\,,\tilde\gamma=0 \right)\,,
\ee 
which is usually denoted ``flipped $SU(5)$''~\cite{DeRujula:1980qc,Barr:1981qv}
embedding because the SM hypercharge is spanned non-trivially on the $SU(5)\otimes U(1)_{Z}$ subgroup\footnote{By definition, a flipped variant of a specific GUT model based on a simple gauge group $G$ is obtained by embedding the SM hypercharge nontrivially into the $G\otimes U(1)$ tensor product.} of $SO(10)$, $Y=\tfrac{1}{5}(Z-Y')$.
Remarkably, from the $SU(3)_{C}\otimes SU(2)_{L}\otimes SU(2)_{R}\otimes U(1)_{B-L}$ perspective this setting corresponds to a sign flip of the $SU(2)_{R}$ Cartan operator $T_{R}^{(3)}$, namely $Y=-T_{R}^{(3)}+\tfrac{1}{2}(B-L)$ which can be viewed as a $\pi$ rotation in the $SU(2)_{R}$ algebra.

A third solution corresponds to
\be\label{flippedSO10abc}
\alpha=0\,,\beta=-\tfrac{1}{4}\,,\gamma=\tfrac{1}{4}\,, \qquad \left(  \tilde\alpha=-\tfrac{1}{5}\,,\tilde\beta=-\tfrac{1}{20}\,,\tilde\gamma=\tfrac{1}{4} \right),
\ee
denoted as ``flipped $SO(10)$''~\cite{Kephart:1984xj,Rizos:1988jn,Barr:1988yj} embedding of the SM hypercharge. Notice, in particular, the fundamental difference between the setting (\ref{flippedSO10abc}) with $\gamma =\tilde\gamma= \tfrac{1}{4}$ and the two previous cases (\ref{standardabc}) and (\ref{flippedSU5abc}) where $U(1)_{X}$ does not play any role.

Analogously to what is found for $Y$, once we consider the additional anomaly-free $U(1)_X$ gauge factor,
there are three SM-compatible ways of embedding the physical $\left( B-L \right)$ into $SO(10) \otimes U(1)_X$.
Using the $SU(5)$ compatible description they are respectively given by (see Ref. \cite{Harada:2003sb} for a complete set of relations)
\bea
\label{BLst}
\left( B-L \right)
&=& \tfrac{1}{5} \left( 4 Y' + Z \right)
\, , \\
\label{BLtw1}
\left( B-L \right)
&=& \tfrac{1}{20} \left( 16 Y' - Z + 5 X \right)
\, , \\
\label{BLtw2}
\left( B-L \right)
&=& - \tfrac{1}{20} \left( 8 Y' - 3 Z - 5 X \right)
\, .
\eea
where the first assignment is the standard $B-L$ embedding in \eq{YPS}.
Out of $3 \times 3$ possible pairs of $Y$ and $\left( B-L \right)$ charges only $6$ do correspond to
the quantum numbers of the SM matter \cite{Harada:2003sb}.
By focussing on the flipped $SO(10)$ hypercharge embedding
in \eq{flippedSO10abc}, the two SM-compatible $\left( B-L \right)$ assignments are those in \eqs{BLtw1}{BLtw2} (they are related by a sign flip in $T^{(3)}_R$).
In what follows we shall employ the $\left( B-L \right)$ assignment in \eq{BLtw2}.

\subsubsection{Spinor and adjoint SM singlets in flipped $SO(10)$}
The active role of the $U(1)_{X}$ generator in the SM hypercharge
(and $B-L$) identification within the flipped $SO(10)$ scenario has relevant consequences for model building. In particular, 
the SM decomposition of the $SO(10)$ representations change so that there are additional SM singlets both in $16_{H}\oplus \overline{16}_{H}$ as well as in $45_{H}$.

The pattern of SM singlet components in flipped $SO(10)$ has a simple and intuitive interpretation from the $SO(10)\otimes U(1)_{X}\subset E_{6}$ perspective, where $16_{+1}\oplus \overline{16}_{-1}$ (with the subscript indicating the $U(1)_{X}$ charge)
are contained in $27\oplus \overline{27}$ while $45_{0}$ is a part of the $E_{6}$ adjoint $78$. The point is that the flipped SM hypercharge assignment makes the various SM singlets within the complete $E_{6}$ representations ``migrate'' among their different $SO(10)$ sub-multiplets; namely, the two SM singlets in the $27$ of $E_{6}$ that in the standard embedding (\ref{standardabc}) reside in the $SO(10)$ singlet $1$ and spinorial $16$ components both happen to fall into just the single $16\subset 27$ in the flipped $SO(10)$ case.

Similarly, there are two additional SM singlet directions in $45_{0}$ in the flipped $SO(10)$ scenario, that, in the standard $SO(10)$ embedding, belong to the $16_{-3}\oplus \overline{16}_{+3}$ components of the $78$ of  $E_{6}$, thus accounting for a total of four adjoint SM singlets.

In Tables \ref{tab:10decomp}, \ref{tab:16decomp} and \ref{tab:45decomp}
we summarize the decomposition of the $10_{-2}$, $16_{+1}$ and $45_{0}$ representations of $SO(10)\otimes U(1)_{X}$ under the SM subgroup,
in both the standard and the flipped $SO(10)$ cases (and in both the LR
and $SU(5)$ descriptions). The pattern of the SM singlet components
is emphasized in boldface.

\renewcommand{\arraystretch}{1.3}
\begin{table}[ht]
\centering
\begin{tabular}{llll}
\hline \hline
\multicolumn{2}{c}{LR} & \multicolumn{2}{c}{$SU(5)$}
\\
\hline
$SO(10)$
& $SO(10)_f$
& $SO(10)$
& $SO(10)_f$
\\
\hline
$(3,1;-\tfrac{1}{3})_{6}$
& $(3,1;-\tfrac{1}{3})_{6}$
& $(3,1;-\tfrac{1}{3})_{5}$
& $(3,1;-\tfrac{1}{3})_{5}$
\\
\null
$(\overline{3},1;+\tfrac{1}{3})_{6}$
& $(\overline{3},1;-\tfrac{2}{3})_{6}$
& $(1,2;+\tfrac{1}{2})_{5}$
& $(1,2;-\tfrac{1}{2})_{5}$
\\
\null
$(1,2;+\tfrac{1}{2})_{1^+}$
& $(1,2;-\tfrac{1}{2})_{1^+}$
& $(\overline{3},1;+\tfrac{1}{3})_{\overline{5}}$
& $(\overline{3},1;-\tfrac{2}{3})_{\overline{5}}$
\\
\null
$(1,2;-\tfrac{1}{2})_{1^-}$
& $(1,2;-\tfrac{1}{2})_{1^-}$
& $(1,2;-\tfrac{1}{2})_{\overline{5}}$
& $(1,2;-\tfrac{1}{2})_{\overline{5}}$
\\
\hline \hline
\end{tabular}
\mycaption{Decomposition of the fundamental 10-dimensional representation under $SU(3)_C \otimes SU(2)_L \otimes U(1)_Y$,
for standard $SO(10)$ and flipped $SO(10) \otimes U(1)_X$ ($SO(10)_f$) respectively.
In the first two columns (LR) the subscripts keep track of the
$SU(4)_C$ origin of the multiplets (the extra symbols $\pm$ correspond to the eigenvalues of the $T^{(3)}_{R}$ Cartan generator)
while in the last two columns the $SU(5)$ content is shown.}
\label{tab:10decomp}
\end{table}

\renewcommand{\arraystretch}{1.3}
\begin{table}[ht]
\centering
\begin{tabular}{llll}
\hline \hline
\multicolumn{2}{c}{LR} & \multicolumn{2}{c}{$SU(5)$}
\\
\hline
$SO(10)$
& $SO(10)_f$
& $SO(10)$
& $SO(10)_f$
\\
\hline
$(3,2;+\tfrac{1}{6})_{4}$
& $(3,2;+\tfrac{1}{6})_{4}$
& $(\overline{3},1;+\tfrac{1}{3})_{\overline{5}}$
& $(\overline{3},1;+\tfrac{1}{3})_{\overline{5}}$
\\
\null
$(1,2; -\tfrac{1}{2})_{4}$
& $(1,2; +\tfrac{1}{2})_{4}$
& $(1,2; -\tfrac{1}{2})_{\overline{5}}$
& $(1,2; +\tfrac{1}{2})_{\overline{5}}$
\\
\null
$(\overline{3},1;+\tfrac{1}{3})_{\overline{4}^+}$
& $(\overline{3},1;+\tfrac{1}{3})_{\overline{4}^+}$
& $(3,2;+\tfrac{1}{6})_{10}$
& $(3,2;+\tfrac{1}{6})_{10}$
\\
\null
$(\overline{3},1;-\tfrac{2}{3})_{\overline{4}^-}$
& $(\overline{3},1;+\tfrac{1}{3})_{\overline{4}^-}$
& $(\overline{3},1;-\tfrac{2}{3})_{10}$
& $(\overline{3},1;+\tfrac{1}{3})_{10}$
\\
\null
$(1,1;+1)_{\overline{4}^+}$
& $\mathbf{(1,1;0)_{\overline{4}^+}}$
& $(1,1;+1)_{10}$
& $\mathbf{(1,1;0)_{10}}$
\\
\null
$\mathbf{(1,1;0)_{\overline{4}^-}}$
& $\mathbf{(1,1;0)_{\overline{4}^-}}$
& $\mathbf{(1,1;0)_{1}}$
& $\mathbf{(1,1;0)_{1}}$
\\
\hline \hline
\end{tabular}
\mycaption{The same as in Table \ref{tab:10decomp} for the spinor $16$-dimensional representation.
The SM singlets are emphasized in boldface and shall be denoted, in the $SU(5)$ description, as $e \equiv (1,1;0)_{10}$ and $\nu \equiv (1,1;0)_{1}$.
The LR decomposition shows that $e$ and $\nu$ belong to an $SU(2)_R$ doublet.}
\label{tab:16decomp}
\end{table}

\renewcommand{\arraystretch}{1.3}
\begin{table}[ht]
\centering
\begin{tabular}{llll}
\hline \hline
\multicolumn{2}{c}{LR} & \multicolumn{2}{c}{$SU(5)$}
\\
\hline
$SO(10)$
& $SO(10)_f$
& $SO(10)$
& $SO(10)_f$
\\
\hline
$\mathbf{(1,1;0)_{1^0}}$
& $\mathbf{(1,1;0)_{1^0}}$
& $\mathbf{(1,1;0)_{1}}$
& $\mathbf{(1,1;0)_{1}}$
\\
\null
$\mathbf{(1,1;0)_{15}}$
& $\mathbf{(1,1;0)_{15}}$
& $\mathbf{(1,1;0)_{24}}$
& $\mathbf{(1,1;0)_{24}}$
\\
\null
$(8,1;0)_{15}$
& $(8,1;0)_{15}$
& $(8,1;0)_{24}$
& $(8,1;0)_{24}$
\\
\null
$(3,1;+\tfrac{2}{3})_{15}$
& $(3,1;-\tfrac{1}{3})_{15}$
& $(3,2;-\tfrac{5}{6})_{24}$
& $(3,2;+\tfrac{1}{6})_{24}$
\\
\null
$(\overline{3},1;-\tfrac{2}{3})_{15}$
& $(\overline{3},1;+\tfrac{1}{3})_{15}$
& $(\overline{3},2;+\tfrac{5}{6})_{24}$
& $(\overline{3},2;-\tfrac{1}{6})_{24}$
\\
\null
$(1,3;0)_{1}$
& $(1,3;0)_{1}$
& $(1,3;0)_{24}$
& $(1,3;0)_{24}$
\\
\null
$(3,2;+\tfrac{1}{6})_{6^+}$
& $(3,2;+\tfrac{1}{6})_{6^+}$
& $(3,2;+\tfrac{1}{6})_{10}$
& $(3,2;+\tfrac{1}{6})_{10}$
\\
\null
$(\overline{3},2;+\tfrac{5}{6})_{6^+}$
& $(\overline{3},2;-\tfrac{1}{6})_{6^+}$
& $(\overline{3},1;-\tfrac{2}{3})_{10}$
& $(\overline{3},1;+\tfrac{1}{3})_{10}$
\\
\null
$(1,1;+1)_{1^+}$
& $\mathbf{(1,1;0)_{1^+}}$
& $(1,1;+1)_{10}$
& $\mathbf{(1,1;0)_{10}}$
\\
\null
$(\overline{3},2;-\tfrac{1}{6})_{6^-}$
& $(\overline{3},2;-\tfrac{1}{6})_{6^-}$
& $(\overline{3},2;-\tfrac{1}{6})_{\overline{10}}$
& $(\overline{3},2;-\tfrac{1}{6})_{\overline{10}}$
\\
\null
$(3,2;-\tfrac{5}{6})_{6^-}$
& $(3,2;+\tfrac{1}{6})_{6^-}$
& $(3,1;+\tfrac{2}{3})_{\overline{10}}$
& $(3,1;-\tfrac{1}{3})_{\overline{10}}$
\\
\null
$(1,1;-1)_{1^-}$
& $\mathbf{(1,1;0)_{1^-}}$
& $(1,1;-1)_{\overline{10}}$
& $\mathbf{(1,1;0)_{\overline{10}}}$
\\
\hline \hline
\end{tabular}
\mycaption{{The same} as in Table \ref{tab:10decomp} for the $45$ representation.
The SM singlets are given in boldface and labeled throughout the text as $\omega_{B-L} \equiv (1,1;0)_{15}$,
$\omega^+ \equiv (1,1;0)_{1^+}$, $\omega_R \equiv (1,1;0)_{1^0}$ and $\omega^- \equiv (1,1;0)_{1^-}$ where again the LR notation has been used.
The LR decomposition also shows that $\omega^+$, $\omega_R$ and $\omega^-$ belong to an $SU(2)_R$ triplet, while $\omega_{B-L}$ is a $B-L$ singlet.}
\label{tab:45decomp}
\end{table}

\subsubsection{The supersymmetric flipped $SO(10)$ model}

The presence of additional SM singlets (some of them transforming non-trivially under $SU(5)$) in the lowest-dimensional representations of the flipped realisation of the $SO(10)$ gauge symmetry provides the
ground for obtaining a viable symmetry breaking with a significantly simplified renormalizable Higgs sector. Naively, one may guess that the pair of VEVs in $16_{H}$ (plus another conjugated pair in $\overline{16}_{H}$ to maintain the required $D$-flatness)
might be enough to break the GUT symmetry entirely, since
one component transforms as a $10$ of $SU(5)\subset SO(10)$, while the other one is identified with the $SU(5)$ singlet (cf.~Table~\ref{tab:16decomp}). Notice that even in the presence of an additional four-dimensional vacuum manifold of the adjoint Higgs multiplet, the little group is determined by the $16_H$ VEVs since, due to the simple form of the renormalizable superpotential $F$-flatness makes the VEVs of $45_{H}$ align with those of ${16}_{H}\overline{16}_{H}$, providing just enough freedom for them to develop non-zero values.

{Unfortunately, this is still not enough to support the desired symmetry breaking pattern.} The two VEV directions in $16_{H}$ are equivalent to one and a residual $SU(5) \otimes U(1)$ symmetry is always preserved by $\vev{16}_{H}$~\cite{Buccella:1980qb}.
Thus, even in the flipped $SO(10)\otimes U(1)$ setting the Higgs model spanned on ${16}_{H}\oplus\overline{16}_{H}\oplus 45_{H}$ suffers from
an $SU(5)\otimes U(1)$ lock analogous to the one of the standard SUSY $SO(10)$ models with the same Higgs sector.
This can be understood by taking into account the freedom in choosing the basis in the $SO(10)$ algebra so that the pair of VEVs within $16$ can be ``rotated'' onto a single component, which can be then viewed as the direction of the singlet in the decomposition of $16=\overline{5}\oplus 10\oplus 1$ with respect to an $SU(5)$ subgroup of the original $SO(10)$ gauge symmetry.

On the other hand, with a pair of interacting $16_{H}\oplus \overline{16}_{H}$'s the vacuum directions in the two $16_{H}$'s need not be aligned and the intersection of the two different invariant subalgebras (e.g.~, standard and flipped $SU(5)$ for a specific VEV configuration) leaves as a little group the $SU(3)_{C}\otimes SU(2)_{L}\otimes U(1)_{Y}$ of the SM.
$F$-flatness makes then the adjoint VEVs ($45_H$ is the needed carrier
of $16_H$ interaction at the renormalizable level) aligned to the SM vacuum.
Hence, as we will show in the next section, $2\times (16_{H}+ \overline{16}_{H})\oplus 45_{H}$ defines the minimal renormalizable Higgs setting for the SUSY flipped $SO(10)\otimes U(1)_{X}$ model.
For comparison, let us reiterate that in the standard renormalizable $SO(10)$ setting the SUSY vacuum is always $SU(5)$ regardless of how many copies of $16_{H}\oplus \overline{16}_{H}$ are employed together with at most a pair of adjoints.

\subsubsection{The matter sector}

Due to the flipped hypercharge assignment, the SM matter can no longer be fully embedded into the 16-dimensional $SO(10)$ spinor, as in the standard case. By inspecting Table~\ref{tab:16decomp} one can see that in the flipped setting the pair of the SM sub-multiplets of $16$ transforming as $u^{c}$ and $e^{c}$ is traded for an extra $d^{c}$-like state and an extra SM singlet.
The former pair is instead found in the $SO(10)$ vector and the singlet (the lepton doublet as well appears in the vector multiplet). Thus, flipping spreads each of the SM matter generations across $16\oplus 10\oplus 1$ of $SO(10)$, which, by construction, can be viewed as the complete 27-dimensional fundamental representation of $E_{6}\supset SO(10)\otimes U(1)_{X}$. This brings in a set of additional degrees of freedom, in particular $(1,1,0)_{{16}}$, $(\overline{3},1,+\tfrac{1}{3})_{{16}}$,   $(1,2,+\tfrac{1}{2})_{{16}}$,  $(3,1,-\tfrac{1}{3})_{{10}}$ and $(1,2,-\tfrac{1}{2})_{{10}}$,  where the subscript indicates their ${SO(10)}$ origin. Notice, however, that these SM ``exotics'' can be grouped into superheavy vector-like pairs and thus no extra states appear in the low energy spectrum. Furthermore, the $U(1)_{X}$ anomalies associated with each of the $SO(10)\otimes U(1)_{X}$ matter multiplets cancel when summed over the entire reducible representation $16_{1}\oplus 10_{-2}\oplus 1_{4}$. An elementary discussion of the matter spectrum in this scenario is deferred to \sect{TRflavor}.

\subsection{Supersymmetric vacuum}
\label{vacuumFSO10}

The most general renormalizable Higgs superpotential, made of the representations $45 \oplus 16_1 \oplus \overline{16}_1 \oplus 16_2  \oplus \overline{16}_2$ is given by
\be
\label{WHFSO10}
W_H = \frac{\mu}{2} \, \Tr 45^2 + \rho_{ij} 16_{i} \overline{16}_{j} + \tau_{ij} 16_{i} 45  \overline{16}_{j} \, ,
\ee
where $i,j = 1,2$ and the notation is explained in \app{flippedSO10notation}.
Without loss of generality we can take $\mu$ real by a global phase redefinition, while $\tau$ (or $\rho$) can be diagonalized by a bi-unitary transformation acting on the flavor indices of the $16$ and the $\overline{16}$. Let us choose, for instance, $\tau_{ij} = \tau_i \delta_{ij}$, with $\tau_i$ real.
We label the SM-singlets contained in the $16$'s in the following way:
$e \equiv (1,1;0)_{10}$ (only for flipped $SO(10)$) and $\nu \equiv (1,1;0)_{1}$
(for all embeddings).

By plugging in the SM-singlet VEVs $\omega_R$, $\omega_{B-L}$, $\omega^+$, $\omega^-$, $e_{1,2}$, $\overline{e}_{1,2}$, $\nu_{1,2}$ and $\overline{\nu}_{1,2}$
(cf.~\app{flippedSO10notation}), the superpotential on the vacuum reads
\begin{align}
\label{vevWHFSO10}
\vev{W_H} &= \mu \left(2 \omega _R^2+3 \omega_{B-L}^2+4 \omega ^- \omega ^+\right) \nn \\
& + \rho _{11} \left(e_1 \overline{e}_1+\nu _1 \overline{\nu }_1\right)+\rho _{21} \left(e_2
   \overline{e}_1+\nu _2 \overline{\nu }_1\right) 
   + \rho _{12} \left(e_1 \overline{e}_2+\nu _1 \overline{\nu }_2\right)+\rho _{22} \left(e_2 \overline{e}_2+\nu _2 \overline{\nu }_2\right) \nn \\
   &+ \tau _1 \left[ - \omega ^- e_1 \overline{\nu }_1- \omega ^+ \nu _1 \overline{e}_1 - \frac{\omega _R}{\sqrt{2}} \left(e_1 \overline{e}_1-\nu _1 \overline{\nu }_1\right) 
   +  \frac{3}{2} \frac{\omega_{B-L}}{\sqrt{2}} \left(e_1 \overline{e}_1
   +\nu _1 \overline{\nu }_1\right) \right] \nn \\
   &+ \tau _2 \left[- \omega ^- e_2 \overline{\nu }_2- \omega ^+ \nu _2  \overline{e}_2 - \frac{\omega _R}{\sqrt{2}} \left(e_2
   \overline{e}_2-\nu _2 \overline{\nu }_2\right) + \frac{3}{2} \frac{\omega_{B-L}}{\sqrt{2}} \left( e_2 \overline{e}_2+ \nu _2 \overline{\nu }_2\right)\right] \, .
\end{align}
In order to retain SUSY down to the TeV scale
we must require
that the GUT gauge symmetry breaking preserves supersymmetry.
In \app{DFtermFSO10} we work out the relevant $D$- and $F$-term equations.
We find that the existence of a nontrivial vacuum requires $\rho$ (and $\tau$ for consistency) to be hermitian matrices.
This is a consequence of the fact that $D$-term flatness for the flipped $SO(10)$ embedding implies
$\vev{16_i} = \vev{\overline{16}_i}^*$ ({see \eq{complexDtermsFSO10} and the discussion next to it}).
With this restriction the vacuum manifold is given by
{
\begin{align}
\label{vacmanifoldFSO10}
8 \mu\, \omega^+ &= \tau _1 r_1^2 \sin{2\alpha_1} e^{i (\phi_{e_1} - \phi_{\nu_1})} + \tau _2 r_2^2 \sin{2\alpha_2} e^{i (\phi_{e_2} - \phi_{\nu_2})} \, , \nn \\
8 \mu\, \omega^- &= \tau _1  r_1^2 \sin{2\alpha_1} e^{-i (\phi_{e_1} - \phi_{\nu_1})} + \tau _2 r_2^2 \sin{2\alpha_2} e^{-i (\phi_{e_2} - \phi_{\nu_2})} \, , \nn \\
4 \sqrt{2} \mu\, \omega_R &= \tau _1 r_1^2 \cos{2\alpha_1} + \tau _2 r_2^2 \cos{2\alpha_2} \, , \nn \\
4 \sqrt{2} \mu\, \omega_{B-L} &= -\tau _1 r_1^2 - \tau _2 r_2^2 \, , \nn \\
e_{1,2} &= r_{1,2} \cos \alpha_{1,2} \ e^{i \phi_{e_{1,2}}} \, , \nn \\
\nu_{1,2} &= r_{1,2} \sin \alpha_{1,2}\ e^{i \phi_{\nu_{1,2}}} \, , \nn \\
\overline{e}_{1,2} &= r_{1,2} \cos \alpha_{1,2}\ e^{- i \phi_{e_{1,2}}} \, , \nn \\
\overline{\nu}_{1,2} &= r_{1,2} \sin \alpha_{1,2}\ e^{- i \phi_{\nu_{1,2}}} \, ,
\end{align}
}
where $r_{1,2}$ and $\alpha^{\pm} \equiv \alpha_1 \pm \alpha_2$ are fixed in terms of the superpotential parameters,
\begin{align}
\label{r1sqFSO10}
& r_{1}^{2} = -\frac{2 \mu \left( \rho _{22} \tau _1-5 \rho _{11} \tau _2\right)}{3 \tau _1^2 \tau _2} \, , \\
\label{r2sqFSO10}
& r_{2}^{2} = -\frac{2 \mu \left(\rho _{11} \tau _2-5 \rho _{22} \tau _1\right)}{3 \tau _1 \tau _2^2} \, , \\
\label{cosalpha1m2FSO10}
& \cos{\alpha^{-}} = \xi \ \frac{\sin \Phi_\nu - \sin \Phi_e}{\sin \left(\Phi_\nu - \Phi_e \right)} \, , \\
\label{cosalpha1p2FSO10}
& \cos{\alpha^{+}} = \xi \ \frac{\sin \Phi_\nu + \sin \Phi_e}{\sin \left(\Phi_\nu - \Phi_e \right)}  \, ,
\end{align}
with
\be
\label{xiFSO10}
\xi = \frac{6 |\rho _{12}|}{\sqrt{-\frac{5 \rho _{11}^2 \tau _2}{\tau _1}-\frac{5 \rho _{22}^2 \tau _1}{\tau _2}+26 \rho _{22} \rho _{11}}} \, .
\ee
The phase factors $\Phi_\nu$ and $\Phi_e$ are defined as
\be
\label{PhienuFSO10}
\Phi_\nu \equiv \phi_{\nu_1}-\phi_{\nu_2}+\phi_{\rho _{12}} \, , \quad \Phi_e \equiv \phi_{e_1}-\phi_{e_2}+\phi_{\rho _{12}} \, ,
\ee
in terms of the relevant phases $\phi_{\nu_{1,2}}$, $\phi_{e_{1,2}}$ and $\phi_{\rho_{12}}$.
\eqs{cosalpha1m2FSO10}{cosalpha1p2FSO10} imply that for $\Phi_{\nu} = \Phi_{e} = \Phi$, \eq{cosalpha1m2FSO10} reduces
to $\cos{\alpha^{-}} \rightarrow \xi \cos{\Phi}$ while $\alpha^{+}$ is undetermined (thus parametrizing an orbit of isomorphic vacua).

In order to determine the little group of the vacuum manifold
we explicitly compute the corresponding gauge boson spectrum in \app{gaugespectrum}.
We find that, for $\alpha^- \neq 0$ and/or $\Phi_\nu \neq \Phi_e$, the vacuum in
\eq{vacmanifoldFSO10} does preserve the SM algebra.

As already mentioned in the introduction this result is a consequence
of the misalignement of the spinor VEVs, that is made possible
at the renormalizable level by the interaction with the $45_H$.
If we choose to align the $16_1 \oplus \overline{16}_1$ and $16_2 \oplus \overline{16}_2$ VEVs ($\alpha^- = 0$ and $\Phi_{\nu} = \Phi_{e}$)
or equivalently, to decouple one of the Higgs spinors from the vacuum ($r_2 = 0$ for instance)
the little group is $SU(5)\otimes U(1)$.

This result can be easily understood by observing that in the case with just one pair of $16_{H} \oplus \overline{16}_{H}$
(or with two pairs of $16_{H} \oplus \overline{16}_{H}$ aligned) the two SM-singlet directions, $e_H$ and $\nu_H$, are connected by
an $SU(2)_R$ transformation. This freedom can be used to rotate one of the VEVs to zero, so that the little group is standard or flipped $SU(5) \otimes U(1)$, depending on which of the two VEVs is zero.

In this respect, the Higgs adjoint plays the role of a renormalizable agent that prevents the two pairs of spinor vacua from aligning with each other along the $SU(5) \otimes U(1)$ direction.
Actually, by decoupling the adjoint Higgs, $F$-flatness makes the (aligned) $16_i \oplus \overline{16}_i$ vacuum trivial, as one verifies by inspecting the $F$-terms in \eq{FtermsFSO10} of~\app{DFtermFSO10} for $\vev{45_H}=0$ and $\det{\rho}\neq 0$.

The same result with just two pairs of $16_{H} \oplus \overline{16}_{H}$ Higgs multiplets is obtained by adding non-renormalizable spinor interactions,
at the cost of
introducing a potentially critical GUT-scale threshold hierarchy.
In the flipped $SO(10)$ setup here proposed the GUT symmetry breaking is driven by the renormalizable part of the Higgs superpotential,
thus allowing naturally for a one-step matching with the minimal supersymmetric extension of the SM (MSSM).

Before addressing the possible embedding of the model in a
unified $E_6$ scenario, we comment in brief on the naturalness
of the doublet-triplet mass splitting in flipped embeddings.

\subsection{Doublet-Triplet splitting in flipped models}
\label{DTsplittingflipped}

Flipped embeddings offers a rather economical way to implement the Doublet-Triplet (DT) splitting through the so called Missing Partner (MP) mechanism \cite{Antoniadis:1987dx,Barr:2010tv}. In order to show the relevat features let us consider first the
flipped $SU(5) \otimes U(1)_Z$.

In order to implement the MP mechanism in the flipped $SU(5) \otimes U(1)_Z$ the Higgs superpotential is required to have the couplings
\be
\label{MPflippedSU5}
W_H \supset 10_{+1} 10_{+1} 5_{-2} + \overline{10}_{-1} \overline{10}_{-1} \overline{5}_{+2} \, ,
\ee
where the subscripts correspond to the $U(1)_Z$ quantum numbers,
but not the $5_{-2} \overline{5}_{+2}$ mass term.
From \eq{MPflippedSU5}
we extract the relevant terms that lead to a mass for the Higgs triplets
\be
\label{MPflippedSU5vev}
W_H \supset \vev{(1,1;0)_{10}} (\overline{3},1;+\tfrac{1}{3})_{10} (3,1;-\tfrac{1}{3})_{5}
+ \vev{(1,1;0)_{\overline{10}}} (3,1;-\tfrac{1}{3})_{\overline{10}} (\overline{3},1;+\tfrac{1}{3})_{\overline{5}} \, .
\ee
On the other hand, the Higgs doublets, contained in the $5_{-2} \oplus \overline{5}_{+2}$ remain massless since they have no partner in the
$10_{+1} \oplus \overline{10}_{-1}$ to couple with.

The MP mechanism cannot be implemented in standard $SO(10)$.
The relevant interactions, analogue of \eq{MPflippedSU5}, are contained into the $SO(10)$ invariant term
\be
W_H \supset 16\, 16\, 10 + \overline{16}\, \overline{16}\, 10 \, ,
\ee
which, however, gives a mass to the doublets as well, via the superpotential terms
\be
W_H \supset \vev{(1,1;0)_{1_{16}}} (1,2; -\tfrac{1}{2})_{\overline{5}_{16}} (1,2;+\tfrac{1}{2})_{5_{10}}
+ \vev{(1,1;0)_{1_{\overline{16}}}} (1,2; +\tfrac{1}{2})_{5_{\overline{16}}} (1,2;-\tfrac{1}{2})_{\overline{5}_{10}} \, .
\ee

Flipped $SO(10) \otimes U(1)_X$, on the other hand,  offers again the possibility of implementing the MP mechanism.
The prize to pay is the necessity of avoiding a large number of terms, both bilinear and trilinear, in the Higgs superpotential.
In particular, the analogue of \eq{MPflippedSU5} is given by the non-renormalizable term~\cite{Maekawa:2003wm}
\be
\label{MPflippedSO10}
W_H \supset \frac{1}{M_P} \overline{16}_1 16_2 16_2 \overline{16}_1 + \frac{1}{M_P} 16_1 \overline{16}_2 \overline{16}_2 16_1 \, .
\ee
By requiring that $16_1$ ($\overline{16_1}$) takes a VEV in the $1_{16}$ ($1_{\overline{16}}$) direction while $16_2$ ($\overline{16_2}$) in the $10_{16}$ ($\overline{10}_{\overline{16}}$) component, one gets
\be
W_H \supset \frac{1}{M_P} \vev{1_{\overline{16}_1}} \vev{{10}_{16_2}} 10_{16_2} 5_{\overline{16}_1}
+ \frac{1}{M_P} \vev{1_{16_1}} \vev{{\overline{10}}_{\overline{16}_2}} \overline{10}_{\overline{16}_2} \overline{5}_{16_1} \, ,
\ee
which closely resembles \eq{MPflippedSU5}, leading to massive triplets and massless doublets.
In order to have minimally one pair of electroweak doublets, one must further require that the $2 \times 2$
mass matrix of the $16$'s has rank equal to one.
Due to the active role of non-renormalizable operators,
the Higgs triplets turn out to be two orders of magnitude below the flipped $SO(10) \otimes U(1)_X$ scale,
reintroducing the issues discussed as in \sect{littlehierarchy}.

An alternative possibility for naturally implementing the DT splitting in $SO(10)$ is the Dimopoulos-Wilczek (DW) (or the missing VEV) mechanism \cite{Dimopoulos:1981xm}.
In order to explain the key features it is convenient to decompose the relevant $SO(10)$ representations in terms of the $SU(4)_C \otimes SU(2)_L \otimes SU(2)_R$ group
\bea
\label{SO10decomp}
&& 45 \equiv (1,1,3) \oplus (15,1,1) \oplus \ldots \nn \\
&& 16 \equiv (4,2,1) \oplus (\overline{4},1,2) \, , \nn \\
&& \overline{16} \equiv (\overline{4},2,1) \oplus (4,1,2) \, , \nn \\
&& 10 \equiv (6,1,1) \oplus (1,2,2) \, ,
\eea
where $\omega_R \equiv \vev{(1,1,3)}$ and $\omega_{B-L} \equiv \vev{(15,1,1)}$.
In the standard $SO(10)$ case (see \cite{Babu:1994dq,Barr:1997hq} and \cite{Babu:2010ej} for a recent discussion)
one assumes that the $SU(2)_L$ doublets are contained in two vector
multiplets ($10_1$ and $10_2$).
From the decompositions in \eq{SO10decomp} it's easy to see that the interaction $10_1 45\, 10_2$
(where the antisymmetry of $45$ requires the presence of two $10$'s)
leaves the $SU(2)_L$ doublets massless provided that $\omega_R = 0$.
For the naturalness of the setting
other superpotential terms must not appear, as a direct mass term for one of the $10$'s
and the interaction term $16\, 45 \, \overline{16}$. The latter aligns the SUSY vacuum in the $SU(5)$ direction ($\omega_R = \omega_{B-L}$),
thus destabilizing the DW solution.

On the other hand, the absence of the $16\, 45 \, \overline{16}$
interaction enlarges the global symmetries of the
scalar potential with the consequent appearance of a set of light pseudo-Goldstone bosons in the
spectrum.
To avoid that the adjoint and the spinor sector must be coupled in an indirect way by adding extra fields and symmetries (see for instance \cite{Babu:1994dq,Barr:1997hq,Babu:2010ej}).

Our flipped $SO(10) \otimes U(1)_X$ setting offers the rather economical possibility of embedding the electroweak doublets directly into the spinors without the need of $10_H$
(see \sect{TRflavor}).
As a matter of fact, there exists a variant of the DW mechanism
where the $SU(2)_L$ doublets, contained in the $16{_H} \oplus \overline{16}{_H}$, are kept massless
by the condition $\omega_{B-L} = 0$ (see e.g.~\cite{Dvali:1996wh}).
However, in order to satisfy in a natural way the $F$-flatness for the configuration $\omega_{B-L} = 0$,
again a contrived superpotential is required, when compared to that in \eq{WHFSO10}.
In conclusion, we cannot implement in our simple setup any of the natural mechanisms so far proposed 
and we have to resort to the standard minimal fine-tuning.

\section{Minimal $E_6$ embedding}
\label{minimalE6Higgs}

The natural and minimal unified embedding of the flipped $SO(10)\otimes U(1)$ model is $E_6$ with one $78_H$ and two pairs of $27_H \oplus \overline{27}_H$
in the Higgs sector. The three matter families are contained in three $27_F$ chiral superfields.
The decomposition of the $27$ and $78$ representations under the SM quantum numbers is detailed
in Tables \ref{tab:27decompSU(5)} 
and \ref{tab:78decompSU(5)}, 
according to
the different hypercharge embeddings.

\renewcommand{\arraystretch}{1.3}
\begin{table}[ht]
\centering
\begin{small}
\begin{tabular}{lll}
\hline \hline
$SU(5)$ \ \
& $SU(5)_f$\ \
& $SO(10)_f$
\\
\hline
$(\overline{3},1;+\tfrac{1}{3})_{\overline{5}_{16}}$
& $(\overline{3},1;-\tfrac{2}{3})_{\overline{5}_{16}}$
& $(\overline{3},1;+\tfrac{1}{3})_{\overline{5}_{16}}$
\\
\null
$(1,2; -\tfrac{1}{2})_{\overline{5}_{16}}$
& $(1,2; -\tfrac{1}{2})_{\overline{5}_{16}}$
& $(1,2; +\tfrac{1}{2})_{\overline{5}_{16}}$
\\
\null
$(3,2;+\tfrac{1}{6})_{10_{16}}$
& $(3,2;+\tfrac{1}{6})_{10_{16}}$
& $(3,2;+\tfrac{1}{6})_{10_{16}}$
\\
\null
$(\overline{3},1;-\tfrac{2}{3})_{10_{16}}$
& $(\overline{3},1;+\tfrac{1}{3})_{10_{16}}$
& $(\overline{3},1;+\tfrac{1}{3})_{10_{16}}$
\\
\null
$(1,1;+1)_{10_{16}}$
& $(1,1;0)_{10_{16}}$
& $(1,1;0)_{10_{16}}$
\\
\null
$(1,1;0)_{1_{16}}$
& $(1,1;+1)_{1_{16}}$
& $(1,1;0)_{1_{16}}$
\\
\hline
$(3,1;-\tfrac{1}{3})_{5_{10}}$
& $(3,1;-\tfrac{1}{3})_{5_{10}}$
& $(3,1;-\tfrac{1}{3})_{5_{10}}$
\\
\null
$(1,2;+\tfrac{1}{2})_{5_{10}}$
& $(1,2;-\tfrac{1}{2})_{5_{10}}$
& $(1,2;-\tfrac{1}{2})_{5_{10}}$
\\
\null
$(\overline{3},1;+\tfrac{1}{3})_{\overline{5}_{10}}$
& $(\overline{3},1;+\tfrac{1}{3})_{\overline{5}_{10}}$
& $(\overline{3},1;-\tfrac{2}{3})_{\overline{5}_{10}}$
\\
\null
$(1,2;-\tfrac{1}{2})_{\overline{5}_{10}}$
& $(1,2;+\tfrac{1}{2})_{\overline{5}_{10}}$
& $(1,2;-\tfrac{1}{2})_{\overline{5}_{10}}$
\\
\hline
$(1,1;0)_{1_{1}}$
& $(1,1;0)_{1_{1}}$
& $(1,1;+1)_{1_{1}}$
\\
\hline \hline
\end{tabular}
\end{small}
\mycaption{Decomposition of the fundamental representation $27$ of $E_6$ under $SU(3)_C \otimes SU(2)_L \otimes U(1)_Y$, according to the three SM-compatible different embeddings of the hypercharge ($f$ stands for flipped). The numerical subscripts keep track of the $SU(5)$ and $SO(10)$ origin.}
\label{tab:27decompSU(5)}
\end{table}

\renewcommand{\arraystretch}{1.3}
\begin{table}[ht]
\centering
\begin{small}
\begin{tabular}{lll}
\hline \hline
$SU(5)$ \ \
& $SU(5)_f$\ \
& $SO(10)_f$
\\
\hline
$(1,1;0)_{1_{1}}$
& $(1,1;0)_{1_{1}}$
& $(1,1;0)_{1_{1}}$
\\
\hline
$(1,1;0)_{1_{45}}$
& $(1,1;0)_{1_{45}}$
& $(1,1;0)_{1_{45}}$
\\
\null
$(8,1;0)_{24_{45}}$
& $(8,1;0)_{24_{45}}$
& $(8,1;0)_{24_{45}}$
\\
\null
$(3,2;-\tfrac{5}{6})_{24_{45}}$
& $(3,2;+\tfrac{1}{6})_{24_{45}}$
& $(3,2;+\tfrac{1}{6})_{24_{45}}$
\\
\null
$(\overline{3},2;+\tfrac{5}{6})_{24_{45}}$
& $(\overline{3},2;-\tfrac{1}{6})_{24_{45}}$
& $(\overline{3},2;-\tfrac{1}{6})_{24_{45}}$
\\
\null
$(1,3;0)_{24_{45}}$
& $(1,3;0)_{24_{45}}$
& $(1,3;0)_{24_{45}}$
\\
\null
$(1,1;0)_{24_{45}}$
& $(1,1;0)_{24_{45}}$
& $(1,1;0)_{24_{45}}$
\\
\null
$(3,2;+\tfrac{1}{6})_{10_{45}}$
& $(3,2;-\tfrac{5}{6})_{10_{45}}$
& $(3,2;+\tfrac{1}{6})_{10_{45}}$
\\
\null
$(\overline{3},1;-\tfrac{2}{3})_{10_{45}}$
& $(\overline{3},1;-\tfrac{2}{3})_{10_{45}}$
& $(\overline{3},1;+\tfrac{1}{3})_{10_{45}}$
\\
\null
$(1,1;+1)_{10_{45}}$
& $(1,1;-1)_{10_{45}}$
& $(1,1;0)_{10_{45}}$
\\
\null
$(\overline{3},2;-\tfrac{1}{6})_{\overline{10}_{45}}$
& $(\overline{3},2;+\tfrac{5}{6})_{\overline{10}_{45}}$
& $(\overline{3},2;-\tfrac{1}{6})_{\overline{10}_{45}}$
\\
\null
$(3,1;+\tfrac{2}{3})_{\overline{10}_{45}}$
& $(3,1;+\tfrac{2}{3})_{\overline{10}_{45}}$
& $(3,1;-\tfrac{1}{3})_{\overline{10}_{45}}$
\\
\null
$(1,1;-1)_{\overline{10}_{45}}$
& $(1,1;+1)_{\overline{10}_{45}}$
& $(1,1;0)_{\overline{10}_{45}}$
\\
\hline
$(\overline{3},1;+\tfrac{1}{3})_{\overline{5}_{16}}$
& $(\overline{3},1;-\tfrac{2}{3})_{\overline{5}_{16}}$
& $(\overline{3},1;-\tfrac{2}{3})_{\overline{5}_{16}}$
\\
\null
$(1,2; -\tfrac{1}{2})_{\overline{5}_{16}}$
& $(1,2; -\tfrac{1}{2})_{\overline{5}_{16}}$
& $(1,2; -\tfrac{1}{2})_{\overline{5}_{16}}$
\\
\null
$(3,2;+\tfrac{1}{6})_{10_{16}}$
& $(3,2;+\tfrac{1}{6})_{10_{16}}$
& $(3,2;-\tfrac{5}{6})_{10_{16}}$
\\
\null
$(\overline{3},1;-\tfrac{2}{3})_{10_{16}}$
& $(\overline{3},1;+\tfrac{1}{3})_{10_{16}}$
& $(\overline{3},1;-\tfrac{2}{3})_{10_{16}}$
\\
\null
$(1,1;+1)_{10_{16}}$
& $(1,1;0)_{10_{16}}$
& $(1,1;-1)_{10_{16}}$
\\
\null
$(1,1;0)_{1_{16}}$
& $(1,1;+1)_{1_{16}}$
& $(1,1;-1)_{1_{16}}$
\\
\hline
$(3,1;-\tfrac{1}{3})_{5_{\overline{16}}}$
& $(3,1;+\tfrac{2}{3})_{5_{\overline{16}}}$
& $(3,1;+\tfrac{2}{3})_{5_{\overline{16}}}$
\\
\null
$(1,2; +\tfrac{1}{2})_{5_{\overline{16}}}$
& $(1,2; +\tfrac{1}{2})_{5_{\overline{16}}}$
& $(1,2; +\tfrac{1}{2})_{5_{\overline{16}}}$
\\
\null
$(\overline{3},2;-\tfrac{1}{6})_{\overline{10}_{\overline{16}}}$
& $(\overline{3},2;-\tfrac{1}{6})_{\overline{10}_{\overline{16}}}$
& $(\overline{3},2;+\tfrac{5}{6})_{\overline{10}_{\overline{16}}}$
\\
\null
$(3,1;+\tfrac{2}{3})_{\overline{10}_{\overline{16}}}$
& $(3,1;-\tfrac{1}{3})_{\overline{10}_{\overline{16}}}$
& $(3,1;+\tfrac{2}{3})_{\overline{10}_{\overline{16}}}$
\\
\null
$(1,1;-1)_{\overline{10}_{\overline{16}}}$
& $(1,1;0)_{\overline{10}_{\overline{16}}}$
& $(1,1;+1)_{\overline{10}_{\overline{16}}}$
\\
\null
$(1,1;0)_{1_{\overline{16}}}$
& $(1,1;-1)_{1_{\overline{16}}}$
& $(1,1;+1)_{1_{\overline{16}}}$
\\
\hline \hline
\end{tabular}
\end{small}
\mycaption{{The same} as in Table \ref{tab:27decompSU(5)} for the $78$ representation.}
\label{tab:78decompSU(5)}
\end{table}

In analogy with the flipped $SO(10)$ discussion, we shall label the SM-singlets contained in the $27$ as
$e \equiv \left(1,1;0\right)_{1_{1}}$ and $\nu \equiv \left(1,1;0\right)_{1_{16}}$.

As we are going to show, the little group of $\vev{78\oplus 27_1\oplus 27_2\oplus \overline{27}_1\oplus \overline{27}_2}$ is 
SUSY-$SU(5)$ in the renormalizable case.
This is just a consequence of the larger $E_6$ algebra.
In order to obtain a SM vacuum, we need to resort to a non-renormalizable scenario
that allows for a disentanglement of the $\vev{78_H}$ directions, and, consistently, for a flipped $SO(10)\otimes U(1)$ intermediate stage.
We shall make the case for an $E_6$ gauge symmetry broken near the
Planck scale, leaving an effective flipped $SO(10)$ scenario down
to the $10^{16}$ GeV.

\clearpage

\subsection{$Y$ and $B-L$ into $E_6$}
\label{YandBLintoE6}

Interpreting the different possible definitions of the SM hypercharge in terms of the $E_6$ maximal subalgebra $SU(3)_C \otimes SU(3)_L \otimes SU(3)_R$, one finds that the three assignments in \eqs{standardabc}{flippedSO10abc}
are each orthogonal to the three possible ways of embedding $SU(2)_I$ (with $I=R,R',E$) into $SU(3)_R$ \cite{Harada:2003sb}.
Working in the Gell-Mann basis (cf.~\app{app:SU33formalism}) the $SU(3)_R$ Cartan generators read
\begin{align}
& T^{(3)}_R = \tfrac{1}{2} \left( T^{1'}_{1'} - T^{2'}_{2'} \right) \, \label{T3R}
, \\
& T^{(8)}_R = \tfrac{1}{2\sqrt{3}} \left( T^{1'}_{1'} + T^{2'}_{2'} - 2 T^{3'}_{3'} \right) \, ,
\label{T8R}
\end{align}
which defines the $SU(2)_R$ embedding.
The $SU(2)_{R'}$ and $SU(2)_E$ embeddings are obtained from \eqs{T3R}{T8R} by flipping
respectively $2' \leftrightarrow 3'$ and $3' \leftrightarrow 1'$.
Considering the standard and flipped $SO(10)$ embeddings of the hypercharge in \eq{standardabc}
and \eq{flippedSO10abc}, in the $SU(3)^3$ notation they are respectively given by
\be
\label{Ystandard}
Y = \tfrac{1}{\sqrt{3}} T^{(8)}_L + T^{(3)}_R + \tfrac{1}{\sqrt{3}} T^{(8)}_R 
= \tfrac{1}{\sqrt{3}} T^{(8)}_L - \tfrac{2}{\sqrt{3}} T^{(8)}_E \, ,
\ee
and
\be
\label{Yso10}
Y = \tfrac{1}{\sqrt{3}} T^{(8)}_L - \tfrac{2}{\sqrt{3}} T^{(8)}_R
= \tfrac{1}{\sqrt{3}} T^{(8)}_L + T^{(3)}_E + \tfrac{1}{\sqrt{3}} T^{(8)}_E \, .
\ee
Analogously, the three SM-compatible assignments of
$B-L$ in \eqs{BLst}{BLtw2} are as well orthogonal to the three possible
ways of embedding $SU(2)_I$ into $SU(3)_R$.
However, once we fix the embedding of the hypercharge we have
only two consistent choices for $B-L$ available. They correspond to the pairs where $Y$ and $B-L$
are not orthogonal to the same $SU(2)_I$ \cite{Harada:2003sb}.

For the standard hypercharge embedding, the $B-L$ assignment in \eq{BLst} reads
\be
\label{BLstSU3}
B-L = \tfrac{2}{\sqrt{3}} \left( T^{(8)}_L + T^{(8)}_R \right)
= \tfrac{2}{\sqrt{3}} T^{(8)}_L - T^{(3)}_E - \tfrac{1}{\sqrt{3}} T^{(8)}_E \, ,
\ee
while the $B-L$ assignment in \eq{BLtw2}, consistent
with the flipped $SO(10)$ embedding of the hypercharge, reads
\be
\label{BLtw2SU3}
B-L = \tfrac{2}{\sqrt{3}} T^{(8)}_L - T^{(3)}_R - \tfrac{1}{\sqrt{3}} T^{(8)}_R
= \tfrac{2}{\sqrt{3}} \left( T^{(8)}_L + T^{(8)}_E \right) \, .
\ee

\subsection{The $E_6$ vacuum manifold}
\label{vacuumE6}

The most general renormalizable Higgs superpotential, made of the representations $78 \oplus 27_1 \oplus 27_2 \oplus \overline{27}_1 \oplus \overline{27}_2$, is given by
\be
\label{WHE6}
W_H = \frac{\mu}{2} \Tr 78^2 + \rho_{ij} 27_i \overline{27}_j + \tau_{ij} 27_i 78 \overline{27}_j
+ \alpha_{ijk} 27_i 27_j 27_k + \beta_{ijk} \overline{27}_i \overline{27}_j \overline{27}_k \, ,
\ee
where $i,j = 1,2$. The couplings $\alpha_{ijk}$ and $\beta_{ijk}$ are totally symmetric in $ijk$, so that each one of them contains four complex parameters.
Without loss of generality we can take $\mu$ real by a phase redefinition of the superpotential,
while $\tau$ can be diagonalized by a bi-unitary transformation acting on the indices of the $27$ and the $\overline{27}$.
We take, $\tau_{ij} = \tau_i \delta_{ij}$, with $\tau_i$ real.
Notice that $\alpha$ and $\beta$ are not relevant for the present study, since the corresponding invariants vanish on the SM orbit.

In the standard hypercharge embedding of \eq{Ystandard}, the SM-preserving vacuum directions are parametrized by
\be
\label{vev78}
\vev{78}= a_1 T^{3'}_{2'} + a_2 T^{2'}_{3'} + \frac{a_3}{\sqrt{6}} (T^{1'}_{1'} + T^{2'}_{2'} - 2T^{3'}_{3'})
+ \frac{a_4}{\sqrt{2}} (T^{1'}_{1'} - T^{2'}_{2'}) + \frac{b_3}{\sqrt{6}} (T^{1}_{1} + T^{2}_{2} - 2T^{3}_{3}) \, ,
\ee
and
\bea
\label{vev2712}
\vev{27_i} &=&  (e_i) v^{3}_{3'} + (\nu_i) v^{3}_{2'} \, , \\
\label{vev27bar12}
\vev{\overline{27}_i} &=& (\overline{e}_i) u^{3'}_{3} + (\overline{\nu}_i) u^{2'}_{3} \, .
\eea
where $a_1$, $a_2$, $a_3$, $a_4$, $b_3$, $e_{1,2}$, $\overline{e}_{1,2}$, $\nu_{1,2}$ and $\overline{\nu}_{1,2}$
are 13 SM-singlet VEVs (see \app{app:SU33formalism} for notation).
Given the $B-L$ expression in \eq{BLstSU3} and the fact that we can rewrite the Cartan part of $\vev{78}$ as
\be
\sqrt{2} a_4 T^{(3)}_R + \tfrac{1}{\sqrt{2}} (a_3 + b_3) \left( T^{(8)}_R + T^{(8)}_L \right)
+ \tfrac{1}{\sqrt{2}} (a_3 - b_3) \left( T^{(8)}_R - T^{(8)}_L \right) \, ,
\ee
we readily identify the standard $SO(10)$ VEVs used in the previous section with the present $E_6$ notation as
$\omega_R \propto a_4$, $\omega_{B-L} \propto a_3 + b_3$, while
$\Omega \propto a_3 - b_3$ is the $SO(10) \otimes U(1)_X$ singlet VEV in $E_6$ ($T_X\propto T^{(8)}_R - T^{(8)}_L$).

We can also write the vacuum manifold in such a way that it is manifestly invariant under the flipped $SO(10)$
hypercharge in \eq{Yso10}. This can be obtained by flipping $1' \leftrightarrow 3'$ in \eqs{vev78}{vev27bar12}, yielding
\begin{multline}
\label{vev78flip}
\vev{78} = a_1 T^{1'}_{2'} + a_2 T^{2'}_{1'} + \sqrt{2} a'_4 T^{(3)}_E 
+ \tfrac{1}{\sqrt{2}} (a'_3 + b_3) \left( T^{(8)}_E + T^{(8)}_L \right) \\
+ \tfrac{1}{\sqrt{2}} (a'_3 - b_3) \left( T^{(8)}_E - T^{(8)}_L \right) \, , 
\end{multline}
\begin{align}
\label{vev2712flip}
& \vev{27_i} =  (e_i) v^{3}_{1'} + (\nu_i) v^{3}_{2'} \, , \\
\label{vev27bar12flip}
& \vev{\overline{27}_i} = (\overline{e}_i) u^{1'}_{3} + (\overline{\nu}_i) u^{2'}_{3} \, ,
\end{align}
where we recognize the $B-L$ generator defined in \eq{BLtw2SU3}.
Notice that the Cartan subalgebra is actually invariant both under the standard and the flipped $SO(10)$ form of $Y$.
We have
\be
a'_3 T^{(8)}_E + a'_4 T^{(3)}_E
= a_3 T^{(8)}_R + a_4 T^{(3)}_R \, ,
\label{RvsEcartans}
\ee
with
\begin{align}
2 a'_3 &= - a_3 - \sqrt{3} a_4 \, , \\
2 a'_4 &= - \sqrt{3} a_3 + a_4 \,
\label{a34prime}
\end{align}
thus making the use of $a_{3,4}$ or $a_{3,4}'$ directions in the flipped or standard vacuum manifold completely equivalent.
We can now complete the identification of the notation used for $E_6$ with that of the flipped $SO(10) \otimes U(1)_X$ model studied in \sect{sect:minimalflippedSO10}, by $\omega^{\pm}\propto a_{1,2}$.

From the $E_6$ stand point, the analyses of the standard and flipped vacuum manifolds given, respectively,
in \eqs{vev78}{vev27bar12} and \eqs{vev78flip}{vev27bar12flip}, lead,
as expected, to the same results with the roles of standard and flipped hypercharge interchanged (see \app{app:E6vacuum}). In order to determine the vacuum little group we may therefore proceed with the explicit discussion of the standard setting.

By writing the superpotential in \eq{WHE6} on the SM-preserving vacuum in \eqs{vev78}{vev27bar12}, we find
\begin{align}
\label{vevWHE6}
& \vev{W_H} = \mu \left(a_1 a_2+\frac{a_3^2}{2}+\frac{a_4^2}{2}+\frac{b_3^2}{2}\right) \\
& + \rho _{11} \left(e _1 \overline{e }_1+\nu_1 \overline{\nu}_1\right)+\rho _{21} \left(e _2\overline{e }_1+\nu_2 \overline{\nu}_1\right)
+ \rho _{12} \left(e _1 \overline{e }_2+\nu_1 \overline{\nu}_2\right)+\rho _{22} \left(e _2 \overline{e }_2+\nu_2 \overline{\nu}_2\right) \nn \\
&+ \tau _1 \left[-a_1 e _1 \overline{\nu}_1-a_2 \nu_1 \overline{e }_1 + \sqrt{\frac{2}{3}} a_3 \left(e _1 \overline{e }_1 - \frac{1}{2} \nu_1 \overline{\nu}_1\right)  
+  \frac{a_4 \nu_1 \overline{\nu}_1}{\sqrt{2}}-\sqrt{\frac{2}{3}} b_3 \left(e _1 \overline{e }_1 + \nu_1 \overline{\nu}_1 \right)\right] \nn \\
&+ \tau _2 \left[-a_1 e _2 \overline{\nu}_2-a_2 \nu_2 \overline{e}_2 + \sqrt{\frac{2}{3}} a_3 \left( e _2 \overline{e }_2 - \frac{1}{2} \nu_2 \overline{\nu}_2 \right)
+ \frac{a_4 \nu_2 \overline{\nu}_2}{\sqrt{2}}-\sqrt{\frac{2}{3}} b_3 (e _2 \overline{e }_2 + \nu_2\overline{\nu}_2)\right] \nn \, .
\end{align}
When applying the constraints coming from $D$- and $F$-term equations, a nontrivial vacuum exists if $\rho$ and $\tau$ are hermitian, as in the flipped $SO(10)$ case.
This is a consequence of the fact that $D$-flatness implies
$\vev{27_i} = \vev{\overline{27}_i}^*$ (see \app{DFtermE6} for details).

After imposing all the constraints due to $D$- and $F$-flatness, the $E_6$ vacuum manifold can be finally written as
\begin{align}
\label{vacmanifoldE6}
2 \mu a_1 &= \tau _1 r_1^2 \sin{2\alpha_1}\ e^{i(\phi_{\nu_1}-\phi_{e_1})} 
+ \tau _2 r_2^2 \sin{2\alpha_2}\ e^{i(\phi_{\nu_2}-\phi_{e_2})} \, , \nn \\
2 \mu a_2 &= \tau _1 r_1^2 \sin{2\alpha_1}\ e^{-i(\phi_{\nu_1}-\phi_{e_1})}
+ \tau _2 r_2^2 \sin{2\alpha_2}\ e^{-i(\phi_{\nu_2}-\phi_{e_2})} \, , \nn \\
2 \sqrt{6} \mu a_3 &= - \tau _1 r_1^2 (3 \cos{2\alpha_1} + 1) - \tau _2 r_2^2 (3 \cos{2\alpha_2} + 1) \, , \nn \\
\sqrt{2} \mu a_4 &= -\tau _1 r_1^2 \sin^2{\alpha_1} - \tau _2 r_2^2 \sin^2{\alpha_2} \, , \nn \\
\sqrt{3} \mu b_3 &= \sqrt{2} \tau _1 r_1^2 + \sqrt{2} \tau _2 r_2^2 \, , \nn \\
e_{1,2} &= r_{1,2} \cos \alpha_{1,2}\ e^{i \phi_{e_{1,2}}} \, , \nn \\
\nu_{1,2} &= r_{1,2} \sin \alpha_{1,2}\ e^{i \phi_{\nu_{1,2}}} \, , \nn \\
\overline{e}_{1,2} &= r_{1,2} \cos \alpha_{1,2}\ e^{- i \phi_{e_{1,2}}} \, , \nn \\
\overline{\nu}_{1,2} &= r_{1,2} \sin \alpha_{1,2}\ e^{- i \phi_{\nu_{1,2}}} \, ,
\end{align}
where $r_{1,2}$ and $\alpha^{\pm} \equiv \alpha_1 \pm \alpha_2$ are fixed in terms of superpotential parameters, as follows
\bea
\label{r1sq}
&& r_{1}^{2} = - \frac{\mu (\rho _{22} \tau _1 -4 \rho _{11} \tau _2) }{5 \tau _1^2 \tau _2} \, , \\
\label{r2sq}
&& r_{2}^{2} = - \frac{\mu (\rho _{11} \tau _2 -4 \rho _{22} \tau _1) }{5 \tau _1 \tau _2^2} \, , \\
\label{cosalpha1m2}
&& \cos{\alpha^{-}} = \xi \ \frac{\sin \Phi_\nu - \sin \Phi_e}{\sin \left(\Phi_\nu - \Phi_e \right)} \, , \\
\label{cosalpha1p2}
&& \cos{\alpha^{+}} = \xi \ \frac{\sin \Phi_\nu + \sin \Phi_e}{\sin \left(\Phi_\nu - \Phi_e \right)} \, ,
\eea
with
\be
\label{xiE6}
\xi = \frac{5 |\rho _{12}|}{\sqrt{-\frac{4 \rho _{11}^2 \tau _2}{\tau _1}-\frac{4 \rho _{22}^2 \tau _1}{\tau _2}+17 \rho _{22} \rho _{11}}} \, .
\ee
The phase factors $\Phi_\nu$ and $\Phi_e$ are defined as
\be
\Phi_\nu \equiv \phi_{\nu_1}-\phi_{\nu_2}+\phi_{\rho _{12}} \, , \quad \Phi_e \equiv \phi_{e_1}-\phi_{e_2}+\phi_{\rho _{12}} \, .
\ee

In \app{formalproof} we show that the little group of the
the vacuum manifold in \eq{vacmanifoldE6} is $SU(5)$.

It is instructive to look at the configuration in which one pair of $27_H$, let us say $27_2 \oplus \overline{27}_2$, is decoupled.
This case can be obtained by setting
$\tau_2=\rho_{12}=\rho_{22}=0$ in the relevant equations.
In agreement with Ref.~\cite{Buccella:1987kc}, we find that $\alpha_{1}$ turns out to be undetermined by the $F$-term constraints,
thus parametrizing a set of isomorphic solutions.
We may therefore take in \eq{vacmanifoldE6} $\alpha_1=\alpha_2=0$
and show that the little group corresponds in this case to $SO(10)$ (see \app{formalproof}), thus recovering the result of Ref.~\cite{Buccella:1987kc}.

The same result is obtained in the case in which the vacua of the two copies of $27_{H} \oplus \overline{27}_{H}$ are
aligned, i.e. $\alpha^-=0$ and $\Phi_\nu =\Phi_e$.
Analogously to the discussion in Sect.~\ref{vacuumFSO10},
$\alpha^+$ is in this case undetermined and it can be set to zero,
that leads us again to the one $27_{H} \oplus \overline{27}_{H}$ case,
with $SO(10)$ as the preserved algebra.

These results are intuitively understood by considering that in case there is just one pair of $27_{H} \oplus \overline{27}_{H}$
(or the vacua of the two pairs of $27_i \oplus \overline{27}_i$ are aligned) the SM-singlet directions $e$ and $\nu$ are connected by
an $SU(2)_R$ transformation which can be used to rotate one of the VEVs to zero, so that the little group is locked to an $SO(10)$ configuration.
On the other hand, two misaligned $27_{H} \oplus \overline{27}_{H}$ VEVs in the $e-\nu$ plane lead (just by inspection of the VEV quantum numbers) to an $SU(5)$ little group.

In analogy with the flipped $SO(10)$ case, the Higgs adjoint plays the role of a renormalizable agent that prevents the two pairs of $\vev{27_i \oplus \overline{27}_i}$ from aligning within each other along the $SO(10)$ vacuum.
Actually, by decoupling the adjoint Higgs, $F$-flatness makes the (aligned) $27_i\oplus 27_i$ vacuum trivial, as one verifies by inspecting the $F$-terms in \eq{FtermsE6} of~\app{DFtermE6} for $\vev{78_H}=0$ and $\det{\rho}\neq 0$.

In conclusion, due to the larger $E_6$ algebra, the vacuum little group remains $SU(5)$, never landing to the SM.
In this respect we guess that the authors of Ref. \cite{Kawase:2010na}, who advocate a $78{_H} \oplus 2 \times \left( 27{_H} \oplus \overline{27}{_H} \right)$ Higgs sector, implicitly refer to a non-renormalizable setting.

\subsection{Breaking the residual $SU(5)$ via effective interactions}
\label{NRopsE6}

In this section we consider the
possibility of breaking the residual $SU(5)$ symmetry
of the renormalizable $E_6$ vacuum through the inclusion of effective
adjoint Higgs interactions near the Planck scale $M_P$.
We argue that an effective flipped $SO(10)\otimes U(1)_X\equiv SO(10)_f$ may survive down to the $M_f\approx 10^{16}$ GeV scale, with thresholds spread in between $M_P$ and $M_f$ in such a way not to affect proton stability and lead to realistic neutrino masses.

The relevant part of the non-renormalizable superpotential at the $E_6$ scale $M_E < M_P$ can be written as
\be
\label{WHE6NR}
W^{\text{NR}}_H = \frac{1}{M_P} \left[ \lambda_1 \left( \Tr 78^2 \right)^2 + \lambda_2 \Tr 78^4 + \ldots \right] \, ,
\ee
where the ellipses stand for terms which include powers of the $27$'s representations and $D \ge 5$ operators.
Projecting \eq{WHE6NR} along the SM-singlet vacuum directions in \eqs{vev78}{vev27bar12}
we obtain
\begin{multline}
\label{WHE6NRvev}
\vev{W^{\text{NR}}_H} = \frac{1}{M_P} \left\{ \lambda_1 \left( 2 a_1 a_2 + a_3^2 + a_4^2 + b_3^2 \right)^2 \right. \\
+ \left. \lambda_2 \left[ 2 a_1 a_2 \left( a_1^2 a_2^2 + a_3^2 + a_4^2 + \tfrac{1}{\sqrt{3}} a_3 a_4 \right) 
+ \tfrac{1}{2} \left( a_3^2 + a_4^2 \right)^2 + \tfrac{1}{2} b_3^4 \right] + \ldots \right\} \, .
\end{multline}
One verifies that including the non-renormalizable contribution in the $F$-term equations
allows for a disentanglement of the $\vev{78}$ and
$\vev{27_1 \oplus \overline{27}_1 \oplus 27_2 \oplus \overline{27}_2}$ VEVs, so that the breaking to the SM is achieved.
In particular, the SUSY vacuum allows for an intermediate
$SO(10)_f$ stage (that is prevented by the simple renormalizable
vacuum manifold in \eq{vacmanifoldE6}).
By including \eq{WHE6NRvev} in the $F$-term equations, we can consistently neglect
all VEVs but the $SO(10) \otimes U(1)$ singlet $\Omega$,
that reads
\be
\Omega^2 = - \frac{\mu M_P}{5 \lambda_1 + \frac{1}{2} \lambda_2} \, .
\ee
It is therefore possible to envisage a scenario where the $E_6$ symmetry is broken at a scale $M_E < M_P$ leaving an effective
flipped $SO(10) \otimes U(1)_X$ scenario down to the $10^{16}$ GeV,
as discussed in \sect{sect:minimalflippedSO10}.
All remaining SM singlet VEVs are contained in
$45 \oplus 16_1 \oplus \overline{16}_1 \oplus 16_2 \oplus \overline{16}_2$ that are the only Higgs multiplets required
to survive at the $M_f\ll M_E$ scale.
It is clear that this is a plausibility argument and that a detailed
study of the $E_6$ vacuum and related thresholds is needed to ascertain
the feasibility of the scenario.

The non-renormalizable breaking of $E_6$ through an intermediate $SO(10)_f$ stage driven by $\Omega\gg M_f$,
while allowing (as we shall discuss next) for a consistent unification pattern, avoids the issues arising within a one-step breaking.
As a matter of fact, the colored triplets responsible for $D=5$ proton decay live naturally at the $\Omega^2/M_P > M_f$ scale,
while the masses of the SM-singlet neutrino states which enter the "extended" type-I seesaw formula
are governed by the $\vev{27}\sim M_f$ (see the discussion in \sect{TRflavor}).

\subsection{A unified $E_6$ scenario}
\label{UnifiedE6}

{Let us} examine the plausibility of the two-step gauge unification scenario
discussed in the previous subsection. We consider here just a simplified description that neglects thresholds effects.
As a first quantitative estimate of the running effects on the
$SO(10)_f$ couplings let us introduce the quantity
\be
\Delta (M_f) \equiv \frac{\alpha_{\hat{X}}^{-1} (M_f) - \alpha_{10}^{-1} (M_f)}{\alpha_{E}^{-1}} = 
\frac{1}{\alpha_{E}^{-1}} \frac{b_{\hat{X}} - b_{10}}{2 \pi} \log \frac{M_E}{M_f} \, ,
\ee
where $M_{E}$ is the $E_6$ unification scale and $\alpha_{E}$ is the $E_6$ gauge coupling. The $U(1)_X$ charge has been properly normalized to $\hat{X}=X/\sqrt{24}$.
The one-loop beta coefficients for the superfield content
$45_H \oplus 2 \times \left( 16_H \oplus \overline{16}_H \right)
\oplus 3 \times (16_F \oplus 10_F \oplus 1_F) \oplus 45_G$
are found to be $b_{10} = 1$ and $b_{\hat{X}} = 67/24$.

Taking, for the sake of an estimate, a typical MSSM value for the GUT coupling
$\alpha_{E}^{-1} \approx 25$, for $M_E/M_f <  10^2$
one finds
$\Delta (M_f) <  5\%$.

In order to match the $SO(10)_f$ couplings with the measured SM couplings, we consider as a typical setup
the two-loop MSSM gauge running with a 1 TeV SUSY scale. The (one-loop) matching of the non abelian gauge couplings (in dimensional reduction) at the scale $M_f$ reads
\be
\alpha_{10}^{-1} (M_f) = \alpha_2^{-1} (M_f) = \alpha_3^{-1} (M_f) \, ,
\ee
while for the properly normalized hypercharge $\hat{Y}$ one obtains
\be
\label{alphaXext}
 \alpha_{\hat{Y}}^{-1} (M_f) =
\left( \hat{\alpha}^2 + \hat{\beta}^2 \right)  \alpha_{10}^{-1} (M_f)
+ \hat{\gamma}^2 \alpha_{\hat{X}}^{-1} (M_f) \, .
\ee
Here we have implemented the relation among the properly normalized U(1) generators (see \eq{flippedSO10abc})
\be
\label{runY}
\hat{Y} = \hat{\alpha} \hat{Y}' + \hat{\beta} \hat{Z} + \hat{\gamma} \hat{X} \, ,
\ee
with $\{\hat{\alpha}, \hat{\beta}, \hat{\gamma}\} = \{-\frac{1}{5}, -\frac{1}{5}\sqrt{\frac{3}{2}}, \frac{3}{\sqrt{10}}\}$.

\begin{figure}[h]
\centering
\includegraphics[width=8.5cm]{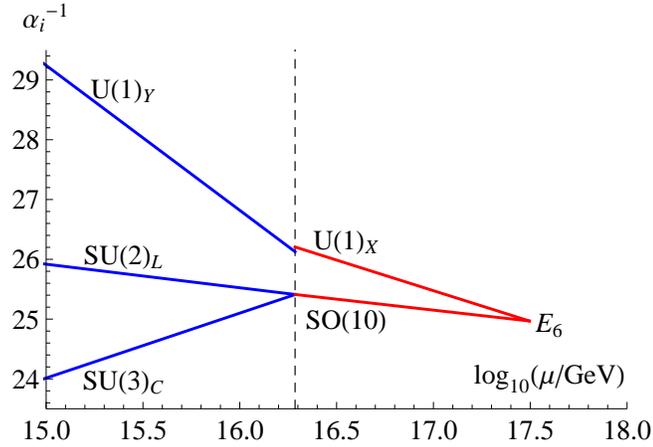}
\caption{\label{E6unification}
Sample picture of the gauge coupling unification in the $E_6$-embedded
$SO(10)\otimes U(1)_X$ model.
}
\end{figure}

The result of this simple exercise is depicted in \fig{E6unification}.
Barring detailed threshold effects,
it is interesting to see that the qualitative behavior
of the relevant gauge couplings is, indeed, consistent with
the basic picture of  the flipped $SO(10)\otimes U(1)_{X}$
embedded into a genuine $E_{6}$ GUT emerging below the Planck scale.

\section{Towards a realistic flavor}
\label{TRflavor}

The aim of this section is to provide an elementary discussion of the main features and of the possible issues arising in the Yukawa sector of the 
flipped $SO(10) \otimes U(1)_X$ model under consideration.
In order to keep the discussion simple we shall consider a basic Higgs contents with just one pair of $16_H \oplus \overline{16}_H$.
{As a
complement} of the tables given in Sect. \ref{sect:minimalflippedSO10}, we summarize  the SM-decomposition of the
representations relevant to the Yukawa sector in Table \ref{tab:stanVSflipYukawa}.

\renewcommand{\arraystretch}{1.3}
\begin{table*}[h]
\begin{small}
\centering
\begin{tabular}{lll}
\hline \hline
\ \
& $SO(10)$
& $SO(10)_f$
\\
\hline
$16_F$ &
$\left( D^c \oplus L \right)_{\overline{5}} \oplus \left( U^c \oplus Q \oplus E^c \right)_{10} \oplus (N^c)_1$ \quad\quad &
$\left( D^c \oplus \Lambda^c \right)_{\overline{5}} \oplus \left( \Delta^c \oplus Q \oplus S \right)_{10} \oplus (N^c)_1$
\\
\null
$10_F$ &
$\left( \Delta \oplus \Lambda^c \right)_{5} \oplus \left( \Delta^c \oplus \Lambda \right)_{\overline{5}}$ &
$\left( \Delta \oplus L \right)_{5} \oplus \left( U^c \oplus \Lambda \right)_{\overline{5}}$
\\
\null
$1_F$ &
$(S)_1$ &
$(E^c)_1$
\\
$\vev{16_H}$ &
$\left( 0 \oplus \vev{H_d} \right)_{\overline{5}} \oplus \left( 0 \oplus 0 \oplus 0 \right)_{10} \oplus (\nu_H)_1$ &
$\left( 0 \oplus \vev{H_u} \right)_{\overline{5}} \oplus \left( 0 \oplus 0 \oplus s_H \right)_{10} \oplus (\nu_H)_1$
\\
\null
$\vev{\overline{16}_H}$ \quad\quad &
$\left( 0 \oplus \vev{H_u} \right)_{5} \oplus \left( 0 \oplus 0 \oplus 0 \right)_{\overline{10}} \oplus (\nu_H)_1$ &
$\left( 0 \oplus \vev{H_d} \right)_{5} \oplus \left( 0 \oplus 0 \oplus s_H \right)_{\overline{10}} \oplus (\nu_H)_1$
\\
\hline \hline
\end{tabular}
\end{small}
\mycaption{
SM decomposition of $SO(10)$ representations relevant for the Yukawa sector in the standard and flipped hypercharge embedding.
In the $SO(10)_f$ case $B-L$ is assigned according to \eq{BLtw2}.
A self-explanatory SM notation is used, with the outer subscripts labeling the $SU(5)$ origin.
The $SU(2)_L$ doublets decompose as $Q = (U, \ D)$, $L = (N, \ E)$, $\Lambda = (\Lambda^0, \ \Lambda^-)$ and
$\Lambda^{c} = (\Lambda^{c+}, \ \Lambda^{c0})$.
Accordingly, $\vev{H_u} = (0, \ v_u)$ and $\vev{H_d} = (v_d, \ 0)$.
The $D$-flatness constraint on the SM-singlet VEVs, $s_H$ and $\nu_H$, is taken into account.
}
\label{tab:stanVSflipYukawa}
\end{table*}

For what follows, we refer to
\cite{Nandi:1985uh,Frank:2004vg,Malinsky:2007qy,Heinze:2010du} and references therein where the basic features of models with extended matter sector are discussed in the $E_{6}$ and the standard $SO(10)$ context. For a scenario employing flipped $SO(10) \otimes U(1)$ (with an additional anomalous $U(1)$) see Ref. \cite{Maekawa:2003wm}.

\subsection{Yukawa sector of the flipped $SO(10)$ model}
\label{YukawaFSO10}

Considering for simplicity just one pair of spinor Higgs multiplets and imposing a $Z_2$ matter-parity (negative for matter and positive for Higgs superfields) the Yukawa superpotential (up to $d= 5$ operators) reads
\be
\label{YukFSO10}
W_Y =  Y_{U} 16_F 10_F 16_H 
+ \frac{1}{M_P} \left[ Y_{E} 10_F 1_F \overline{16}_H \overline{16}_H
+  Y_{D} 16_F 16_F \overline{16}_H \overline{16}_H \right] \, ,
\ee
where family indexes are understood.
Notice (cf.~Table~\ref{tab:invdec}) that due to the flipped
embedding the up-quarks receive mass at the renormalizable level,
while all the other fermion masses need Planck-suppressed effective contributions in order to achieve a realistic texture.

\renewcommand{\arraystretch}{1.3}
\begin{table}[h]
\centering
\begin{scriptsize}
\begin{tabular}{lll}
\hline \hline
$16_F 10_F \vev{16_{H}}$
& $10_F 1_F \vev{\overline{16}_H} \vev{\overline{16}_H}$
& $16_F 16_F \vev{\overline{16}_{H}} \vev{\overline{16}_{H}}$
\\
\hline
$(1)\ 10_F \overline{5}_F \vev{\overline{5}_H} \supset (Q U^c + S \Lambda) \vev{H_u}$
& $(2)\  \overline{5}_F 1_F \vev{5_H} \vev{\overline{1}_H} \supset \Lambda E^c \vev{H_{d}} \nu_H$
& $(1)\ 1_F 1_F \vev{\overline{1}_H} \vev{\overline{1}_H} \supset N^c N^c \nu_H^2$
\\
\null
$(1)\ 1_F 5_F \vev{\overline{5}_H} \supset N^c L \vev{H_u}$
& $(2)\ 5_F 1_F \vev{\overline{10}_H} \vev{5_H} \supset L E^c \vev{H_{d}} s_H $
& $(1)\ 10_F 10_F \vev{\overline{10}_H} \vev{\overline{10}_H} \supset S S s_H^2$
\\
\null
$(1)\ \overline{5}_F 5_F \vev{1_H} \supset (D^c \Delta + \Lambda^c L) \nu_H$
& 
& $(4)\ 10_F 1_F \vev{\overline{10}_H} \vev{\overline{1}_H} \supset S N^c s_H \nu_H$
\\
\null
$(1)\ \overline{5}_F \overline{5}_F \vev{10_H} \supset \Lambda^c \Lambda s_H$
&
& $(1)\ \overline{5}_F \overline{5}_F \vev{5_H} \vev{5_H} \supset \Lambda^c \Lambda^c \vev{H_d} \vev{H_d}$
\\
\null
$(1)\ 10_F 5_F \vev{10_H} \supset \Delta^c \Delta s_H$
&
& $(4)\ 10_F \overline{5}_F \vev{\overline{10}_H} \vev{5_H} \supset ( \Lambda^c S + Q D^c ) \vev{H_d} s_H$
\\
\null
&
& $(2)\ 10_F 10_F \vev{5_H} \vev{\overline{1}_H} \supset Q \Delta^c \vev{H_d} \nu_H$
\\
\null
&
& $(4)\ \overline{5}_F 1_F \vev{5_H} \vev{\overline{1}_H} \supset \Lambda^c N^c \vev{H_d} \nu_H$
\\
\hline \hline
\end{tabular}
\mycaption{Decomposition
of the invariants in \eq{YukFSO10} according to flipped $SU(5)$ and SM. The number in the round brackets stands for the multiplicity of the invariant. The contractions
$\overline{5}_{10_F} 1_{1_F} \vev{\overline{10}_{H}} \vev{\overline{10}_{H}}$ and
$\overline{5}_{16_F} 1_{16_F} \vev{\overline{10}_{H}} \vev{\overline{10}_{H}}$ yield no SM invariant.
}
\label{tab:invdec}
\end{scriptsize}
\end{table}

\subsubsection{Mass matrices}
\label{GUTmassmatrices}

In order to avoid the recursive $1/M_P$ factors we introduce the following notation
for the relevant VEVs (see Table \ref{tab:stanVSflipYukawa}): $\hat{v}_d \equiv v_d / M_P$, $\hat{\nu}_H \equiv \nu_H / M_P$ and $\hat{s}_H \equiv s_H / M_P$.
The $M_f$-scale mass matrices for the matter fields
sharing the same unbroken $SU(3)_C \otimes U(1)_Q$ quantum numbers
can be extracted readily
by inspecting the SM decomposition of the relevant $1+10+16$ matter multiplets in the flipped SO(10) setting:
\bea
\label{MuGUT}
&& M_u = Y_U v_u \, , \nn \\[0.2ex]
\label{MdGUT}
&& M_d =
\left(
\begin{array}{cc}
 Y_D \hat{\nu}_H v_d & Y_D \hat{s}_H v_d \\
 Y_U s_H & Y_U \nu_H \\[0.2ex]
\end{array}
\right) \, , \nn \\[1ex]
\label{MeGUT}
&& M_e =
\left(
\begin{array}{cc}
 Y_E \hat{\nu}_H v_d  & Y_U s_H \\
 Y_E \hat{s}_H v_d & Y_U \nu_H
\end{array}
\right) \, ,
\eea
\be
\label{MnuGUT}
M_\nu = 
\left(
\begin{array}{ccccc}
 0 & 0 & Y_U s_H & 0 & Y_U v_u \\
 0 & 0 & Y_U \nu_H & Y_U v_u & 0 \\
 Y_U s_H & Y_U \nu_H & Y_D \hat{v}_d v_d & 2 Y_D \hat{v}_d \nu_H & 2 Y_D \hat{v}_d s_H \\
 0 & Y_U v_u & 2 Y_D \hat{\nu}_H v_d & Y_D \hat{\nu}_H \nu_H & 2 Y_D \hat{\nu}_H s_H \\
 Y_U v_u & 0 & 2 Y_D \hat{s}_H v_d & 2 Y_D \hat{s}_H \nu_H & Y_D \hat{s}_H s_H
\end{array}
\right) ,
\ee
where, for convenience, we redefined $Y_D \rightarrow Y_D / 2$ and $Y_E \rightarrow Y_E / 2$.
The basis
$(U)(U^c)$ is used for $M_u$, $(D, \Delta)(\Delta^c, D^c)$ for $M_d$ and
$(\Lambda^-, E)(E^c, \Lambda^{c+})$ for $M_e$.
The Majorana mass matrix $M_\nu$ is written in the basis $(\Lambda^0, N, \Lambda^{c0}, N^c, S)$.

\subsubsection{Effective mass matrices}
\label{Effmassmatrices}

Below the $M_{f}\sim s_H \sim \nu_H $ scale,
the exotic (vector) part of the matter spectrum decouples
{and one is left with} the three standard MSSM families.
In what follows, we shall use the calligraphic symbol $\mathcal{M}$
for the $3\times 3$ effective MSSM fermion mass matrices in order to distinguish them from the mass matrices in \eqs{MuGUT}{MnuGUT}.
\vskip 0.2mm
\emph{i) Up-type quarks:}
The effective up-quark mass matrix coincides with the mass matrix in \eq{MuGUT}
\be
\label{Mueff}
\mathcal{M}_u = Y_U v_u \, .
\ee
\indent
\emph{ii) Down-type quarks and charged leptons:}
The $6 \times 6$ mass matrices in \eqs{MdGUT}{MeGUT} can be brought into a convenient
form by means of the transformations
\be
\label{Mderot}
M_d \rightarrow M_d U^{\dag}_d \equiv M'_d \, , \quad
M_e \rightarrow U^\ast_e M_e \equiv M'_e \, ,
\ee
where $U_{d,e}$ are $6 \times 6$ unitary matrices such that $M'_d$ and $M'_e$ are block-triangular
\be
\label{Mtriangular}
M'_d =
\mathcal{O} \left(
\begin{array}{cc}
 v & v \\
 0 & M_f
\end{array}
\right)
\, , \quad
M'_e =
\mathcal{O}
\left(
\begin{array}{cc}
 v & 0 \\
 v & M_f
\end{array}
\right)
\, .
\ee
Here $v$ denotes weak scale entries.
This corresponds to the change of basis
\be
\label{changeofbasis}
\left(
\begin{array}{c}
 d^c \\
 \tilde{\Delta}^c
\end{array}
\right)
\equiv U_d
\left(
\begin{array}{c}
 \Delta^c \\
 D^c
\end{array}
\right)
\, , \quad
\left(
\begin{array}{c}
 e \\
 \tilde{\Lambda}^-
\end{array}
\right)
\equiv U_e
\left(
\begin{array}{c}
 \Lambda^- \\
 E
\end{array}
\right)
\, ,
\ee
in the right-handed (RH) down quark and left-handed (LH) charged lepton sectors, respectively.
The upper components of the rotated vectors ($d^c$ and $e$) correspond
to the light MSSM degrees of freedom.
Since the residual rotations acting on the LH down quark and RH charged lepton components, that transform
the $M'_{d,e}$ matrices into fully block-diagonal forms, are extremely tiny (of $\mathcal{O}(v / M_f)$), the $3 \times 3$ upper-left blocks (ULB) in \eq{Mtriangular} can
be identified with the effective light down-type quark and charged lepton mass matrices,
i.e., $\mathcal{M}_d \equiv \left( M'_d \right)_{ULB}$ and $\mathcal{M}_e \equiv \left( M'_e \right)_{ULB}$.

{It is instructive} to work out the explicit form of the unitary matrices $U_d$ and $U_e$. For the sake of simplicity, in what follows we shall stick to the
single family case and assume the reality of all the relevant parameters. Dropping same order Yukawa factors as well, one
writes \eqs{MdGUT}{MeGUT} as
\be
M_d =
\left(
\begin{array}{cc}
 v_{\nu} & v_{s} \\
 s_H & \nu_H
\end{array}
\right) \, ,
\quad
M_e =
\left(
\begin{array}{cc}
 v_{\nu} & s_H \\
 v_{s} & \nu_H
\end{array}
\right) \, ,
\ee
and the matrices $U_d$ and $U_e$ are explicitly given by
\be
\label{Ude2by2}
U_{d,e} =
\left(
\begin{array}{cc}
 \cos{\alpha} & - \sin{\alpha} \\
 \sin{\alpha} & \cos{\alpha}
\end{array}
\right) \, .
\ee

By applying \eq{Mderot} we get that $M'_d$ and $M'_e$ have the form in
\eq{Mtriangular} provided that $\tan \alpha = s_H / \nu_H$.
In particular, with a specific choice of the global phase, we can write
\be
\cos{\alpha} = \frac{\nu_H}{\sqrt{s^2_H + \nu^2_H}} \, , \quad \sin{\alpha} = \frac{s_H}{\sqrt{s^2_H + \nu^2_H}} \, ,
\ee
so that the mass eigenstates (up to $\mathcal{O}(v/M_f)$ effects) are finally given by (see \eq{changeofbasis})
\be
\label{deigenstates}
\left(
\begin{array}{c}
 d^c \\
 \tilde{\Delta}^c
\end{array}
\right)
=
\frac{1}{\sqrt{s^2_H + \nu^2_H}}
\left(
\begin{array}{c}
 \nu_H \Delta^c - s_H D^c \\
 s_H \Delta^c + \nu_H D^c
\end{array}
\right)
\, ,
\ee
and
\be
\label{eeigenstates}
\left(
\begin{array}{c}
 e \\
 \tilde{\Lambda}^-
\end{array}
\right)
=
\frac{1}{\sqrt{s^2_H + \nu^2_H}}
\left(
\begin{array}{c}
 \nu_H \Lambda^-  - s_H E  \\
 s_H \Lambda^- + \nu_H E
\end{array}
\right)
\, ,
\ee
where the upper (SM) components have mass of $\mathcal{O}(v_{\nu,s})$ and the lower (exotic) ones of $\mathcal{O}(M_f)$.

\emph{iii) Neutrinos:}
{Working} again in the same approximation,
the lightest eigenvalue of $M_\nu$ in \eq{MnuGUT} is given by
\be
\label{lighteigen}
m_{\nu}
\sim \frac{(\nu_H^2 + s_H^2)^2 + 2 s_H^2 \nu_H^2}{3 s_H^2 \nu_H^2  (s_H^2 + \nu_H^2)} M_P v_u^2 \, .
\ee
{For} $s_H\sim \nu_H\sim M_f\sim 10^{16}$ GeV
$M_P \sim 10^{18} \ \text{GeV}$ and $v_u \sim 10^2 \ \text{GeV}$
one obtains
\be
m_{\nu} \sim \frac{ v_u^2}{M_f^2/M_P} \sim 0.1 \ \text{eV} \, ,
\ee
which is within the ballpark of the current lower bounds on the light neutrino masses set by the oscillation experiments.

It is also useful to examine the composition of the lightest neutrino eigenstate $\nu$.
At the leading order, the light neutrino eigenvector obeys the equation $M_\nu \nu =0$
which, in the components $\nu=(x_1,x_2,x_3,x_4,x_5)$, reads
\bea
\label{eqsimplv1}
&& s_H x_3 = 0 \, , \\
\label{eqsimplv2}
&& \nu_H x_3 = 0 \, , \\
\label{eqsimplv3}
&& s_H x_1 + \nu_H x_2 = 0 \, , \\
\label{eqsimplv4}
&& \hat{\nu}_H \nu_H x_4 + 2 \hat{\nu}_H s_H x_5 = 0 \, , \\
\label{eqsimplv5}
&& 2 \hat{s}_H \nu_H x_4 + \hat{s}_H s_H x_5 = 0 \, .
\eea
By inspection, \eqs{eqsimplv4}{eqsimplv5} are compatible only if $x_4 = x_5 = 0$, while \eqs{eqsimplv1}{eqsimplv2} imply $x_3 = 0$.
Thus, the non-vanishing components of the neutrino eigenvector
are just $x_1$ and $x_2$. From
\eq{eqsimplv3}, up to a phase factor, we obtain
\be
\label{nueigenstate}
\nu = \frac{\nu_H}{\sqrt{\nu_H^2 + s_H^2}} \Lambda^{0} + \frac{- s_H}{\sqrt{\nu_H^2 + s_H^2}} N \, .
\ee
Notice that the lightest neutrino eigenstate $\nu$
and the lightest charged lepton show the same admixtures
of the corresponding electroweak doublet components.
Actually, this can be easily understood by taking the limit $v_u=v_d=0$ in which the preserved $SU(2)_L$
gauge symmetry imposes the same $U_e$ transformation
on the $(\Lambda^0, N)$ components.
Explicitly, given the form of $U_e$ in \eq{Ude2by2}, one obtains
in the rotated basis
\be
\label{Mnublockdiag}
M'_\nu =
\left(
\begin{array}{ccccc}
 0 & 0 & 0 & 0 & 0 \\
 0 & 0 & M_f & 0 & 0 \\
 0 & M_f & 0 & 0 & 0 \\
 0 & 0 & 0 & \frac{M_f^2}{M_P} & 2 \frac{M_f^2}{M_P} \\
 0 & 0 & 0 & 2 \frac{M_f^2}{M_P} & \frac{M_f^2}{M_P}
\end{array}
\right) \, ,
\ee
where we have taken $s_H \sim \nu_H\sim M_f$.
$M'_\nu$ is defined on the basis $(\nu, \tilde{\Lambda}^0, \Lambda^{c0}, N^c, S)$,
where
\be
\label{nueigenstates}
\left(
\begin{array}{c}
 \nu \\
 \tilde{\Lambda}^0
\end{array}
\right)
=
\frac{1}{\sqrt{2}}
\left(
\begin{array}{c}
 \Lambda^0 - N \\
 \Lambda^0 + N
\end{array}
\right)
\, .
\ee
In conclusion, we see that the "light" eigenstate $\nu$
decouples from the heavy spectrum,
\bea
\label{heavyneutrinos}
&& m_{\nu_{\text{M}_1}} \sim - M_f^2 / M_P \quad\ \, \nu_{\text{M}_1} \sim \tfrac{1}{\sqrt{2}} (N^c - S) \, , \\
&& m_{\nu_{\text{M}_2}} \sim 3 \cdot M_f^2 / M_P \quad \nu_{\text{M}_2} \sim \tfrac{1}{\sqrt{2}} (N^c + S) \, , \\
&& m_{\nu_{\text{PD}_1}} \sim - M_f \quad\quad\quad\, \nu_{\text{PD}_1} \sim \tfrac{1}{\sqrt{2}} (\tilde{\Lambda}^0 - \Lambda^{c0}) \, , \\
&& m_{\nu_{\text{PD}_2}} \sim M_f \quad\quad\quad\ \ \ \nu_{\text{PD}_2} \sim \tfrac{1}{\sqrt{2}} (\tilde{\Lambda}^0 + \Lambda^{c0}) \, ,
\eea
where $\nu_{\text{M}_1}$ and $\nu_{\text{M}_2}$ are two Majorana neutrinos of intermediate mass, $O(10^{14})$ GeV,
while the states $\nu_{\text{PD}_1}$ and $\nu_{\text{PD}_2}$ form a pseudo-Dirac neutrino of mass of $O(10^{16})$ GeV.

Notice finally that the
charged current $W_L \bar\nu_L e_L$ coupling is unaffected (cf.~\eq{nueigenstate} with~\eq{eeigenstates}),
contrary to the claim in Refs. \cite{Nandi:1985uh} and \cite{Frank:2004vg},
that are based on the unjustified assumption that the physical electron $e$ is predominantly made of $E$.

\chapter*{Outlook: the quest for the minimal nonsupersymmetric $SO(10)$ theory}
\addcontentsline{toc}{chapter}{Outlook: the quest for the minimal nonsupersymmetric $SO(10)$ theory}
\markboth{\textsc{Outlook: the quest for the minimal nonsupersymmetric $SO(10)$ theory}}{\textsc{Outlook: the quest for the minimal nonsupersymmetric $SO(10)$ theory}}
\label{SO10w126}

In the previous chapters we argued that an Higgs sector based on $10_H \oplus 45_H \oplus 126_H$ has all 
the ingredients to be the minimal nonsupersymmetric $SO(10)$ theory.  
We are going to conclude this thesis by mentioning some preliminary results of ongoing work and future developments. 

The first issue to be faced is the minimization of the scalar potential. 
Though there exist detailed studies of the scalar spectrum of nonsupersymmetric $SO(10)$ Higgs sectors based on 
$10_H \oplus 54_H \oplus 126_H$~\cite{Buccella:1984ft,Buccella:1986hn}, 
such a survey is missing in the $10_H \oplus 45_H \oplus 126_H$ case. 
The reason can be simply attributed the tree level no-go which was plaguing the class of models 
with just the adjoint governing the first stage of the GUT breaking~\cite{Yasue:1980fy,Yasue:1980qj,Anastaze:1983zk,Babu:1984mz}. 
On the other hand the results obtained in Chapter~\ref{thequantumvac} show that the situation is drastically changed at the quantum level, 
making the study of the $10_H \oplus 45_H \oplus 126_H$ scalar potential worth of a detailed investigation.

We have undertaken such a computation in the case of the $45_H \oplus 126_H$ scalar potential 
and some preliminary results are already available~\cite{BDLM1}.  
The first technical trouble in such a case has to do with the group-theoretical treatment of the 
$126_H$, especially as far as concerns the $126_H^4$ invariants. 
The presence of several invariants in the scalar potential is reflected in the fact that 
there are many SM sub-multiplets into the $45_H \oplus 126_H$ reducible representation and each one of them feels the
$SO(10)$ breaking in a different way.
Indeed the number of real parameters is 16 and apparently, if compared with the 9 of the $45_H \oplus 16_H$ system (cf.~e.g.~\eqs{potentialV45}{potentialV4516}), 
one would think that predictivity is compromised. 
However, out of these 16 couplings, 3 are fixed by the stationary equations, 
3 contribute only to the mass of SM-singlet states and 3 do not contribute at all to the scalar masses. 
Thus we are left with 7 real parameters governing the 22 scalar states that transform non-trivially under the SM gauge group. 
After imposing the gauge hierarchy $\vev{45_H} \gg \vev{126_H}$, required by gauge unification, 
the GUT-scale spectrum is controlled just by 4 real parameters while the intermediate-scale spectrum is controlled 
by the remaining 3. 
Notice also that these couplings are not completely free since they must fulfill the vacuum constraints, like e.g.~the positivity of the scalar spectrum. 

The message to take home is that in spite of the complexity of the $45_H \oplus 126_H$ system one cannot move the scalar states at will. 
This can be considered a nice counterexample to the criticism developed in~\cite{Dixit:1989ff} about the futility of high-precision $SO(10)$ calculations.  

Actually the knowledge of the scalar spectrum is a crucial information in view of the (two-loop) study of gauge coupling unification. 
The analysis of the intermediate scales performed in Chapter~\ref{intermediatescales} was based on the ESH~\cite{delAguila:1980at}:
at every stage of the symmetry breaking only those scalars
are present that develop a VEV
at the current or the subsequent levels of the spontaneous symmetry breaking, while 
all the other states are clustered at the GUT 
scale\footnote{With the spectrum at hand one can verify explicitly that this assumption is 
equivalent to the requirement of the minimal number of
fine-tunings to be imposed onto the scalar potential, as advocated in full generality by~\cite{Mohapatra:1982aq}.}. 
In this respect the two-loop values obtained for $M_{B-L}$, $M_U$ and $\alpha_U^{-1}$ in 
the case of the two phenomenologically allowed breaking chains were
\begin{itemize}
\item $SO(10)\chain{M_U}{\vev{45_H}} 3_C 2_L 2_R 1_{B-L}\chain{M_{B-L}}{\vev{126_H}} \mbox{SM}$
$$
M_{B-L} = 3.2 \times 10^{9} \ \text{GeV} \, , \qquad M_U = 1.6 \times 10^{16} \ \text{GeV} \, , \qquad \alpha^{-1}_U=45.5 \, ,
$$
\item $SO(10)\chain{M_U}{\vev{45_H}} 4_C 2_L 1_R\chain{M_{B-L}}{\vev{126_H}} \mbox{SM}$
$$
M_{B-L} = 2.5 \times 10^{11} \ \text{GeV} \, , \qquad M_U = 2.5 \times 10^{14} \ \text{GeV} \, , \qquad \alpha^{-1}_U=44.1 \, .
$$
\end{itemize}
Taken at face value both the scenarios are in trouble either because of a too small $M_{B-L}$ ($3_C 2_L 2_R 1_{B-L}$ and $4_C 2_L 1_R$ case) 
or a too small $M_U$ ($4_C 2_L 1_R$ case). Strictly speaking the lower bound on the $B-L$ breaking scale depends from the details 
of the Yukawa sector, but it would be natural to require $M_{B-L} \gtrsim 10^{13 \div 14} \ \text{GeV}$. On the other hand the lower bound on 
the unification scale is sharper since it comes from the $d=6$ gauge induced proton decay. 
This constraint yields something like $M_U \gtrsim 2.3 \times 10^{15} \ \text{GeV}$. 

Thus in order to restore the agreement with the phenomenology one has to go beyond the ESH 
and consider thresholds effects, i.e.~states which are not exactly clustered at the GUT scale and that can contribute to the running. 
Let us stress that whenever we pull down a state from the GUT scale the consistence with the vacuum constraints must 
be checked and it is not obvious a priori that we can do it.
 
For definiteness let us analyze the $3_C 2_L 2_R 1_{B-L}$ case.  
A simple one-loop analytical survey of the gauge running equations yields the following closed solutions for $M_{B-L}$ and $M_U$ 
\begin{small}
\begin{align}
& \frac{M_{B-L}}{M_Z} =
\exp \left(\frac{2 \pi  \left(\left(\alpha _2^{-1}-\alpha _3^{-1}\right) \left(\frac{2}{5} a_{B-L}^{3221}+ \frac{3}{5} a_R^{3221}\right)+
   \left(\alpha _1^{-1}-\alpha _2^{-1}\right) a_C^{3221} + \left(\alpha _3^{-1}-\alpha _1^{-1}\right)
   a_L^{3221}\right)}
   {\Delta}\right) \, , \nn \\
& \frac{M_U}{M_{B-L}} =  
\label{MU3221}
\exp \left(\frac{2 \pi  \left(
   \left( \alpha _2^{-1}-\alpha _3^{-1}\right) a_Y^{\text{SM}}
   +\left(\alpha _1^{-1}-\alpha _2^{-1}\right) a_C^{\text{SM}}
   +\left(\alpha _3^{-1}-\alpha _1^{-1}\right) a_L^{\text{SM}}
   \right)}
   {-\Delta}\right) \, ,
\end{align}
\end{small}
with
\be
\label{defDelta}
\Delta = 
   \left(a_L^{\text{SM}}- a_C^{\text{SM}}\right) \left( \frac{2}{5} a_{B-L}^{3221}+\frac{3}{5} a_R^{3221}\right) 
   + \left( a_Y^{\text{SM}}- a_L^{\text{SM}}\right) a_C^{3221} 
   + \left(a_C^{\text{SM}}- a_Y^{\text{SM}}\right) a_L^{3221} \, ,
\ee
where $\alpha_{1,2,3}$ are the properly normalized gauge couplings at the $M_Z$ scale, 
while $a^{\text{SM}}_{C,L,Y}$ and $a^{3221}_{C,L,R,B-L}$ are respectively the one-loop beta-functions for the SM and the 
$3_C 2_L 2_R 1_{B-L}$ gauge groups. 

The values of the gauge couplings are such that 
$\left(\alpha _2^{-1}-\alpha _3^{-1}\right) \sim 21.1$,  
$\left(\alpha _1^{-1}-\alpha _2^{-1}\right) \sim 29.4$, 
$\left(\alpha _3^{-1}-\alpha _1^{-1}\right) \sim -50.5$ and,
assuming the field content of the ESH (cf.~e.g.~\Table{tab:submultiplets}), we have $\Delta < 0$. 
Then as long as $\Delta$ remains negative when lowering new states below the GUT scale, 
the fact that the matter fields contribute positively to the beta-functions leads us to conclude  
that $M_{B-L}$ is increased (reduced) by the states charged under $SU(2)_L$ 
($SU(3)_C$ or $SU(2)_R$ or $U(1)_{B-L}$). 

Thus, in order maximize the raise of $M_{B-L}$, we must select among the $3_C 2_L 2_R 1_{B-L}$ sub-multiplets 
of $45_H \oplus 126_H$ those fields with $a^{3221}_L > a^{3221}_{C,R,B-L}$. 
The best candidate turns out to be the scalar multiplet $(6,3,1,+\tfrac{1}{3}) \subset 126_H$. 
By pulling this color sextet down to the scale 
$M_{B-L}$, we get at one-loop
$$
M_{B-L} = 8.6 \times 10^{12} \ \text{GeV} \, , \qquad M_U = 5.5 \times 10^{15} \ \text{GeV} \, , \qquad \alpha^{-1}_U = 41.3 \, ,
$$
which is closer to a phenomenologically reasonable benchmark. 
In order for the color sextet to be lowered we have to impose 
a fine-tuning which goes beyond that needed for the gauge hierarchy. 
It is anyway remarkable that the vacuum dynamics allows such a configuration. 
Another allowed threshold that helps in increasing $M_{B-L}$ is given by the 
scalar triplet $(1,3,0)$ which can be eventually pulled down till to the TeV scale.
A full treatment of the threshold patterns is still ongoing~\cite{BDLM1}. 

What about the addition of a $10_H$ in the scalar potential? Though it brings in many new couplings it does not change the bulk of the 
$45_H \oplus 126_H$ spectrum.  
The reason is simply because the $10_H$ can develop only electroweak VEVs which are negligible 
when compared with the GUT (intermediate) scale one of the $45_H$ ($126_H$). 
Thus we expect that adding a $10_H$ will not invalidate the conclusions about the vacuum of the $45_H \oplus 126_H$ scalar potential, 
including the threshold patterns.  
Of course that will contribute to the mass matrices of the isospin doublets and color triplets which are crucial for other issues like the doublet-triplet 
splitting and the scalar induced $d=6$ proton decay. 

The other aspect of the theory to be addressed is the Yukawa sector. 
Such a program has been put forward 
in Ref.~\cite{Bajc:2005zf}. 
The authors focus on renormalizable models with combinations of 
$126_H^*$ and $10_H$ (or $120_H$) in the Yukawa sector. 
They work out, neglecting the first generation masses, some interesting analytic correlations 
between the neutrino and the charged fermion sectors. 

In a recent paper~\cite{Joshipura:2011nn} the full three generation study of such settings has been numerically addressed.  
The authors claim that the model with $120_H \oplus 126^*_H$ cannot 
fit the fermions, while the setting with $10_H$ and $126^*_H$ yields an excellent fit in the case of type-I seesaw dominance.

A subtle feature, as pointed out in~\cite{Bajc:2005zf}, is that the $10_H$ must be complex. 
The reason being that in the real case one predicts $m_t \sim m_b$ 
(at least when working in the two heaviest generations limit and with real parameters). 
A complex $10_H$ implies then the presence of one additional Yukawa coupling. In turn this entails  
a loss of predictivity in the Yukawa sector when compared to the supersymmetric case. 
The proposed way out advocated by the authors of Ref.~\cite{Bajc:2005zf} was to consider 
a PQ symmmetry, relevant for dark matter and the strong CP problem, 
which forbids that extra Yukawa. 

Sticking to a pure $SO(10)$ approach, 
some predictivity could be also recovered working with three Yukawas but requiring only one Higgs doublet in the 
effective theory, as a preliminary numerical study with three generations shows~\cite{Ketan}. 

The comparison between the $10_H \oplus 45_H \oplus 126_H$ scenario and the next-to-minimal one with a 
$54_H$ in place of a $45_H$ is also worth a comment. 
At first sight the $54_H$ seems a good option as well in view of the two-loop values emerging from the unification analysis of 
Chapter~\ref{intermediatescales}: $M_{B-L} = 4.7 \times 10^{13} \ \text{GeV}$ and $M_U = 1.2 \times 10^{15} \ \text{GeV}$. 
However the choice between the $54_H$ and the $45_H$ leads to crucially distinctive features. 

The first issue has to do with the nature of the light Higgs. 
In this respect the $126^*_H$ plays a fundamental role in the Yukawa sector 
where it provides the necessary breaking of
the down-quark/charged-lepton mass degeneracy (cf.~\eqs{Md3Yuk}{Me3Yuk}).
For this to work one needs a reasonably large admixture
between the bi-doublets $(1,2,2) \subset 10_H$ and $(15,2,2) \subset 126^*_H$. 
In the model with the $45_H$ this mixing is guaranteed by the interaction $10_H 126^*_H 45_H 45_H$, but there is not such a similar invariant in the case of the $54_H$. 
Though there always exists a mixing term of the type $10_H 126^*_H 126_H 126_H$, 
this yields a suppressed mixing due to the unification constraint $\vev{45_H} \gg \vev{126_H}$. 

The other peculiar difference between the models with $45_H$ and $54_H$ has to do with the interplay between type-I and type-II seesaw. 
As already observed in~\sect{typeIvstypeII} one expects that 
in theories in which the breaking of the D-parity is decoupled from that of $SU(2)_R$
the type-II seesaw is naturally suppressed by a factor $(M_{B-L}/M_U)^2$ with respect to the 
type-I. Whilst the $45_H$ leads to this last class of models, 
the $54_H$ preserves the D-parity which is subsequently broken by the $126_H$ together with $SU(2)_R$. 
The dominance of type-I seesaw in the case of the $45_H$ has a double role: it makes the Yukawa sector more predictive 
and it does not lead to $b$-$\tau$ unification, which is badly violated without supersymmetry.

So where do we stand at the moment? 
In order to say something sensible one has to test the consistency of 
the $10_H \oplus 45_H \oplus 126_H$ vacuum against gauge unification and the SM fermion spectrum. 
If the vacuum turned out to be compatible with the phenomenological requirements  
it would be then important to perform an accurate estimate of the proton decay braching ratios. 
As a matter of fact nonsupersymmetric GUTs offer the possibility of making definite predictions for proton 
decay, especially in the presence of symmetric Yukawa matrices, as in the $10_H \oplus 45_H \oplus 126_H$ case, 
where the main theoretical uncertainty lies in the mass of the leptoquark vector bosons, subject to gauge unification constraints. 

Though the path is still long we hope to have contributed to a little step towards the quest for the minimal $SO(10)$ theory.

\appendix



\chapter{One- and Two-loop beta coefficients}
\label{app:2Lbeta}

In this appendix we report the one- and two-loop $\beta$-coefficients
used in the numerical analysis of Chapter~\ref{intermediatescales}. The calculation of the $U(1)$ mixing coefficients
and of the Yukawa contributions to the gauge coupling renormalization is detailed in Apps. \ref{app:U1mix} and
\ref{app:Yukawa} respectively.


\begin{table}[h]
\centering
\begin{scriptsize}
\begin{tabular}{ccc}
\hline
\multicolumn{3}{c}{SM ($M_Z \rightarrow M_{1}$)} \\
\hline
{\rm Chain}  &  {\rm $a_i$}  &  {\rm $b_{ij}$}
\\
\hline
\sepB
{\rm All}  &
$(-7, -\frac{19}{6}, \frac{41}{10})$ &
$\left(
\begin{array}{ccc}
 -26 & \frac{9}{2} & \frac{11}{10} \\
 12 & \frac{35}{6} & \frac{9}{10} \\
 \frac{44}{5} & \frac{27}{10} & \frac{199}{50}
\end{array}
\right)$
\\
\hline
\end{tabular}
\end{scriptsize}
\mycaption{The $a_i$ and $b_{ij}$ coefficients are given for the $3_C 2_L 1_Y$ (SM) gauge running. The scalar sector includes one Higgs doublet.}
\label{tab:beta-SM}
\end{table}

\begin{table*}[h]
\centering
\begin{scriptsize}
\begin{tabular}{cccccc}
\hline
\multicolumn{6}{c}{G1 ($M_{1} \rightarrow M_{2}$)} \\
\hline
{\rm Chain}  &  {\rm $a_i$}  &  {\rm $b_{ij}$}  & {\rm Chain}  &  {\rm $a_i$}  &  {\rm $b_{ij}$} \\
\hline
\\[-2ex]
\sepB
{\rm Ia}  &
$(-7, -3, -\frac{7}{3}, \frac{11}{2})$ &
$\left(
\begin{array}{cccc}
 -26 & \frac{9}{2} & \frac{9}{2} & \frac{1}{2} \\
 12 & 8 & 3 & \frac{3}{2} \\
 12 & 3 & \frac{80}{3} & \frac{27}{2} \\
 4 & \frac{9}{2} & \frac{81}{2} & \frac{61}{2} 
\end{array}
\right)$
&
{\rm  Ib}  &
$(-7, -3, -\frac{17}{6}, \frac{17}{4})$ &
$\left(
\begin{array}{cccc}
 -26 & \frac{9}{2} & \frac{9}{2} & \frac{1}{2} \\
 12 & 8 & 3 & \frac{3}{2} \\
 12 & 3 & \frac{61}{6} & \frac{9}{4} \\
 4 & \frac{9}{2} & \frac{27}{4} & \frac{37}{8} 
\end{array}
\right)$
\\

\sepB
{\rm  IIa}  &
$(-7, -\frac{7}{3}, -\frac{7}{3}, 7)$ &
$\left(
\begin{array}{cccc}
 -26 & \frac{9}{2} & \frac{9}{2} & \frac{1}{2} \\
 12 & \frac{80}{3} & 3 & \frac{27}{2} \\
 12 & 3 & \frac{80}{3} & \frac{27}{2} \\
 4 & \frac{81}{2} & \frac{81}{2} & \frac{115}{2} 
\end{array}
\right)$
&
{\rm  IIb}  &
$(-7, -\frac{17}{6}, -\frac{17}{6}, \frac{9}{2})$ &
$\left(
\begin{array}{cccc}
 -26 & \frac{9}{2} & \frac{9}{2} & \frac{1}{2} \\
 12 & \frac{61}{6} & 3 & \frac{9}{4} \\
 12 & 3 & \frac{61}{6} & \frac{9}{4} \\
 4 & \frac{27}{4} & \frac{27}{4} & \frac{23}{4} 
\end{array}
\right)$
\\

\sepB
{\rm IIIa}  &
$(-7, -3, -\frac{7}{3}, \frac{11}{2})$ &
$\left(
\begin{array}{cccc}
 -26 & \frac{9}{2} & \frac{9}{2} & \frac{1}{2} \\
 12 & 8 & 3 & \frac{3}{2} \\
 12 & 3 & \frac{80}{3} & \frac{27}{2} \\
 4 & \frac{9}{2} & \frac{81}{2} & \frac{61}{2} 
\end{array}
\right)$
&
{\rm  IIIb}  &
$(-7, -3, -\frac{17}{6}, \frac{17}{4})$ &
$\left(
\begin{array}{cccc}
 -26 & \frac{9}{2} & \frac{9}{2} & \frac{1}{2} \\
 12 & 8 & 3 & \frac{3}{2} \\
 12 & 3 & \frac{61}{6} & \frac{9}{4} \\
 4 & \frac{9}{2} & \frac{27}{4} & \frac{37}{8} 
\end{array}
\right)$
\\

\sepB
{\rm IVa}  &
$(-7, -3, -\frac{7}{3}, \frac{11}{2})$ &
$\left(
\begin{array}{cccc}
 -26 & \frac{9}{2} & \frac{9}{2} & \frac{1}{2} \\
 12 & 8 & 3 & \frac{3}{2} \\
 12 & 3 & \frac{80}{3} & \frac{27}{2} \\
 4 & \frac{9}{2} & \frac{81}{2} & \frac{61}{2} 
\end{array}
\right)$
 &
{\rm   IVb}  &
$(-7, -3, -\frac{17}{6}, \frac{17}{4})$ &
$\left(
\begin{array}{cccc}
 -26 & \frac{9}{2} & \frac{9}{2} & \frac{1}{2} \\
 12 & 8 & 3 & \frac{3}{2} \\
 12 & 3 & \frac{61}{6} & \frac{9}{4} \\
 4 & \frac{9}{2} & \frac{27}{4} & \frac{37}{8} 
\end{array}
\right)$
\\

\sepA
{\rm  Va}  &
$(-\frac{29}{3}, -\frac{19}{6}, \frac{15}{2})$ &
$\left(
\begin{array}{ccc}
 -\frac{101}{6} & \frac{9}{2} & \frac{27}{2} \\
 \frac{45}{2} & \frac{35}{6} & \frac{1}{2} \\
 \frac{405}{2} & \frac{3}{2} & \frac{87}{2} 
\end{array}
\right)$
&
{\rm  Vb}  &
$(-\frac{21}{2}, -\frac{19}{6}, \frac{9}{2})$ &
$\left(
\begin{array}{ccc}
 -\frac{295}{4} & \frac{9}{2} & 2 \\
 \frac{45}{2} & \frac{35}{6} & \frac{1}{2} \\
 30 & \frac{3}{2} & \frac{9}{2} 
\end{array}
\right)$
\\

\sepA
{\rm  VIa}  &
$(-\frac{29}{3}, -\frac{19}{6}, \frac{15}{2})$ &
$\left(
\begin{array}{ccc}
 -\frac{101}{6} & \frac{9}{2} & \frac{27}{2} \\
 \frac{45}{2} & \frac{35}{6} & \frac{1}{2} \\
 \frac{405}{2} & \frac{3}{2} & \frac{87}{2} 
\end{array}
\right)$
&
{\rm  VIb}  &
$(-\frac{21}{2}, -\frac{19}{6}, \frac{9}{2})$ &
$\left(
\begin{array}{ccc}
 -\frac{295}{4} & \frac{9}{2} & 2 \\
 \frac{45}{2} & \frac{35}{6} & \frac{1}{2} \\
 30 & \frac{3}{2} & \frac{9}{2} 
\end{array}
\right)$
\\

\sepA
{\rm  VIIa}  &
$(-\frac{23}{3}, -3, \frac{11}{3})$ &
$\left(
\begin{array}{ccc}
 \frac{643}{6} & \frac{9}{2} & \frac{153}{2} \\
 \frac{45}{2} & 8 & 3 \\
 \frac{765}{2} & 3 & \frac{584}{3} 
\end{array}
\right)$
&
{\rm  VIIb}  &
$(-\frac{31}{3}, -3, -\frac{7}{3})$ &
$\left(
\begin{array}{ccc}
 -\frac{206}{3} & \frac{9}{2} & \frac{15}{2} \\
 \frac{45}{2} & 8 & 3 \\
 \frac{75}{2} & 3 & \frac{50}{3} 
\end{array}
\right)$
\\

\hline
\end{tabular}
\end{scriptsize}
\mycaption{The $a_{i}$ and $b_{ij}$ coefficients due to gauge interactions are reported for the G1 chains I to VII with $\overline{126}_{H}$ (left) and $\overline{16}_{H}$ (right) respectively.
The two-loop contributions induced by Yukawa couplings are given
in Appendix~\ref{app:Yukawa}}
\label{tab:beta-G1}
\end{table*}

\begin{table*}[h]
\centering
\begin{scriptsize}
\begin{tabular}{cccccc}
\hline
\multicolumn{6}{c}{G1 ($M_{1} \rightarrow M_{2}$)} \\
\hline
{\rm Chain}  &  {\rm $a_i$}  &  {\rm $b_{ij}$}  & {\rm Chain}  &  {\rm $a_i$}  &  {\rm $b_{ij}$} \\
\hline

\sepC
$\begin{array}{c}
{\rm  VIIIa} \\
\vdots \\
{\rm  XIIa}
\end{array}$
 &
\;\;$(-7, -\frac{19}{6}, \frac{9}{2}, \frac{9}{2})$ &
$\left(
\begin{array}{cccc}
 -26 & \frac{9}{2} & \frac{3}{2} & \frac{1}{2} \\
 12 & \frac{35}{6} & \frac{1}{2} & \frac{3}{2} \\
 12 & \frac{3}{2} & \frac{15}{2} & \frac{15}{2} \\
 4 & \frac{9}{2} & \frac{15}{2} & \frac{25}{2} 
\end{array}
\right)$\;\;
&
\sepB
$\begin{array}{c}
{\rm  VIIIb} \\
\vdots \\
{\rm  XIIb}
\end{array}$
&
\;\;$(-7, -\frac{19}{6},  \frac{17}{4}, \frac{33}{8})$&
$\left(
\begin{array}{cccc}
 -26 & \frac{9}{2} & \frac{3}{2} & \frac{1}{2} \\
 12 & \frac{35}{6} & \frac{1}{2} & \frac{3}{2} \\
 12 & \frac{3}{2} & \frac{15}{4} & \frac{15}{8} \\
 4 & \frac{9}{2} & \frac{15}{8} & \frac{65}{16} 
\end{array}
\right)$
\\

\hline
\end{tabular}
\end{scriptsize}
\mycaption{The $a_{i}$ and $b_{ij}$ coefficients due to purely gauge interactions for the
G1 chains VIII to XII are reported.
For comparison with previous studies the $\beta$-coefficients
are given neglecting systematically one- and two-loops $U(1)$ mixing effects
(while all diagonal $U(1)$ contributions
to abelian and non-abelian gauge coupling renormalization are included).
The complete (and correct) treatment of $U(1)$ mixing
is detailed in Appendix~\ref{app:U1mix}.}
\label{tab:beta-U1nomix}
\end{table*}

\begin{table*}[h]
\centering
\begin{scriptsize}
\begin{tabular}{lcclcc}
\hline
\multicolumn{6}{c}{G2 ($M_{2}$ $\rightarrow$ $M_{U}$)} \\
\hline
{\rm Chain}  &  {\rm $a_j$}  &  {\rm $b_{ij}$}  & {\rm Chain}  &  {\rm $a_j$}  &  {\rm $b_{ij}$} \\
\hline

\sepB
{\rm  Ia}  &
$(-7, -3, \frac{11}{3})$ &
$\left(
\begin{array}{ccc}
 \frac{289}{2} & \frac{9}{2} & \frac{153}{2} \\
 \frac{45}{2} & 8 & 3 \\
 \frac{765}{2} & 3 & \frac{584}{3}
\end{array}
\right)$
&
{\rm  Ib}  &
$(-\frac{29}{3}, -3, -\frac{7}{3})$ &
$\left(
\begin{array}{ccc}
 -\frac{94}{3} & \frac{9}{2} & \frac{15}{2} \\
 \frac{45}{2} & 8 & 3 \\
 \frac{75}{2} & 3 & \frac{50}{3} 
\end{array}
\right)$
\\

{\rm  IIa}  &
$({-4, \frac{11}{3}, \frac{11}{3}})$ &
$\left(
\begin{array}{ccc}
 \frac{661}{2} & \frac{153}{2} & \frac{153}{2} \\
 \frac{765}{2} & \frac{584}{3} & 3 \\
 \frac{765}{2} & 3 & \frac{584}{3} 
\end{array}
\right)$
&
{\rm  IIb}  &
$({-\frac{28}{3}, -\frac{7}{3}, -\frac{7}{3}})$ &
$\left(
\begin{array}{ccc}
 -\frac{127}{6} & \frac{15}{2} & \frac{15}{2} \\
 \frac{75}{2} & \frac{50}{3} & 3 \\
 \frac{75}{2} & 3 & \frac{50}{3} 
\end{array}
\right)$
\\

\sepB
{\rm  IIIa}  &
$({-4, \frac{11}{3}, \frac{11}{3}})$ &
$\left(
\begin{array}{ccc}
 \frac{661}{2} & \frac{153}{2} & \frac{153}{2} \\
 \frac{765}{2} & \frac{584}{3} & 3 \\
 \frac{765}{2} & 3 & \frac{584}{3} 
\end{array}
\right)$
&
{\rm  IIIb}  &
$({-\frac{28}{3}, -\frac{7}{3}, -\frac{7}{3}})$ &
$\left(
\begin{array}{ccc}
 -\frac{127}{6} & \frac{15}{2} & \frac{15}{2} \\
 \frac{75}{2} & \frac{50}{3} & 3 \\
 \frac{75}{2} & 3 & \frac{50}{3} 
\end{array}
\right)$
\\

\sepB
{\rm IVa}  &
$({-7, -\frac{7}{3}, -\frac{7}{3}, 7})$ &
$\left(
\begin{array}{cccc}
 -26 & \frac{9}{2} & \frac{9}{2} & \frac{1}{2} \\
 12 & \frac{80}{3} & 3 & \frac{27}{2} \\
 12 & 3 & \frac{80}{3} & \frac{27}{2} \\
 4 & \frac{81}{2} & \frac{81}{2} & \frac{115}{2} 
\end{array}
\right)$
&
{\rm  IVb}  &
$(-7, -\frac{17}{6}, -\frac{17}{6}, \frac{9}{2})$ &
$\left(
\begin{array}{cccc}
 -26 & \frac{9}{2} & \frac{9}{2} & \frac{1}{2} \\
 12 & \frac{61}{6} & 3 & \frac{9}{4} \\
 12 & 3 & \frac{61}{6} & \frac{9}{4} \\
 4 & \frac{27}{4} & \frac{27}{4} & \frac{23}{4} 
\end{array}
\right)$
\\

\sepA
{\rm  Va}  &
$(-\frac{23}{3}, -3, 4)$ &
$\left(
\begin{array}{ccc}
 \frac{643}{6} & \frac{9}{2} & \frac{153}{2} \\
 \frac{45}{2} & 8 & 3 \\
 \frac{765}{2} & 3 & 204 
\end{array}
\right)$
&
{\rm  Vb}  &
$(-\frac{31}{3}, -3, -2)$ &
$\left(
\begin{array}{ccc}
 -\frac{206}{3} & \frac{9}{2} & \frac{15}{2} \\
 \frac{45}{2} & 8 & 3 \\
 \frac{75}{2} & 3 & 26 
\end{array}
\right)$
\\

\sepA
{\rm  VIa}  &
$(-\frac{14}{3}, 4, 4)$ &
$\left(
\begin{array}{ccc}
 \frac{1759}{6} & \frac{153}{2} & \frac{153}{2} \\
 \frac{765}{2} & 204 & 3 \\
 \frac{765}{2} & 3 & 204 
\end{array}
\right)$
&
{\rm  VIb}  &
$(-10, -2, -2)$ &
$\left(
\begin{array}{ccc}
 -\frac{117}{2} & \frac{15}{2} & \frac{15}{2} \\
 \frac{75}{2} & 26 & 3 \\
 \frac{75}{2} & 3 & 26 
\end{array}
\right)$
\\

\sepA
{\rm  VIIa}  &
$(-\frac{14}{3}, \frac{11}{3}, \frac{11}{3})$ &
$\left(
\begin{array}{ccc}
 \frac{1759}{6} & \frac{153}{2} & \frac{153}{2} \\
 \frac{765}{2} & \frac{584}{3} & 3 \\
 \frac{765}{2} & 3 & \frac{584}{3} 
\end{array}
\right)$
&
{\rm  VIIb}  &
$(-10, -\frac{7}{3}, -\frac{7}{3})$ &
$\left(
\begin{array}{ccc}
 -\frac{117}{2} & \frac{15}{2} & \frac{15}{2} \\
 \frac{75}{2} & \frac{50}{3} & 3 \\
 \frac{75}{2} & 3 & \frac{50}{3} 
\end{array}
\right)$
\\

\sepB
{\rm  VIIIa}  &
$(-7, -3, -2, \frac{11}{2})$ &
$\left(
\begin{array}{cccc}
 -26 & \frac{9}{2} & \frac{9}{2} & \frac{1}{2} \\
 12 & 8 & 3 & \frac{3}{2} \\
 12 & 3 & 36 & \frac{27}{2} \\
 4 & \frac{9}{2} & \frac{81}{2} & \frac{61}{2} 
\end{array}
\right)$
&
{\rm  VIIIb}  &
$(-7, -3, -\frac{5}{2}, \frac{17}{4})$ &
$\left(
\begin{array}{cccc}
 -26 & \frac{9}{2} & \frac{9}{2} & \frac{1}{2} \\
 12 & 8 & 3 & \frac{3}{2} \\
 12 & 3 & \frac{39}{2} & \frac{9}{4} \\
 4 & \frac{9}{2} & \frac{27}{4} & \frac{37}{8} 
\end{array}
\right)$
\\

\sepB
{\rm  IXa}  &
$(-7, -2, -2, 7)$ &
$\left(
\begin{array}{cccc}
 -26 & \frac{9}{2} & \frac{9}{2} & \frac{1}{2} \\
 12 & 36 & 3 & \frac{27}{2} \\
 12 & 3 & 36 & \frac{27}{2} \\
 4 & \frac{81}{2} & \frac{81}{2} & \frac{115}{2} 
\end{array}
\right)$
&
{\rm  IXb}  &
$(-7, -\frac{5}{2}, -\frac{5}{2}, \frac{9}{2})$ &
$\left(
\begin{array}{cccc}
 -26 & \frac{9}{2} & \frac{9}{2} & \frac{1}{2} \\
 12 & \frac{39}{2} & 3 & \frac{9}{4} \\
 12 & 3 & \frac{39}{2} & \frac{9}{4} \\
 4 & \frac{27}{4} & \frac{27}{4} & \frac{23}{4} 
\end{array}
\right)$
\\

\sepB
{\rm  Xa}  &
$(-\frac{17}{3}, -3, \frac{26}{3})$ &
$\left(
\begin{array}{ccc}
 \frac{1315}{6} & \frac{9}{2} & \frac{249}{2} \\
 \frac{45}{2} & 8 & 3 \\
 \frac{1245}{2} & 3 & \frac{1004}{3} 
\end{array}
\right)$
&
{\rm  Xb}  &
$(-\frac{25}{3}, -3, \frac{8}{3})$ &
$\left(
\begin{array}{ccc}
 \frac{130}{3} & \frac{9}{2} & \frac{111}{2} \\
 \frac{45}{2} & 8 & 3 \\
 \frac{555}{2} & 3 & \frac{470}{3} 
\end{array}
\right)$
\\

{\rm  XIa}  &
$(-\frac{2}{3}, \frac{26}{3}, \frac{26}{3})$ &
$\left(
\begin{array}{ccc}
 \frac{3103}{6} & \frac{249}{2} & \frac{249}{2} \\
 \frac{1245}{2} & \frac{1004}{3} & 3 \\
 \frac{1245}{2} & 3 & \frac{1004}{3} 
\end{array}
\right)$
&
{\rm  XIb}  &
$(-6, \frac{8}{3}, \frac{8}{3})$ &
$\left(
\begin{array}{ccc}
 \frac{331}{2} & \frac{111}{2} & \frac{111}{2} \\
 \frac{555}{2} & \frac{470}{3} & 3 \\
 \frac{555}{2} & 3 & \frac{470}{3} 
\end{array}
\right)$
\\

\sepB
{\rm  XIIa}  &
$(-9, -\frac{19}{6}, \frac{15}{2})$ &
$\left(
\begin{array}{ccc}
 \frac{41}{2} & \frac{9}{2} & \frac{27}{2} \\
 \frac{45}{2} & \frac{35}{6} & \frac{1}{2} \\
 \frac{405}{2} & \frac{3}{2} & \frac{87}{2} 
\end{array}
\right)$
&
{\rm  XIIb}  &
$(-\frac{59}{6}, -\frac{19}{6}, \frac{9}{2})$ &
$\left(
\begin{array}{ccc}
 -\frac{437}{12} & \frac{9}{2} & 2 \\
 \frac{45}{2} & \frac{35}{6} & \frac{1}{2} \\
 30 & \frac{3}{2} & \frac{9}{2} 
\end{array}
\right)$
\\

\hline
\end{tabular}
\end{scriptsize}
\mycaption{The $a_{i}$ and $b_{ij}$ coefficients due to pure gauge interactions are
reported for
the G2 chains with $\overline{126}_{H}$ (left) and ${16}_{H}$ (right) respectively.
The two-loop contributions induced by Yukawa couplings are given
in Appendix \ref{app:Yukawa}}
\label{tab:beta-G2}
\end{table*}


\begin{table}[h]
\centering
\begin{scriptsize}
\begin{tabular}{lcc}
\hline
{\rm Chain}
&
$\tilde{b}_{ij}$
&
{\rm Eq. in Ref. \cite{Chang:1984qr}\ }
\\
\hline
\\
{\rm All/SM}
&
$\left(
\begin{array}{ccc}
 \frac{199}{205} & -\frac{81}{95} & -\frac{44}{35} \\
 \frac{9}{41} & -\frac{35}{19} & -\frac{12}{7} \\
 \frac{11}{41} & -\frac{27}{19} & \frac{26}{7}
\end{array}
\right)$
&
{\rm A7}
\\
\\
{\rm VIIIa/G1}
&
$\left(
\begin{array}{cccc}
 \frac{25}{9} & \frac{5}{3} & -\frac{27}{19} & -\frac{4}{7} \\
 \frac{5}{3} & \frac{5}{3} & -\frac{9}{19} & -\frac{12}{7} \\
 \frac{1}{3} & \frac{1}{9} & -\frac{35}{19} & -\frac{12}{7} \\
 \frac{1}{9} & \frac{1}{3} & -\frac{27}{19} & \frac{26}{7}
\end{array}
\right)$
&
{\rm A10}
\\
\\
{\rm VIIIa/G2}
&
$\left(
\begin{array}{cccc}
 \frac{61}{11} & -\frac{3}{2} & -\frac{81}{4} & -\frac{4}{7} \\
 \frac{3}{11} & -\frac{8}{3} & -\frac{3}{2} & -\frac{12}{7} \\
 \frac{27}{11} & -1 & -18 & -\frac{12}{7} \\
 \frac{1}{11} & -\frac{3}{2} & -\frac{9}{4} & \frac{26}{7}
\end{array}
\right)$
&
{\rm A13}
\\
\\
{\rm Ia/G2}
&
$\left(
\begin{array}{ccc}
 -\frac{8}{3} & \frac{9}{11} & -\frac{45}{14} \\
 -1 & \frac{584}{11} & -\frac{765}{14} \\
 -\frac{3}{2} & \frac{459}{22} & -\frac{289}{14}
\end{array}
\right)$
&
{\rm A14}
\\
\\
{\rm Va/G1}
&
$\left(
\begin{array}{ccc}
 -\frac{35}{19} & \frac{1}{15} & -\frac{135}{58} \\
 -\frac{9}{19} & \frac{29}{5} & -\frac{1215}{58} \\
 -\frac{27}{19} & \frac{9}{5} & \frac{101}{58}
\end{array}
\right)$
&
{\rm A15}
\\
\\
{\rm XIIa/G2}
&
$\left(
\begin{array}{ccc}
 -\frac{35}{19} & \frac{1}{15} & -\frac{5}{2} \\
 -\frac{9}{19} & \frac{29}{5} & -\frac{45}{2} \\
 -\frac{27}{19} & \frac{9}{5} & -\frac{41}{18}
\end{array}
\right)$
&
{\rm A18}
\\
\\
\hline
\end{tabular}
\end{scriptsize}
\mycaption{The rescaled two-loop $\beta$-coefficients $\tilde{b}_{ij}$
computed in this work are shown together with the corresponding equations in Ref.~\cite{Chang:1984qr}.
For the purpose of comparison
Yukawa contributions are neglected and no $U(1)$ mixing is included in chain VIIIa/G1.
Care must be taken of the different ordering between abelian and non-abelian gauge
group factors in Ref.~\cite{Chang:1984qr}.
We report those cases where disagreement is found in some of the entries, while
we fully agree with the Eqs. A9, A11 and A16.
}
\label{tab:betaChang}
\end{table}

\begin{table}[h]
\centering
\begin{scriptsize}
\begin{tabular}{ccc}
\hline
{$\phi^{126}$}  &  {\rm $a_i$}  &  {\rm $b_{ij}$}
\\
\hline
\\
{\rm $(15,2,2)$}  &
$(\frac{16}{3}, 5, 5)$ &
$\left(
\begin{array}{ccc}
 \frac{896}{3} & 48 & 48 \\
 240 & 65 & 45 \\
 240 & 45 & 65 
\end{array}
\right)$
\\
\\
{\rm $(15, 2,+\frac{1}{2})$}  &
$(\frac{8}{3}, \frac{5}{2}, \frac{5}{2})$ &
$\left(
\begin{array}{ccc}
 \frac{448}{3} & 24 & 8 \\
 120 & \frac{65}{2} & \frac{15}{2} \\
 120 & \frac{45}{2} & \frac{15}{2} 
\end{array}
\right)$
\\
\\
\hline
\end{tabular}
\end{scriptsize}
\mycaption{One- and two-loop additional contributions to the $\beta$-coefficients
related to the presence of the $\phi^{126}$ scalar multiplets
in the $4_C 2_L 2_R$ (top) and $4_C 2_L 1_R$ (bottom) stages.}
\label{tab:phi126betas}
\end{table}

\clearpage

\section{Beta-functions with $U(1)$ mixing}
\label{app:U1mix}

The basic building blocks of the one- and two-loop $\beta$-functions for the abelian couplings
with $U(1)$ mixing, cf.~\eqs{bU12loops}{bpU12loops}, can be conveniently
written as
\be
g_{ka}g_{kb}=g_{sa}\Gamma^{(1)}_{sr}g_{rb}
\label{1loopterm}
\ee
and
\be
g_{ka}g_{kb}g^2_{kc}=g_{sa}\Gamma^{(2)}_{sr}g_{rb}
\label{2loopterm}\,,
\ee
where $\Gamma^{(1)}$ and $\Gamma^{(2)}$ are functions of the abelian charges $Q_{k}^{a}$
and, at two loops, also of the gauge couplings.
In the case of interest, i.e. for two abelian charges $U(1)_A$ and $U(1)_B$,
one obtains
\begin{align}
\Gamma^{(1)}_{AA}&=(Q_k^A)^2\,,
\nn \\[1ex]
\Gamma^{(1)}_{AB}&=\Gamma^{(1)}_{BA}=Q_k^AQ_k^B\,,
\label{1loopgammatilde}
\\[1ex]
\Gamma^{(1)}_{BB}&=(Q_k^B)^2\,, \nn
\end{align}
and
\begin{small}
\begin{align}
\Gamma^{(2)}_{AA}&=(Q_k^A)^4(g^2_{AA}+g^2_{AB})+2(Q_k^A)^3Q_k^B(g_{AA}g_{BA}+g_{AB}g_{BB})+
(Q_k^A)^2(Q_k^B)^2(g^2_{BA}+g^2_{BB})\,,
\nn \\[2ex]
\Gamma^{(2)}_{AB}&=\Gamma^{(2)}_{BA}=(Q_k^A)^3Q_k^B(g^2_{AA}+g^2_{AB})+2(Q_k^A)^2(Q_k^B)^2(g_{AA}g_{BA}+g_{AB}g_{BB})+
Q_k^A(Q_k^B)^3(g^2_{BA}+g^2_{BB})\,, \nn
\label{2loopgammatilde}
\\[2ex]
\Gamma^{(2)}_{BB}&=(Q_k^A)^2(Q_k^B)^2(g^2_{AA}+g^2_{AB})+2Q_k^A(Q_k^B)^3(g_{AA}g_{BA}+g_{AB}g_{BB})+
(Q_k^B)^4(g^2_{BA}+g^2_{BB}) \,.
\end{align}
\end{small}
All other contributions
in \eq{bU12loops} and \eq{bpU12loops} can be easily obtained
from \eqs{1loopgammatilde}{2loopgammatilde} by including the appropriate group factors.
It is worth mentioning that for complete $SO(10)$ multiplets,
$(Q_k^A)^n(Q_k^B)^m = 0$ for n and m odd
(with $n+m=2$ at one-loop and $n+m=4$ at two-loop level).

By evaluating \eqs{1loopgammatilde}{2loopgammatilde} for the particle content relevant to
the $3_C2_L1_R1_{B-L}$ stages in chains VIII-XII, and by substituting into \eqs{bU12loops}{bpU12loops},
one finally obtains

\begin{itemize}
\item Chains VIII-XII with $\overline{126}_{H}$ in the Higgs sector:
\begin{small}
\begin{align}
\label{gammaXR_1}
& \gamma_C = -7+\frac{1}{(4\pi)^2}\left[\frac{3}{2}(g_{R,R}^2+g_{R,B-L}^2)+
\frac{1}{2}(g_{B-L,R}^2+g_{B-L,B-L}^2)+\frac{9}{2}g^2_L-26g^2_C\right] \,, \\[1ex]
& \gamma_L = -\frac{19}{6}+\frac{1}{(4\pi)^2}\left[\frac{1}{2}(g_{R,R}^2+g_{R,B-L}^2)+
\frac{3}{2}(g_{B-L,R}^2+g_{B-L,B-L}^2)+\frac{35}{6}g^2_L+12g^2_C\right] \,,\nn \\[1ex]
& \gamma_{R,R} = \frac{9}{2} +\frac{1}{(4\pi)^2}\left[\frac{15}{2}(g_{R,R}^2+g_{R,B-L}^2)-4\sqrt{6}(g_{R,R}g_{B-L,R}+g_{R,B-L}g_{B-L,B-L}) \right. \nn \\[1ex]
& \qquad\qquad + \left. \frac{15}{2}(g_{B-L,R}^2+g_{B-L,B-L}^2)+\frac{3}{2}g^2_L+12g^2_C\right]\,, \nn \\[1ex]
& \gamma_{R,B-L} = \gamma_{B-L,R}= -\frac{1}{\sqrt{6}} +\frac{1}{(4\pi)^2}\left[-2{\sqrt{6}}(g_{R,R}^2+g_{R,B-L}^2) \right. \nn \\[1ex]
& \qquad\qquad + \left. 15(g_{R,R}g_{B-L,R}+g_{R,B-L}g_{B-L,B-L}) - 3\sqrt{6}(g_{B-L,R}^2+g_{B-L,B-L}^2)\right]\,, \nn \\[1ex]
& \gamma_{B-L,B-L} = \frac{9}{2} +\frac{1}{(4\pi)^2}\left[\frac{15}{2}(g_{R,R}^2+g_{R,B-L}^2)-6\sqrt{6}(g_{R,R}g_{B-L,R}+g_{R,B-L}g_{B-L,B-L}) \right. \nn \\[1ex]
& \qquad\qquad + \left. \frac{25}{2}(g_{B-L,R}^2+g_{B-L,B-L}^2)+\frac{9}{2}g^2_L+4g^2_C\right] \,; \nn
\end{align}
\end{small}
\item Chains VIII-XII with $\overline{16}_{H}$ in the Higgs sector:
\begin{small}
\begin{align}
\label{gammaXR_2}
& \gamma_C = -7+\frac{1}{(4\pi)^2}\left[\frac{3}{2}(g_{R,R}^2+g_{R,B-L}^2)+
\frac{1}{2}(g_{B-L,R}^2+g_{B-L,B-L}^2)+\frac{9}{2}g^2_L-26g^2_C\right] \\[1ex]
& \gamma_L = -\frac{19}{6}+\frac{1}{(4\pi)^2}\left[\frac{1}{2}(g_{R,R}^2+g_{R,B-L}^2)+
\frac{3}{2}(g_{B-L,R}^2+g_{B-L,B-L}^2)+\frac{35}{6}g^2_L+12g^2_C\right]\,, \nn \\[1ex]
& \gamma_{R,R} = \frac{17}{4} +\frac{1}{(4\pi)^2}\left[\frac{15}{4}(g_{R,R}^2+g_{R,B-L}^2)-\frac{1}{2}\sqrt{\frac{3}{2}}(g_{R,R}g_{B-L,R}+g_{R,B-L}g_{B-L,B-L}) \right. \nn \\[1ex]
& \qquad\qquad + \left. \frac{15}{8}(g_{B-L,R}^2+g_{B-L,B-L}^2)+\frac{3}{2}g^2_L+12g^2_C\right]\,, \nn \\[1ex]
& \gamma_{R,B-L} = \gamma_{B-L,R}=-\frac{1}{4\sqrt{6}} +\frac{1}{(4\pi)^2}\left[-\frac{1}{4}\sqrt{\frac{3}{2}}(g_{R,R}^2+g_{R,B-L}^2)
\right. \nn \\[1ex] & \qquad\qquad + \left. \frac{15}{4}(g_{RR}g_{B-L,R}+g_{R,B-L}g_{B-L,B-L}) - \frac{3}{8}\sqrt{\frac{3}{2}}(g_{B-L,R}^2+g_{B-L,B-L}^2)\right]\,, \nn \\[1ex]
& \gamma_{B-L,B-L} = \frac{33}{8}  +\frac{1}{(4\pi)^2}\left[\frac{15}{8}(g_{R,R}^2+g_{R,B-L}^2)-\frac{3}{4}\sqrt{\frac{3}{2}}(g_{RR}g_{B-L,R}+g_{R,B-L}g_{B-L,B-L}) 
\right. \nn \\[1ex] &  \qquad\qquad + \left. \frac{65}{16}(g_{B-L,R}^2+g_{B-L,B-L}^2)+\frac{9}{2}g^2_L+4g^2_C\right] \nn \,. 
\end{align}
\end{small}
\end{itemize}

By setting  $\gamma_{B-L,R}=\gamma_{R,B-L}=0$ and $g_{B-L,R}=g_{R,B-L}=0$ in \eqs{gammaXR_1}{gammaXR_2}
one obtains the one- and two-loop $\beta$-coefficients in the diagonal approximation, as reported
in \Table{tab:beta-U1nomix}. The latter are used in \figs{fig:1to12a}{fig:1to12b}
for the only purpose of exhibiting the effect of the abelian mixing in the gauge coupling
renormalization.

\section{Yukawa contributions}
\label{app:Yukawa}

The Yukawa couplings enter the gauge $\beta$-functions first at the two-loop level,
cf.~\eq{Gp2loops} and \eq{bU12loops}. Since the notation adopted in
\eqs{Y4}{Yukawa} is rather concise we shall detail the structure
of \eq{Y4}, paying particular attention to the calculation of the $y_{pk}$
coefficients in \eq{Ycpk}.

The trace on the RHS of \eq{Y4} is taken over all indices of the fields
entering the Yukawa interaction in \eq{Yukawa}. Considering for instance the up-quark
Yukawa sector of the SM the term
$ \overline{Q_{L}}Y_{U}U_{R}\tilde{h}+h.c.$
(with $\tilde h=i\sigma_{2}h^{*}$) can be explicitly written as
\begin{equation}
\label{Yukawaguts}
Y^{ab}_{U}{\varepsilon}^{kl}{\delta_{3}}_{j}^{i}\overline{Q_{L}^{a}}_{ik}U^{bj}_{R}h^{*}_{l}+h.c.\,,
\end{equation}
where $\{a,b\}$, $\{i,j\}$ and $\{k,l\}$ label flavour, $SU(3)_{C}$ and
$SU(2)_{L}$ indices respectively, while $\delta_{n}$ denotes the $n$-dimensional Kronecker $\delta$ symbol.
Thus, the Yukawa coupling entering \eq{Y4} is a 6-dimensional
object with the index structure
$Y^{ab}_{U}{\varepsilon}^{kl}{\delta_{3}}_{j}^{i}$.
The contribution of \eq{Yukawaguts} to the three $y_{pU}$ coefficients
(conveniently separated into two terms corresponding to the fermionic representations
$Q_{L}$ and $U_{R}$) can then be written as
\be
y_{pU} = \frac{1}{d(G_{p})}
\left[C_{2}^{(p)}(Q_{L})+C_{2}^{(p)}(U_{R})\right] 
\sum_{ab,ij,kl}Y^{ab}_{U}{\varepsilon}^{kl}{\delta_{3}}_{j}^{i}Y^{ab*}_{U}
{\varepsilon}_{kl}{\delta_{3}}_{i}^{j}
\label{ypU}
\ee
The sum can be factorized into the flavour space part
$\sum_{ab}Y^{ab*}_{U}Y^{ab}_{U}={\rm Tr}[Y_{U}Y_{U}^{\dagger}]$
times the trace over the gauge contractions ${\rm Tr}[\Delta\Delta^{\dagger}]$
where $\Delta\equiv{\varepsilon}^{kl}{\delta_{3}}_{j}^{i}$.
For the SM gauge group (with the properly normalized hypercharge) one then obtains
$y_{1U}=\frac{17}{10}$,
$y_{2U}=\frac{3}{2}$
and
$y_{3U}=2$,
that coincide with the values given in the first column of the matrix (B.5) in 
Refs.~\cite{Machacek:1983tz,Machacek:1983fi,Machacek:1984zw}.

All of the $y_{pk}$ coefficients as well as the structures of the
relevant $\Delta$-tensors are reported
in Table \ref{tab:Yukawas}.

\renewcommand{\arraystretch}{1.2}
\renewcommand\tabcolsep{1pt}
\begin{table}[h]
\centering
\begin{scriptsize}
\begin{tabular}{ccccccc}
\hline
{$G_p$}  &  $y_{pk}$  &  {$k$} &  {\rm Gauge structure} &  {\rm Higgs rep.} &  Tensor $\Delta$  & Tr$[\Delta\Delta^{\dagger}]$
\\
\hline
\\
$\begin{array}{l}
3_C \\
2_L \\
1_Y
\end{array}$
&
$\left(
\begin{array}{lcc}
 2 & 2 & 0 \\
 \frac{3}{2} & \frac{3}{2} & \frac{1}{2} \\
 \frac{17}{10} & \frac{1}{2} & \frac{3}{2}
\end{array}
\right)$  &
$\begin{array}{c}
{\rm U} \\
{\rm D} \\
{\rm E}
\end{array}$
&
$\begin{array}{c}
\overline{Q_{L}}_{kj}U_{R}^{i}\tilde h_{l} \\
\overline{Q_{L}}_{kj}D_{R}^{i}h^{l} \\
\overline{L_{L}}_{k}E_{R}^{i}h^{l} \\
\end{array}$
&
$h^{l}: (+\frac{1}{2},2,1)$
&
$\begin{array}{c}
\epsilon^{kl}{\delta_3}_{i}^{j} \\
{\delta_{2}}^{k}_{l}{\delta_3}_{i}^{j}  \\
{\delta_{2}}^{k}_{l} \\
\end{array}$
&
$\begin{array}{c}
6 \\
6 \\
2 \\
\end{array}$
\\
\\
$\begin{array}{l}
3_{C}\\[0.3ex]
2_L\\[0.3ex]
1_{R,R} \\[0.3ex]
1_{R,B-L}\\[0.3ex]
1_{B-L,R}\\[0.3ex]
1_{B-L,B-L}
\end{array}$ &
$\left(
\begin{array}{cccc}
 2 & 2 & 0 & 0 \\
 \frac{3}{2} & \frac{3}{2} & \frac{1}{2} & \frac{1}{2} \\
 \frac{3}{2} & \frac{3}{2} & \frac{1}{2} & \frac{1}{2} \\
 \frac{1}{2}\sqrt{\frac{3}{2}} & -\frac{1}{2}\sqrt{\frac{3}{2}} & -\frac{1}{2}\sqrt{\frac{3}{2}} & \frac{1}{2}\sqrt{\frac{3}{2}} \\
 \frac{1}{2}\sqrt{\frac{3}{2}} & -\frac{1}{2}\sqrt{\frac{3}{2}} & -\frac{1}{2}\sqrt{\frac{3}{2}} & \frac{1}{2}\sqrt{\frac{3}{2}} \\
 \frac{1}{2} & \frac{1}{2} & \frac{3}{2} & \frac{3}{2} 
\end{array}
\right)$  &
$\begin{array}{c}
{\rm U} \\
{\rm D} \\
{\rm N} \\
{\rm E}
\end{array}$ &
$\begin{array}{c}
\overline{Q_{L}}_{kj}U_{R}^{i}\tilde h_{l} \\
\overline{Q_{L}}_{kj}D_{R}^{i}h^{l} \\
\overline{L_{L}}_{k}N_{R}\tilde h_{l} \\
\overline{L_{L}}_{k}E_{R}h^{l} \\
\end{array}$
&
$h^{l}:(2,+\frac{1}{2},0,1)$
&
$\begin{array}{c}
\epsilon^{kl}{\delta_3}_{i}^{j} \\
{\delta_{2}}^{k}_{l}{\delta_3}_{i}^{j}  \\
\epsilon^{kl}\\
{\delta_{2}}^{k}_{l} \\
\end{array}$
&
$\begin{array}{c}
6 \\
6 \\
2 \\
2 \\
\end{array}$
\\
\\
$\begin{array}{l}
3_{C} \\
2_L\\
2_{R} \\
1_{B-L}
\end{array}$
&
$\left(
\begin{array}{ccc}
 4 & 0 \\
 3 & 1 \\
 3 & 1 \\
 1 & 3 
\end{array}
\right)$  &
$\begin{array}{c}
{\rm Q} \\
{\rm L}
\end{array}$ &
$\begin{array}{c}
Q_{L}^{ik}Q^{c\,m}_{Lj}\phi^{ln} \\
L_{L}^{k}L^{c\,m}_{L}\phi^{ln} \\
\end{array}$
&
$\phi^{ln}:(2,2,0,1)$
&
$\begin{array}{c}
\epsilon_{kl}\epsilon_{mn}{\delta_3}_{i}^{j} \\
\epsilon_{kl}\epsilon_{mn} \\
\end{array}$
&
$\begin{array}{c}
12 \\
4 \\
\end{array}$
\\
\\
$\begin{array}{l}
4_{C} \\
2_L\\
1_{R}
\end{array}$
&
$\left(
\begin{array}{cc}
 2 & 2 \\
 2 & 2 \\
 2 & 2
\end{array}
\right)$  &
$\begin{array}{c}
{\rm F^{U}} \\
{\rm F^{D}}
\end{array}$
&
$\begin{array}{c}
\overline{F_{L}}_{kj}F^{Ui}_{R}\tilde h_{l} \\
\overline{F_{L}}_{kj}F^{Di}_{R}h^{l}
\end{array}$
&
$h^{l}:(2,+\frac{1}{2},1)$
&
$\begin{array}{c}
\epsilon^{kl}{\delta_4}_{i}^{j} \\
{\delta_{2}}^{k}_{l}{\delta_4}_{i}^{j}  \\
\end{array}$
&
$\begin{array}{c}
8 \\
8 \\
\end{array}$
\\
\\
$\begin{array}{c}
4_{C} \\
2_L\\
2_{R}
\end{array}$
&
$\left(
\begin{array}{c}
 4 \\
 4 \\
 4
\end{array}
\right)$  &
{\rm F} &
$F_{L}^{ik}F^{c\,m}_{Lj}\phi^{ln}$
&
$\phi^{ln}:(2,2,1)$
&
$\begin{array}{c}
\epsilon_{kl}\epsilon_{mn}{\delta_4}_{i}^{j} \\
\end{array}$
&
$\begin{array}{c}
16 \\
\end{array}$
\\
\\
\\
$\begin{array}{l}
4_{C} \\
2_L\\
1_{R}
\end{array}$
&
$\left(
\begin{array}{cc}
\frac{15}{4} & \frac{15}{4} \\
\frac{15}{4} & \frac{15}{4} \\
\frac{15}{4} & \frac{15}{4}
\end{array}
\right)$  &
$\begin{array}{c}
{\rm F^{U}} \\
{\rm F^{D}}
\end{array}$ &
$\begin{array}{c}
\overline{F_{L}}_{kj}F^{Ui}_{R}\tilde H^{a}_{l} \\
\overline{F_{L}}_{kj}F^{Di}_{R}H^{la}
\end{array}$
 &
$H^{la}:(2,+\frac{1}{2},15)$
&
$\begin{array}{c}
\epsilon^{kl}(T_{a})_{i}^{j}\\
\delta^{k}_{l}(T_{a})_{i}^{j}\\
\end{array}$
&
$\begin{array}{c}
15 \\
15 \\
\end{array}$
\\
\\
$\begin{array}{c}
4_{C}\\
2_L\\
2_{R}
\end{array}$
&
$\left(
\begin{array}{c}
\frac{15}{2} \\
\frac{15}{2} \\
\frac{15}{2}
\end{array}
\right)$  &
{\rm F} &
$F_{L}^{ik}F^{c\,m}_{Lj}\Phi^{lna}$
&
$\Phi^{lna}:(2,2,15)$
&
$\epsilon_{kl}\epsilon_{mn}(T_{a})_{i}^{j}$
&
$\begin{array}{c}
30 \\
\end{array}$
\\
\\
\hline
\end{tabular}
\end{scriptsize}
\mycaption{The two-loop Yukawa contributions to the gauge sector $\beta$-functions
in \eq{Ycpk} are detailed. The index $p$ in $y_{pk}$ labels the gauge groups while
$k$ refers to flavour.
In addition to the Higgs bi-doublet from the 10-dimensional representation
(whose components are denoted according to the relevant gauge symmetry by $h$ and $\phi$)
extra bi-doublet components in $\overline{126}_{H}$ (denoted by $H$ and $\Phi$)
survives from unification
down to the Pati-Salam breaking scale as required by a realistic SM fermionic spectrum.
The $T_{a}$ factors are the generators of $SU(4)_{C}$ in the standard normalization.
As a consequence of minimal fine tuning, only one linear combination of $10_H$
and $\overline{126}_H$ doublets survives below the $SU(4)_{C}$ scale.
The $U(1)_{R,B-L}$ mixing in the case $3_C2_L1_R1_{B-L}$ is explicitly displayed.}
\label{tab:Yukawas}
\end{table}



\chapter{$SO(10)$ algebra representations}
\label{app:so10algebra}

We briefly collect here the conventions for the $SO(10)$ algebra representations adopted in Chapter~\ref{thequantumvac}.

\section{Tensorial representations}
\label{app:tensorrep}

The hermitian and antisymmetric generators of the fundamental representation of $SO(10)$ are given by
\begin{equation}
(\epsilon_{ij})_{ab}=-i(\delta_{a[i}\delta_{bj]}) \, ,
\end{equation}
where $a,b,i,j=1,..,10$ and the square bracket stands for anti-symmetrization.
They satisfy the $SO(10)$ commutation relations
\be
\label{so10commrel}
\left[ \epsilon_{ij}, \epsilon_{kl} \right] = -i ( \delta_{jk}\epsilon_{il} - \delta_{ik}\epsilon_{jl} - \delta_{jl}\epsilon_{ik} + \delta_{il}\epsilon_{jk}) \, ,
\ee
with normalization
\be
\Tr \epsilon_{ij}\epsilon_{kl} = 2\ \delta_{i[k}\delta_{jl]} \, .
\ee
The fundamental (vector) representation $\phi_a$ $ (a=1,...,10)$ transforms
as
\be
\label{10transf}
\phi_a \rightarrow \phi_a - \frac{i}{2} \lambda_{ij} (\epsilon_{ij}\phi)_a \, ,
\ee
where $\lambda_{ij}$ are the infinitesimal parameters of the transformation.

The adjoint representation is then obtained as the
antisymmetric part of the 2-index $10_a\otimes 10_b$ tensor $\phi_{ab}$ $(a,b=1,..,10)$
and transforms as
\be
\phi_{ab}\rightarrow  \phi_{ab}-\frac{i}{2}\lambda_{ij}\left[\epsilon_{ij},\phi\right]_{ab} \, .
\ee
Notice that
$\left[\epsilon_{ij},\phi\right]^T = - \left[\epsilon_{ij},\phi\right]$
and $\left[\epsilon_{ij},\phi\right]^\dag = \left[\epsilon_{ij},\phi\right]$.

\section{Spinorial representations}
\label{app:spinorrep}

Following the notation of Ref.~\cite{Babu:1984mz}, the $SO(10)$ generators $S_{ij}$ ($i,j=0,..,9$) acting on the 32-dimensional
spinor $\Xi$ are defined as
\begin{equation}
\label{spingen32}
S_{ij}=\frac{1}{4i}\left[\Gamma_i,\Gamma_j\right] \, ,
\end{equation}
where the $\Gamma_i$'s satisfy the Clifford algebra 
\be
\left\{\Gamma_i,\Gamma_j\right\}=2\delta_{ij} \, .
\ee
An explicit representation given by~\cite{Georgi:1979dq}
\begin{equation}
\Gamma_0=\left(
\begin{array}{cc}
 0 & I_{16}  \\
 I_{16} & 0
\end{array}
\right) \, , \qquad
\Gamma_p=\left(
\begin{array}{cc}
 0     & is_p  \\
 -is_p & 0
\end{array}
\right) \, , \qquad  p=1,...,9 \; ,
\label{cliffalgebra}
\end{equation}
where the $s_p$ matrices are defined as ($k=1,..,3$)
\begin{equation}
s_k=\eta_k\rho_3 \, , \qquad s_{k+3}=\sigma_k\rho_1 \, , \qquad s_{k+6}=\tau_k\rho_2 \, .
\end{equation}
The matrices $\sigma_k$, $\tau_k$, $\eta_k$ and $\rho_k$,
are given by the following tensor products of $2\times 2$ matrices
\begin{align}
\label{sigmatauetaro}
\sigma_k&=I_{2}\otimes I_{2}\otimes I_{2}\otimes \Sigma_k \, , \nn\\
\tau_k  &=I_{2}\otimes I_{2}\otimes \Sigma_k\otimes I_{2} \, , \\
\eta_k  &=I_{2}\otimes \Sigma_k\otimes I_{2}\otimes I_{2} \, , \nn\\
\rho_k  &=\Sigma_k\otimes I_{2}\otimes I_{2}\otimes I_{2} \, \nn,
\end{align}
where $\Sigma_k$ stand for the ordinary Pauli matrices.
Defining
\begin{equation}
s_{pq}=\frac{1}{2i}\left[s_p,s_q\right]
\end{equation}
for $p,q=1,..,9$, the algebra (\ref{spingen32}) is represented by
\begin{equation}
S_{p0}=\frac{1}{2}\left(
\begin{array}{cc}
 s_p & 0  \\
 0 & -s_p
\end{array}
\right)
\, , \qquad
S_{pq}=\frac{1}{2}\left(
\begin{array}{cc}
 s_{pq} & 0  \\
 0 & s_{pq}
\end{array}
\right) \, .
\label{32spingen}
\end{equation}
The Cartan subalgebra is spanned over $S_{03}$, $S_{12}$, $S_{45}$, $S_{78}$ and $S_{69}$.
One can construct a chiral projector $\Gamma_{\chi}$, that splits the 32-dimensional spinor $\Xi$ into a pair of irreducible 16-dimensional components:
 \begin{equation}
\Gamma_{\chi}=2^{-5} S_{03}S_{12}S_{45}S_{78}S_{69}=\left(
\begin{array}{cc}
 -I_{16} & 0  \\
 0 & I_{16}
\end{array}
\right)
\, .
\end{equation}
It is readily verified that $\Gamma_\chi$ has the following properties:
$\Gamma_\chi^2=I_{32}$, $\left\{ \Gamma_\chi, \Gamma_i \right\}=0$ and hence
$\left[ \Gamma_\chi, S_{ij} \right]=0$.
Introducing the chiral projectors $P_\pm=\frac{1}{2}(I_{32}\mp\Gamma_{\chi})$,
the irreducible chiral spinors are defined as
\be
\chi_+ = P_+\Xi\equiv
\left(
\begin{array}{l}
 \chi  \\
 0
\end{array}
\right) \, , \qquad
\chi_- = P_-\Xi\equiv
\left(
\begin{array}{l}
 0  \\
 \chi^{c}
\end{array}
\right) \, ,
\ee
where $\chi^{c}\equiv C\chi^{\ast}$ and $C$ is the $SO(10)$ charge conjugation matrix (see next subsection).
Analogously, we can use the chiral projectors to write $S_{ij}$ as
\be
\label{genprojected}
S_{ij}=P_+S_{ij}P_+ + P_-S_{ij}P_-
\equiv
\frac{1}{2}
\left(
\begin{array}{cc}
 \sigma_{ij} & 0  \\
 0 & \tilde{\sigma}_{ij}
\end{array}
\right) \, ,
\ee
where the properties $\left[ P_\pm, S_{ij} \right]=0$,
$P_\pm^2=P_\pm$ and $P_+ + P_- = I_{32}$ were used.

Finally, matching \eq{genprojected} with \eq{32spingen}, one
identifies the hermitian generators $\sigma_{ij}/2$ and $\tilde{\sigma}_{ij}/2$ acting
on the $\chi$ and $\chi^c$ spinors, respectively, as
\be
\sigma_{p0} = s_p\,,\quad \sigma_{pq}=s_{pq} \,,\quad
\tilde{\sigma}_{p0} = -s_p\,,\quad \tilde{\sigma}_{pq}=s_{pq} \, ,
\ee
with normalization
\be
\label{Dynkinspinor}
\tfrac{1}{4}\Tr \sigma_{ij}\sigma_{kl} =
\tfrac{1}{4}\Tr \tilde{\sigma}_{ij}\tilde{\sigma}_{kl} =
4\ \delta_{i[k}\delta_{jl]} \, .
\ee

It is convenient to trace out the $\sigma$-matrices in the invariants built off the adjoint representation in the natural basis $\Phi\equiv\sigma_{ij}\phi_{ij}/4$.
From the traces of two and four $\sigma$-matrices one obtains
\begin{align}
\label{tr2sig}
&\Tr\Phi^2= - 2 \Tr\phi^2 \, , \\[1ex]
\label{tr4sig}
&\Tr\Phi^4= \tfrac{3}{4} \left(\Tr\phi^2\right)^2 - \Tr\phi^4
\, .
\end{align}

In order to maintain a consistent notation, from now on
we shall label the indices of the spinorial generators from
1 to 10, and use the following mapping from the basis of Ref.~\cite{Babu:1984mz} into the basis of Ref.~\cite{Bajc:2004xe}
for both vectors and tensors:
$\{0 3 1 2 4 5 7 8 6 9\} \rightarrow \{1 2 3 4 5 6 7 8 9 10\}$.

\section{The charge conjugation $C$}
\label{app:spinorandC}


According to the notation of the previous subsection,
the spinor $\chi$ and its complex conjugate $\chi^{\ast}$ transform as
\be
\chi\rightarrow \chi - \frac{i}{4} \lambda_{ij}\sigma_{ij} \chi \, , \quad
\chi^{\ast}\rightarrow \chi^{\ast} +\frac{i}{4} \lambda_{ij}\sigma_{ij}^{T} \chi^{\ast} \, .
\ee
The charge conjugated spinor $\chi^{c}\equiv C\chi^{\ast}$ obeys
\be
\chi^{c}\rightarrow \chi^{c} - \frac{i}{4} \lambda_{ij}\tilde{\sigma}_{ij} \chi^{c} \, ,
\ee
 and thus $C$ satisfies
\be
\label{Cproperty}
C^{-1}\tilde{\sigma}_{ij}C=-\sigma_{ij}^{T} \, .
\ee
Taking into account \eq{sigmatauetaro}, a formal solution reads
\be
\label{Crelations}
C=\sigma_2\tau_2\eta_2\rho_2 \, ,
\ee
which in our basis yields
\begin{multline}
C = \text{antidiag}(+1,-1,-1,+1,-1,+1,+1,-1, \\
                    -1,+1,+1,-1,+1,-1,-1,+1) ,
\end{multline}
and hence $
C=C^{\ast}=C^{-1}=C^{T}=C^{\dag}$.

\section{The Cartan generators}
\label{app:explicitgenerators}

It is convenient to write the five $SO(10)$ Cartan generators in the $3_C 2_L 2_R 1_{B-L}$ basis,
where the generator $T_{B-L}$ is $(B-L)/2$. For the spinorial
representation we have
\bea
& T^{3}_R=\tfrac{1}{4}(\sigma_{12}+\sigma_{34}) \, , \quad
\widetilde{T}^{3}_R=\tfrac{1}{4}(-\sigma_{12}+\sigma_{34}) \, , &
\nn\\
&T^{3}_L=\tfrac{1}{4}(\sigma_{34}-\sigma_{12}) \, , \quad
\widetilde{T}^{3}_L=\tfrac{1}{4}(\sigma_{34}+\sigma_{12}) \, , &
\nn\\
& T^{3}_c=\widetilde{T}^{3}_c=\tfrac{1}{4}(\sigma_{56}-\sigma_{78}) \nn \, ,
\\
&T^{8}_c=\widetilde{T}^{8}_c=\tfrac{1}{4\sqrt{3}}(\sigma_{56}+\sigma_{78}-2\sigma_{910}) \, ,&
\nn\\
&T_{B-L}=\widetilde{T}_{B-L}=-\tfrac{2}{3}(\sigma_{56}+\sigma_{78}+\sigma_{910}) \, .&
\label{cartanT}
\eea
While the $T$'s act on $\chi$, the $\widetilde{T}$'s (characterized by a sign flip in $\sigma_{1i}$)
act on $\chi^c$.
The normalization of the Cartan generators is chosen according to the usual SM convention.
A GUT-consistent normalization across all generators is obtained by rescaling $T_{B-L}$ (and $\widetilde{T}_{B-L}$) by $\sqrt{3/2}$.

In order to obtain the physical generators acting on the fundamental representation it is enough to replace $\sigma_{ij}/2$ in \eq{cartanT} by $\epsilon_{ij}$.

With this information at hand, one can identify the spinor components of $\chi$ and $\chi^{c}$
\be
\label{chiembedding}
\chi = (\nu,u_1,u_2,u_3,l,d_1,d_2,d_3, 
           -d^c_3,d^c_2,d^c_1,-l^c,u^c_3,-u^c_2,-u^c_1,\nu^c) \, ,
\ee
and
\be
\label{chicembedding}
{\chi^c} = (\nu^c,u^c_1,u^c_2,u^c_3,l^c,d^c_1,d^c_2,d^c_3, 
               -d_3,d_2,d_1,-l,u_3,-u_2,-u_1,\nu)^{\ast} \, ,
\ee
where a self-explanatory SM notation has been naturally extended into the scalar sector.
In particular, the relative signs in \eqs{chiembedding}{chicembedding} arise from the charge conjugation of the
$SO(6)\sim SU(4)_C$ and $SO(4)\sim SU(2)_L\otimes SU(2)_R$
components of $\chi$ and $\chi^c$.

The standard and flipped embeddings of $SU(5)$
commute with two different Cartan generators, $Z$ and $Z'$ respectively:
\be
Z =-4T^{(3)}_R + 6T_X \, , \qquad
Z' =4T^{(3)}_R + 6T_X \, .
\ee
Given the relation $\Tr (T^{3}_R)^2= \frac{3}{2}\Tr T_{B-L}^2$
one obtains
\be
\Tr (YZ)=0 \, , \qquad
\Tr (YZ')\neq 0 \, ,
\ee
where $Y = T^{3}_R + T_{B-L}$
is the weak hypercharge generator.

As a consequence, the standard $SU(5)$ contains the SM group,
while $SU(5)'$ has a subgroup
$SU(3)_C\otimes SU(2)_L\otimes U(1)_{Y'}$, with
\be
\label{defZ}
Y' =-T^{3}_R+T_X \, .
\ee
In terms of $Z'$ and of $Y'$ the weak hypercharge reads
\be
\label{deffwhyp}
Y=\tfrac{1}{5}(Z'-Y') \, .
\ee
Using the explicit form of the Cartan generators in the vector
representation one finds
\begin{align}
Z'& \propto \text{diag}(-1,-1,+1,+1,+1)\otimes \Sigma_2 \, ,\\[1ex]
Z &\propto \text{diag}(+1,+1,+1,+1,+1)\otimes \Sigma_2 \, .
\end{align}
The vacuum configurations
$\omega_R = -\omega_{B-L}$ and $\omega_R = \omega_{B-L}$ in \eq{vacua}
are aligned with the $Z'$ and the $Z$ generator respectively,
thus preserving $SU(5)'\otimes U(1)_{Z'}$ and $SU(5)\otimes U(1)_{Z}$, respectively.

\chapter{Vacuum stability}
\label{boundedpot}

The boundedness of the scalar potential is needed
in order to ensure the global stability of the vacuum.
The requirement that the potential is bounded from below sets non trivial constraints
on the quartic interactions.
We do not provide a fully general analysis for the whole field space,
but limit ourselves to the constraints obtained for
the given vacuum directions.

\begin{itemize}
\item{ 
$(\omega_R$, $\omega_{B-L}$, $\chi_R) \neq 0$

From the quartic part of the scalar potential $V_0^{(4)}$
one obtains
\begin{multline}
4a_1(2\omega_R^2 +3\omega_{B-L}^2)^2
 +\frac{a_2}{4}(8\omega_R^4 + 21\omega_{B-L}^4 + 36\omega_R^2\omega_{B-L}^2)
+ \frac{\lambda_1}{4}\chi_R^4
+ 4\alpha\chi_R^2(2\omega_R^2 +3\omega_{B-L}^2) \\
+\frac{\beta}{4}\chi_R^2(2\omega_R + 3\omega_{B-L})^2
 -\frac{\tau}{2}\chi_R^2(2\omega_R + 3\omega_{B-L}) > 0
\end{multline}
Notice that the $\lambda_2$ term vanishes along the
$16_H$ vacuum direction.
}


\item{$\omega_R = \omega_{B-L}=0$, $\chi_R\neq 0$

Along this direction the quartic potential $V_0^{(4)}$ reads
\be
\label{V4only16}
V_0^{(4)}=\tfrac{1}{4} \lambda _1 \chi _R^4 \, ,
\ee
which implies
\be
\label{boundonly16}
\lambda_1 > 0 \, .
\ee
}
\end{itemize}
From now on, we focus on the $\chi_R=0$ case, cf.~\sect{sec:chi0limit}.

\begin{itemize}


\item{$\omega = \omega_R = -\omega_{B-L}$, $\chi_R=0$

On this orbit the quartic part of the scalar potential reads
\be
\label{V4only45}
V_0^{(4)}=\tfrac{5}{4} \omega ^4 (80 a_1+13 a_2) \, .
\ee
Taking into account that the scalar mass spectrum implies
$a_2<0$, we obtain
\be
\label{bound45only}
a_1 > -\tfrac{13}{80} a_2 \, .
\ee
}


\item{$\omega_R=0$, $\omega_{B-L} \neq 0$, $\chi_R=0$

At the tree level this VEV configuration does not correspond
to a minimum of the potential.
It is nevertheless useful to inspect the stability conditions along this direction.
Since
\be
\label{V4-3211}
V_0^{(4)}=  \tfrac{3}{4} (48 a_1+7 a_2) \omega_{B-L}^4 \ ,
\ee
boundedness is obtained, independently on the sign of $a_2$, when
\be
\label{boundhierarch}
a_1 > - \tfrac{7}{48} a_2 \ .
\ee
}


\item{$\omega_R \neq 0$, $\omega_{B-L} = 0$, $\chi_R=0$

In analogy with the previous case we have
\be
V_0^{(4)}= 2 (8 a_1+a_2) \omega _R^4
\, ,
\label{V4-421}
\ee
which implies the constraint
\be
a_1 > - \tfrac{1}{8} a_2
\, .
\ee
In the case $a_2<0$ the constraint in \eq{bound45only} provides the global lower bound on $a_1$.
}

\end{itemize}


\chapter{Tree level mass spectra}
\label{app:Treemasses}

\section{Gauge bosons}
\label{app:gaugespectrum}

Given the covariant derivatives of the scalar fields
\begin{align}
&(D_{\mu}\phi)_{ab}=\partial_{\mu}\phi_{ab} -i
\frac{1}{2}g(A_\mu)_{ij}\left[\epsilon_{ij},\phi\right]_{ab} \, ,
\\ \nn \\
&(D_{\mu}\chi)_{\alpha}=\partial_{\mu}\chi_{\alpha} -i
\frac{1}{4}g(A_\mu)_{ij}(\sigma_{ij})_{\alpha\beta} \, \chi_{\beta} \, ,
\\ \nn \\
&(D_{\mu}\chi^c)_{\alpha}=\partial_{\mu}\chi^c_{\alpha} -i
\frac{1}{4}g(A_\mu)_{ij}(\tilde{\sigma}_{ij})_{\alpha\beta} \, \chi^c_{\beta} \, ,
\end{align}
and the canonically normalizaed kinetic terms
\be
\label{gauge45piece}
\frac{1}{4}\Tr(D_{\mu}\phi)^{\dag}(D^{\mu}\phi) \, ,
\ee
and
\be
\label{gauge16piece}
\frac{1}{2}(D_{\mu}\chi)^{\dag} (D^{\mu}\chi) \,
+ \frac{1}{2}(D_{\mu}\chi^c)^{\dag} (D^{\mu}\chi^c) \, ,
\ee
one may write the field dependent mass matrices for the gauge bosons
as
\begin{align}
\label{fielddepmass45}
\mathcal{M}_A^2(\phi)_{(ij)(kl)}
&= \frac{g^2}{2} \Tr [\epsilon_{(ij)}, \phi][\epsilon_{(kl)}, \phi]     \, ,
\\[0ex]
\label{fielddepmass16}
\mathcal{M}_A^2(\chi)_{(ij)(kl)}
&=\frac{g^2}{4} \chi^\dag\{\sigma_{(ij)},\sigma_{(kl)}\}\chi \, .
\end{align}
where $(ij),\, (kl)$ stand for ordered pairs of indices, and $\epsilon_{ij}$ ($\sigma_{ij}/2$) with $i,j=1,..,10$
are the generators of the fundamental (spinor) representation (see \app{app:so10algebra}).

\eqs{fielddepmass45}{fielddepmass16}, evaluated on the generic ($\omega_{R,B-L}\neq 0,\ \chi_{R}\neq 0$) vacuum, yield
the following contributions to the tree level gauge boson masses:

\subsection{Gauge bosons masses from 45}
\label{gaugespectrum45}

Focusing on \eq{gauge45piece} one obtains
\begin{align}
&\mathcal{M}_A^2(1,1,+1)=4 g^2 \omega _R^2 \, ,
\nn \\[0ex]
&\mathcal{M}_A^2(\overline{3},1,-\tfrac{2}{3})=4 g^2 \omega_{B-L}^2 \, ,
\nn\\[0ex]
&\mathcal{M}_A^2(1,3,0)=0 \, ,
\nn\\[0ex]
&\mathcal{M}_A^2(8,1,0)=0 \, ,
\\[0ex]
&\mathcal{M}_A^2(3,2,-\tfrac{5}{6})=g^2 \left(\omega _R-\omega_{B-L}\right)^2 \, ,
\nn\\[0ex]
&\mathcal{M}_A^2(3,2,+\tfrac{1}{6})=g^2 \left(\omega _R+\omega_{B-L}\right)^2 \, ,
\nn\\[0ex]
&\mathcal{M}_A^2(1,1,0)=
\left(
\begin{array}{cc}
0 & 0 \\
0 & 0
\end{array}
\right) \, ,\nn
\end{align}
where the SM singlet matrix is defined on the basis ($\psi^{45}_{15}$, $\psi^{45}_{1}$),
with the superscript referring to the original $SO(10)$ representation and
the subscript to the $SU(4)_C$ origin (see \Table{tab:45decomp}).

Note that, in the limits of standard $5\, 1_{Z}$ $(\omega_R = \omega_{B-L})$, flipped $5'\, 1_{Z'}$ $(\omega_R = -\omega_{B-L})$,
$3_C 2_L 2_R 1_{B-L}$ ($\omega_R=0$) and $4_C 2_L 1_R$ ($\omega_{B-L}=0$) vacua, we have respectively
25, 25, 15 and 19 massless gauge bosons, as expected.

\subsection{Gauge bosons masses from 16}

The contributions from \eq{gauge16piece} read
\begin{align}
&\mathcal{M}_A^2(1,1,+1)=g^2 \chi _R^2 \, ,
 \nn\\[0ex]
&\mathcal{M}_A^2(\overline{3},1,-\tfrac{2}{3})=g^2 \chi _R^2 \, ,
\nn\\[0ex]
&\mathcal{M}_A^2(1,3,0)=0 \, ,
\nn\\[0ex]
&\mathcal{M}_A^2(8,1,0)=0 \, ,
\\[0ex]
&\mathcal{M}_A^2(3,2,-\tfrac{5}{6})=0 \, ,
\nn\\[0ex]
&\mathcal{M}_A^2(3,2,+\tfrac{1}{6})=g^2 \chi _R^2 \, ,
\nn\\[0ex]
&\mathcal{M}_A^2(1,1,0)=
\left(
\begin{array}{cc}
 \frac{3}{2}  & \sqrt{\frac{3}{2}}  \\
 \sqrt{\frac{3}{2}}  & 1
\end{array}
\right)g^2 \chi _R^2 \, ,
\nn
\end{align}
where the last matrix is again spanned over ($\psi^{45}_{15}$, $\psi^{45}_{1}$), yielding
\begin{align}
& \text{Det}\, \mathcal{M}_A^2(1,1,0)=0 \, , \\[0ex]
& \Tr \mathcal{M}_A^2(1,1,0)=\tfrac{5}{2} g^2 \chi _R^2 \, .
\end{align}
The number of vanishing entries corresponds to the dimension
of the $SU(5)$ algebra
preserved by the $16_H$ VEV $\chi_R$.

Summing together the $45_H$ and $16_H$ contributions, we recognize 12 massless states, that correspond to the SM gauge bosons.

\section{Anatomy of the scalar spectrum}
\label{app:scalarspectrum}

In order to understand the dependence of the scalar masses on
the various parameters in the Higgs potential
we detail the scalar mass spectrum in the relevant limits of the scalar couplings, according to
the discussion on the accidental global symmetries in \sect{sect:understanding}.

\subsection{45 only}
\label{45only}

Applying the stationary conditions in \eqs{eqstatmu}{eqstat0},
to the flipped $5'\, 1_{Z'}$ vacuum with $\omega=\omega_R=-\omega_{B-L}$, we find
\begin{align}
&M^2(24,0)= -4 a_2 \omega ^2 \, , \nn\\[0.5ex]
&M^2(10,-4) = 0 \, , \\[0.5ex]
&M^2(1,0) = 2 \left(80 a_1 + 13 a_2\right) \omega ^2 \, ,\nn
\end{align}
and, as expected, the spectrum exhibits 20 WGB and 24 PGB whose
mass depends on $a_2$ only.
The required positivity of the scalar masses gives the constraints
\be
\label{fsu5only45cond}
a_2 < 0 \qquad \text{and} \qquad a_1 > -\tfrac{13}{80} a_2 \, ,
\ee
where the second equation coincides with the constraint coming from the stability of the scalar
potential (see \eq{bound45only} in \app{boundedpot}).

\subsection{16 only}
\label{16only}

When only the $16_H$ part of the scalar potential is considered the symmetry is spontaneously broken to
the standard $SU(5)$ gauge group.
Applying the stationary \eq{eqstatnu} we find
\begin{align}
&M^2(\overline{5})= 2 \lambda _2 \chi _R^2 \, ,\nn
\\[0.5ex]
&M^2(10)= 0 \, ,
\label{1016only}\\[0.5ex]
&M^2(1)=
\left(
\begin{array}{cc}
 1 & 1 \\
 1 & 1
\end{array}
\right)
\frac{1}{2}\lambda_1\chi_R^2 \, ,\nn
\end{align}
in the ($\psi^{16}_{1}$, $\psi^{16}_{1^\ast}$) basis, with the subscripts referring to the standard $SU(5)$ origin, that yields
\begin{align}
&\text{Det}\ M^2(1)= 0 \, ,
\nn \\[0ex]
&\Tr M^2(1)=\lambda_1\chi_R^2 \, ,
\end{align}
and as expected we count 21 WGB and 10 PGB modes whose mass depends
on $\lambda_2$ only.
The required positivity of the scalar masses leads to
\be
\label{su5only16cond}
\lambda_2 > 0 \qquad \text{and} \qquad \lambda_1 > 0 \, ,
\ee
where the second equation coincides with the constraint coming from the stability of the scalar potential
(see \eq{boundonly16} in \app{boundedpot}).

\subsection{Mixed 45-16 spectrum ($\chi_R \neq 0$)}
\label{4516spectrum}

In the general case the unbroken symmetry is the SM group. Applying
first the two stationary conditions in \eq{eqstatmu} and \eq{eqstatnu} we find the spectrum below.
The $2\times 2$ matrices are spanned over the ($\psi^{45}$, $\psi^{16}$) basis whereas the $4\times 4$ SM singlet matrix is given in the
($\psi^{45}_{15}$, $\psi^{45}_{1}$, $\psi^{16}_{1}$, $\psi^{16}_{1^\ast}$) basis.
\begin{align}
\label{mass1114516}
&M^2(1,1,+1)=
\left(
\begin{array}{cc}
 \beta  \chi _R^2+2 a_2 \omega_{B-L} \left(\omega _R+\omega_{B-L}\right) & \chi _R \left(\tau -3 \beta  \omega_{B-L}\right) \\
 \chi _R \left(\tau -3 \beta  \omega_{B-L}\right) & 2 \omega _R \left(\tau -3 \beta  \omega_{B-L}\right)
\end{array}
\right) \, ,
\nn\\[0ex]
&M^2(\overline{3},1,-\tfrac{2}{3})=
\left(
\begin{array}{cc}
 \beta  \chi _R^2+2 a_2 \omega _R \left(\omega _R+\omega_{B-L}\right) & \chi _R \left(\tau - \beta  (2\omega _R+\omega_{B-L})\right) \\
 \chi _R \left(\tau - \beta  (2\omega _R+\omega_{B-L})\right) & 2 \omega_{B-L} \left(\tau - \beta  (2\omega _R+\omega_{B-L})\right)
\end{array}
\right) \, ,
\\[1ex]
&M^2(1,3,0)=
2 a_2 (\omega_{B-L} - \omega _R) (\omega_{B-L} + 2 \omega _R) \, , \nn \\[0ex]
\label{mass8104516}
&M^2(8,1,0)=
2 a_2 (\omega _R - \omega_{B-L}) (\omega _R + 2 \omega_{B-L}) \, , \\[0ex]
&M^2(3,2,-\tfrac{5}{6})=
0 \, ,
\nn\\[1ex]
&M^2(3,2,+\tfrac{1}{6})=
\left(
\begin{array}{cc}
 \beta  \chi _R^2+4 a_2 \omega _R \omega_{B-L} & \chi _R \left(\tau -\beta  (\omega _R+2\omega_{B-L})\right) \\
 \chi _R \left(\tau  -\beta  (\omega _R+2\omega_{B-L})\right) & \left(\omega _R+\omega_{B-L}\right) \left(\tau  -\beta  (\omega _R+2\omega_{B-L})\right)
\end{array}
\right) \, ,
\nn\\[0ex]
\label{mass12m124516}
&M^2(1,2,-\tfrac{1}{2})=
\left(\omega _R+3 \omega_{B-L}\right) \left(\tau -\beta  \omega _R\right)+2 \lambda _2 \chi _R^2
 \, , \\[0ex]
\label{mass3bar1134516}
&M^2(\overline{3},1,+\tfrac{1}{3})=
2 \left(\omega _R+\omega_{B-L}\right) \left(\tau -\beta  \omega_{B-L}\right)+2 \lambda _2 \chi _R^2
 \, \nn.
\end{align}
\begin{multline}
M^2(1,1,0)= 
\left(
\begin{array}{c}
 \frac{1}{2} \left(3 \beta  \chi _R^2+4 \left(a_2 \omega _R^2+a_2 \omega_{B-L} \omega _R+(48 a_1+7 a_2) \omega_{B-L}^2\right)\right) \\
 \sqrt{6} \left(\frac{\beta  \chi _R^2}{2}+2 (16 a_1+3 a_2) \omega _R \omega_{B-L}\right) \\
 -\frac{1}{2} \sqrt{3} \chi _R \left(\tau -2 \beta  \omega _R-(16 \alpha +3 \beta ) \omega_{B-L}\right) \\
 -\frac{1}{2} \sqrt{3} \chi _R \left(\tau -2 \beta  \omega _R-(16 \alpha +3 \beta ) \omega_{B-L}\right) 
 \end{array}
\right.  \\[1ex]
\begin{array}{cc}
   \sqrt{6} \left(\frac{\beta  \chi _R^2}{2}+2 (16 a_1+3 a_2) \omega _R \omega_{B-L}\right) & 
      -\frac{1}{2} \sqrt{3} \chi _R \left(\tau -2 \beta \omega _R-(16 \alpha +3 \beta ) \omega_{B-L}\right) \\
 \beta  \chi _R^2+2 \left(4 (8 a_1+a_2) \omega _R^2+a_2 \omega_{B-L} \omega _R+a_2 \omega_{B-L}^2\right) & 
  \frac{\chi _R \left(-\tau +2 (8 \alpha +\beta ) \omega _R+3 \beta  \omega_{B-L}\right)}{\sqrt{2}} \\
 \frac{\chi _R \left(-\tau +2 (8 \alpha +\beta ) \omega _R+3 \beta  \omega_{B-L}\right)}{\sqrt{2}} & 
  \frac{1}{2} \lambda _1 \chi _R^2 \\
  \frac{1}{2} \lambda _1 \chi _R^2 & 
  \frac{\chi _R \left(-\tau +2 (8 \alpha +\beta ) \omega _R+3 \beta  \omega_{B-L}\right)}{\sqrt{2}}
 \end{array} \\[1ex]
 \left.
\begin{array}{c}
   -\frac{1}{2} \sqrt{3} \chi _R \left(\tau -2 \beta  \omega _R-(16 \alpha +3 \beta ) \omega_{B-L}\right) \\
 \frac{\chi _R \left(-\tau +2 (8 \alpha +\beta ) \omega _R+3 \beta  \omega_{B-L}\right)}{\sqrt{2}} \\
 \frac{1}{2} \lambda _1 \chi _R^2 \\
 \frac{1}{2} \lambda _1 \chi _R^2
\end{array}
\right) \, .
\end{multline}
By applying the remaining stationary condition in \eq{eqstat0}
one obtains
\begin{align}
&\text{Det}\ M^2(1,1,+1)=0 \, ,
\nn\\[0ex]
&\Tr M^2(1,1,+1)=
\frac{\left(\chi _R^2+4 \omega _R^2\right) \left(\tau -3 \beta  \omega_{B-L}\right)}{2 \omega _R} \, ,
\nn\\[0ex]
&\text{Det}\ M^2(\overline{3},1,-\tfrac{2}{3})=0 \, , \nn\\[0ex]
\label{massTr31m134516}
&\Tr M^2(\overline{3},1,-\tfrac{2}{3})=
\frac{\left(\chi _R^2+4 \omega_{B-L}^2\right) \left(\tau - \beta (2 \omega _R +  \omega_{B-L}) \right)}{2 \omega_{B-L}} \, , \\[0ex]
&\text{Det}\ M^2(3,2,+\tfrac{1}{6})=0 \, , \nn\\[0ex]
&\Tr M^2(3,2,+\tfrac{1}{6})=\beta  \chi _R^2 + 4 a_2 \omega _R \omega_{B-L}+
\left(\omega _R+\omega_{B-L}\right) \left(\tau - \beta  (\omega _R + 2  \omega_{B-L}) \right) \, ,\nn \\[0ex]
&\text{Rank}\ M^2(1,1,0)= 3 \, , \nn\\[0ex]
&\Tr M^2(1,1,0)=2 \left((32 a_1+5 a_2) \omega _R^2
+8 (6 a_1+a_2) \omega_{B-L}^2+2 a_2 \omega _R \omega_{B-L}\right)+\chi _R^2 \left(\tfrac{5}{2}\beta +\lambda _1\right) \, .\nn
\end{align}
In \eqs{mass1114516}{massTr31m134516} we recognize the 33 WGB with the quantum numbers of the coset $SO(10)/SM$ algebra.

In using the stationary condition in \eq{eqstat0},
we paid attention not to divide by ($\omega_R + \omega_{B-L}$), since the flipped vacuum $\omega=\omega_R=-\omega_{B-L}$
is an allowed configuration. On the other hand, we can freely put
 $\omega_R$ and $\omega_{B-L}$ into the denominators, as
the vacua $\omega_R=0$ and $\omega_{B-L}=0$ are excluded at the tree level.
The coupling $a_2$ in \eq{massTr31m134516} is understood to obey the constraint
\be
\label{laststatcond}
4a_2(\omega_R+\omega_{B-L})\omega_R\omega_{B-L}+\beta\chi_R^2(2\omega_R + 3\omega_{B-L})
-\tau\chi_R^2 = 0 \, .
\ee

\subsection{A trivial 45-16 potential ($a_2=\lambda_2=\beta=\tau=0$)}
\label{4516justnorms}

It is interesting to study the global symmetries of the scalar potential when only the moduli of $45_H$ and $16_H$ appear in the scalar potential.
In order to correctly count the corresponding PGB, the $(1,1,0)$ mass matrix in the limit of $a_2=\lambda_2=\beta=\tau=0$ needs to be scrutinized. We find in the ($\psi^{45}_{15}$, $\psi^{45}_{1}$, $\psi^{16}_{1}$, $\psi^{16}_{1^\ast}$) basis,
\begin{multline}
\label{}
M^2(1,1,0)= \\
\left(
\begin{array}{cccc}
 96 a_1 \omega_{B-L}^2 & 32 \sqrt{6} a_1 \omega _R \omega_{B-L} & 8 \sqrt{3} \alpha  \chi _R \omega_{B-L} & 8 \sqrt{3} \alpha  \chi _R \omega_{B-L} \\
 32 \sqrt{6} a_1 \omega _R \omega_{B-L} & 64 a_1 \omega _R^2 & 8 \sqrt{2} \alpha  \chi _R \omega _R & 8 \sqrt{2} \alpha  \chi _R \omega _R \\
 8 \sqrt{3} \alpha  \chi _R \omega_{B-L} & 8 \sqrt{2} \alpha  \chi _R \omega _R & \frac{1}{2} \lambda _1 \chi _R^2 & \frac{1}{2} \lambda _1 \chi _R^2 \\
 8 \sqrt{3} \alpha  \chi _R \omega_{B-L} & 8 \sqrt{2} \alpha  \chi _R \omega _R & \frac{1}{2} \lambda _1 \chi _R^2 & \frac{1}{2} \lambda _1 \chi _R^2
\end{array}
\right)
\, ,
\label{110simplified}
\end{multline}
with the properties
\begin{align}
&\text{Rank}\ M^2(1,1,0)= 2 \, ,
\nn \\[0ex]
&\Tr M^2(1,1,0)=64 a_1 \omega _R^2+96 a_1 \omega_{B-L}^2+\lambda _1 \chi _R^2 \, .
\end{align}
As expected from the discussion in \sect{sect:understanding},  \eqs{mass1114516}{110simplified} in the $a_2=\lambda_2=\beta=\tau=0$ limit exhibit 75 massless
modes out of which 42 are PGB.

\subsection{A trivial 45-16 interaction ($\beta=\tau=0$)}
\label{4516trivialinteraction}

In this limit, the interaction part of the potential consists only of the $\alpha$ term,
which is the product of $45_H$ and  $16_H$ moduli.
Once again, in order to correctly count the massless modes
we specialize the $(1,1,0)$ matrix to the
$\beta=\tau=0$ limit. In the ($\psi^{45}_{15}$, $\psi^{45}_{1}$, $\psi^{16}_{1}$, $\psi^{16}_{1^\ast}$) basis, we find
\begin{multline}
M^2(1,1,0)= \\[1ex]
\left(
\begin{array}{cc}
 2 \left(a_2 \omega _R^2+a_2 \omega_{B-L} \omega _R+(48 a_1+7 a_2) \omega_{B-L}^2\right) & 2 \sqrt{6} (16 a_1+3 a_2) \omega _R
   \omega_{B-L}  \\
 2 \sqrt{6} (16 a_1+3 a_2) \omega _R \omega_{B-L} & 2 \left(4 (8 a_1+a_2) \omega _R^2+a_2 \omega_{B-L} \omega _R+a_2 \omega
   _Y^2\right) \\
 8 \sqrt{3} \alpha  \chi _R \omega_{B-L} & 8 \sqrt{2} \alpha  \chi _R \omega _R  \\
 8 \sqrt{3} \alpha  \chi _R \omega_{B-L} & 8 \sqrt{2} \alpha  \chi _R \omega _R
\end{array}
\right.
\\[0ex]
\left.
\begin{array}{cc}
8 \sqrt{3} \alpha  \chi _R \omega_{B-L} & 8 \sqrt{3} \alpha  \chi _R \omega_{B-L} \\
8 \sqrt{2} \alpha  \chi _R \omega _R & 8 \sqrt{2} \alpha  \chi _R \omega _R \\
 \frac{1}{2} \lambda _1 \chi _R^2 & \frac{1}{2} \lambda _1 \chi _R^2 \\
  \frac{1}{2} \lambda _1 \chi _R^2 & \frac{1}{2} \lambda _1 \chi _R^2
\end{array}
\right)
\, ,
\end{multline}
with the properties
\begin{align}
&\text{Rank}\ M^2(1,1,0)= 3  \, , \nn\\[0.5ex]
&\Tr M^2(1,1,0) = 2 \left((32 a_1 + 5 a_2) \omega _R^2 
 +8 (6 a_1+a_2) \omega_{B-L}^2 
+ 2 a_2 \omega _R \omega_{B-L}\right)+\lambda _1 \chi _R^2 \, .
\end{align}
According to the discussion in \sect{sect:understanding}, upon inspecting
\eqs{mass1114516}{massTr31m134516} in the $\beta=\tau=0$ limit,
one finds 41 massless scalar modes of which 8 are PGB.

\subsection{The 45-16 scalar spectrum for $\chi_R = 0$ }
\label{4516spectrumchi0}

The application of the stationary conditions in \eqs{eqstatmu}{eqstat0}
(for $\chi_R =0$, \eq{eqstatnu} is trivially satisfied) leads to four different spectra according to the four vacua:
standard $5 \, 1_{Z}$,
flipped $5' \, 1_{Z'}$, $3_C 2_L 2_R 1_{B-L}$ and $4_C 2_L 1_R$.
We specialize our discussion to the last three cases.

The mass eigenstates are conveniently labeled according to
the subalgebras of $SO(10)$ left invariant by each vacuum.
With the help of \Tables{tab:16decomp}{tab:45decomp} one can easily recover the decomposition in the SM components.
In the limit $\chi_R=0$ the states $45_H$ and $16_H$ do not mix.
All of the WGB belong to the $45_H$, since for $\chi_R=0$ the $16_H$ preserves $SO(10)$.

Consider first the case: $\omega = \omega_R = -\omega_{B-L}$ (which preserves the flipped $5' \, 1_{Z'}$ group). For the $45_H$ components we obtain:
\begin{align}
&M^2(24,0)= -4 a_2 \omega ^2 \, , \nn\\[0.5ex]
&M^2(10,-4) = 0 \, , \\[0.5ex]
&M^2(1,0) = 2 \left(80 a_1 + 13 a_2\right) \omega ^2 \, .\nn
\end{align}
Analogously, for the $16_H$ components we get:
\begin{align}
&M^2(10,+1) = \tfrac{1}{4} \left(\omega^2 (80 \alpha +\beta )+2 \tau  \omega -2 \nu^2 \right) \, , \nn\\[0.5ex]
&M^2(\bar 5,-3) = \tfrac{1}{4} \left(\omega^2 (80 \alpha +9 \beta )-6 \tau  \omega -2 \nu^2 \right) \, , \\[0.5ex]
&M^2(1,+5) = \tfrac{1}{4} \left(5 \omega^2 (16 \alpha +5 \beta )+10 \tau  \omega -2 \nu^2 \right) \, .\nn
\end{align}
Since the unbroken group is the flipped $5' \, 1_{Z'}$ we recognize, as
expected, 45-25=20 WGB.
When only trivial $45_H$ invariants (moduli) are considered
the global symmetry of the scalar potential is $O(45)$,
broken spontaneously by $\omega$ to $O(44)$.
This leads to 44 GB in the scalar spectrum.
Therefore 44-20=24 PGB are left in the spectrum.
On general grounds, their masses should receive contributions
from all of the explicitly breaking terms $a_2$, $\beta$ and $\tau$.
As it is directly seen from the spectrum, only the
$a_2$ term contributes at the tree level to $M(24,0)$.
By choosing $a_2<0$ one may obtain a consistent minimum of the scalar
potential. Quantum corrections are not relevant in this case.

Consider then the case $\omega_R=0$ and $\omega_{B-L} \neq 0$ which preserves the $3_C 2_L 2_R 1_{B-L}$ gauge group.
For the $45_H$ components we obtain:
\begin{align}
&M^2(1,3,1,0)= 2 a_2 \omega_{B-L}^2 \, , \nn\\[0.5ex]
&M^2(1,1,3,0)= 2 a_2 \omega_{B-L}^2 \, , \nn\\[0.5ex]
&M^2(8,1,1,0)= -4 a_2 \omega_{B-L}^2 \, , \nn\\[0.5ex]
&M^2(3,2,2,-\tfrac{1}{3})= 0 \, , \\[0.5ex]
&M^2(\overline{3},1,1,-\tfrac{2}{3})= 0 \, , \nn\\[0.5ex]
&M^2(1,1,1,0)= 2 \left(48 a_1 + 7 a_2\right) \omega_{B-L}^2 \, .\nn
\end{align}

Analogously, for the $16_H$ components we get:
\begin{align}
&M^2(3,2,1,+\tfrac{1}{6})= \tfrac{1}{4} \left( \omega_{B-L}^2(48 \alpha +\beta )
-2 \tau \omega_{B-L}-2 \nu ^2\right) \, , \nn\\[0.5ex]
&M^2(\overline{3},1,2,-\tfrac{1}{6})= \tfrac{1}{4} \left( \omega_{B-L}^2(48 \alpha +\beta )
+2 \tau \omega_{B-L}-2 \nu ^2\right) \, , \nn\\[0.5ex]
&M^2(1,2,1,-\tfrac{1}{2})= \tfrac{1}{4} \left(\omega_{B-L}^2 (48 \alpha +9 \beta )+6 \tau  \omega_{B-L}-2 \nu^2\right)
\, , \nn\\[0.5ex]
&M^2(1,1,2,+\tfrac{1}{2})= \tfrac{1}{4} \left( \omega_{B-L}^2(48 \alpha +9 \beta )-6 \tau  \omega_{B-L}-2 \nu ^2\right) \, .
\end{align}
Worth of a note is the mass degeneracy of
the $(1,3,1,0)$ and $(1,1,3,0)$ multiplets which is due to the fact that
for $\omega_R=0$ D-parity is conserved by even $\omega_{B-L}$ powers.
On the contrary, in the $16_H$ components the D-parity is broken by the $\tau$ term that is linear in $\omega_{B-L}$.

Since the unbroken group is $3_C 2_L 2_R 1_{B-L}$ there are 45-15=30 WGB, as it appears from the explicit pattern of the scalar spectrum.
When only trivial invariants (moduli terms) of $45_H$ are considered
the global symmetry of the scalar potential is $O(45)$,
broken spontaneously to $O(44)$, thus
leading to 44 GB in the scalar spectrum.
As a consequence 44-30=14 PGB are left in the spectrum.
On general grounds, their masses should receive contributions
from all of the explicitly breaking terms $a_2$, $\beta$ and $\tau$.
As it is directly seen from the spectrum, only the
$a_2$ term contributes at the tree level to the mass of the 14 PGB,
leading unavoidably to a tachyonic spectrum. This feature is naturally lifted at the quantum level.

Let us finally consider the case $\omega_R \neq 0$ and $\omega_{B-L} = 0$ (which preserves the $4_C 2_L 1_R$ gauge symmetry).
For the $45_H$ components we find:
\begin{align}
&M^2(15,1,0)= 2 a_2 \omega _R^2 \, , \nn\\[0.5ex]
&M^2(1,3,0)= -4 a_2 \omega _R^2 \, , \nn\\[0ex]
&M^2(6,2,+\tfrac{1}{2})= 0 \, , \\[0.5ex]
&M^2(6,2,-\tfrac{1}{2})= 0 \, , \nn\\[0ex]
&M^2(1,1,+1)= 0 \, , \nn\\[0ex]
&M^2(1,1,0)= 8 \left(8 a_1+a_2\right) \omega _R^2 \, .\nn
\end{align}
For the $16_H$ components we obtain:
\begin{align}
&M^2(4,2,0)= 8 \alpha  \omega _R^2 - \tfrac{1}{2}\nu ^2 \, , \nn\\[0.5ex]
&M^2(\overline{4},1,+\tfrac{1}{2})=
\omega _R^2 (8 \alpha +\beta )+\tau  \omega _R - \tfrac{1}{2}\nu ^2 \, , \\[0.5ex]
 &M^2(\overline{4},1,-\tfrac{1}{2})=
\omega _R^2 (8 \alpha +\beta )-\tau  \omega _R - \tfrac{1}{2}\nu ^2 \, .\nn
\end{align}
The unbroken gauge symmetry in this case corresponds to $4_C 2_L 1_R$.
Therefore, one can recognize 45-19=26 WGB in the scalar spectrum.
When only trivial (moduli) $45_H$ invariants are considered
the global symmetry of the scalar potential is $O(45)$,
which is broken spontaneously by $\omega _R$ to $O(44)$.
This leads globally to 44 massless states in the scalar spectrum.
As a consequence, 44-26=18 PGB are left in the $45_H$ spectrum,
that should receive mass contributions
from the explicitly breaking terms $a_2$, $\beta$ and $\tau$.
At the tree level only the $a_2$ term is present,
leading again to a tachyonic spectrum.
This is an accidental tree level feature that is naturally lifted at the quantum level.

\chapter{One-loop mass spectra}
\label{app:1Lmasses}
We have checked explicitly that the one-loop corrected stationary equation (\ref{eqstat0}) maintains
in the $\chi_R=0$ limit the four tree level solutions, namely,
$\omega_R = \omega_{B-L}$,
$\omega_R = -\omega_{B-L}$, $\omega_R=0$ and $\omega_{B-L} = 0$,
corresponding respectively to the
standard $5\, 1_Z$, flipped $5'\, 1_{Z'}$, $3_C 2_L 2_R 1_{B-L}$ and $4_C 2_L 1_R$ vacua.

In what follows we list, for the last three cases,
the leading one-loop corrections,
arising from the gauge and scalar sectors,
to the critical PGB masses. For all other states
the loop corrections provide only sub-leading perturbations of the tree-level masses, and as such irrelevant
to the present discussion.

\section{Gauge contributions to the PGB mass}
\label{gaugePGBoneloop}

Before focusing to the three relevant vacuum configurations,
it is convenient to write the gauge contribution to the
$(1,3,0)$ and $(8,1,0)$ states
in the general case.
\begin{multline}
\Delta M^{2}(1,3,0)=
\frac{g^4 \left(16 \omega _R^2+\omega_{B-L} \omega _R+19 \omega_{B-L}^2\right)}{4 \pi ^2} \\
+ \frac{3 g^4}{4 \pi ^2\left(\omega _R-\omega_{B-L}\right)}
\left[  2 \left(\omega _R-\omega_{B-L}\right){}^3 \log \left(\frac{g^2 \left(\omega _R-\omega_{B-L}\right){}^2}{\mu ^2}\right) \right. \\[1ex]
+\left(4 \omega
   _R-5 \omega_{B-L}\right) \left(\omega _R+\omega_{B-L}\right){}^2 \log \left(\frac{g^2 \left(\omega _R+\omega_{B-L}\right){}^2}{\mu ^2}\right) \\[1ex]
 \left. -4 \omega_R^3 \log \left(\frac{4 g^2 \omega _R^2}{\mu ^2}\right)+8 \omega_{B-L}^3 \log \left(\frac{4 g^2 \omega_{B-L}^2}{\mu ^2}\right) \right] \, ,
\end{multline}
\begin{multline}
\Delta M^{2}(8,1,0)=
\frac{g^4\left(13 \omega _R^2+\omega_{B-L} \omega _R+22 \omega_{B-L}^2\right)}{4 \pi ^2} \\[1ex]
 +\frac{3 g^4}{8 \pi ^2 \left(\omega _R-\omega_{B-L}\right)}
\left[ \left(\omega _R-\omega_{B-L}\right){}^3 \log \left(\frac{g^2 \left(\omega _R-\omega_{B-L}\right){}^2}{\mu ^2}\right) \right. \\[1ex]
+\left(5 \omega
   _R-7 \omega_{B-L}\right) \left(\omega _R+\omega_{B-L}\right){}^2 \log \left(\frac{g^2 \left(\omega _R+\omega_{B-L}\right){}^2}{\mu ^2}\right) \\[1ex]
 \left. +4 \omega
   _Y^2 \left(3 \omega _R+\omega_{B-L}\right) \log \left(\frac{4 g^2 \omega_{B-L}^2}{\mu ^2}\right)-8 \omega _R^3 \log \left(\frac{4 g^2 \omega
   _R^2}{\mu ^2}\right)\right] \, .
\end{multline}
One can easily recognize the (tree-level) masses of the
gauge bosons in the log's arguments and cofactors (see \app{gaugespectrum45}). Note that
only the massive states do contribute to the one-loop correction.
(see \sect{sec:1loopspectrum}).

Let's now specialize to the three relevant vacua. First, for the flipped $5' \, 1_{Z'}$ case $\omega = \omega_R = -\omega_{B-L}$ one has:
\be
\Delta M^{2}(24,0)
= \frac{17 g^4 \omega ^2}{2 \pi ^2}
+\frac{3 g^4 \omega ^2 }{2 \pi ^2}\log \left(\frac{4 g^2 \omega ^2}{\mu ^2}\right) \, .
\ee
Similarly, for $\omega_R=0$ and $\omega_{B-L} \neq 0$\quad ($3_C 2_L 2_R 1_{B-L}$):

\begin{align}
\Delta M^{2}(1,3,1,0) & = \Delta M^{2}(1,1,3,0)  \nn \\
& = \frac{19 g^4 \omega_{B-L}^2}{4 \pi ^2} 
+\frac{21 g^4 \omega_{B-L}^2}{4 \pi ^2} \log \left(\frac{g^2 \omega_{B-L}^2}{\mu ^2}\right)
-\frac{24 g^4 \omega_{B-L}^2}{4 \pi ^2} \log \left(\frac{4 g^2 \omega_{B-L}^2}{\mu^2}\right) \, ,\nn \\[1ex]
\Delta M^{2}(8,1,1,0) & =
\frac{11 g^4 \omega_{B-L}^2}{2 \pi ^2}
+\frac{3 g^4 \omega_{B-L}^2}{2 \pi ^2} \log \left(\frac{g^2 \omega_{B-L}^2}{4 \mu ^2}\right)\, .
\end{align}
Finally, for $\omega_R\neq 0$ and $\omega_{B-L} = 0$\quad ($4_C 2_L 1_R$):

\begin{align}
& \Delta M^{2}(1,3,0)  =
\frac{4 g^4 \omega _R^2}{\pi ^2}
+\frac{3 g^4 \omega _R^2}{2 \pi ^2} \log \left(\frac{g^2 \omega _R^2}{16 \mu ^2}\right)\, , \nn \\[1ex]
& \Delta M^{2}(15,1,0) =
\frac{13 g^4 \omega _R^2}{4 \pi ^2}
+\frac{9 g^4 \omega _R^2}{4 \pi ^2} \log \left(\frac{g^2 \omega _R^2}{\mu ^2}\right)
-\frac{12 g^4 \omega _R^2}{4 \pi ^2}  \log \left(\frac{4 g^2 \omega _R^2}{\mu ^2}\right)\, .
\end{align}

\section{Scalar contributions to the PGB mass}
\label{scalarPGBoneloop}

Since the general formula for the SM vacuum configuration is quite involved,
we give directly the corrections to the PGB masses on the three vacua of our interest.
We consider first the case $\omega = \omega_R = -\omega_{B-L}$ (flipped $5' \, 1_{Z'}$):
\begin{small}
\begin{align}
& \Delta M^{2}(24,0) = 
\frac{\tau ^2 + 5 \beta ^2 \omega ^2}{4 \pi ^2} \\[0ex]
&+ \frac{1}{128 \pi ^2 \omega }\left[
 (-5 \beta  \omega -\tau ) (5 \omega  (16 \alpha  \omega +5 \beta  \omega +2 \tau )-2 \nu^2)
\log \left(\frac{5 \omega^2 (16 \alpha +5 \beta )+10 \tau  \omega -2 \nu^2}{4 \mu ^2}\right) \right. \nn \\[0ex]
& + \left(\omega  \left(3 \tau  \omega  (80 \alpha +3 \beta )+\beta  \omega ^2 (27 \beta -400 \alpha )-10 \tau ^2\right)+\nu^2 (10 \beta  \omega
-6 \tau)\right) \nn \\
& \times \log \left(\frac{\omega^2 (80 \alpha +9 \beta )-6 \tau  \omega -2 \nu^2}{4 \mu ^2}\right) \nn \\[0ex]
& + 2 \left(\omega  \left(\tau  (33 \beta  \omega -80 \alpha  \omega )+\beta \omega ^2 (400 \alpha +17 \beta )+10 \tau ^2\right)+2 \nu^2 (\tau -5 \beta \omega )\right) 
\nn \\
& \left. \times \log \left(\frac{\omega^2 (80 \alpha +\beta )+2 \tau  \omega -2 \nu^2 }{4 \mu ^2}\right) \right].\nn
\end{align}
\end{small}
For $\omega_R=0$ and $\omega_{B-L} \neq 0$ ($3_C 2_L 2_R 1_{B-L}$),
we find:
\begin{small}
\begin{align}
&\Delta M^{2}(1,3,1,0)  = \Delta M^{2}(1,1,3,0)  =
\frac{\tau ^2 + 2 \beta ^2 \omega_{B-L}^2}{4 \pi ^2}  \\[0ex]
&+ \frac{1}{64 \pi ^2 \omega_{B-L} }\left[
-\left(\tau -3 \beta  \omega_{B-L}\right) \left(-3 \omega_{B-L}^2 (16 \alpha +3 \beta )+6 \tau  \omega_{B-L}+2 \nu^2 \right) \right. \nn \\
& \times \log \left(\frac{ \omega_{B-L}^2(48 \alpha +9 \beta )-6 \tau
   \omega_{B-L} -2 \nu ^2 }{4 \mu ^2}\right)  \nn \\[0ex]
&-\left(\beta  \omega_{B-L}+\tau \right) \left(\omega_{B-L}^2 (48 \alpha +\beta )+2 \tau  \omega_{B-L}-2 \nu^2 \right) \log \left(\frac{\omega_{B-L}^2 (48
   \alpha +\beta )+2 \tau  \omega_{B-L}-2 \nu^2 }{4 \mu ^2}\right)
      \nn \\[0ex]
&+ \left(3 \tau  \omega_{B-L}^2 (16 \alpha -11 \beta )+\beta  \omega_{B-L}^3 (240 \alpha +17 \beta )+2 \omega_{B-L} \left(5 \tau ^2-5 \beta \nu^2 \right)-2
    \nu^2  \tau \right) \nn \\ 
    & \times \log \left(\frac{\omega_{B-L}^2 (48 \alpha +\beta )-2 \tau  \omega_{B-L}-2  \nu^2 }{4 \mu ^2}\right)
    \nn \\[0ex]
& +\left(\omega_{B-L}^2 (9 \beta  \tau -48 \alpha  \tau )+3 \beta  \omega_{B-L}^3 (9 \beta -16 \alpha )+2 \omega_{B-L} \left(\beta   \nu^2 -\tau ^2\right)+2
    \nu^2  \tau \right) \nn \\ 
    & \left. \times \log \left(\frac{\omega_{B-L}^2 (48 \alpha +9 \beta )+6 \tau  \omega_{B-L}-2  \nu^2 }{4 \mu ^2}\right) \right]\nn\,,
\end{align}
\begin{align}
&\Delta M^{2}(8,1,1,0) =
\frac{\tau ^2 + 3 \beta ^2 \omega_{B-L}^2}{4 \pi ^2}  \\[0ex]
&+ \frac{1}{64 \pi ^2 \omega_{B-L} } \left[
 - \left(\tau -3 \beta  \omega_{B-L}\right) \left(-3 \omega_{B-L}^2 (16 \alpha +3 \beta )+6 \tau  \omega_{B-L}+2  \nu^2 \right) \right. \nn \\
 & \times \log \left(\frac{\omega_{B-L}^2 (48
   \alpha +9 \beta )-6 \tau  \omega_{B-L}-2 \nu^2 }{4 \mu ^2}\right) \nn \\[0ex]
& + \left(\omega_{B-L}^2 (21 \beta  \tau -48 \alpha  \tau )+\beta  \omega_{B-L}^3 (144 \alpha +11 \beta )+\omega_{B-L} \left(6 \tau ^2-6 \beta  \nu^2 \right)+2
    \nu^2  \tau \right) \nn \\
    & \times \log \left(\frac{\omega_{B-L}^2 (48 \alpha +\beta )+2 \tau  \omega_{B-L}-2  \nu^2 }{4 \mu ^2}\right) \nn \\[0ex]
& - \left(3 \beta  \omega_{B-L}+\tau \right) \left(\omega_{B-L}^2 (48 \alpha +9 \beta )+6 \tau  \omega_{B-L}-2  \nu^2 \right) \log \left(\frac{\omega_{B-L}^2 (48 \alpha
   +9 \beta )+6 \tau  \omega_{B-L}-2  \nu^2 }{4 \mu ^2}\right) \nn \\[0ex]
& + \left(3 \tau  \omega_{B-L}^2 (16 \alpha -7 \beta )+\beta  \omega_{B-L}^3 (144 \alpha +11 \beta )+\omega_{B-L} \left(6 \tau ^2-6 \beta   \nu^2 \right)-2  \nu^2  \tau \right) 
\nn \\ 
& \left. \times \log \left(\frac{\omega_{B-L}^2 (48 \alpha +\beta )-2 \tau  \omega_{B-L}-2  \nu^2 }{4 \mu ^2}\right) \right]\nn.
\end{align}
\end{small}
Finally, for $\omega_R\neq 0$ and $\omega_{B-L} = 0$ ($4_C 2_L 1_R$), we have:
\begin{small}
\begin{align}
& \Delta M^{2}(1,3,0) =
\frac{\tau ^2 + 2 \beta ^2 \omega _R^2}{4 \pi ^2}  \\[0ex]
&+ \frac{1}{64 \pi ^2 \omega_R } \left[
16 \omega _R \left(16 \alpha  \beta  \omega _R^2-\beta   \nu^2 +\tau ^2\right) \log \left(\frac{8 \alpha  \omega _R^2-\frac{\nu ^2}{2}}{\mu ^2}\right) \right.  \nn \\[0ex]
&-4 \left(\tau -2 \beta  \omega _R\right) \left(-2 \omega _R^2 (8 \alpha +\beta )+2 \tau  \omega _R+ \nu^2 \right)
\log \left(\frac{ \omega _R^2 (8 \alpha +\beta )-\tau  \omega _R -\frac{\nu ^2}{2}}{ \mu ^2}\right)   \nn \\[0ex]
&\left. -4 \left(2 \beta  \omega _R+\tau \right) \left(2 \omega _R^2 (8 \alpha +\beta )+2 \tau  \omega _R- \nu^2 \right)
\log \left(\frac{ \omega _R^2 (8 \alpha +\beta )+\tau  \omega _R -\frac{\nu ^2}{2} }{4 \mu ^2}\right) \right]\,, \nn
\\
& \Delta M^{2}(15,1,0) =
\frac{\tau ^2 + \beta ^2 \omega _R^2}{4 \pi ^2} \\[0ex]
&+ \frac{1}{64 \pi ^2 \omega_R } \left[
 8 \omega _R \left(16 \alpha  \beta  \omega _R^2-\beta  \nu^2 +\tau ^2\right) \log \left(\frac{8 \alpha  \omega _R^2-\frac{\nu ^2}{2}}{\mu ^2}\right) \right. \nn \\[0ex]
& -4 \left(2 \beta  \omega _R^3 (8 \alpha -\beta )-16 \alpha  \tau  \omega _R^2+\omega _R \left(\tau ^2-\beta  \nu^2 \right)+ \nu^2  \tau \right)
   \log \left(\frac{ \omega _R^2 (8 \alpha +\beta )-\tau  \omega _R -\frac{\nu ^2}{2}}{\mu ^2}\right) \nn \\[0ex]
& \left. + 4 \left(2 \beta  \omega _R^3 (\beta -8 \alpha )-16 \alpha  \tau  \omega _R^2+\omega _R \left(\beta  \nu^2 -\tau ^2\right)+ \nu^2  \tau \right)
   \log \left(\frac{ \omega _R^2 (8 \alpha +\beta )+\tau  \omega _R -\frac{\nu ^2}{2}}{\mu ^2}\right) \right]\,. \nn
\end{align}
\end{small}
Also in these formulae we recognize  the (tree level)
mass eigenvalues of the $16_H$ states contributing to the one-loop
effective potential
(see \app{4516spectrumchi0}).

Notice that the singlets with respect to each vacuum,
namely $(1,0)$, $(1,1,1,0)$ and $(1,1,0)$, for the flipped $5' \, 1_{Z'}$,
$3_C 2_L 2_R 1_{B-L}$ and $4_C 2_L 1_R$ vacua respectively,
receive a tree level contribution from both $a_1$ as well as $a_2$
(see \app{4516spectrumchi0}).
The $a_1$ term leads the tree level mass and
radiative corrections can be neglected.

One may verify that in the limit of vanishing VEVs
the one-loop masses vanish identically on each of the three vacua, as it should be. This is a non trivial check of the calculation
of the scalar induced corrections.

\chapter{Flipped $SO(10)$ vacuum}
\label{app:flippedSO10vacuum}

\section{Flipped $SO(10)$ notation}
\label{flippedSO10notation}

We work in the basis of Ref. \cite{Rajpoot:1980xy},
where the adjoint is projected along the positive-chirality spinorial generators
\be
45 \equiv 45_{ij} \Sigma^{+}_{ij} \, ,
\ee
with $i,j = 1,..,10$. Here
\be
\left(
\begin{array}{c}
\Sigma^+ \\
\Sigma^-
\end{array}
\right) \equiv \frac{1}{2} \left( I_{32} \pm \Gamma_{\chi} \right) \Sigma \, ,
\ee
where $I_{32}$ is the $32$-dimensional identity matrix and $\Gamma_{\chi}$ is the 10-dimensional analogue of the Dirac $\gamma_{5}$ matrix defined as
\be
\Gamma_{\chi} \equiv - i
\Gamma_{1} \Gamma_{2} \Gamma_{3} \Gamma_{4} \Gamma_{5}
\Gamma_{6} \Gamma_{7} \Gamma_{8} \Gamma_{9} \Gamma_{10} \, .
\ee
The $\Gamma_i$ factors are given by the following tensor products \hyphenation{pro-ducts} of ordinary Pauli matrices $\sigma_{i}$ and the $2$-dimensional identity $I_{2}$:
\bea
&& \Gamma_{1} \equiv \sigma_1 \otimes \sigma_1 \otimes I_2 \otimes I_2 \otimes \sigma_2 \, , \nn \\
&& \Gamma_{2} \equiv \sigma_1 \otimes \sigma_2 \otimes I_2 \otimes \sigma_3 \otimes \sigma_2 \, , \nn \\
&& \Gamma_{3} \equiv \sigma_1 \otimes \sigma_1 \otimes I_2 \otimes \sigma_2 \otimes \sigma_3 \, , \nn \\
&& \Gamma_{4} \equiv \sigma_1 \otimes \sigma_2 \otimes I_2 \otimes \sigma_2 \otimes I_2 \, , \nn \\
&& \Gamma_{5} \equiv \sigma_1 \otimes \sigma_1 \otimes I_2 \otimes \sigma_2 \otimes \sigma_1 \, , \nn \\
&& \Gamma_{6} \equiv \sigma_1 \otimes \sigma_2 \otimes I_2 \otimes \sigma_1 \otimes \sigma_2 \, , \nn \\
&& \Gamma_{7} \equiv \sigma_1 \otimes \sigma_3 \otimes \sigma_1 \otimes I_2 \otimes I_2 \, , \nn \\
&& \Gamma_{8} \equiv \sigma_1 \otimes \sigma_3 \otimes \sigma_2 \otimes I_2 \otimes I_2 \, , \nn \\
&& \Gamma_{9} \equiv \sigma_1 \otimes \sigma_3 \otimes \sigma_3 \otimes I_2 \otimes I_2 \, , \nn \\
&& \Gamma_{10} \equiv \sigma_2 \otimes I_2 \otimes I_2 \otimes I_2 \otimes I_2 \, ,
\eea
{which} satisfy the Clifford algebra
\be
\left\{ \Gamma_i, \Gamma_j \right\} = 2 \delta_{ij} I_{32} \, .
\ee
The spinorial generators, $\Sigma_{ij}$, are then defined as
\be
\Sigma_{ij} \equiv \frac{i}{4} \left[ \Gamma_i , \Gamma_j \right] \, .
\ee

On the flipped $SO(10)$ vacuum the adjoint representation reads
\be
\label{45vevFSO10}
\vev{45} = \left(
\begin{array}{cc}
\vev{45}_{L} & \cdot \\
\cdot & \vev{45}_{R}
\end{array}
\right) \, ,
\ee
where
\be
\vev{45}_{L} =
\text{diag} \left( \lambda_{1}, \lambda_{2}, \lambda_{3}, \lambda_{4}, \lambda_{5}, \lambda_{6}, \lambda_{7}, \lambda_{8} \right) \, ,
\ee
and
\be
\vev{45}_{R} = 
\left(
\begin{array}{cccccccc}
 \lambda_{9} & \cdot & \cdot & \cdot & \omega ^+ & \cdot & \cdot & \cdot \\
 \cdot & \lambda_{10} & \cdot & \cdot & \cdot & \omega ^+ & \cdot & \cdot \\
 \cdot & \cdot & \lambda_{11} & \cdot & \cdot & \cdot & \omega ^+ & \cdot \\
 \cdot & \cdot & \cdot & \lambda_{12} & \cdot & \cdot & \cdot & \omega ^+ \\
 \omega ^- & \cdot & \cdot & \cdot & \lambda_{13} & \cdot & \cdot & \cdot \\
 \cdot & \omega ^- & \cdot & \cdot & \cdot & \lambda_{14} & \cdot & \cdot \\
 \cdot & \cdot & \omega ^- & \cdot & \cdot & \cdot & \lambda_{15} & \cdot \\
 \cdot & \cdot & \cdot & \omega ^- & \cdot & \cdot & \cdot & \lambda_{16}
\end{array}
\right) \, .
\ee
{In the convention defined in section \ref{vacuumFSO10} (cf.~also caption of Table \ref{tab:45decomp}),} the diagonal entries are given by
\bea
\lambda_{1} &=& \lambda_{2} = \lambda_{3} = \lambda_{5} = \lambda_{6} = \lambda_{7} = \frac{\omega_{B-L}}{2 \sqrt{2}} \, , \\
\lambda_{4} &=& \lambda_{8} = -\frac{3 \omega_{B-L}}{2 \sqrt{2}} \, , \nn \\
\lambda_{9} &=& \lambda_{10} = \lambda_{11} = -\frac{\omega_{B-L}}{2 \sqrt{2}}-\frac{\omega _R}{\sqrt{2}} \, ,
\quad \lambda_{12} = \frac{3 \omega_{B-L}}{2 \sqrt{2}}-\frac{\omega _R}{\sqrt{2}} \, , \nn \\
\lambda_{13} &=& \lambda_{14} = \lambda_{15} = -\frac{\omega_{B-L}}{2 \sqrt{2}}+\frac{\omega _R}{\sqrt{2}} \, ,
\quad \lambda_{16} = \frac{3 \omega_{B-L}}{2 \sqrt{2}}+\frac{\omega _R}{\sqrt{2}} \, . \nn
\eea
where $\omega_{B-L}$ and $\omega _R$ are real, while $\omega^+ = \omega^{-*}$.

Analogously, the spinor and the anti-spinor SM-preserving vacuum directions are given by
\bea
\label{16vevFSO10}
&& \vev{16}^T = (\cdot \cdot \cdot \cdot \cdot \cdot \cdot \cdot \cdot \cdot \cdot \ e \ \cdot \cdot \cdot \ -\nu) \, , \\
\label{16barvevFSO10}
&& \vev{\overline{16}}^T = (\cdot \cdot \cdot \ \overline{\nu} \ \cdot \cdot \cdot \ \overline{e} \ \cdot \cdot \cdot \cdot \cdot \cdot \cdot \cdot) \, ,
\eea
where the dots stand for zeros, and the non-vanishing VEVs are generally complex.

It is worth reminding that the shorthand notation $16 \, \overline{16}$ and $16 \, 45 \, \overline{16}$ in \eq{WHFSO10} stands for
$16^T \mathcal{C} \, \overline{16}$ and $16^T 45^T \mathcal{C} \, \overline{16}$, where $\mathcal{C}$ is the ``charge conjugation'' matrix obeying
$(\Sigma^{+})^T \mathcal{C} + \mathcal{C} \, \Sigma^{-} = 0$. In the current convention, $\mathcal{C}$ is given by
\be
\label{CmatrixFSO10}
\mathcal{C} =
\left(
\begin{array}{cccc}
 \cdot & \cdot & \cdot & -I_{4} \\
 \cdot & \cdot & I_{4} & \cdot \\
 \cdot & I_{4} & \cdot & \cdot \\
 -I_{4} & \cdot & \cdot & \cdot
\end{array}
\right) \, ,
\ee
where $I_4$ is the four-dimensional identity matrix.

\section{Supersymmetric vacuum manifold}
\label{DFtermFSO10}

In order for SUSY to survive the spontaneous GUT symmetry breakdown at $M_{U}$
the vacuum manifold must be $D$- and $F$-flat at the GUT scale.
The relevant superpotential $W_H$ given in \eq{WHFSO10}, with the SM-preserving vacuum parametrized by Eq. (\ref{45vevFSO10}) and \eqs{16vevFSO10}{16barvevFSO10}, yields
the following $F$-flatness equations:
\begin{align}
F_{\omega_R} &=
- 4 \mu  \omega _R + \frac{\tau _1}{\sqrt{2}} ( e_1 \overline{e}_1 -\nu _1\overline{\nu }_1 ) + \frac{\tau _2}{\sqrt{2}} ( e_2 \overline{e}_2-\nu _2 \overline{\nu }_2 ) = 0 \, , \nn \\
\tfrac{2}{3}
F_{\omega_{B-L}} \hspace{-0.3em}&=
 4\mu  \omega_{B-L} + \frac{\tau _1}{\sqrt{2}} ( e_1 \overline{e}_1 +\nu _1 \overline{\nu }_1 ) + \frac{\tau _2}{\sqrt{2}}  ( e_2 \overline{e}_2+\nu _2 \overline{\nu }_2 )
= 0 \, , \nn \\
F_{\omega^+} &=
4 \mu  \omega ^- - \tau _1 \nu _1 \overline{e}_1- \tau _2\nu _2 \overline{e}_2 = 0 \, , \nn \\[1ex]
F_{\omega^-} &=
4 \mu  \omega ^+ - \tau _1 e_1 \overline{\nu }_1 - \tau _2 e_2 \overline{\nu }_2 = 0 \, , \nn
\end{align}
\begin{align}
F_{e_1} &=
\tau _1 \left( - \omega ^-
   \overline{\nu }_1 - \frac{ \overline{e}_1 \omega _R}{\sqrt{2}}+\frac{3 \overline{e}_1 \omega_{B-L}}{2 \sqrt{2}} \right) +\rho _{11} \overline{e}_1+\rho _{12} \overline{e}_2 = 0 \, , \nn \\
F_{e_2} &=
\tau _2 \left( - \omega ^-
   \overline{\nu }_2 - \frac{ \overline{e}_2 \omega _R}{\sqrt{2}}+\frac{3 \overline{e}_2 \omega_{B-L}}{2 \sqrt{2}} \right) +\rho _{21} \overline{e}_1+\rho _{22} \overline{e}_2 = 0 \, , \nn \\
F_{\nu_1} &=
\tau _1 \left( - \omega ^+ \overline{e}_1+\frac{\overline{\nu }_1 \omega _R}{\sqrt{2}}+\frac{3 \overline{\nu }_1 \omega_{B-L}}{2 \sqrt{2}} \right) +\rho _{11} \overline{\nu }_1+\rho
   _{12} \overline{\nu }_2 = 0 \, , \nn \\
F_{\nu_2} &=
\tau _2 \left( - \omega ^+ \overline{e}_2+\frac{\overline{\nu }_2 \omega _R}{\sqrt{2}}+\frac{3 \overline{\nu }_2 \omega_{B-L}}{2 \sqrt{2}} \right) +\rho _{21} \overline{\nu }_1+\rho
   _{22} \overline{\nu }_2 = 0 \, , \nn
\end{align}
\begin{align}
F_{\overline{e}_1} &=
\tau _1  \left( - \omega ^+ \nu _1  - \frac{e_1 \omega _R}{\sqrt{2}}+\frac{3 e_1 \omega_{B-L}}{2 \sqrt{2}} \right) + \rho _{11} e_1 + \rho _{21} e_2  = 0 \, , \nn \\[1ex]
F_{\overline{e}_2} &=
\tau _2 \left( - \omega ^+ \nu _2  - \frac{e_2 \omega _R}{\sqrt{2}}+\frac{3 e_2 \omega_{B-L}}{2 \sqrt{2}} \right) + \rho _{12} e_1 + \rho _{22} e_2  = 0 \, , \nn \\[1ex]
F_{\overline{\nu}_1} &=
\tau _1 \left( - \omega ^- e_1 +\frac{\nu _1 \omega _R}{\sqrt{2}}+\frac{3 \nu _1 \omega_{B-L}}{2 \sqrt{2}} \right) + \rho _{11}\nu _1+ \rho _{21}\nu _2 = 0 \, , \nn \\[1ex]
F_{\overline{\nu}_2} &=
\tau _2 \left( - \omega ^- e_2  +\frac{\nu _2 \omega _R}{\sqrt{2}}+\frac{3 \nu _2 \omega_{B-L}}{2 \sqrt{2}} \right) +\rho _{12}\nu _1 + \rho _{22} \nu _2  = 0 \, . 
\label{FtermsFSO10}
\end{align}
One can use the first four equations above to replace $\omega_R$, $\omega_{B-L}$, $\omega^+$ and $\omega^-$ in the remaining eight (complex) relations
which can be rewritten in the form
\begin{multline}
16 \mu F^{\omega}_{e_1} = 16 \mu \left( \rho _{11} \overline{e}_1+\rho _{12} \overline{e}_2 \right) \\ 
- 5 \tau _1^2 \left( \nu _1 \overline{\nu }_1 + e_1 \overline{e}_1\right) \overline{e}_1
- \tau _1 \tau _2 \left( \nu _2 \overline{\nu }_2 \overline{e}_1 + \left( 4 \nu _2 \overline{\nu }_1 + 5 e_2 \overline{e}_1 \right) \overline{e}_2 \right)
= 0 \, , \nn
\end{multline}
\begin{multline}
16 \mu F^{\omega}_{\overline{e}_1} = 16 \mu \left( \rho _{11} e_1 + \rho _{21} e_2 \right) \\
 - 5 \tau _1^2 \left( \overline{\nu }_1 \nu _1 + \overline{e}_1 e_1 \right) e_1
- \tau _1 \tau _2 \left( \overline{\nu }_2 \nu _2 e_1 + \left( 4 \overline{\nu }_2 \nu _1 + 5 \overline{e}_2 e_1 \right) e_2  \right)
= 0 \, , \nn
\end{multline}
\begin{multline}
16 \mu F^{\omega}_{\nu_1} = 16 \mu \left( \rho _{11} \overline{\nu }_1+\rho _{12}
   \overline{\nu }_2 \right) \\
    - 5 \tau _1^2 \left( e_1 \overline{e}_1 + \nu _1 \overline{\nu }_1\right) \overline{\nu }_1
- \tau _1 \tau _2 \left( e_2 \overline{e}_2 \overline{\nu }_1 + \left( 4 e_2 \overline{e}_1 +5 \nu _2 \overline{\nu }_1 \right) \overline{\nu }_2 \right)
= 0 \, , \nn 
\end{multline}
\begin{multline}
16 \mu F^{\omega}_{\overline{\nu}_1} = 16 \mu \left( \rho _{11} \nu _1 + \rho _{21} \nu _2 \right) \\ 
- 5 \tau _1^2 \left( \overline{e}_1 e_1 +  \overline{\nu }_1 \nu _1 \right) \nu _1
- \tau _1 \tau _2 \left( \overline{e}_2 e_2 \nu _1 + \left( 4 \overline{e}_2 e_1 + 5 \overline{\nu }_2 \nu _1 \right) \nu _2  \right)
= 0 \, , \label{Fterms16}
\end{multline}
{where the other four equations are obtained from these by exchanging
$1 \leftrightarrow 2$.}

{There are two classes of $D$-flatness conditions corresponding, respectively, to the VEVs of the $U(1)_{X}$ and the $SO(10)$ generators.
For the $X$-charge one finds}
\begin{multline}
\label{DtermsX}
D_{X} = \vev{45}^{\dag} X \vev{45}
+ \vev{16_{1}}^{\dag} X \vev{16_{1}} + \vev{\overline{16}_{1}}^{\dag} X \vev{\overline{16}_{1}} 
+ \vev{16_{2}}^{\dag} X \vev{16_{2}} + \vev{\overline{16}_{2}}^{\dag} X \vev{\overline{16}_{2}} \\
= |e_1|^2 + |\nu_1|^2 - |\overline{e}_1|^2 - |\overline{\nu}_1|^2
+ |e_2|^2 + |\nu_2|^2 - |\overline{e}_2|^2 - |\overline{\nu}_2|^2 = 0 \, ,
\end{multline}
while for the $SO(10)$ generators one has
\be
\label{DtermsSO10}
D_{ij} \equiv \ D_{ij}^{45} + D_{ij}^{16 \oplus \overline{16}} = 0 \, ,
\ee
where
\be
D_{ij}^{45} = \Tr \vev{45}^{\dag} \left[ \Sigma^{+}_{ij} , \vev{45} \right] \, ,
\ee
and
\be
D_{ij}^{16 \oplus \overline{16}} = \vev{16_{1}}^{\dag} \Sigma^{+}_{ij} \vev{16_{1}} + \vev{\overline{16}_{1}}^{\dag} \Sigma^{-}_{ij} \vev{\overline{16}_{1}} 
+ \vev{16_{2}}^{\dag} \Sigma^{+}_{ij} \vev{16_{2}} + \vev{\overline{16}_{2}}^{\dag} \Sigma^{-}_{ij} \vev{\overline{16}_{2}} \, .
\ee
Given that
\be
\Tr \vev{45}^{\dag} \left[ \Sigma^{+}_{ij} , \vev{45} \right]
= \Tr \Sigma^{+}_{ij} \left[ \vev{45} , \vev{45}^\dag \right] \, ,
\ee
we obtain
\be
\left[ \vev{45} , \vev{45}^\dag \right] =
\left(
\begin{array}{cc}
\cdot & \cdot \\
\cdot & D_{R}
\end{array}
\right) \, ,
\ee
where
\be
D_{R} =
\left(
\begin{array}{cccccccc}
 A & \cdot & \cdot & \cdot & \sqrt{2} B^\ast & \cdot & \cdot & \cdot \\
 \cdot & A & \cdot & \cdot & \cdot & \sqrt{2} B^\ast & \cdot & \cdot \\
 \cdot & \cdot & A & \cdot & \cdot & \cdot & \sqrt{2} B^\ast & \cdot \\
 \cdot & \cdot & \cdot & A & \cdot & \cdot & \cdot & \sqrt{2} B^\ast \\
 \sqrt{2} B & \cdot & \cdot & \cdot & -A & \cdot & \cdot & \cdot \\
 \cdot & \sqrt{2} B & \cdot & \cdot & \cdot & -A & \cdot & \cdot \\
 \cdot & \cdot & \sqrt{2} B & \cdot & \cdot & \cdot & -A & \cdot \\
 \cdot & \cdot & \cdot & \sqrt{2} B & \cdot & \cdot & \cdot & -A
\end{array}
\right) \, ,
\ee
and
\bea
A &=& \left|\omega ^+\right|^2 - \left|\omega ^-\right|^2 \, , \nn \\
B &=& \left(\omega ^+\right)^{\ast} \omega _R - \left(\omega_R\right)^{\ast} \omega ^- \, .
\eea
{Since $\omega_R$ is real and $\omega^+ = (\omega^-)^\ast$, $D_{ij}^{45} =0$ as it should be. Notice that $F_{\omega^\pm}$-flatness
implies}
\be
\tau _1 e_1 \overline{\nu }_1+\tau _2 e_2 \overline{\nu }_2 =
\tau _1 (\nu _1 \overline{e}_1)^* + \tau _2 (\nu _2 \overline{e}_2)^*
\label{Dterms45}
\ee
where the reality of $\tau_{1,2}$ has been taken into account.

For the spinorial contribution in (\ref{DtermsSO10}) we find
\begin{multline}
D_{ij}^{16 \oplus \overline{16}} =
(\Sigma^{+}_{ij})_{12,12}  \left( |e_1|^2 + |e_2|^2 \right) + (\Sigma^{+}_{ij})_{16,16} \left( |\nu_1|^2 + |\nu_2|^2 \right) \\
+ (\Sigma^{-}_{ij})_{4,4} \left( |\overline{\nu}_1|^2 + |\overline{\nu}_2|^2 \right) + (\Sigma^{-}_{ij})_{8,8}  \left( |\overline{e}_1|^2 + |\overline{e}_2|^2 \right) \\
- (\Sigma^{+}_{ij})_{12,16}  \left( e_1^{\ast} \nu_1 + e_2^{\ast} \nu_2 \right) - (\Sigma^{+}_{ij})_{16,12} \left( \nu_1^{\ast} e_1 + \nu_2^{\ast} e_2 \right) \\
+ (\Sigma^{-}_{ij})_{4,8} \left( \overline{\nu}_1^{\ast} \overline{e}_1 + \overline{\nu}_2^{\ast} \overline{e}_2 \right)
+ (\Sigma^{-}_{ij})_{8,4} \left( \overline{e}_1^{\ast} \overline{\nu}_1 + \overline{e}_2^{\ast} \overline{\nu}_2 \right) \, .
\end{multline}
Given $\Sigma^{-} = - \mathcal{C}^{-1} (\Sigma^{+})^T \mathcal{C}$ and the explicit form of $\mathcal{C}$ in \eq{CmatrixFSO10},
one can verify readily that
\bea
&& (\Sigma^{-}_{ij})_{4,4} = - (\Sigma^{+}_{ij})_{16,16} \, , \nn \\
&& (\Sigma^{-}_{ij})_{8,8} = - (\Sigma^{+}_{ij})_{12,12} \, , \nn \\
&& (\Sigma^{-}_{ij})_{4,8} = + (\Sigma^{+}_{ij})_{12,16} \, .
\eea
Thus, $D_{ij}^{16 \oplus \overline{16}}$ can be simplified to
\begin{multline}
\label{DtermsSO10nosigmam}
(\Sigma^{+}_{ij})_{12,12}  ( |e_1|^2 + |e_2|^2 - |\overline{e}_1|^2 - |\overline{e}_2|^2 )
+ (\Sigma^{+}_{ij})_{16,16} ( |\nu_1|^2 + |\nu_2|^2 - |\overline{\nu}_1|^2 - |\overline{\nu}_2|^2 ) \\
- \left[ (\Sigma^{+}_{ij})_{12,16} ( e_1^{\ast} \nu_1 + e_2^{\ast} \nu_2 - \overline{\nu}_1^{\ast} \overline{e}_1 - \overline{\nu}_2^{\ast} \overline{e}_2 )
+ \text{c.c.} \right] = 0 \, ,
\end{multline}
or, with \eq{DtermsX} at hand, to
\begin{multline}
\label{DtermsSO10nosigmamred}
\left[ (\Sigma^{+}_{ij})_{16,16}  - (\Sigma^{+}_{ij})_{12,12} \right] ( |\nu_1|^2 + |\nu_2|^2 - |\overline{\nu}_1|^2 - |\overline{\nu}_2|^2 ) \\
- \left[ (\Sigma^{+}_{ij})_{12,16} ( e_1^{\ast} \nu_1 + e_2^{\ast} \nu_2 - \overline{\nu}_1^{\ast} \overline{e}_1 - \overline{\nu}_2^{\ast} \overline{e}_2 )
+ \text{c.c.} \right] = 0 \, .
\end{multline}

{Taking into account the basic features of the spinorial generators
$\Sigma^{+}_{ij}$ (e.g.~, the bracket [$(\Sigma^{+}_{ij})_{16,16}  - (\Sigma^{+}_{ij})_{12,12}]$ and $(\Sigma^{+}_{ij})_{12,16}$ can never act against each other because at least one of them always vanishes, or the fact that $(\Sigma^{+}_{ij})_{12,16}$ is complex) \eq{DtermsSO10nosigmamred} can be satisfied for all $ij$ if and only if}
\bea
\label{Dterms16s}
|e_1|^2 + |e_2|^2 - |\overline{e}_1|^2 - |\overline{e}_2|^2 & = & 0 \, , \nn \\
|\nu_1|^2 + |\nu_2|^2 - |\overline{\nu}_1|^2 - |\overline{\nu}_2|^2 & = & 0 \, , \nn \\
e_1^{\ast} \nu_1 + e_2^{\ast} \nu_2 - \overline{\nu}_1^{\ast} \overline{e}_1 - \overline{\nu}_2^{\ast} \overline{e}_2 & = & 0 \, ,
\eea
{Combining this with} \eq{Dterms45}, the required $D$- and $F$-flatness
{can be in general maintained only if} $e_{1,2}^{\ast} = \overline{e}_{1,2}$ and $\nu_{1,2}^{\ast} = \overline{\nu}_{1,2}$.
Hence, we can write
\bea
\label{complexDtermsFSO10}
&& e_{1,2} \equiv |e_{1,2}| e^{i \phi_{e_{1,2}}} \, , \qquad \overline{e}_{1,2} \equiv |e_{1,2}| e^{- i \phi_{e_{1,2}}} \, , \nn \\
&& \nu_{1,2} \equiv |\nu_{1,2}| e^{i \phi_{\nu_{1,2}}} \, , \qquad \overline{\nu}_{1,2} \equiv |\nu_{1,2}| e^{- i \phi_{\nu_{1,2}}} \, .
\eea
{With this at hand, one can further simplify the $F$-flatness conditions \eq{Fterms16}.
To this end, it is convenient to define the}
following linear combinations
\bea
\label{LmlincombFSO10}
&& L^{-}_{V} \equiv C_{1}^{V}\cos{\phi_{V}}  - C_{2}^{V}\sin{\phi_{V}}  \, , \\
\label{LplincombFSO10}
&& L^{+}_{V} \equiv C_{1}^{V}\sin{\phi_{V}} + C_{2}^{V}\cos{\phi_{V}}  \, ,
\eea
where
\bea
C_{1}^{V} \equiv \frac{1}{2i} \left( F^{\omega}_{\overline{V}} - F^{\omega}_{V} \right) \, , \qquad C_{2}^{V} \equiv \frac{1}{2} \left( F^{\omega}_{\overline{V}} + F^{\omega}_{V} \right) \, ,\nn
\eea
with $V$ running over the spinorial VEVs $e_1$, $e_2$, $\nu_1$ and $\nu_2$.
{For $\mu$, $\tau_1$ and $\tau_2$ real by definition, the requirement of $L^{\pm}_{V}=0$ for all $V$ is equivalent to}
\begin{multline}
4 \mu \Re L^{-}_{e_1} = \ 
\left|e_2\right| \left(\tau _1 \tau _2 \left|\nu _1\right| \left|\nu _2\right| \sin \left(\phi_{e_1}-\phi_{e_2}-\phi_{\nu _1}+\phi_{\nu _2}\right) \right. \\
\left. -2 \mu  \left(\left|\rho _{21}\right| \sin \left(\phi_{e_1}-\phi_{e_2}-\phi_{\rho _{21}}\right) 
+\left|\rho _{12}\right| \sin \left(\phi_{e_1}-\phi_{e_2}+\phi_{\rho _{12}}\right)\right)\right)
= 0 \, , \nn
\end{multline}
\begin{multline}
4 \mu \Re L^{-}_{\nu_1} = \ 
\left|\nu_2\right| \left(\tau _1 \tau _2 \left|e _1\right| \left|e _2\right| \sin \left(\phi_{\nu_1}-\phi_{\nu_2}-\phi_{e _1}+\phi_{e _2}\right) \right. \\
\left. -2 \mu  \left(\left|\rho _{21}\right| \sin \left(\phi_{\nu_1}-\phi_{\nu_2}-\phi_{\rho _{21}}\right) 
+\left|\rho _{12}\right| \sin \left(\phi_{\nu_1}-\phi_{\nu_2}+\phi_{\rho _{12}}\right)\right)\right)
= 0 \, ,
\end{multline}
\be
\!\!\!\!\!\!\!\!\!\!\!\!\! - 2 \Im L^{-}_{e_1} = \
\left|e_2\right| \left(\left|\rho _{21}\right| \cos \left(\phi_{e_1}-\phi_{e_2}-\phi_{\rho
   _{21}}\right) 
   -\left|\rho _{12}\right| \cos \left(\phi_{e_1}-\phi_{e_2}+\phi_{\rho
   _{12}}\right)\right)
= 0 \, , \nn
\ee
\be
- 2 \Im L^{-}_{\nu_1} = \ 
\left|\nu_2\right| \left(\left|\rho _{21}\right| \cos \left(\phi_{\nu_1}-\phi_{\nu_2}-\phi_{\rho
   _{21}}\right) 
   -\left|\rho _{12}\right| \cos \left(\phi_{\nu_1}-\phi_{\nu_2}+\phi_{\rho
   _{12}}\right)\right)
= 0 \, ,
\ee
and
\begin{multline}
- 16 \mu \Re L^{+}_{e_1} = \ 
- 16 \mu \left|e_1\right| \left|\rho _{11}\right| \cos \left(\phi_{\rho _{11}}\right) +5 \tau _1^2 \left( \left|e_1\right|{}^2 + \left|\nu_1\right|{}^2 \right) \left|e_1\right| \\
- 8 \mu \left|e_2\right| \left(\left|\rho _{21}\right| \cos \left(\phi_{e_1}-\phi_{e_2}-\phi_{\rho
   _{21}}\right)+\left|\rho _{12}\right| \cos \left(\phi_{e_1}-\phi_{e_2}+\phi_{\rho
   _{12}}\right)\right) \\
   +\tau _1 \tau _2 \left( \left(5  \left|e_2\right|{}^2 + \left|\nu _2\right|{}^2\right) \left|e_1\right|
   +4 \left|\nu _1\right| \left|\nu _2\right| \left|e_2\right| \cos \left(\phi_{e_1}-\phi_{e_2}-\phi_{\nu _1}+ \phi_{\nu _2}\right)\right)
    = 0 \, , \nn
\end{multline}
\begin{multline}
- 16 \mu \Re L^{+}_{\nu_1} = \ 
- 16 \mu \left|\nu_1\right| \left|\rho _{11}\right| \cos \left(\phi_{\rho _{11}}\right) +5 \tau _1^2 \left( \left|\nu_1\right|{}^2 + \left|e_1\right|{}^2 \right) \left|\nu_1\right| \\
- 8 \mu \left|\nu_2\right| \left(\left|\rho _{21}\right| \cos \left(\phi_{\nu_1}-\phi_{\nu_2}-\phi_{\rho
   _{21}}\right)+\left|\rho _{12}\right| \cos \left(\phi_{\nu_1}-\phi_{\nu_2}+\phi_{\rho
   _{12}}\right)\right) \\
   +\tau _1 \tau _2 \left( \left(5  \left|\nu_2\right|{}^2 + \left|e _2\right|{}^2\right) \left|\nu_1\right|
   +4 \left|e _1\right| \left|e _2\right| \left|\nu_2\right| \cos \left(\phi_{\nu_1}-\phi_{\nu_2}-\phi_{e _1}+ \phi_{e _2}\right)\right)
    = 0 \, ,
\end{multline}
\begin{multline}
2 \Im L^{+}_{e_1} = \ 
2 \left|e_1\right| \left|\rho _{11}\right| \sin \left(\phi_{\rho _{11}}\right) \\
+\left|e_2\right| \left(\left|\rho
   _{12}\right| \sin \left(\phi_{e_1}-\phi_{e_2}+\phi_{\rho _{12}}\right) 
   -\left|\rho _{21}\right| \sin
   \left(\phi_{e_1}-\phi_{e_2}-\phi_{\rho _{21}}\right)\right)
= 0 \, , \nn
\end{multline}
\begin{multline}
2 \Im L^{+}_{\nu_1} = \ 
2 \left|\nu_1\right| \left|\rho _{11}\right| \sin \left(\phi_{\rho _{11}}\right) \\ 
+\left|\nu_2\right| \left(\left|\rho
   _{12}\right| \sin \left(\phi_{\nu_1}-\phi_{\nu_2}+\phi_{\rho _{12}}\right) 
   -\left|\rho _{21}\right| \sin
   \left(\phi_{\nu_1}-\phi_{\nu_2}-\phi_{\rho _{21}}\right)\right)
= 0 \, ,
\end{multline}
{where, as before, the remaining eight real equations for V=$e_{2},\,\nu_{2}$ are obtained  by swapping $1 \leftrightarrow 2$.}

Focusing first on $L^{-}$, one finds that
$|e_1| L^{-}_{e_1} + |e_2| L^{-}_{e_2} = 0$ and $|\nu_1| L^{-}_{\nu_1} + |\nu_2| L^{-}_{\nu_2} = 0$. Thus, we can consider just $L^{-}_{e_1}$ and $L^{-}_{\nu_1}$ as independent equations. For instance, from $\Im L^{-}_{e_1}=0$ one readily gets
\bea
\label{rhoratioFSO10}
\frac{|\rho_{21}|}{|\rho_{12}|} = \frac{ \cos \left(\phi_{e _1}-\phi_{e _2}+\phi_{\rho _{12}}\right)}
{\cos \left(\phi_{e _1}-\phi_{e_2}-\phi_{\rho _{21}}\right)} \,.
\eea
{On top of that, the remaining ${\rm Re\,} L^{-}_{V}={\rm Im\,}L^{-}_{V}=0$ equations can be solved only for} $\phi_{\rho _{12}} = - \phi_{\rho _{21}}$, which, plugged into \eq{rhoratioFSO10} gives $|\rho_{12}| = |\rho_{21}|$. Thus, we end up with the following condition for the off-diagonal entries of the $\rho$ matrix:
\be
\label{rho21eqrho12stFSO10}
\rho_{21} = \rho_{12}^{\ast} \, .
\ee
{Inserting this into the $\Re L^{-}_{e_1}=0$ and $\Re L^{-}_{\nu_1}=0$ equations, they simplify to}
\bea
\label{rho12constFSO10}
-4 \mu |\rho_{12}| &=& \tau _1 \tau _2 \left|\nu_1\right| \left|\nu_2\right| \sin \left(\Phi_\nu - \Phi_e \right) \csc \Phi_e \, , \;\;\\
\label{rho12constbisFSO10}
4 \mu |\rho_{12}| &=& \tau _1 \tau _2 \left|e _1\right| \left|e _2\right|  \sin \left(\Phi_\nu - \Phi_e \right) \csc \Phi_\nu \, ,\;\;
\eea
{where we have denoted}
\be
\Phi_\nu \equiv \phi_{\nu_1}-\phi_{\nu_2}+\phi_{\rho _{12}} \, , \quad \Phi_e \equiv \phi_{e_1}-\phi_{e_2}+\phi_{\rho _{12}} \, .
\ee
These, taken together, yield
\bea
\label{nu1nu2e1e2constFSO10}
&& |e_1| |e_2| \sin \Phi_e  =  - |\nu_1| |\nu_2| \sin \Phi_\nu \, ,
\eea
and
\bea
\label{nu1nu2e1e2constBISFSO10}
&& |\nu_1| |\nu_2| + |e_1| |e_2| = \frac{4 \mu |\rho_{12}|}{\tau _1 \tau _2} \frac{\sin \Phi_\nu - \sin \Phi_e}
{\sin \left(\Phi_\nu - \Phi_e \right)} \, .
\eea
Notice that in the zero phases limit the constraint (\ref{nu1nu2e1e2constFSO10}) is trivially relaxed, while $\tfrac{\sin \Phi_\nu - \sin \Phi_e}
{\sin \left(\Phi_\nu - \Phi_e \right)} \rightarrow 1$.

{Returning to the $L^{+}_{V}=0$ equations, the constraint (\ref{rho21eqrho12stFSO10}) implies, e.g.}
\begin{align}
\Im L^{+}_{e_1} = \ &
\left|e _1\right| \left|\rho _{11}\right| \sin \left(\phi_{\rho _{11}}\right)
= 0 \, , \nn \\
\Im L^{+}_{e_2} = \ &
\left|e _2\right| \left|\rho _{22}\right| \sin \left(\phi_{\rho _{22}}\right)
= 0 \, , \nn \\
\Im L^{+}_{\nu_1} = \ &
\left|\nu_1\right| \left|\rho _{11}\right| \sin \left(\phi_{\rho _{11}}\right)
= 0 \, , \nn \\
\Im L^{+}_{\nu_2} = \ &
\left|\nu_2\right| \left|\rho _{22}\right| \sin \left(\phi_{\rho _{22}}\right)
= 0 \, .
\end{align}
{For generic VEVs, these relations} require $\phi_{\rho _{11}}$ and $\phi_{\rho _{22}}$ to vanish. In conclusion, a nontrivial vacuum requires $\rho$ (and hence $\tau$ for consistency) to be hermitian. This is a consequence of the fact that $D$-flatness for the flipped $SO(10)$ embedding implies
$\vev{16_i} = \vev{\overline{16}_i}^*$, cf.~\eq{complexDtermsFSO10}.
{Let us also note that} such a setting is preserved by supersymmetric wavefunction renormalization.

Taking $\rho=\rho^\dag$ in the remaining $\Re L^{+}_{V}=0$ equations and trading $|\rho_{12}|$ for $|\nu_1||\nu_2|$
{in $\Re L^+_{e_{1,2}}=0$} by means of \eq{rho12constFSO10} and for $|e_1||e_2|$ in $\Re L^+_{\nu_{1,2}}=0$ via \eq{rho12constbisFSO10}, one obtains
\begin{align}
- 16 \mu \Re L^{+}_{e_1} & = \ 
\left|e _1\right| \left[ -16 \mu \rho _{11} + 5 \tau _1^2 \left( \left|\nu_1\right|{}^2 + \left|e _1\right|{}^2 \right) \right. \nn \\
& \left. + \tau _1 \tau _2 \left( \left|\nu_2\right|{}^2  +5 \left|e _2\right|{}^2 \right) \right] + 4 \tau _1 \tau _2 \left|\nu_1\right| \left|\nu_2\right| \left|e _2\right| \sin \Phi_\nu \csc \Phi_e
= 0 \, , \nn \\ \nn \\
- 16 \mu \Re L^{+}_{e_2} & = \ 
\left|e _2\right| \left[ -16 \mu \rho _{22} + 5 \tau _2^2 \left( \left|\nu_2\right|{}^2 + \left|e _2\right|{}^2 \right) \right. \nn \\
& \left. + \tau _1 \tau _2 \left( \left|\nu_1\right|{}^2  +5 \left|e _1\right|{}^2 \right) \right] + 4 \tau _1 \tau _2 \left|\nu_1\right| \left|\nu_2\right| \left|e _1\right| \sin \Phi_\nu \csc \Phi_e
= 0 \, , \nn \\ \nn \\
- 16 \mu \Re L^{+}_{\nu_1} & = \ 
\left|\nu _1\right| \left[ -16 \mu \rho _{11} + 5 \tau _1^2 \left( \left|e_1\right|{}^2 + \left|\nu _1\right|{}^2 \right) \right. \nn \\
& \left. + \tau _1 \tau _2 \left( \left|e_2\right|{}^2  +5 \left|\nu _2\right|{}^2 \right) \right] + 4 \tau _1 \tau _2 \left|\nu_2\right| \left|e _1\right| \left|e _2\right| \csc \Phi_\nu \sin \Phi_e
= 0 \, , \nn \\ \nn \\
- 16 \mu \Re L^{+}_{\nu_2} & = \ 
\left|\nu _2\right| \left[ -16 \mu \rho _{22} + 5 \tau _2^2 \left( \left|e_2\right|{}^2 + \left|\nu _2\right|{}^2 \right) \right. \nn \\
& \left. + \tau _1 \tau _2 \left( \left|e_1\right|{}^2  +5 \left|\nu _1\right|{}^2 \right) \right] + 4 \tau _1 \tau _2 \left|\nu_1\right| \left|e _1\right| \left|e _2\right| \csc \Phi_\nu \sin \Phi_e
= 0 \, . 
\label{ReLprho12subFSO10}
\end{align}
Since only two out of these four are independent constraints, it is convenient to consider the following linear combinations
\be
C_{3} \equiv |\nu_1|^2 \left(|e_1| \Re L^{+}_{e_1} - |e_2| \Re L^{+}_{e_2}\right) 
- |e_1|^2 \left(|\nu_1| \Re L^{+}_{\nu_1} - |\nu_2| \Re L^{+}_{\nu_2}\right) \, ,
\ee
\be
C_{4} \equiv |\nu_2|^2 \left(|e_1| \Re L^{+}_{e_1} - |e_2| \Re L^{+}_{e_2}\right)
- |e_2|^2 \left(|\nu_1| \Re L^{+}_{\nu_1} - |\nu_2| \Re L^{+}_{\nu_2}\right) \, ,
\ee
which admit for a simple factorized form
\begin{align}
\label{CK3FSO10}
16 \mu C_{3} & = \ \left(\left|\nu_2\right|{}^2 \left|e _1\right|{}^2-\left|\nu_1\right|{}^2 \left|e _2\right|{}^2\right) \nn \\
& \times \left[5 \tau _2^2 \left(\left|\nu_2\right|{}^2+\left|e
   _2\right|{}^2\right)+\tau _1 \tau _2 \left(\left|\nu_1\right|{}^2+\left|e _1\right|{}^2\right)-16 \mu \rho _{22}\right] = 0 \, ,  \\
\label{CK4FSO10}
16 \mu C_{4} & = \ \left(\left|\nu_2\right|{}^2 \left|e _1\right|{}^2-\left|\nu_1\right|{}^2 \left|e _2\right|{}^2\right) \nn \\
& \times \left[5 \tau _1^2 \left(\left|\nu_1\right|{}^2+\left|e
   _1\right|{}^2\right)+\tau _1 \tau _2 \left(\left|\nu_2\right|{}^2+\left|e _2\right|{}^2\right)-16 \mu \rho _{11}\right] = 0 \, .
\end{align}
{These relations can be generically satisfied only if the square brackets are zero, providing}
\begin{align}
16 \mu \rho_{11} &= 
5 \tau _1^2 \left(\left|\nu_1\right|{}^2+\left|e _1\right|{}^2\right) + \tau _1 \tau _2 \left(\left|\nu_2\right|{}^2+\left|e _2\right|{}^2\right) \, ,\nn\\
16 \mu \rho_{22} &=  5 \tau _2^2 \left(\left|\nu_2\right|{}^2+\left|e _2\right|{}^2\right) + \tau _1 \tau _2 \left(\left|\nu_1\right|{}^2+\left|e _1\right|{}^2\right) \, .
\label{absrhoFSO10}
\end{align}
By introducing a pair of symbolic 2-dimensional vectors $\vec{r}_{1}=(|\nu_{1}|,|e_{1}|)$ and $\vec{r}_{2}=(|\nu_{2}|,|e_{2}|)$ one can write
\bea
&& r_{1}^{2} = |\nu_1|^2+|e_1|^2 \, , \nn\\
&& r_{2}^{2} = |\nu_2|^2+|e_2|^2 \, , \nn \\
\label{r1dotr2vevFSO10}
&& \vec{r}_{1}.\vec{r}_{2} = |\nu_1| |\nu_{2}|+|e_1||e_{2}| \, .
\eea
{which, in combination with eqs.
(\ref{nu1nu2e1e2constBISFSO10}) and (\ref{absrhoFSO10}) yields}
\bea
&& r_{1}^{2}  =  - \frac{2 \mu (\rho _{22} \tau _1 -5 \rho _{11} \tau _2) }{3 \tau _1^2 \tau _2} \, , \nn\\
&& r_{2}^{2}  =  - \frac{2 \mu (\rho _{11} \tau _2 -5 \rho _{22} \tau _1) }{3 \tau _1 \tau _2^2} \, ,  \nn\\
\label{r1dotr2supparFSO10}
&& \vec{r}_{1}.\vec{r}_{2} = \frac{4 \mu |\rho _{12}|}{\tau _1 \tau _2} \frac{\sin \Phi_\nu - \sin \Phi_e}{\sin \left(\Phi_\nu - \Phi_e \right)} \, .
\eea
{With this at hand, the vacuum manifold can be conveniently parametrized by means of} two angles $\alpha_1$ and $\alpha_2$
\bea
&& |\nu_1| = r_1 \sin \alpha_1 \, , \qquad |e_1| = r_1 \cos \alpha_1 \, , \nn \\
&& |\nu_2| = r_2 \sin \alpha_2 \, , \qquad |e_2| = r_2 \cos \alpha_2 \, .
\label{alpha12paramFSO10}
\eea
{which are} fixed in terms of the superpotential parameters.
By defining $\alpha^{\pm} \equiv \alpha_1 \pm \alpha_2$, \eqs{r1dotr2vevFSO10}{alpha12paramFSO10} give
\be
\label{cosalphamFSO10}
\cos{\alpha^{-}} = \frac{\vec{r}_{1}.\vec{r}_{2}}{r_{1} r_{2}} =
\xi \ \frac{\sin \Phi_\nu - \sin \Phi_e}{\sin \left(\Phi_\nu - \Phi_e \right)} \, ,
\ee
where
\be
\xi = \frac{6 |\rho _{12}|}{\sqrt{-\frac{5 \rho _{11}^2 \tau _2}{\tau _1}-\frac{5 \rho _{22}^2 \tau _1}{\tau _2}+26 \rho _{22} \rho _{11}}} \, .
\ee
Analogously, \eq{nu1nu2e1e2constFSO10} can be rewritten as
\be
\cos \alpha_1 \cos \alpha_2 \sin \Phi_e  =  - \sin \alpha_1 \sin \alpha_2 \sin \Phi_\nu \, ,
\ee
which gives
\be
\frac{\sin \Phi_e}{\sin \Phi_\nu}  =
\frac{\cos{\alpha^+} - \cos{\alpha^-}}{\cos{\alpha^-} + \cos{\alpha^+}} \, ,
\ee
and thus, using \eq{cosalphamFSO10}, we obtain
\be
\label{cosalphapFSO10}
\cos \alpha^{+} = \xi \ \frac{\sin \Phi_\nu + \sin \Phi_e}{\sin \left(\Phi_\nu - \Phi_e \right)} \, .
\ee
Notice also that in the real case (i.e., $\Phi_\nu = \Phi_e = 0$) $\alpha^{+}$ is undetermined, while $\cos \alpha^{-} = \xi$.

This justifies the shape of the vacuum manifold given in \eq{vacmanifoldFSO10} of \sect{vacuumFSO10}.

\section{Gauge boson spectrum}
\label{gaugespectrum}

In order to determine the residual symmetry corresponding to a specific vacuum configuration we compute explicitly the gauge spectrum.
Given the $SO(10)\otimes U(1)_X$ covariant derivatives for the
scalar components of the Higgs chiral superfields
\bea
\label{covder16}
&& D_{\mu} 16 = \partial_{\mu} 16 - i g (A_{\mu})_{(ij)} \Sigma^+_{(ij)} 16 - i g_X X_{\mu} 16 \, , \nn \\
\label{covder16bar}
&& D_{\mu} \overline{16} = \partial_{\mu} \overline{16} - i g (A_{\mu})_{(ij)} \Sigma^-_{(ij)} \overline{16} + i g_X X_{\mu} \overline{16} \, , \nn \\
\label{covder45}
&& D_{\mu} 45 = \partial_{\mu} 45 - i g (A_{\mu})_{(ij)} \left[ \Sigma^+_{(ij)},  45 \right]  \, ,
\eea
where the indices in brackets $(ij)$ stand for ordered pairs,
and the properly normalized kinetic terms
\be
D_{\mu} 16^{\dag} D_{\mu} 16 \, , \qquad
D_{\mu} \overline{16}^{\dag} D_{\mu} \overline{16} \, , \qquad
\tfrac{1}{4} \Tr D_{\mu} 45^{\dag} D_{\mu} 45 \, , \\
\ee
one can write the 46-dimensional gauge boson mass matrix governing the mass bilinear of the form
\be
\label{masstermgauge}
\frac{1}{2} \left( (A_{\mu})_{(ij)}\ ,  X_{\mu} \right) \mathcal{M}^2 (A,X) \left( (A^{\mu})_{(kl)}\ ,  X^{\mu} \right)^{T} \,
\ee
as
\bea
&& \mathcal{M}^2(A,X) =
\left(
\begin{array}{ll}
 \mathcal{M}_{(ij)(kl)}^2 & \mathcal{M}_{(ij)X}^2  \\
 \mathcal{M}_{X(kl)}^2 & \mathcal{M}_{XX}^2
\end{array}
\right) \, .
\eea
The relevant matrix elements are given by
\begin{multline}
\mathcal{M}_{(ij)(kl)}^2 =
g^2\left(\vev{16}^{\dag} \{\Sigma^+_{(ij)},\Sigma^+_{(kl)}\} \vev{16} +\vev{\overline{16}}^{\dag} \{\Sigma^-_{(ij)},\Sigma^-_{(kl)}\} \vev{\overline{16}} \right. \\
\left. + \frac{1}{2} \Tr  \left[ \Sigma^+_{(ij)},  \vev{45} \right]^\dag \left[ \Sigma^+_{(kl)}, \vev{45} \right]  \right) \, , \nn
\end{multline}
\begin{align}
& \mathcal{M}_{(ij) X}^2 =
2 g g_X \left(\vev{16}^{\dag} \Sigma^+_{(ij)} \vev{16} - \vev{\overline{16}}^{\dag} \Sigma^-_{(ij)} \vev{\overline{16}}\right) \, , \nn \\
& \mathcal{M}_{X (kl)}^2 =
2 g g_X \left(\vev{16}^{\dag} \Sigma^+_{(kl)} \vev{16} - \vev{\overline{16}}^{\dag} \Sigma^-_{(kl)} \vev{\overline{16}}\right) \, , \nn \\
& \mathcal{M}_{XX}^2 =
2 g_X^{2} \left(\vev{16}^{\dag} \vev{16} + \vev{\overline{16}}^{\dag} \vev{\overline{16}}\right) \, .
\end{align}

\subsection{Spinorial contribution}
\label{gaugespectrum16}

Considering first the contribution of the reducible representation $\vev{16_1 \oplus 16_2 \oplus \overline{16}_1 \oplus \overline{16}_2}$ to the gauge boson mass matrix, we find
\begin{align}
\label{Mgauge130_FSO10}
& \mathcal{M}_{{16}}^2 (1,3,0)_{1_{45}} = 0 \, , \\
\label{Mgauge810_FSO10}
& \mathcal{M}_{{16}}^2 (8,1,0)_{15_{45}} = 0 \, ,
\end{align}
\begin{multline}
\label{Mgauge13m13_FSO10}
\mathcal{M}_{{16}}^2 (3,1,-\tfrac{1}{3})_{15_{45}} = \\
g^2 \left( |e_1|^2 + |\nu_1|^2 + |e_2|^2 + |\nu_2|^2
+ |\overline{e}_1|^2 + |\overline{\nu}_1|^2 + |\overline{e}_2|^2 + |\overline{\nu}_2|^2  \right) \, ,
\end{multline}

{In} the $(6^-_{45},6^+_{45})$ basis (see Table \ref{tab:45decomp} for the labelling of the states) we obtain
\begin{multline}
\label{Mgauge32p16_FSO10}
\mathcal{M}_{{16}}^2 (3,2,+\tfrac{1}{6}) = \\
\left(
\begin{array}{cc}
 g^2 \left( |\nu_1|^2 + |\nu_2|^2 + |\overline{\nu}_1|^2 + |\overline{\nu}_2|^2 \right) &
 -i g^2 \left( e_1^\ast \nu_1 + e_2^\ast \nu_2 + \overline{\nu}_1^\ast \overline{e}_1 + \overline{\nu}_2^\ast \overline{e}_2 \right) \\
 i g^2 \left( e_1 \nu_1^\ast + e_2 \nu_2^\ast + \overline{\nu}_1 \overline{e}_1^\ast + \overline{\nu}_2 \overline{e}_2^\ast \right) &
 g^2 \left( |e_1|^2 + |e_2|^2 + |\overline{e}_1|^2 + |\overline{e}_2|^2 \right)
\end{array}
\right)
\, ,
\end{multline}
The five dimensional SM singlet mass matrix in the $\left( 15_{45}, 1^-_{45}, 1^0_{45}, 1^+_{45}, 1_{1}  \right)$ basis reads
{
\begin{multline}
\label{Mgauge110_two16}
\mathcal{M}_{{16}}^2 (1,1,0) = \\
\left(
\begin{array}{ccccc}
 \frac{3}{2} g^2 S_{1} & i \sqrt{3} g^2 S_{3} & -\sqrt{\frac{3}{2}} g^2 S_{2} & -i \sqrt{3} g^2 S_{3}^{\ast}  & -\sqrt{3} g g_X
   S_{1} \\
 -i \sqrt{3} g^2 S_{3}^{\ast}  & g^2 S_{1} & 0 & 0 & 2 i g g_X S_{3} \\
 -\sqrt{\frac{3}{2}} g^2 S_{2} & 0 & g^2 S_{1}  & 0 & \sqrt{2} g g_X S_{2} \\
 i \sqrt{3} g^2 S_{3} & 0 & 0 & g^2 S_{1} & -2 i g g_X S_{3}^{\ast}  \\
 -\sqrt{3} g g_X S_{1} & - 2 i g g_X S_{3}^{\ast} & \sqrt{2} g g_X S_{2} &2 i g g_X S_{3} &
2 g_X^2 S_{1}
\end{array}
\right)
\end{multline}
where $S_{1}\equiv |e_1|^2+|e_2|^2+|\nu _1|^2+|\nu _2|^2+|\overline{e}_1|^2+|\overline{e}_2|^2+|\overline{\nu }_1|^2+|\overline{\nu }_2|^2$, $S_{2}\equiv |e_1|^2+|e_2|^2-|\nu _1|^2-|\nu _2|^2+|\overline{e}_1|^2+|\overline{e}_2|^2-|\overline{\nu }_1|^2-|\overline{\nu }_2|^2$ and $S_{3}\equiv e_1 \nu _1^\ast+e_2 \nu
   _2^\ast+\overline{e}_1^\ast \overline{\nu }_1+\overline{e}_2^\ast \overline{\nu }_2$.}

For generic VEVs $\text{Rank}\ \mathcal{M}_{{16}}^2 (1,1,0) = 4$, and we recover 12 massless gauge bosons
with the quantum numbers of the SM algebra.

We verified that this result is maintained when implementing the
constraints of the flipped vacuum manifold in \eq{vacmanifoldFSO10}.
Since it is, by construction, the smallest algebra that can be preserved by the whole vacuum manifold, it must be maintained when adding the $\vev{45_H}$ contribution. We can therefore claim that the invariant algebra on the generic vacuum is the SM.
On the other hand, the $45_H$ plays already an active role in this result since it allows for a misalignment of the VEV directions in the two $16_H\oplus \overline{16}_H$ spinors
such that the spinor vacuum preserves SM and not $SU(5)\otimes U(1)$. More details shall be given in the next section.

\subsection{Adjoint contribution}
\label{gaugespectrum45}

Considering the contribution of $\vev{45_H}$ to the
gauge spectrum, we find
\bea
\label{Mgauge130_45}
&& \mathcal{M}_{{45}}^2 (1,3,0)_{1_{45}} = 0 \, , \\
\label{Mgauge810_45}
&& \mathcal{M}_{{45}}^2 (8,1,0)_{15_{45}} = 0 \, , \\
\label{Mgauge13m13_45}
&& \mathcal{M}_{{45}}^2 (3,1,-\tfrac{1}{3})_{15_{45}} =
4 g^2 \omega_{B-L}^2 \, .
\eea
Analogously, in the $(6^-_{45},6^+_{45})$ basis, we have
\begin{multline}
\label{Mgauge32p16_45}
\mathcal{M}_{{45}}^2 (3,2,+\tfrac{1}{6}) = \\
\left(
\begin{array}{cc}
  g^2 \left(\left(\omega _R+\omega_{B-L}\right){}^2+2 \omega ^- \omega ^+\right) &  i 2 \sqrt{2} g^2 \omega_{B-L} \omega ^- \\
 - i 2 \sqrt{2} g^2 \omega_{B-L} \omega ^+ &  g^2 \left(\left(\omega _R-\omega_{B-L}\right){}^2+2 \omega ^- \omega ^+\right)
\end{array}
\right)
\, .
\end{multline}
The SM singlet mass matrix in the $\left( 15_{45}, 1^-_{45}, 1^0_{45}, 1^+_{45}, 1_{1}  \right)$ basis reads
\be
\label{Mgauge110_45}
\mathcal{M}_{{45}}^2 (1,1,0) =
\left(
\begin{array}{ccccc}
 0 & 0 & 0 & 0 & 0 \\
 0 & 4 g^2 \left(\omega _R^2+\omega ^- \omega ^+\right) & - i 4 g^2 \omega _R \omega ^- & 4 g^2 \left(\omega ^-\right)^2 & 0 \\
 0 & i 4 g^2 \omega _R \omega ^+ & 8 g^2 \omega ^- \omega ^+ & - i 4 g^2 \omega _R \omega ^- & 0 \\
 0 & 4 g^2 \left(\omega ^+\right)^2 & i 4 g^2 \omega _R \omega ^+ & 4 g^2 \left(\omega _R^2+\omega ^- \omega ^+\right) & 0 \\
 0 & 0 & 0 & 0 & 0
\end{array}
\right) \, .
\ee
For generic VEVs we find $\text{Rank}\ \mathcal{M}_{{45}}^2 (1,1,0) = 2$
leading globally to the 14 massless gauge bosons
of the $SU(3)_C \otimes SU(2)_L \otimes U(1)^3$ algebra.


\subsection{Vacuum little group}
\label{FormalproofFSO10}

{With the results of sections \ref{gaugespectrum16} and \ref{gaugespectrum45} at hand the residual gauge symmetry can be readily identified from the properties of the} complete gauge boson mass matrix. For the sake of simplicity here we shall present the results in the real VEV approximation.

Trading the VEVs for the superpotential parameters,
one can immediately identify the strong and weak gauge bosons of the SM that, as expected, remain massless:
\bea
&& \mathcal{M}^2 (8,1,0)_{15_{45}} = 0 \, , \nn\\
&& \mathcal{M}^2 (1,3,0)_{1_{45}} = 0 \, .
\eea
{Similarly, it is straightforward to obtain}
\begin{multline}
\mathcal{M}^2 (3,1,-\tfrac{1}{3})_{15_{45}} = \\
\frac{4 g^2}{9 \tau _1^2 \tau _2^2}
\left(3 \mu  \left(\rho _{22} \tau _1 \left(5 \tau _1-\tau _2\right)  +\rho _{11} \tau _2 \left(5 \tau _2-\tau
   _1\right)\right) +2 \left(\rho _{22} \tau _1+\rho _{11} \tau _2\right){}^2\right) \, .
\end{multline}
On the other hand, the complete matrices $\mathcal{M}^2 (3,2,+\tfrac{1}{6})$ and $\mathcal{M}^2 (1,1,0)$ turn out to be
quite involved once the vacuum constraints are imposed, and we do not show them here explicitly. Nevertheless, it is sufficient to consider
\be
\Tr \mathcal{M}^2 (3,2,+\tfrac{1}{6}) =
\frac{g^2}{8 \mu ^2} \left[ 16 \mu^2 \left( r_1^2 + r_2^2 \right) + \tau _1^2 r_1^4 + \tau _2^2 r_2^4 
+ \tau _1 \tau _2 r_1^2 r_2^2
\left( 1 + \cos 2 \alpha^-\right)\right]
\ee
and
\begin{multline}
\label{det3216r2alpham}
\det \mathcal{M}^2 (3,2,+\tfrac{1}{6}) =
\frac{g^4 r_1^2 r_2^2}{128 \mu ^4}
\left[512 \mu ^4 + 32 \mu ^2 \left( \tau _1^2 r_1^2 + \tau _2^2 r_2^2 \right)
\right. \\ \left.
+ \tau _1^2 \tau _2^2 r_1^2 r_2^2 \left( 1 - \cos 2 \alpha^-\right) \right] \sin ^2 \alpha^- \,
\end{multline}
to see that for a generic non-zero value of $\sin\alpha^-$ one gets Rank $\mathcal{M}^2 (3,2,+\tfrac{1}{6})=2$.
On the other hand, when $\alpha^- = 0$ (i.e., $\vev{16_{1}}\propto\vev{16_{2}}$) or $r_2 = 0$ (i.e., $\vev{16_{2}}=0$),
$\text{Rank}\ \mathcal{M}^2 (3,2,+\tfrac{1}{6}) = 1$ and one is left with an additional massless $(3,2,+\tfrac{1}{6})\oplus (\overline{3},2,-\tfrac{1}{6})$ gauge boson, corresponding to an enhanced residual symmetry.

In the case of the $5$-dimensional matrix $\mathcal{M}^2 (1,1,0)$ it is sufficient to notice that for a generic non-zero $\sin\alpha^{-}$
\be
\label{rank110r2alpham}
\text{Rank}\ \mathcal{M}^2 (1,1,0) = 4 \,,
\ee
on the vacuum manifold, which leaves a massless $U(1)_Y$ gauge boson, thus completing the SM algebra.
As before, for $\alpha^- = 0$ or for $r_2 = 0$,
we find $\text{Rank}\ \mathcal{M}^2 (1,1,0) = 3$.
Taking into account the massless states in the $(3,2,+\tfrac{1}{6})\oplus (\overline{3},2,-\tfrac{1}{6})$ sector, we recover, as expected, the flipped $SU(5) \otimes U(1)$ algebra.

\chapter{$E_6$ vacuum}
\label{app:E6vacuum}

\section{The $SU(3)^3$ formalism}
\label{app:SU33formalism}

Following closely the notation of Refs.~\cite{Buccella:1987kc,Miele}, we decompose the adjoint and fundamental representations of $E_6$ under its $SU(3)_C \otimes SU(3)_L \otimes SU(3)_R$ maximal subalgebra as
\begin{align}
\label{adjointE6}
78 &\equiv (8,1,1) \oplus (1,8,1) \oplus (1,1,8) \oplus (\overline{3},3,3) \oplus (3,\overline{3},\overline{3}) \nn \\
& \subset T^{\alpha}_{\beta} \oplus T^{i}_{j} \oplus T^{i'}_{j'} \oplus Q^{\alpha}_{i j'} \oplus Q^{i j'}_{\alpha} \, , \\
\label{fundE6}
27 &\equiv (3,3,1) \oplus (1,\overline{3},3) \oplus (\overline{3},1,\overline{3}) 
\equiv v_{\alpha i} \oplus v^{i}_{j'} \oplus v^{\alpha j'} \, , \\
\label{antifundE6}
\overline{27} &\equiv (\overline{3},\overline{3},1) \oplus (1,3,\overline{3}) \oplus (3,1,3)
\equiv u^{\alpha i} \oplus u^{j'}_{i} \oplus u_{\alpha j'} \, ,
\end{align}
where the greek, latin and primed-latin indices, corresponding to $SU(3)_C$, $SU(3)_L$ and $SU(3)_R$, respectively, run from $1$ to $3$.
As far as the $SU(3)$ algebras in \eq{adjointE6} are concerned,
the generators follow the standard Gell-Mann convention
\begin{align}
\label{gellmannbasis}
& T^{(1)} = \tfrac{1}{2} (T^{1}_{2} + T^{2}_{1}) \, , \qquad
T^{(2)} = \tfrac{i}{2} (T^{1}_{2} - T^{2}_{1})  \, , \nn \\
& T^{(3)} = \tfrac{1}{2} (T^{1}_{1} - T^{2}_{2})  \, , \qquad
T^{(4)} = \tfrac{1}{2} (T^{1}_{3} + T^{3}_{1})  \, ,  \\
& T^{(5)} = \tfrac{i}{2} (T^{1}_{3} - T^{3}_{1}) \, , \qquad
T^{(6)} = \tfrac{1}{2} (T^{2}_{3} + T^{3}_{2}) \, , \nn \\
& T^{(7)} = \tfrac{i}{2} (T^{2}_{3} - T^{3}_{2}) \, , \qquad
T^{(8)} = \tfrac{1}{2\sqrt{3}} (T^{1}_{1} + T^{2}_{2} - 2 T^{3}_{3}) \, ,\nn
\end{align}
with  $(T^{a}_{b})^{k}_{l} = \delta^{k}_{b} \delta^{a}_{l}$, so they are all normalized so that
$\Tr\ T^{(a)} T^{(b)} = \frac{1}{2} \delta^{ab}$.

Taking into account \eqs{adjointE6}{gellmannbasis},
the $E_6$ algebra can be written as
\bea
&& [ T^{\alpha}_{\beta}, T^{\gamma}_{\eta} ] = \delta^{\alpha}_{\eta} T^{\gamma}_{\beta} - \delta^{\gamma}_{\beta} T^{\alpha}_{\eta} \nn \\
&& [ T^{i}_{j}, T^{k}_{l} ] = \delta^{i}_{l} T^{k}_{j} - \delta^{k}_{j} T^{i}_{l} \nn \\
&& [ T^{i'}_{j'}, T^{k'}_{l'} ] = \delta^{i'}_{l'} T^{k'}_{j'} - \delta^{k'}_{j'} T^{i'}_{l'} \nn \\
&& [ T^{\alpha}_{\beta}, T^{i}_{j} ] = [ T^{\alpha}_{\beta}, T^{i'}_{j'} ] = [ T^{i}_{j}, T^{i'}_{j'} ] = 0 \, ,
\label{E6action78_1}
\eea
\bea
&& [ Q^{\gamma}_{i j'}, T^{\alpha}_{\beta} ] = \delta^{\gamma}_{\beta} Q^{\alpha}_{i j'} \nn \\
&& [ Q^{i j'}_{\gamma}, T^{\alpha}_{\beta} ] = - \delta^{\alpha}_{\gamma} Q^{i j'}_{\beta} \nn \\
&& [ Q^{\gamma}_{i j'}, T^{k}_{l} ] = - \delta^{k}_{i} Q^{\gamma}_{l j'} \nn \\
&& [ Q^{i j'}_{\gamma}, T^{k}_{l} ] = \delta^{i}_{l} Q^{k j'}_{\gamma} \nn \\
&& [ Q^{\gamma}_{i j'}, T^{k'}_{l'} ] = - \delta^{k'}_{j'} Q^{\gamma}_{i l'} \nn \\
&& [ Q^{i j'}_{\gamma}, T^{k'}_{l'} ] = \delta^{j'}_{l'} Q^{i k'}_{\gamma} \, ,
\label{E6action78_2}
\eea
\bea
&& [ Q^{\alpha}_{i j'}, Q^{k l'}_{\beta} ] = - \delta^{\alpha}_{\beta} \delta^{k}_{i} T^{l'}_{j'} - \delta^{\alpha}_{\beta} \delta^{l'}_{j'} T^{k}_{i} + \delta^{k}_{i} \delta^{l'}_{j'} T^{\alpha}_{\beta} \nn \\
&& [ Q^{\alpha}_{i j'}, Q^{\beta}_{k l'} ] = \epsilon^{\alpha \beta \gamma} \epsilon_{i k p} \epsilon_{j' l' q'} Q^{p q'}_{\gamma} \nn \\
&& [ Q^{i j'}_{\alpha}, Q^{k l'}_{\beta} ] = - \epsilon_{\alpha \beta \gamma} \epsilon^{i k p} \epsilon^{j' l' q'} Q^{\gamma}_{p q'} \, ,
\label{E6action78_3}
\eea
The action of the algebra on the fundamental $27$ representation reads
\bea
&& T^{\beta}_{\gamma} v_{\alpha i} = \delta^{\beta}_{\alpha} v_{\gamma i} \nn \\
&& T^{k}_{l} v_{\alpha i} = \delta^{k}_{i} v_{\alpha l} \nn \\
&& T^{k'}_{l'} v_{\alpha i} = 0 \nn \\
&& Q^{\beta}_{p q'} v_{\alpha i} = \delta^{\beta}_{\alpha} \epsilon_{p i k} v^{k}_{q'} \nn \\
&& Q^{p q'}_{\beta} v_{\alpha i} = \delta^{p}_{i} \epsilon_{\beta \alpha \gamma} v^{\gamma q'} \, ,
\label{E6action27a}
\eea
\bea
&& T^{\beta}_{\gamma} v^{i}_{j'} = 0 \nn \\
&& T^{k}_{l} v^{i}_{j'} = - \delta^{i}_{l} v^{k}_{j'} \nn \\
&& T^{k'}_{l'} v^{i}_{j'} = \delta^{k'}_{j'} v^{i}_{l'} \nn \\
&& Q^{\beta}_{p q'} v^{i}_{j'} = - \delta^{i}_{p} \epsilon_{q' j' k'} v^{\beta k'} \nn \\
&& Q^{p q'}_{\beta} v^{i}_{j'} = \delta^{q'}_{j'} \epsilon^{p i k} v_{\beta k} \, ,
\label{E6action27b}
\eea
\bea
&& T^{\beta}_{\gamma} v^{\alpha j'} = - \delta^{\alpha}_{\gamma} v^{\beta j'} \nn \\
&& T^{k}_{l} v^{\alpha j'} = 0 \nn \\
&& T^{k'}_{l'} v^{\alpha j'} = - \delta^{j'}_{l'} v^{\alpha k'} \nn \\
&& Q^{\beta}_{p q'} v^{\alpha j'} = - \delta^{j'}_{q'} \epsilon^{\beta \alpha \gamma} v_{\gamma p} \nn \\
&& Q^{p q'}_{\beta} v^{\alpha j'} = - \delta^{\alpha}_{\beta} \epsilon^{q' j' k'} v^{p}_{k'} \, ,
\label{E6action27c}
\eea
and accordingly on $\overline{27}$
\bea
&& T^{\beta}_{\gamma} u^{\alpha i} = - \delta^{\alpha}_{\gamma} u^{\beta i} \nn \\
&& T^{k}_{l} u^{\alpha i} = - \delta^{i}_{l} u^{\alpha k} \nn \\
&& T^{k'}_{l'} u^{\alpha i} = 0 \nn \\
&& Q^{\beta}_{p q'} u^{\alpha i} = - \delta^{i}_{p} \epsilon^{\beta \alpha \gamma} u_{\gamma q'} \nn \\
&& Q^{p q'}_{\beta} u^{\alpha i} = - \delta^{\alpha}_{\beta} \epsilon^{p i k} u^{q'}_{k} \, ,
\label{E6action27bara}
\eea
\bea
&& T^{\beta}_{\gamma} u^{j'}_{i} = 0 \nn \\
&& T^{k}_{l} u^{j'}_{i} = \delta^{k}_{i} u^{j'}_{l} \nn \\
&& T^{k'}_{l'} u^{j'}_{i} = - \delta^{j'}_{l'} u^{k'}_{i} \nn \\
&& Q^{\beta}_{p q'} u^{j'}_{i} = - \delta^{j'}_{q'} \epsilon_{p i k} u^{\beta k} \nn \\
&& Q^{p q'}_{\beta} u^{j'}_{i} = \delta^{p}_{i} \epsilon^{q' j' k'} u_{\beta k'} \, ,
\label{E6action27barb}
\eea
\bea
&& T^{\beta}_{\gamma} u_{\alpha j'} = \delta^{\beta}_{\alpha} u_{\gamma j'} \nn \\
&& T^{k}_{l} u_{\alpha j'} = 0 \nn \\
&& T^{k'}_{l'} u_{\alpha j'} = \delta^{k'}_{j'} u_{\alpha l'} \nn \\
&& Q^{\beta}_{p q'} u_{\alpha j'} = \delta^{\beta}_{\alpha} \epsilon_{q' j' k'} u^{k'}_{p} \nn \\
&& Q^{p q'}_{\beta} u_{\alpha j'} = \delta^{q'}_{j'} \epsilon_{\beta \alpha \gamma} u^{\gamma p} \, .
\label{E6action27barc}
\eea
Given the SM hypercharge definition
\be
\label{SMhyperchargeE6}
Y = \frac{1}{\sqrt{3}} T^{(8)}_L + T^{(3)}_R + \frac{1}{\sqrt{3}} T^{(8)}_R \, ,
\ee
the SM-preserving vacuum direction corresponds to \cite{Buccella:1987kc}
\be
\label{78vacE6}
\vev{78} = a_1 T^{3'}_{2'} + a_2 T^{2'}_{3'} + \frac{a_3}{\sqrt{6}} (T^{1'}_{1'} + T^{2'}_{2'} - 2T^{3'}_{3'})
+ \frac{a_4}{\sqrt{2}} (T^{1'}_{1'} - T^{2'}_{2'}) + \frac{b_3}{\sqrt{6}} (T^{1}_{1} + T^{2}_{2} - 2T^{3}_{3}) \, ,
\ee
\be
\label{27vacE6}
\vev{27} =  e v^{3}_{3'} + \nu v^{3}_{2'} \, , \quad \vev{\overline{27}} = \overline{e} u^{3'}_{3} + \overline{\nu} u^{2'}_{3} \, ,
\ee
where $a_1$, $a_2$, $a_3$, $a_4$, $b_3$, $e$, $\overline{e}$, $\nu$ and $\overline{\nu}$ are SM-singlet VEVs.
This can be checked by means of \eqs{E6action78_1}{E6action27barc}. Notice that the adjoint VEVs $a_3,\ a_4$ and $b_3$ are real, while $a_1=a_2^*$. The VEVs of $27\oplus \overline{27}$ are generally complex.

\section{$E_6$ vacuum manifold}
\label{DFtermE6}

Working out the $D$-flatness equations, one finds that the nontrivial constraints are given by
\begin{align}
& D_{E_\alpha} = \left( \frac{3 a_3}{\sqrt{6}} -\frac{a_4}{\sqrt{2}} \right) a_2^\ast - a_1 \left( \frac{3 a_3^\ast}{\sqrt{6}} - \frac{a_4^\ast}{\sqrt{2}} \right)
+ e_1^\ast \nu_1 - \overline{e}_1 \overline{\nu}_1^\ast + e_2^\ast \nu_2 - \overline{e}_2 \overline{\nu}_2^\ast = 0 \, , \nn \\[1ex]
& D_{T^{(8)}_R} = 3 \left( |a_1|^2 - |a_2|^2 \right)\! +\! 2 \left( |\overline{e}_1|^2 - |e_1|^2 \right)\! +\! 2 \left( |\overline{e}_2|^2 - |e_2|^2 \right) 
+ |\nu_1|^2 - |\overline{\nu}_1|^2 + |\nu_2|^2 - |\overline{\nu}_2|^2 = 0 \, , \nn \\[1ex]
& D_{T^{(3)}_R} = |a_2|^2 - |a_1|^2 + |\overline{\nu}_1|^2 - |\nu_1|^2 + |\overline{\nu}_2|^2 - |\nu_2|^2 = 0 \, , \nn \\
& D_{T^{(8)}_L} = |e_1|^2 + |\nu_1|^2 + |e_2|^2 + |\nu_2|^2 - |\overline{e}_1|^2 - |\overline{\nu}_1|^2 - |\overline{e}_2|^2 - |\overline{\nu}_2|^2 = 0 \, ,
\label{E6Dterms}
\end{align}
where $D_{E_\alpha}$ is the ladder operator from the $(1,1,8)$ sub-multiplet of $78$.
Notice that the relations corresponding to $D_{T^{(8)}_R}$, $D_{T^{(3)}_R}$ and $D_{T^{(8)}_L}$ are linearly dependent, since the linear combination associated to the SM hypercharge in Eq. (\ref{SMhyperchargeE6}) vanishes.

The superpotential $W_H$ in \eq{WHE6} evaluated on the vacuum manifold (\ref{78vacE6})-(\ref{27vacE6}) yields \eq{vevWHE6}.
Accordingly, one finds the following $F$-flatness equations
\begin{align}
F_{a_1} = \ & \mu a_2 - \tau _1 e _1 \overline{\nu}_1- \tau _2 e _2 \overline{\nu}_2
= 0 \, , \nn \\
F_{a_2} = \ & \mu a_1 - \tau _1 \nu_1 \overline{e }_1- \tau _2 \nu_2 \overline{e }_2
= 0 \, , \nn \\
F_{a_3} = \ & \mu a_3 - \frac{1}{\sqrt{6}} \left( \tau _1 ( \nu_1 \overline{\nu}_1 - 2 e _1 \overline{e }_1 ) + \tau _2 ( \nu_2 \overline{\nu}_2 - 2 e _2 \overline{e }_2 ) \right)
= 0 \, , \nn \\
F_{a_4} = \ & \mu a_4 + \frac{1}{\sqrt{2}} \left( \tau _1 \nu_1 \overline{\nu}_1+ \tau _2 \nu_2 \overline{\nu}_2 \right)
= 0 \, , \nn \\
F_{b_3} = \ & \mu b_3 -\sqrt{\frac{2}{3}} \left( \tau _1 (\nu_1 \overline{\nu}_1+e _1 \overline{e }_1) + \tau _2 (\nu_2 \overline{\nu}_2+e _2 \overline{e }_2) \right)
= 0 \, , \nn \\
3 F_{e_1} = \ & 3 ( \rho _{11} \overline{e }_1+\rho _{12} \overline{e }_2 ) 
-  \tau _1 \left(\sqrt{6}  \left(b_3-a_3\right) \overline{e }_1 +3 a_1 \overline{\nu}_1\right)
= 0 \, , \nn \\
3 F_{e_2} = \ & 3 ( \rho _{21} \overline{e }_1+\rho _{22} \overline{e }_2 ) 
-  \tau _2 \left(\sqrt{6}  \left(b_3-a_3\right) \overline{e }_2 +3 a_1 \overline{\nu}_2\right)
= 0 \, , \nn \\
6 F_{\nu_1} = \ & 6 ( \rho _{11} \overline{\nu}_1+\rho _{12} \overline{\nu}_2 ) 
-  \tau _1 \left( \sqrt{2}( \sqrt{3} a_3 -3 a_4 +2 \sqrt{3} b_3 ) \overline{\nu}_1+6 a_2 \overline{e }_1\right)
= 0 \, , \nn \\
6 F_{\nu_2} = \ & 6 ( \rho _{21} \overline{\nu}_1+\rho _{22} \overline{\nu}_2 )
- \tau _2 \left( \sqrt{2}( \sqrt{3} a_3 -3 a_4 +2 \sqrt{3} b_3 ) \overline{\nu}_2+6 a_2 \overline{e }_2\right)
= 0 \, , \nn \\
3 F_{\overline{e}_1} = \ & 3 ( \rho _{11} e _1+\rho _{21} e _2 ) 
- \tau _1 \left(\sqrt{6} \left(b_3-a_3\right) e _1 +3 a_2 \nu_1\right)
= 0 \, , \nn \\
3 F_{\overline{e}_2} = \ & 3 ( \rho _{12} e _1+ \rho _{22} e _2 ) 
- \tau _2 \left(\sqrt{6} \left(b_3-a_3\right) e _2 +3 a_2 \nu_2\right)
= 0 \, , \nn \\
6 F_{\overline{\nu}_1} = \ & 6 ( \rho _{11} \nu_1+ \rho _{21}\nu_2 ) 
- \tau _1 \left( \sqrt{2} ( \sqrt{3} a_3 -3 a_4 +2 \sqrt{3} b_3 ) \nu_1 +6 a_1 e _1\right)
= 0 \, , \nn \\
6 F_{\overline{\nu}_2} = \ & 6 ( \rho _{12} \nu_1+ \rho _{22} \nu_2 ) 
- \tau _2 \left( \sqrt{2} ( \sqrt{3} a_3 -3 a_4 +2 \sqrt{3} b_3 ) \nu_2+6 a_1 e _2\right)
= 0 \, . 
\label{FtermsE6}
\end{align}
{Following the strategy of Appendix \ref{DFtermFSO10} one can solve the first five equations above for   $a_1$, $a_2$, $a_3$, $a_4$ and $b_3$}:
\begin{align}
\mu a_1 &= \tau _1 \nu_1 \overline{e }_1+ \tau _2\nu_2 \overline{e }_2 \nn \, , \\
\mu a_2 &= \tau _1 e _1 \overline{\nu}_1+ \tau _2e _2 \overline{\nu}_2 \, , \nn \\
\sqrt{6} \mu a_3 &= \tau _1 \left(\nu_1 \overline{\nu}_1-2 e _1 \overline{e }_1\right)+\tau _2 \left(\nu_2 \overline{\nu}_2-2 e _2 \overline{e }_2\right) \, ,  \nn \\
\sqrt{2} \mu a_4 &= - \tau _1 \nu_1 \overline{\nu}_1- \tau _2 \nu_2 \overline{\nu}_2 \, , \nn \\
\sqrt{3} \mu b_3 &= \sqrt{2} \left( \tau _1 \left(\nu_1 \overline{\nu}_1+ e _1 \overline{e }_1\right)+ \tau _2 \left(\nu_2 \overline{\nu}_2+e _2 \overline{e
   }_2\right) \right) \, . 
\label{a1tob3asfunof27}   
\end{align}
{Since $a_{1}=a_{2}^{*}$ and $\tau_1$ and $\tau_2$ can be taken real
without loss of generality (see \sect{vacuumE6}), the first two equations above} imply
\be
\tau _1 \nu_1 \overline{e }_1+ \tau _2\nu_2 \overline{e }_2 =
\tau _1 (e _1 \overline{\nu}_1)^* + \tau _2 (e _2 \overline{\nu}_2)^*
\, ,
\label{Dterma1a2}
\ee
Using~\eq{a1tob3asfunof27} the remaining $F$-flatness conditions
in~\eq{FtermsE6} can be rewritten in the form
\begin{multline}
3 \mu F^{a}_{e_1} = 3 \mu ( \rho _{11} \overline{e }_1+\rho _{12} \overline{e }_2 ) - 4 \tau _1^2 \left( \nu_1 \overline{\nu}_1 + e _1 \overline{e }_1\right)\overline{e }_1
\\ - \tau _1 \tau _2 \left(3\nu_2 \overline{\nu}_1 \overline{e }_2+ ( \nu_2 \overline{\nu}_2 +4 e _2 \overline{e }_2 ) \overline{e }_1\right)
= 0 \, , \nn 
\end{multline}
\begin{multline}
3 \mu F^{a}_{\overline{e}_1} = 3 \mu ( \rho _{11} e _1 + \rho _{21} e _2 ) - 4 \tau _1^2 \left( \overline{\nu}_1 \nu_1  + \overline{e }_1 e _1 \right)e _1 
\\ - \tau _1 \tau _2 \left(3 \overline{\nu}_2 \nu_1 e _2 + ( \overline{\nu}_2 \nu_2 +4  \overline{e }_2 e _2  ) e _1 \right)
= 0 \, , \nn 
\end{multline}
\begin{multline}
3 \mu F^{a}_{\nu_1} = 3 \mu ( \rho _{11} \overline{\nu}_1+\rho _{12} \overline{\nu}_2 ) - 4 \tau _1^2 \left( e _1  \overline{e }_1+ \nu_1 \overline{\nu}_1\right)\overline{\nu}_1
\\ - \tau _1 \tau _2 \left(3 e _2 \overline{e}_1 \overline{\nu}_2 + ( e _2  \overline{e }_2+4 \nu_2  \overline{\nu}_2 ) \overline{\nu}_1\right)
= 0 \, , \nn 
\end{multline}
\begin{multline}
3 \mu F^{a}_{\overline{\nu}_1} = 3 \mu ( \rho _{11} \nu_1+ \rho _{21} \nu_2  ) - 4 \tau _1^2 \left( \overline{e }_1 e _1 + \overline{\nu}_1 \nu_1 \right) \nu_1
\\ - \tau _1 \tau _2 \left(3 \overline{e}_2 e _1 \nu_2 + ( \overline{e }_2 e _2 +4  \overline{\nu}_2 \nu_2  ) \nu_1\right)
= 0 \, , 
\label{Faterms}
\end{multline}
{and the additional four relations can be again obtained} by exchanging
$1 \leftrightarrow 2$.
Similarly, the triplet of linearly independent $D$-flatness conditions in \eq{E6Dterms} can be brought to the form
\begin{align}
D_{E_\alpha} = \ & e_1^\ast \nu_1 - \overline{e}_1 \overline{\nu}_1^\ast + e_2^\ast \nu_2 - \overline{e}_2 \overline{\nu}_2^\ast = 0 \, , \nn \\
D_{T^{(3)}_R} = \ & |\overline{\nu}_1|^2 - |\nu_1|^2 + |\overline{\nu}_2|^2 - |\nu_2|^2 = 0 \, , \nn \\
D_{T^{(8)}_L} = \ & |e_1|^2 + |\nu_1|^2 + |e_2|^2 + |\nu_2|^2
- |\overline{e}_1|^2 - |\overline{\nu}_1|^2 - |\overline{e}_2|^2 - |\overline{\nu}_2|^2 = 0 \, .
\end{align}
Combining these with \eq{Dterma1a2}, the $D$-flatness is ensured if and only if $e_{1,2}^{\ast} = \overline{e}_{1,2}$ and $\nu_{1,2}^{\ast} = \overline{\nu}_{1,2}$.
Hence, in complete analogy with the flipped $SO(10)$ case \eq{complexDtermsFSO10}, one can write
\bea
\label{complexDterms}
&& e_{1,2} \equiv |e_{1,2}| e^{i \phi_{e_{1,2}}} \, , \qquad \overline{e}_{1,2} \equiv |e_{1,2}| e^{- i \phi_{e_{1,2}}} \, , \nn \\
&& \nu_{1,2} \equiv |\nu_{1,2}| e^{i \phi_{\nu_{1,2}}} \, , \qquad \overline{\nu}_{1,2} \equiv |\nu_{1,2}| e^{- i \phi_{\nu_{1,2}}} \, .
\eea

From now on, the discussion of the vacuum manifold follows very closely that for the flipped $SO(10)$ in \sect{DFtermFSO10} and we shall not repeat it here. In particular the existence of a nontrivial vacuum requires the hermiticity  of the $\rho$ and $\tau$ couplings. This is related to the fact that $D$- and $F$-flatness require $\vev{27_i}=\vev{\overline{27}_i}^*$.
The detailed shape of the resulting vacuum manifold so obtained is
{given} in \eq{vacmanifoldE6} of \sect{vacuumE6}.

\section{Vacuum little group}
\label{formalproof}

In order to find the algebra left invariant by the vacuum configurations in \eq{vacmanifoldE6}, we need to
compute the action of the $E_6$ generators
on the $\vev{78 \oplus 27_1 \oplus 27_2 \oplus \overline{27}_1 \oplus \overline{27}_2}$ VEV.
From \eqs{E6action78_1}{E6action78_2} one obtains
\begin{align}
\label{E6actionvev78}
T^{\alpha}_{\beta} \vev{78} &= 0 \nn \\
T^{i}_{j} \vev{78} &= \frac{b_3}{\sqrt{6}} ( \delta^{i}_{1} T^{1}_{j} - \delta^{1}_{j} T^{i}_{1} + \delta^{i}_{2} T^{2}_{j} - \delta^{2}_{j} T^{i}_{2}
- 2 \delta^{i}_{3} T^{3}_{j} + 2 \delta^{3}_{j} T^{i}_{3} ) \nn \\
T^{i'}_{j'} \vev{78} &= a_1 ( \delta^{i'} _{2'} T^{3'}_{j'} - \delta^{3'} _{j'} T^{i'}_{2'} ) + a_2 ( \delta^{i'} _{3'} T^{2'}_{j'} - \delta^{2'} _{j'} T^{i'}_{3'} ) 
+ \frac{a_4}{\sqrt{2}} ( \delta^{i'}_{1'} T^{1'}_{j'} - \delta^{1'}_{j'} T^{i'}_{1'} - \delta^{i'}_{2'} T^{2'}_{j'} + \delta^{2'}_{j'} T^{i'}_{2'} ) \nn \\
&+ \frac{a_3}{\sqrt{6}} ( \delta^{i'}_{1'} T^{1'}_{j'} - \delta^{1'}_{j'} T^{i'}_{1'} + \delta^{i'}_{2'} T^{2'}_{j'} - \delta^{2'}_{j'} T^{i'}_{2'} 
- 2 \delta^{i'}_{3'} T^{3'}_{j'} + 2 \delta^{3'}_{j'} T^{i'}_{3'} ) \nn \\
Q^{\alpha}_{i j'} \vev{78} &= - a_1 ( \delta^{3'}_{j'} Q^{\alpha}_{i 2'} ) - a_2 ( \delta^{2'}_{j'} Q^{\alpha}_{i 3'} )
- \frac{a_3}{\sqrt{6}} (\delta^{1'}_{j'} Q^{\alpha}_{i 1'} + \delta^{2'}_{j'} Q^{\alpha}_{i 2'} - 2 \delta^{3'}_{j'} Q^{\alpha}_{i 3'}) \nn \\
& - \frac{a_4}{\sqrt{2}} (\delta^{1'}_{j'} Q^{\alpha}_{i 1'} - \delta^{2'}_{j'} Q^{\alpha}_{i 2'}) 
- \frac{b_3}{\sqrt{6}} (\delta^{1}_{i} Q^{\alpha}_{1 j'} + \delta^{2}_{i} Q^{\alpha}_{2 j'} - 2 \delta^{3}_{i} Q^{\alpha}_{3 j'}) \nn \\
Q^{i j'}_{\alpha} \vev{78} &= a_1 ( \delta^{j'}_{2'} Q^{i 3'}_{\alpha} ) + a_2 ( \delta^{j'}_{3'} Q^{i 2'}_{\alpha} )
+ \frac{a_3}{\sqrt{6}} (\delta^{j'}_{1'} Q^{i 1'}_{\alpha} + \delta^{j'}_{2'} Q^{i 2'}_{\alpha} - 2 \delta^{j'}_{3'} Q^{i 3'}_{\alpha}) \nn \\
&+ \frac{a_4}{\sqrt{2}} (\delta^{j'}_{1'} Q^{i 1'}_{\alpha} - \delta^{j'}_{2'} Q^{i 2'}_{\alpha}) 
+ \frac{b_3}{\sqrt{6}} (\delta^{i}_{1} Q^{1 j'}_{\alpha} + \delta^{i}_{2} Q^{2 j'}_{\alpha} - 2 \delta^{i}_{3} Q^{3 j'}_{\alpha}) \, ,
\end{align}
on the adjoint vacuum.
For $\vev{27_1 \oplus 27_2}$ one finds
\begin{align}
\label{E6actionvev271}
T^{\alpha}_{\beta} \vev{27_1 \oplus 27_2} &= 0 \nn \\
T^{i}_{j} \vev{27_1 \oplus 27_2} &=
- (e_1+e_2) [ \delta^{3}_{j} v^{i}_{3'} ] - (\nu_1+\nu_2) [ \delta^{3}_{j} v^{i}_{2'} ] \nn \\
T^{i'}_{j'} \vev{27_1 \oplus 27_2} &=
(e_1+e_2) [ \delta^{i'}_{3'} v^{3}_{j'}  ] + (\nu_1+\nu_2) [ \delta^{i'}_{2'} v^{3}_{j'}  ] \nn \\
Q^{\alpha}_{i j'} \vev{27_1 \oplus 27_2} &=
- (e_1+e_2) [ \delta^{3}_{i} \epsilon_{j' 3' k'} v^{\alpha k'} ] - (\nu_1+\nu_2) [ \delta^{3}_{i} \epsilon_{j' 2' k'} v^{\alpha k'}  ] \nn \\
Q^{i j'}_{\alpha} \vev{27_1 \oplus 27_2} &=
(e_1+e_2) [ \delta^{j'}_{3'} \epsilon^{i 3 k} v_{\alpha k}  ] + (\nu_1+\nu_2) [ \delta^{j'}_{2'} \epsilon^{i 3 k} v_{\alpha k}  ] \, ,
\end{align}
and, accordingly, for $\vev{\overline{27}_1 \oplus \overline{27}_2}$
\begin{align}
\label{E6actionvev272}
T^{\alpha}_{\beta} \vev{\overline{27}_1 \oplus \overline{27}_2} &= 0 \nn \\
T^{i}_{j} \vev{\overline{27}_1 \oplus \overline{27}_2} &=
(\overline{e}_1+\overline{e}_2) [ \delta^{i}_{3} u^{3'}_{j} ] + (\overline{\nu}_1+\overline{\nu}_2) [ \delta^{i}_{3} u^{2'}_{j} ] \nn \\
T^{i'}_{j'} \vev{\overline{27}_1 \oplus \overline{27}_2} &=
- (\overline{e}_1+\overline{e}_2) [ \delta^{3'}_{j'} u^{i'}_{3} ] - (\overline{\nu}_1+\overline{\nu}_2) [ \delta^{2'}_{j'} u^{i'}_{3} ] \nn \\
Q^{\alpha}_{i j'} \vev{\overline{27}_1 \oplus \overline{27}_2} &=
- (\overline{e}_1+\overline{e}_2) [ \delta^{3'}_{j'} \epsilon_{i 3 k} u^{\alpha k} ] - (\overline{\nu}_1+\overline{\nu}_2) [ \delta^{2'}_{j'} \epsilon_{i 3 k} u^{\alpha k} ] \nn \\
Q^{i j'}_{\alpha} \vev{\overline{27}_1 \oplus \overline{27}_2} &=
(\overline{e}_1+\overline{e}_2) [ \delta^{i}_{3} \epsilon^{j' 3' k'} u_{\alpha k'} ] + (\overline{\nu}_1+\overline{\nu}_2) [ \delta^{i}_{3} \epsilon^{j' 2' k'} u_{\alpha k'} ] \, .
\end{align}

{On the vacuum manifold in Eq. (\ref{vacmanifoldE6})} one finds that the generators generally preserved by the VEVs of $78 \oplus 27_1 \oplus 27_2 \oplus \overline{27}_1 \oplus \overline{27}_2$ are
\begin{align}
& T_{C}^{(1)}\ T_{C}^{(2)}\ T_{C}^{(3)}\ T_{C}^{(4)}\ T_{C}^{(5)}\ T_{C}^{(6)}\ T_{C}^{(7)}\ T_{C}^{(8)}: \ (8,1,0) \, , \nn \\
& T_{L}^{(1)}\ T_{L}^{(2)}\ T_{L}^{(3)}: \ (1,3,0) \, , \nn \\
& Y : \ (1,1,0) \, , \nn \\
& Q^{\alpha}_{1 1'} \ Q^{\alpha}_{2 1'} \ Q^{1 1'}_{\alpha} \ Q^{2 1'}_{\alpha}: \ (\overline{3},2,+\tfrac{5}{6}) \oplus (3,2,-\tfrac{5}{6}) \, ,
\label{SU5generators}
\end{align}
which generate an $SU(5)$ algebra. As an example showing the nontrivial constraints enforced by the vacuum manifold in Eq. (\ref{vacmanifoldE6}),
let us inspect the action of one of the lepto-quark generators, say $Q^{\alpha}_{1 1'}$:
\begin{align}
& Q^{\alpha}_{1 1'} \vev{78} = - \tfrac{1}{\sqrt{6}} \left( a_3 + \sqrt{3} a_4 + b_3 \right) Q^{\alpha}_{1 1'} \, , \label{Q11action}\\
& Q^{\alpha}_{1 1'} \vev{27_1 \oplus 27_2} = 0 \, , \nn\\
& Q^{\alpha}_{1 1'} \vev{\overline{27}_1 \oplus \overline{27}_2} = 0 \, . \nn
\end{align}
It is easy to check that $a_3 + \sqrt{3} a_4 + b_3$ vanishes on the whole vacuum manifold in Eq. (\ref{vacmanifoldE6}) and, thus, $Q^\alpha_{11'}$ is preserved. Let us also remark that the $U(1)_Y$ charges above correspond to the standard $SO(10)$ embedding (see the discussion in sect.~\ref{vacuumE6}). In the flipped $SO(10)$ embedding, the $(\overline{3},2) \oplus (3,2)$ generators in \eq{SU5generators} carry  hypercharges $\mp \tfrac{1}{6}$, respectively.

Considering instead the vacuum manifold invariant with respect to the flipped $SO(10)$ hypercharge (see \eqs{vev78flip}{vev27bar12flip}),
the preserved generators, in addition to those of the SM, are
$Q^{\alpha}_{1 3'} \ Q^{\alpha}_{2 3'} \ Q^{1 3'}_{\alpha} \ Q^{2 3'}_{\alpha}$.
These, for the standard hypercharge embedding of \eq{Ystandard}, transform as
$(\overline{3},2,-\tfrac{1}{6}) \oplus (3,2,+\tfrac{1}{6})$, whereas with the
flipped hypercharge assignment in \eq{Yso10}, the same transform as $(\overline{3},2,+\tfrac{5}{6}) \oplus (3,2,-\tfrac{5}{6})$.
{Needless to say, one finds again $SU(5)$ as the vacuum little group.}

It is interesting to consider the configuration $\alpha_1=\alpha_2=0$, which can be chosen without loss of generality once a pair, let us say $27_2 \oplus \overline{27}_2$, is decoupled or when the two copies of $27_H \oplus \overline{27}_H$ are aligned. According to \eq{vacmanifoldE6} this implies all VEVs equal to zero but $a_3 = -b_3$ and $e_1$ ($e_2$).
Then, from \eqs{E6actionvev78}{E6actionvev272}, one verifies that the preserved generators are (see \eq{gellmannbasis} for notation)
\begin{align}
& T_{C}^{(1)}\ T_{C}^{(2)}\ T_{C}^{(3)}\ T_{C}^{(4)}\ T_{C}^{(5)}\ T_{C}^{(6)}\ T_{C}^{(7)}\ T_{C}^{(8)}: \ (8,1,0) \, , \nn \\
& T_{L}^{(1)}\ T_{L}^{(2)}\ T_{L}^{(3)}: \ (1,3,0) \, , \nn \\
& T_{R}^{(1)}\ T_{R}^{(2)}\ T_{R}^{(3)}: \ (1,1,-1) \oplus (1,1,0) \oplus (1,1,+1) \, , \nn \\
& T_{L}^{(8)} + T_{R}^{(8)}: \ (1,1,0) \, , 
\end{align}
\begin{align}
& Q^{\alpha}_{1 1'} \ Q^{\alpha}_{2 1'} \ Q^{1 1'}_{\alpha} \ Q^{2 1'}_{\alpha}: \ (\overline{3},2,+\tfrac{5}{6}) \oplus (3,2,-\tfrac{5}{6}) \, , \nn \\
& Q^{\alpha}_{1 2'} \ Q^{\alpha}_{2 2'} \ Q^{1 2'}_{\alpha} \ Q^{2 2'}_{\alpha}: \ (\overline{3},2,-\tfrac{1}{6}) \oplus (3,2,+\tfrac{1}{6}) \, , \nn \\
& Q^{\alpha}_{3 3'} \ Q^{3 3'}_{\alpha}: \ (\overline{3},1,-\tfrac{2}{3}) \oplus (3,1,+\tfrac{2}{3}) \, ,
\end{align}
which support an $SO(10)$ algebra.
In particular, $a_3 = -b_3$ preserves $SO(10) \otimes U(1)$, where the extra $U(1)$ generator,
which commutes with all $SO(10)$ generators, is proportional to $T_{L}^{(8)} - T_{R}^{(8)}$.
On the other hand, the VEV $e_1$ breaks $T_{L}^{(8)} - T_{R}^{(8)}$ (while preserving the sum).
We therefore recover the result of Ref. \cite{Buccella:1987kc} for
the $E_6$ setting with $78_H \oplus 27_H \oplus \overline{27}_H$.

\bibliography{bibthesis}{}
\bibliographystyle{hieeetr}


\chapter*{Acknowledgments}

I warmly thank:
Stefano Bertolini for his attitude (would you ever bet there was something to dig out ?),
Michal Malinsk\'y for teaching me a lot, 
Goran Senjanovi\'c for inspiration,  
Ketan Patel for useful communications, 
Lorenzo Seri still looking forward for his first symphony, 
Enzo Vitagliano for showing us it was possible to go barefoot all the summer,
Goffredo Chirco as an example of how big things a small man can do, 
Adriano Contillo for his communicative skills,
Marco Nardecchia because he is an happy person, 
Giorgio Arcadi for being always sober, 
Robert Ziegler for relaxed discussions about physics,
Chetan Krishnan because he is not cynic at all, 
Jarah Evslin for a discussion on spinors and a proof of the roundedness of the earth, 
Chica and Smilla two most beautiful creatures,  
Paolo Antonelli for sharing some good time here in Trieste, 
Antonella Garzilli for being a bit lunatic,  
Angus Prain for initiating me to the spirit of the game, 
Daniele De Martino for telling us the importance of being present,
Fabio Caccioli because it was nice to look at him listening to people, 
Taskin Deniz for the power of obsession, 
Maya Sundukova, Pierpaolo Vivo and Luca Tubiana for standing me and my hairs,
Ugo Marzolino for being around me in the last 15 years and DFW for the footnotes.

\end{document}